%% file: arxiv.tex
\documentclass[11pt]{article} 

\usepackage{amsfonts}
\usepackage{xcolor}
\usepackage{amsthm}

\usepackage{amsmath}
\usepackage{amssymb}
\usepackage[authoryear,sort&compress]{natbib}

\usepackage{graphicx}
\usepackage{url}
\usepackage{algorithm2e}
\PassOptionsToPackage{algo2e,ruled}{algorithm2e}
\usepackage{amsthm}
\usepackage[margin=1.1in]{geometry}
\usepackage{hyperref}
\usepackage[capitalize]{cleveref}
\hypersetup{colorlinks=true,linkcolor=blue,citecolor=blue}

\usepackage{tikz}
\usetikzlibrary{graphs}
\usetikzlibrary{patterns}
\usetikzlibrary{calc}
\usetikzlibrary{decorations.pathreplacing}
\usepackage{booktabs,tabularx,makecell,multirow}

\usepackage{pgfplots}
\pgfplotsset{compat=1.17}
\usepackage{caption}
\usepackage{subcaption}
\usepackage[mathscr]{eucal}

\newtheorem{theorem}{Theorem}[section]
\newtheorem{lemma}[theorem]{Lemma}
  
  \newtheorem{proposition}[theorem]{Proposition}
  \newtheorem{remark}[theorem]{Remark}
  \newtheorem{corollary}[theorem]{Corollary}

  \newtheorem{fact}{Fact}[section]
  \newtheorem{openpb}{Open Question}

\newtheorem{assumption}[theorem]{Assumption}

\title{Near-Optimal Mechanisms for Resource Allocation Without Monetary Transfers}

\author{
  \textbf{Mo\"ise Blanchard}\\
  Georgia Institute of Technology\\
  \small{\texttt{mblanchard41@gatech.edu}}
  \and
  \textbf{Patrick Jaillet}\\
  Massachusetts Institute of Technology\\
  \small{\texttt{jaillet@mit.edu}}
}
\date{}

\usepackage{thmtools}
\usepackage{thm-restate}
\usepackage[mathscr]{eucal}
\usepackage{bbm}
\usepackage{latexsym,tikz,graphicx}
\usetikzlibrary{calc}
\usetikzlibrary{decorations.pathreplacing}
\usetikzlibrary{patterns}
\usetikzlibrary{math}

\renewenvironment{proof}[1][]{\par\noindent{\bf Proof #1\ }}{\hfill$\blacksquare$ \vspace{4mm}}

\input{shortcuts}

\begin{document}

\maketitle

\begin{abstract}
    We study the problem in which a central planner sequentially allocates a single resource to multiple strategic agents using their utility reports at each round, but without using any monetary transfers. We consider general agent utility distributions and two standard settings: a finite horizon $T$ and an infinite horizon with $\gamma$ discounts.
    We provide general tools to characterize the convergence rate between the optimal mechanism for the central planner and the first-best allocation if true agent utilities were available. This heavily depends on the utility distributions, yielding rates anywhere between $1/\sqrt T$ and $1/T$ for the finite-horizon setting, and rates faster than $\sqrt{1-\gamma}$, including exponential rates for the infinite-horizon setting as agents are more patient $\gamma\to 1$. On the algorithmic side, we design mechanisms based on the promised-utility framework to achieve these rates and leverage structure on the utility distributions. Intuitively, the more flexibility the central planner has to reward or penalize any agent while incurring little social welfare cost, the faster the convergence rate. In particular, discrete utility distributions typically yield the slower rates $1/\sqrt T$ and $\sqrt{1-\gamma}$, while smooth distributions with density typically yield faster rates $1/T$ (up to logarithmic factors) and $1-\gamma$.
\end{abstract}

\tableofcontents

\section{Introduction}

Resource allocation or auction design is one of the foundational problems in mechanism design and has been thoroughly studied across economics, operations research, and computer science. In this problem, a planner aims to allocate resources to $n$ strategic agents while maximizing some notion of global welfare, using reports/bids from the agents on their utility for the resources. Importantly, agent reports are strategic; hence, the planner only has limited or partial information about the agents' preferences.  To achieve positive results, fundamental tools in mechanism design include money transfers between the agents and the planner. The large majority of approaches heavily rely on the use of such monetary payments to align incentives between the strategic users and the central planner. For instance, the celebrated Vickrey-Clarke-Groves (VCG) mechanism \citep{v61,c71,gr73} ensures incentive-compatibility while implementing optimal allocations by carefully selecting payment mechanisms.
In many applications, however, money transfers may be impractical. Enforcing payments from beneficiaries to the central planner is typically undesirable when allocating public services: from designing antipoverty programs \citep{bardhan2005decentralizing}, choosing school assignments \citep{russell2011high,ashlagi2016optimal}, allocating scientific equipment \citep{nro22}, to distributing food to food banks. Cloud computing is another application in which memory or computing resources may be scarce; hence, enforcing a scheduling policy is necessary, but monetary transfers may be unpopular/inadequate, especially when allocating internal resources \citep{ng2011online}.

\paragraph{Resource allocation without monetary transfers.}
Following these motivations, we consider the setting in which monetary transfers are not allowed and focus on single resource allocation for simplicity. The reports of the agents are strategic and hence form a perfect Bayesian equilibrium. In a full-information setting where the planner has direct access to all agent utilities, the optimal policy is to allocate the resource to an agent with maximum utility, also known as the \emph{first-best} allocation. Unfortunately, prohibiting the use of money severely limits the ability to approximate the first-best allocation in light of negative results arising from Myerson-Satterthwaite's impossibility theorem \citep{myerson1983efficient,arrow2012social}. For a single allocation, whenever there are at least $3$ alternative allocations, it is impossible to design a non-dictatorial incentive-compatible mechanism \citep{gi73,s75}. Indeed, without penalization from monetary transfers, agents can report as large utilities as possible for the resource; utility reports from the agents carry virtually no information. In this case, agent reports can at most signal whether agents have negative, zero, or positive utility for the resource. We note that mechanism design without monetary transfers can be successful in specific settings, e.g., if agent preferences are single-peaked, when matching users pairwise in a bipartite graph (see \cite{sv07} for an overview of classical results), or in interdomain routing \citep{feigenbaum2007distributed,levin2008interdomain}; these exploit specific properties of the problem that are difficult to generalize.

In a repeated allocation setting, however, the central planner can use future allocations to align incentives by favoring or disfavoring agents. We consider two classical repeated settings in the literature: an infinite-horizon setting with a fixed utility time discount factor $\gamma\in {[}0,1)$ and a finite-horizon $T$ setting without discounts. To illustrate the benefit of repeated allocations, the central planner can, for instance, fix budgets for each agent that serve as virtual currency. This is routinely used in practice \citep{walsh2014allocation}, including for allocating computing resources in distributed systems \citep{ng2011online} or food to food banks \citep{prendergast2022allocation}.

\paragraph{Characterizing the convergence to first-best allocation.} From the folk theorem \citep{friedman1971non,fudenberg2009folk}, one can approximate the optimal first-best allocation arbitrarily well as agents become more patient $\gamma\to 1$, or when the horizon grows $T\to\infty$. However, this result is not quantitative in nature and does not provide direct insights into which allocation mechanisms might be optimal.

The goal of this paper is to provide \emph{quantitative} answers to the following set of questions: 
\begin{enumerate}
    \item What is the welfare gap between non-monetary and monetary mechanisms? Equivalently, what is the welfare gap between strategic and non-strategic agents for non-monetary mechanisms? More formally, can we quantify the \emph{non-asymptotic} social welfare gap between the optimal first-best allocation (which can be achieved via monetary mechanisms) and non-monetary resource allocation mechanisms for general utility distributions? Conceptually, we aim to understand how this gap depends on the utility distributions of the agents, and on the patience of agents (either through the discount factor $\gamma$ or the horizon $T$). \label{item:first_set}
    \item On the algorithmic front, what are (near-)optimal allocation mechanisms without monetary transfers? Since the optimal allocation strategy may depend on the specific problem at hand, through utility distributions and horizon and agents' patience, we aim to provide a generic non-monetary allocation mechanism strategy whose parameters can easily be adjusted for any scenario to achieve near-optimal social welfare.\label{item:second_set}
\end{enumerate}

From a practical perspective, understanding the main drivers (\cref{item:first_set}) for the social welfare gap can (1) help identify which applications are best suited for non-monetary mechanisms, or (2) inform decisions that may impact utility distributions. As a preview, our results show that the social welfare depends heavily on the geometry of utility distributions and identify a main driver for achieving small social welfare gaps: favoring (approximate) utility ties between agents is most beneficial for non-monetary mechanisms. In addition, characterizing the social welfare gap has algorithmic implications (\cref{item:second_set})---this will be used in the parameter choice for the generic allocation strategy.

\subsection{Overview of the contributions}

To answer these questions, we use a geometric approach, in a similar spirit to \citet{bgs19}. Specifically, instead of solely focusing on approaching the social-welfare maximizing utilities, one can consider the complete region of achievable utilities for all $n$ agents by non-monetary mechanisms, which we denote $\Ucal$: it is the collection of utility vectors $(U_i)_{i\in[n]}$ for which one can achieve expected per-round $U_i$ utility for all agents $i$, using some non-monetary mechanism. In this geometric perspective, the gap between non-monetary and monetary mechanisms precisely corresponds to the gap between the achievable region $\Ucal$ and the target region $\Ucal^\star$ of utilities that could be allocated if agents were non-strategic.
At the conceptual level, our results show that the local smoothness of this target region $\Ucal^\star$ at welfare-maximizing points, precisely determines the gap between monetary and non-monetary mechanisms.

\paragraph{Near-optimal allocation mechanisms.}
On the positive side, we construct allocation mechanisms (see \cref{alg:main_mechanism}) that can achieve any utility inside a pre-specified ball within the target region $\Ucal^\star$ which has sufficient margin $\delta>0$ to its boundary. This mechanism builds upon the so-called \emph{promised utility} framework \citep{spear1987repeated,abreu1990toward,thomas1990income} and tools from \citet{bgs19}, which reformulate the problem in a dynamic programming form. Crucially, the larger the radius of this ball, the smaller this margin $\delta$ is required to be. Geometrically, for a target region $\Ucal^\star$ with smoother boundary, we can employ our mechanisms to achieve utility vectors closer to social-welfare optimal boundary. In words, our results show that non-monetary mechanisms perform best when agent utility distributions significantly overlap: situations with close-ties are particularly advantageous for a central planner since they have flexibility to choose who to allocate the resource to without incurring significant social welfare cost. In fact, when exact ties arise frequently, our results prove exponentially small social welfare gaps between monetary and non-monetary mechanisms.

We complement these positive results with corresponding negative results, which essentially show that the local smoothness of the target region $\Ucal^\star$ indeed characterizes the social welfare gap between non-monetary and monetary mechanisms. This effectively transforms the multi-horizon multi-agent game-theoretic resource allocation problem into a purely geometric question of understanding the local smoothness of $\Ucal^\star$ (so that our mechanism \cref{alg:main_mechanism} can be implemented with an appropriate ball parameter).

\paragraph{Unified perspective on $\gamma$-discounted and horizon-$T$ settings.} Up to minor modification of the algorithms, we also show that the same tools described above can be used for both the infinite-horizon setting with discount factor, and the finite horizon $T$ setting. This unifies the analysis and the algorithmic approach for both settings.
Formally, the same results can be achieved up to logarithmic factors by setting $1-\gamma\approx 1/T$ (see \cref{lemma:upper_bound_finite_horizon,lemma:ball_in_region_finite_horizon}).

\paragraph{Characterizations of the non-monetary/monetary social welfare gap.} Using this general methodology, we can characterize the social welfare gap between non-monetary and monetary mechanisms for general utility distributions, beyond prior results.

First, the non-monetary/monetary social welfare gap is always at most $\Ocal(\sqrt{1-\gamma})$ (resp. $\Ocal(1/\sqrt T)$) for the $\gamma$-discounted infinite-horizon (resp. finite-horizon $T$) setting (\cref{thm:universal_lower_bound_simplified}). In particular, this holds for any number of agents and any utility distributions with bounded support. Beyond the universal asymptotic convergence guarantee from the folk theorem \citep{friedman1971non,fudenberg2009folk}, this shows a universal convergence rate upper bound.

While this universal upper bound can be tight\footnote{this is typically the case for discrete utility distributions whose boundary of $\Ucal^\star$ is piece-wise affine (non-smooth), see \cref{thm:discrete_case_full}.}, the convergence rate can be faster in general, and depends on the local geometry of $\Ucal^\star$ as discussed above. Specifically, while the distributional assumptions from \cite{bgs19} prohibited faster rates than $1-\gamma$, we give non-trivial examples of exponential convergence rates as fast as $(1-\gamma)e^{-c/\sqrt{1-\gamma}}$ (see \cref{thm:extra_fast_rates} which essentially corresponds to the case when any agent $i\in[n]$ can have ties for the first-best allocation with non-zero probability). From a practical standpoint, this shows that when ties between several agents are possible, our mechanisms which leverage this additional geometrical structure can achieve significantly (exponentially) smaller social welfare gap.
For the finite-horizon case, faster rates than $1/\sqrt T$ are possible but they cannot converge faster than $1/T$ (\cref{lemma:no_faster_1/T}) as a result of the boundary at time $T$. In particular, the exponential rates from the infinite-horizon setting do not carry to the finite-horizon case. To the best of our knowledge, this is the first example showing exponential gaps between the discounted and finite-horizon settings for non-monetary resource allocation. 

Our results provide general upper and lower bounds on the optimal social welfare gap for arbitrary distributions (\cref{thm:full_characterization})
While the bounds do not match in general, both share the same form and give insights into (1) the dependency of the convergence rate in terms of the utility distributions and (2) corresponding optimal allocation mechanisms. Intuitively, the main driver is whether the central planner has enough flexibility to reward or penalize any agent in future rounds, without incurring large social welfare costs. Accordingly, our mechanisms use this flexibility to align incentives: the larger the flexibility, the closer one can approach the first-best allocation.

\paragraph{Special cases and examples.}
In particular, we can give necessary (\cref{thm:no_faster_rate_1-beta}) and sufficient (\cref{thm:faster_rates_1-beta}) conditions to reach the faster rate $1-\gamma$ from \cite{bgs19} in the infinite-horizon setting, significantly generalizing their results (the necessary and sufficient conditions are both satisfied under their distributional assumptions). In the finite-horizon setting, this corresponds to the fastest rate $1/T$ up to logarithmic factors. At the high-level, we show that these faster rates can be achieved if any agent $i\in[n]$ shares some common utility support with some other agent $j\neq i$: the mechanism can easily then implement future penalties (resp. rewards) on agent $i$ by allocating to agent $j$ (resp. $i$) when their utilities are sufficiently close in following rounds. Using this perspective, we introduce a sufficient condition for fast convergence rates in terms of a graph structure between agents, which can be efficiently tested given their utility distributions. 


The previous example covers the case of smooth distributions (\cref{cor:super_smooth_case,cor:super_smooth_case_finite_horizon}). On the other extreme, we also take the example of discrete utility distributions (\cref{thm:discrete_case_full,cor:discrete_case_full_finite_horizon}). In this case, the optimal convergence rate is usually $\sqrt{1-\gamma}$ (resp. $1/\sqrt T$) for the infinite-horizon (resp. finite-horizon) setting.


In comparison, \cite{gorokh2021monetary} showed that for discrete distributions, one can reach a convergence rate $1/T$ to the first-best allocation with artificial currencies, but in a weaker form of Bayesian equilibrium in which incentive-compatibility constraints are only satisfied up to $\sqrt{\log T/T}$ terms. Our results show that the optimal rate at the perfect Bayesian equilibrium is, in general, exactly $\Theta(1/\sqrt T)$, when incentive-compatibility constraints are perfectly enforced. Intuitively, the incentive-compatibility gap from \cite{gorokh2021monetary} can be traded for the social welfare gap from the central planner. Unlike \cite{gorokh2021monetary}, our upper bound holds for arbitrary distributions.

A summary of these social welfare characterizations for various utility distribution classes is given in \cref{tab:swg_summary_compact}: $(i)$ without any assumptions on the utility distributions one can always achieve $O(\sqrt{1-\gamma})$ convergence rate, $(ii)$ this is tight for discrete distributions in general, $(iii)$ but may be improved to fast rates $\Theta(1-\gamma)$, for instance when distributions have density bounded above and below. $(iv)$ Intermediate situations are also possible and we provide such an example in \cref{fig:construction_H}, with convergence rate $\Theta((1-\gamma)^{3/4})$.  Finally, $(v)$ when there are significant ties between agents, exponential rates are possible in the discounted setting; however these do not translate to the finite-horizon $T$ setting.

\begin{table*}[t]
\centering
\fontsize{6.5pt}{7.5pt}\selectfont
\setlength{\tabcolsep}{3.5pt}

\begin{tabularx}{\textwidth}{
l
>{\centering\arraybackslash\hsize=0.9\hsize}X
>{\centering\arraybackslash\hsize=1.1\hsize}X
>{\centering\arraybackslash\hsize=0.9\hsize}X
>{\centering\arraybackslash\hsize=0.7\hsize}X
>{\centering\arraybackslash\hsize=1\hsize}X
>{\centering\arraybackslash\hsize=0.8\hsize}X
>{\centering\arraybackslash\hsize=1.2\hsize}X
>{\centering\arraybackslash\hsize=0.8\hsize}X
>{\centering\arraybackslash\hsize=1.8\hsize}X
>{\centering\arraybackslash\hsize=0.8\hsize}X
}
\toprule
& \multicolumn{2}{c}{\makecell{\textbf{General distributions} \\ \textbf{(Universal bounds)}}} 
& \multicolumn{2}{c}{\makecell{\textbf{Density bounded} \\ \textbf{above \& below}}}
& \multicolumn{2}{c}{\makecell{\textbf{Discrete}\\ \textbf{(Generic case)}}} 
& \multicolumn{2}{c}{\makecell{\textbf{Intermediary}\\  \textbf{example of \cref{fig:construction_H}}}} 
& \multicolumn{2}{c}{\makecell{\textbf{Significant ties between} \\ \textbf{agents (see Thm \ref{thm:extra_fast_rates})}}} \\
\cmidrule(lr){2-3}\cmidrule(lr){4-5}\cmidrule(lr){6-7}\cmidrule(lr){8-9}\cmidrule(lr){10-11}
\textbf{Setting}
& \(\gamma\)-disc.
& $T$-hor.
& \(\gamma\)-disc.
& $T$-hor.
& \(\gamma\)-disc.
& $T$-hor.
& \(\gamma\)-disc.
& $T$-hor.
& \(\gamma\)-disc.
& $T$-hor. \\
\midrule

\textbf{SWG}
& \(\Ocal(\sqrt{1-\gamma})\)
& $\Ocal\paren{\frac{1}{\sqrt{T}}}$ and $\Omega\paren{\frac{1}{T}}$
& \(\Theta(1-\gamma)\)
& \(\widetilde{\Theta}\!\left(\frac{1}{T}\right)\)
& \(\Theta(\sqrt{1-\gamma})\)
& \(\Theta\!\left(\frac{1}{\sqrt{T}}\right)\)
& \(\Theta((1-\gamma)^{\frac{3}{4}})\)
& \(\widetilde{\Theta}\paren{\frac{1}{T^{\frac{3}{4}}}}\)
& \(\Ocal\paren{(1-\gamma)e^{-\frac{c}{\sqrt{1-\gamma}}}}\)
& \(\widetilde{\Theta}\!\left(\frac{1}{T}\right)\) \\

Ref.
& \multicolumn{2}{c}{Thm~\ref{thm:universal_lower_bound_simplified}
+Lem~\ref{lemma:no_faster_1/T}\&\ref{lemma:ball_in_region_finite_horizon}}
& Cor~\ref{cor:super_smooth_case}
& Cor~\ref{cor:super_smooth_case_finite_horizon}
& Thm~\ref{thm:discrete_case_full}
& Cor~\ref{cor:discrete_case_full_finite_horizon}
& \multicolumn{2}{c}{Prop~\ref{prop:2_agent_case_example} +Lem~\ref{lemma:upper_bound_finite_horizon}\&\ref{lemma:ball_in_region_finite_horizon}}
& \multicolumn{2}{c}{Thm~\ref{thm:extra_fast_rates}
+Lem~\ref{lemma:no_faster_1/T}\&\ref{lemma:ball_in_region_finite_horizon}} \\
\midrule

\makecell[l]{\textbf{Prior}\\ \textbf{results}}
& ---
&\parbox[c]{1.2\linewidth}{\centering\(\Ocal(1/T)\) for \\ $\widetilde\Ocal(1/\sqrt T)$-approx.\ IC}
& \parbox[c]{1.1\linewidth}{\raggedleft\(\Theta(1-\gamma)\) \\ for \(n=2\)}
& ---
& ---
& ---
& ---
& ---
& ---
& --- \\

Ref.
& \parbox[c]{5\linewidth}{\centering\cite{gorokh2021monetary} \;\;\cite{bgs19}} 
& 
& 
&
&  
& 
& 
& 
&  
& \\

\bottomrule
\end{tabularx}

\caption{Examples of social welfare gap ($\SWG$) for various utility distribution classes for $\gamma$-discounted and finite $T$-horizon settings. The first column gives universal bounds---always valid---while others are consequences of our general distribution-dependent characterization (\cref{thm:full_characterization}). This includes distributions with well-behaved densities, discrete distributions without significant ties (see \cref{thm:discrete_case_full,thm:extra_fast_rates}), an example with intermediary convergence rates, and scenarios with significant ties.}
\label{tab:swg_summary_compact}
\end{table*}

\subsection{Related works}

\paragraph{Single-time allocation without money.} 
We first discuss the literature on one-shot resource allocation without money. When there are two resources and agents share the same utility distribution, \cite{miralles2012cardinal} characterizes optimal mechanisms and shows that these can be constructed as combinations of lotteries and auction-like strategies. For several resources and additive utilities for the agents, \cite{guo2010strategy,han2011strategy,cole2013positive} give characterizations on the competitive ratio of strategy-proof mechanisms compared to the first-best allocation, without making prior assumptions on the agent utility distributions. Slightly different from our setting without money, the line of work including \cite{hartline2008optimal,hoppe2009theory,condorelli2012money,chakravarty2013optimal} studies the setting in which the central planner can impose payments from the agents but cannot redistribute these between agents. This setting is often referred to \emph{money burning}: any utility signal in the form of payment to the central planner is lost.
As discussed above, impossibility results due to Myerson-Satterthwaite's impossibility theorem generally prohibit the central planner from converging to the first-best allocation for a single allocation, while our focus is on designing mechanisms that converge to this optimal allocation as fast as possible.

\paragraph{Repeated allocation without money.}
Closest to our work, we now discuss the literature on the repeated allocation setting without transfers. While the folk theorem \cite{fudenberg2009folk} guarantees the existence of a mechanism that asymptotically converges to the first-best allocation, \cite{guo2009competitive} explicitly provides a mechanism that achieves at least $75\%$ of the optimal welfare in the infinite-horizon setting, provided agents are sufficiently patient. On the other hand, their mechanism may not converge to the optimal first-best allocation as $\gamma\to 1$. \cite{guo2015dynamic} characterize the optimal allocation in the stylized setting in which the central planner has a fixed cost and decides at each iteration whether to allocate the resource to a single agent whose utilities evolve according to a two-state Markov chain.

We previously introduced budget-based mechanisms, also called \emph{artificial currency} or \emph{script} strategies \cite{fhk06,kash2007optimizing,kash2015equilibrium,jss14}, in which each agent is given an initial budget that can be used as currency to bid on resources in each round. These have the advantage of being very simple and interpretable, and their implementation in several real-world applications has motivated the study of their theoretical performance  \citep{budish2011combinatorial,jackson2007overcoming,gorokh2021monetary}. For instance, \cite{gorokh2021monetary} shows that in the finite-horizon $T$ setting and for discrete utility distributions their convergence to the first-best allocation decreases as $1/T$. However, their results hold under a weaker notion of approximate Nash equilibrium in which incentive-compatibility constraints are only satisfied approximately up to $\sqrt{\log T/T}$ terms. In a similar spirit to artificial currencies, \cite{ebcdf22} recently proposed the use of karma mechanisms and discussed its applications in ride-hailing platforms, in which agents can receive an additional budget reward for accepting to give the resource to other agents.

The work of \cite{bgs19} is closest to the present paper. They consider the discounted infinite-horizon setting where agent utility distributions admit densities bounded above and below by some fixed constant. With these distributional assumptions and for two agents, they provide allocation mechanisms that converge to the first-best allocation as $\Theta(1-\gamma)$ for $\gamma\to 1$ and show that this is the best possible among all mechanisms up to constants. In comparison, they show that budget-based mechanisms have a convergence rate of at most $\Omega(\sqrt{1-\gamma})$; hence, budget-based strategies are suboptimal under these utility distribution assumptions. For $n\geq 3$ agents, they prove a $\Ocal((1-\gamma)^{1/(n+4)})$ convergence rate for their mechanism, leaving a significant gap compared to their $\Omega(1-\gamma)$ lower bound. 

Compared to these previous works, we give tools to prove convergence rates for \emph{general} utility distributions. As an example, with the specific assumptions from \cite{bgs19}, we provide an allocation mechanism that achieves the optimal convergence rate $\Theta(1-\gamma)$ for any number of agents $n$. More precisely, the upper (resp. lower) bound holds whenever the utility distributions have densities lower (resp. upper) bounded by a constant. Our work also aims to reconcile the analysis for the finite-horizon and infinite-horizon settings by using the same techniques for both.

Last, we make some comments about potential extensions of our current approach. In many operational scenarios, the utility distributions of agents may evolve over time following a periodic structure---say every $p$ rounds---for instance, due to seasonality in food banks. We expect that our current approach can be adapted to this non-i.i.d.\ setting by grouping $p$ consecutive times for the allocation function---effectively replacing $p$ distributions with an appropriate mixture on each length-$p$ group. Beyond seasonality, utility distributions may also be dependent on the past history of allocations. We leave for further work whether our results can have implications for such adaptive utility distributions.

\subsection{Organization of the paper}

We define the model and the framework used for our constructions in \cref{sec:preliminaries}. In \cref{sec:generic_mechanism}, we describe our general resource allocation mechanism and then present our main geometric tool to derive social welfare lower bound gaps in \cref{sec:main_tool_lower_bound}. Using these two geometric tools, we derive convergence rates for social welfare in the infinite-horizon setting in \cref{sec:infinite_horizon_paper}. In \cref{sec:links_to_finite_horizon}, we show how the tools developed for the infinite-horizon setting can be used for the finite-horizon setting as well, with only minor modifications. We provide examples of the results that can be obtained with our characterizations for utility distributions of interest in \cref{sec:examples}. Lastly, we conclude in \cref{sec:conclusion}.

\subsection{Notations}

For $n\geq 1$, we denote $[n]:=\{1,\ldots,n\}$. We also write $\Rbb_+:=[0,\infty)$ and $\mb 1=(1,\ldots,1)\in\Rbb^n$. By default, we use $\| \cdot\|$ for the Euclidean norm $\|\cdot\|_2$, otherwise, we will always specify the norm within the paper.
For any $\mb x\in\Rbb^n$ and $r>0$, we denote $B(\mb x,r)=\{\mb y\in\Rbb^n:\|\mb y-\mb x\|\leq r\}$ and $B^\circ(\mb x,r)=\{\mb y\in\Rbb^n:\|\mb y-\mb x\| < r\}$, the closed and open balls centered at $\mb x$ respectively. We let $S_{n-1}=\{\mb y\in\Rbb^n:\|\mb y\|=1\}$ be the unit $n$-dimensional sphere. For a compact $C\subset\Rbb^n$ and any $\mb x\in\Rbb^n$, we define $d(\mb x,C)=\max_{\mb y\in C}\|\mb y-\mb x\|$, where the maximum is attained since $C$ is compact. Next, for any subset $S\subseteq [n]$ and $\mb x\in\Rbb^n$, we denote $\mb x_S:=(x_i)_{i\in S}$. For $i\in[n]$ we also denote $\mb x_{-i}=(x_j)_{j\neq i}$. We denote $\Delta_n:=\{\mb x\in\Rbb_+^n,\sum_{i\in[n]}x_i\leq 1\}$ the probability simplex (augmented with a do-nothing option). We denote by $\mb a\odot\mb b=(a_ib_i)_i$ the component-wise product of two vectors $\mb a$ and $\mb b$.

We use standard Landau notations. For two functions $f$, and $g$, we write $f(x)=\Ocal(g(x))$ if there exists a constant $C$ such that $f(x)\leq Cg(x)$ for all $x$ in the support of $f$. We may specify $f(x)\underset{x\to a}{=}\Ocal(g(x))$ when the inequality $f(x)\leq Cg(x)$ only holds in a neighborhood of $a$ within the support of $f$. We write $f(x)=\Omega(g(x))$ when $g(x) = \Ocal(f(x))$, and $f(x)=\Theta(g(x))$ if both $f(x)=\Ocal(g(x))$ and $f(x)=\Omega(g(x))$ hold. We write $f(x)\underset{x\to a}{=}\omega(g(x))$ if $0\leq h(x)g(x)\leq f(x)$ where $h(x)\to\infty$ as $x\to a$. Last, the notations $\widetilde\Ocal(\cdot)$ and $\widetilde{\Theta}(\cdot)$ hide logarithmic factors in the horizon $T$.

\section{Problem formulation and preliminaries}
\label{sec:preliminaries}

We first detail the model and give useful preliminaries about the promised utility framework. Except for characterizations in \cref{subsec:trivial_cases}, most statements in this section can be considered standard in the literature.

\subsection{Model definitions}

We consider the sequential problem in which a central planner repeatedly allocates a single resource between $n$ agents. At every round $t\geq 1$, each strategic agent $i\in [n]$ observes their private realized utility $u_i(t)$, then reports a value $v_i(t)$ to the central planner only, and last, having observed the reported values $\mb v(t)=(v_i(t))_{i\in[n]}$ the central planner allocates the single resource to one or none of the agents, which we represent by the index $\hat i(t)\in\{0,\ldots n\}$ where $\hat i(t)=0$ signifies that the resource was not allocated at time $t$. Alternatively, the central planner selects a probabilistic allocation $\mb p(t)=(p_i(t))_{i\in[n]}\in\Delta_n=\{\mb y\geq \mb 0:\sum_{i\in[n]}y_i\leq 1\}$. Last, at the end of the round, the reported values $\mb v(t)$ and allocation $\mb p(t)$ are made public to all agents.

Crucially, we suppose that the true utilities are independent between agents and independent and identically distributed (i.i.d.) for each agent. That is, for any $i\in [n]$, we have $(u_i(t))_{t\geq 1}\overset{iid}{\sim} \Dcal_i$. We denote by $\bar F_i$ the complementary cumulative function of $\Dcal_i$, that is, $\bar F_i(x)=\Pbb_{u\sim\Dcal_i}[u\geq x]$. We assume that the fixed distributions $\Dcal_1,\ldots,\Dcal_n$ have bounded support in $[0,\bar v]$ for some fixed value $\bar v>0$, and are publicly known to the central planner and to all agents $i\in[n]$. Last, we consider the case of myopic agents, that is, at time $t$, agent $i$ does not have access to their realized utilities at future times $u_i(t')$ for $t'>t$. Formally, write the history available at time $t$ as $\mb h(t)=(v_j(t'),p_j(t'),j\in[n],t'<t)$. At time $t$, the central planner strategy is given by a function $S_t:[0,\bar v]^{n(t-1)}\times \Delta_n^{t-1}\times [0,\bar v]^n\to\Delta_n$ and the allocation is given by $\mb p(t) = S_t(\mb h(t),\mb v(t))$, while the strategy of agent $i\in[n]$ is given by a function $\sigma_{i,t}: [0,\bar v]^{n(t-1)}\times \Delta_n^{t-1} \times [0,\bar v] \to [0,\bar v]$ and their report is given by $v_i(t)=\sigma_{i,t}(\mb h(t),u_i(t))$.
For convenience, throughout the paper, $\mb u$ represents a random utility vector with independent coordinates with $u_i\sim\Dcal_i$.

The goal of the central planner is to maximize the social welfare or equivalently, to minimize the regret compared to the optimal allocation in hindsight $i^\star(t) \in\argmax_{i\in[n]}u_i(t)$. We consider two different settings.
\begin{enumerate}
    \item In the \emph{$\gamma$-discounted infinite-horizon} setting, the central planner and all agents share the same discount factor $\gamma\in[0,1)$. That is, the central planner aims to minimize the expected regret
    \begin{equation*}
         (1-\gamma) \Ebb\sqb{ \sum_{t\geq 1} \gamma^{t-1} \paren{\max_{i\in[n]} u_i(t) - u_{\hat i(t)}(t)}  }.
    \end{equation*}
    while agent $i\in[n]$ aims to maximize their total discounted utility $(1-\gamma)\Ebb\sqb{\sum_{t\geq 1} \gamma^{t-1} u_i(t)p_i(t)}$.
    
    \item In the \emph{finite horizon $T$} setting, the central planner aims to minimize the average regret
    \begin{equation*}
       \frac{1}{T} \Ebb\sqb{\sum_{t=1}^T \max_{i\in[n]} u_i(t) - u_{\hat i(t)}(t)},
    \end{equation*}
    while agent $i\in[n]$ aims to maximize their total utility $\frac{1}{T}\Ebb\sqb{\sum_{t\in[T]} u_i(t)p_i(t) }$.
\end{enumerate}

The reports of the strategic agent form a perfect Bayesian equilibrium, as detailed below. Using the same notations as in \cite{bgs19}, we let $V_{i,t}$ be the expected remaining utility of agent $i$ starting from time $t$, given its own utility $u_i(t)$ and the history $\mb h(t)$, with the allocation strategy $\mb S=(S_{t'})_{t'\geq 1}$, all other agents employing their strategy $\mb \sigma_{-i}=(\sigma_{j,t'})_{j\neq i,t'\geq 1}$ and agent $i$ using strategy $\tilde\sigma_i=(\tilde\sigma_{i,t'})_{t'\geq 1}$. Formally, we define $V_{i,t}(\mb S,\mb\sigma_{-i},\tilde\sigma_i\mid u_i(t),\mb h(t))$ as 
\begin{equation*}
(1-\gamma)\Ebb\sqb{ \sum_{t'\geq t}\gamma^{t'-t} u_i(t') p_i(t') \mid u_i(t),\mb h(t)} 
\end{equation*}
for the $\gamma$-discounted setting, and
\begin{equation*}
\frac{1}{T-t+1}\Ebb\sqb{\sum_{t'=t}^T u_i(t') p_i(t') \mid u_i(t),\mb h(t)} 
\end{equation*}
for the finite horizon $T$ setting. In both definitions, $\mb p(t')$ is induced by the central planner strategy $S_{t'}$ for $t'\geq t$ starting from the history $\mb h(t)$, using the reports of agents $j\neq i$ according to $\sigma_j$ and the reports of agent $i$ according to $\tilde\sigma_i$.
The perfect Bayesian equilibrium constraints then write
\begin{align}\label{eq:perfect_bayesian_equilibrium}
    \forall i,t,u_i(t),\mb h(t),\tilde\sigma_i,\quad V_{i,t}(\mb S,\mb\sigma_{-i},\sigma_i\mid u_i(t),\mb h(t)) \notag \\
    \geq V_{i,t}(\mb S,\mb\sigma_{-i},\tilde\sigma_i\mid u_i(t),\mb h(t)). \tag{PBE}
\end{align}

\subsection{Achievable utility sets}

As an alternative reformulation, letting $\mb U=(U_i)_{i\in[n]}$ be the vector of realized utilities where $U_i$ is the total expected utility gained by agent $i$, the goal is to maximize the total utility $\mb 1^\top \mb U$ among realizable utility vectors $\mb U$. For our purposes we will consider the more general problem in which agents can have weights $\mb\alpha\in\Rbb_+^n\setminus\{\mb 0\}$ and the goal of the central planner becomes maximizing the total utility $\mb\alpha^\top\mb U$. Equivalently, letting $V_i(\mb S,\mb\sigma):=\Ebb_{u_i(1)}[V_{i,1}(\mb S,\mb\sigma\mid u_i(1))]$, we aim to characterize the region of achievable utilities:
\begin{equation*}
    \set{\mb U = (V_i(\mb S,\mb\sigma))_{i\in[n]} \in\Rbb_+^n: \mb S,\mb\sigma \text{ satisfy }\eqref{eq:perfect_bayesian_equilibrium}},
\end{equation*}
which we denote $\Ucal_\gamma$ for the $\gamma$-discounted setting, and $\Vcal_T$ for the finite horizon $T$ setting.

\paragraph{Full-information set.} We first characterize the optimal allocation baseline, which corresponds to the \emph{full-information} setting in which the central planner has direct access to all agent utilities $(u_i(t))_{i\in[n]}$ before deciding its allocation. This also corresponds to the setting in which agents are non-strategic. The set of achievable utilities in this full-information setting is
\begin{equation*}
    \Ucal^\star = \set{ (\Ebb[u_i p_i(\mb u)])_{i\in[n]} , \; \mb p:[0,\bar v]^n\to\Delta_n  }.
\end{equation*}
We can also easily characterize the Pareto-optimal points in this set of achievable utilities. This boundary is parametrized by weights $\mb \alpha \in\Rbb_+^n\setminus \{\mb 0\}$ with the corresponding maximizing utility vectors $\argmax_{\mb U\in\Ucal^\star}\mb \alpha ^\top \mb U= \Ebb \sqb{\max_{i\in[n]}\alpha_i u_i }$. Such $\mb\alpha$-optimal vectors exactly correspond to first-best allocation functions that always allocate the resource to $i\in\argmax_{j\in[n]}\alpha_j u_j$. By default, we say that $\mb x\in\Ucal^\star$ is optimal if it is $\mb 1$-optimal. 


Recall that the central planner aims to maximize the social welfare, corresponding to equalitarian weights $\mb\alpha=\mb 1$. Hence, the optimal regret in the $\gamma$-discounted setting, which corresponds to the social welfare gap between strategic and non-strategic agents is
\begin{equation}\label{eq:def_SWG}
    \SWG(\gamma):= \max_{\mb U\in\Ucal^\star} \mb 1^\top \mb U- \max_{\mb U\in\Ucal_\gamma} \mb 1^\top \mb U.
\end{equation}
By abuse of notation, we denote by $\SWG(T)$ the social welfare gap in the finite-horizon-$T$ setting, that is, replacing $\Ucal_\gamma$ with $\Vcal_T$ in \cref{eq:def_SWG}.

\begin{remark}
    While the central planner mainly focuses on equalitarian weights $\mb\alpha = \mb 1$, it turns out that the geometry of the boundary of $\Ucal^\star$---or equivalently, maximizing utility vectors for weights $\mb\alpha\neq \mb 1$---is crucial to characterize the optimal social welfare gap and construct corresponding mechanisms. 
\end{remark}

\paragraph{No-information set.}
We next define the set of utilities that can be achieved \emph{without reports}, that is without information on the utility realizations. This corresponds to a setting in which the central planner selects a fixed allocation $\mb p\in\Delta_n$, which gives the following frontier
\begin{equation*}
    \Ucal^{NI} = \set{ (p_i \Ebb_{u_i\sim \Dcal_i}[u_i])_{i\in[n]}, \mb p\in\Delta_n}.
\end{equation*}
For intuition, if $\Pbb(u_i=0)=0$ for all $i\in[n]$, we have $\Ucal^{NI}=\Ucal_0=\Vcal_1$. We will see, however, that if this does not hold we may have $\Ucal^{NI}\subsetneq \Ucal_0=\Vcal_1$, see \cref{lemma:trivial_case}, scenario 3(b).\footnote{Specifically, the optimal vector $\mb U$ constructed therein is outside $\Ucal^{NI}=\text{Conv}(\mb 0;\{\Ebb[u_i]\mb e_i, i\in[n]\})$.} 
Using mixed allocation strategies, we can easily check the following.

\begin{fact}\label{fact:convexity}
    The sets $\Ucal^\star$, $\Ucal^{NI}$, $\Ucal_\gamma$, and $\Vcal_T$ are convex sets of $\Rbb_+^n$ for all $\gamma\in[0,1)$ and $T\geq 1$.
\end{fact}

\subsection{The promised utility framework}

Before giving our results, we present a convenient reformulation of the sequential allocation problem in terms of dynamic programming, referred to as the \emph{promised utility framework} \citep{spear1987repeated,abreu1990toward,thomas1990income,bgs19}. We give below an introduction and summary of the framework for the $\gamma$-discounted infinite-horizon setting as detailed in \cite{bgs19}.

We focus on the $\gamma$-discounted infinite-horizon setting for now. In its direct form, maximizing social welfare requires solving an infinite-horizon perfect Bayes equilibrium. However, the symmetry of the problem in time allows to drastically simplify the formulation. In light of the revelation principle, we can without loss of generality suppose that the mechanism is incentive-compatible hence agents report truthfully.
Say we want to design an allocation strategy to ensure that the final utility vector is $\mb U$. The first decision is to decide on the allocation function for the first time step $t=1$ which we denote $\mb p(\cdot\mid \mb U):[0,\bar v]^n \to\Delta_n$. It takes as input the agent reports $\mb v(t=1)=\mb u(1)$ and outputs the allocation distribution $\mb p(1)$. We then denote by $\mb W(\cdot\mid \mb U):[0,\bar v]^n\to\Rbb^n$ the remaining utility to be gathered by the agents at future times $t\geq 2$, that is,
\begin{equation*}
    W_i(\mb v\mid \mb U) = (1-\gamma) \Ebb\sqb{\sum_{t\geq 2} \gamma^{t-2} u_i \1_{\hat i_t=i} \mid \mb u(1)=\mb v}.
\end{equation*}
Note that the history also includes the allocation $\mb p(1)$ but this is completely subsumed by the knowledge of $\mb u(1)=\mb v$ via $\mb p(1)= \mb p(\mb v\mid\mb U) $.
The problem at time $t=2$ is now perfectly equivalent to the original problem except that now the central planner aims to ensure that the utility vector to realize is $\mb W(\mb u(1)\mid \mb U)$.

This motivates focusing only on the allocation function and the promised utility function: to achieve a region $\Rcal\subseteq\Ucal^\star$, it suffices to specify two functions $\mb p(\cdot\mid \mb U):[0,\bar v]^n\to \Delta_n$ and $\mb W(\cdot\mid \mb U):[0,\bar v]^n \to \Rbb^n$ for all $\mb U\in\Rcal$, where the later is viewed as a utility promise to the agents. These should satisfy the following constraints. The \emph{promise-keeping} constraint checks that the target utility $\mb U$ is met: for all $i\in[n]$ and $\mb U\in\Rcal$,
\begin{equation}\label{eq:target_met}
    U_i = \Ebb[(1-\gamma) u_i p_i(\mb u \mid \mb U) + \gamma W_i(\mb u \mid \mb U)].
\end{equation}
Second, the central planner should ensure that the promised utilities can be fulfilled for any possible reports of the agent. That is,
\begin{equation}\label{eq:valid_promise}
    \forall \mb U\in\Rcal,\forall \mb v\in [0,\bar v]^n,\quad \mb W (\mb v\mid \mb U) \in\Rcal.
\end{equation}
The last constraint is to ensure that \eqref{eq:perfect_bayesian_equilibrium} is satisfied, which can be simplified since we supposed that the mechanism is incentive-compatible. Having defined the interim allocation $P_i(v\mid \mb U):=\Ebb_{\mb u_{-i}}[p_i(v,\mb u_{-i}\mid \mb U)]$ and the interim promised utility $W_i(v\mid \mb U)=\Ebb_{\mb u_{-i}}[W_i(v,\mb u_{-i}\mid \mb U)]$ for all $i\in[n]$, the perfect Bayesian incentive-compatibility equations write for all $\mb U\in\Rcal$, $i\in[n]$, and $u\in[0,\bar v]$,
\begin{equation}\label{eq:incentive_compatibility}
    u\in\argmax_{v\in[0,\bar v]}  (1-\gamma)u P_i(v\mid \mb U) + \gamma W_i(v\mid \mb U).
\end{equation}

\begin{fact}\label{fact:promised_utility_1}
    A region $\Rcal'\subseteq\Ucal^\star$ can be achieved in the infinite-horizon $\gamma$-discounted setting if and only there exists a superset $\Rcal\supseteq \Rcal'$, functions $\mb p(\cdot\mid\mb U)$ and $\mb W(\cdot\mid \mb U)$ for all $\mb U\in\Rcal$ satisfying \cref{eq:target_met,eq:valid_promise,eq:incentive_compatibility}. 
\end{fact}

When $\gamma>0$, we can further characterize the form of the allocation and promised utility functions following a standard argument from \cite{myerson1981optimal}. Given the incentive-compatibility constraint \cref{eq:incentive_compatibility}, without loss of generality, we can suppose that the functions $P_i(\cdot \mid \mb U)$ are non-decreasing. For convenience, we say that a choice of allocation function $\mb p(\cdot\mid\mb U)$ is non-decreasing if the corresponding interim allocation function $P_i(\cdot\mid\mb U)$ is non-decreasing for all $i\in[n]$. Then, \cref{eq:incentive_compatibility} implies
\begin{align}\label{eq:formula_interim_initial}
    &W_i(v_i\mid \mb U) - W_i(0\mid \mb U) \notag \\
    &=\frac{1-\gamma}{\gamma} \paren{ \int_0^{v_i} P_i(v\mid \mb U)dv - P_i(v_i\mid\mb U) v_i}.
\end{align} 
Combining this with the constraint from \cref{eq:target_met}, we obtain for all $\mb U\in\Rcal$, $i\in[n]$, and $v_i\in[0,\bar v]$,
\begin{align}\label{eq:formula_interim_promise}
    &W_i(v_i\mid \mb U) = \frac{1}{\gamma}\Bigl[U_i + (1-\gamma)\Bigl(\int_0^{v_i} P_i(v\mid \mb U)dv \notag \\
    &- P_i(v_i\mid\mb U) v_i - \int_0^{\bar v} \bar F_i(v) P_i(v\mid\mb U)dv \Bigr) \Bigr].
\end{align}
This formula can be found in \cite{bgs19} as Eq~(10).
This equation effectively replaces the promise-keeping constraint \cref{eq:target_met} and the incentive-compatibility constraint \cref{eq:incentive_compatibility}. Further, we can check that $W_i(v_i\mid \mb U)$ is non-increasing in $v_i$ by re-writing $\int_0^{v_i} P_i(v\mid \mb U)dv - P_i(v_i\mid\mb U) v_i  = \int_0^{v_i} (P_i(v\mid\mb U) - P_i(v_i\mid \mb U) )dv$.
The promised utility function only needs to satisfy the following validity equation for the interim promise: for all $\mb U\in\Rcal$, $ i\in[n]$, and $v\in[0,\bar v]$,
\begin{equation}\label{eq:valid_interim_promise}
    W_i(v\mid \mb U)=\Ebb_{\mb u_{-i}}[W_i(v,\mb u_{-i}\mid \mb U)].
\end{equation}
As a summary, we have the following characterization.

\begin{fact}\label{fact:promised_utility_2}
    Let $\gamma\in(0,1)$. A region $\Rcal'\subseteq\Ucal^\star$ can be achieved in the infinite-horizon $\gamma$-discounted setting if and only if there exists a superset $\Rcal\supseteq \Rcal'$, functions $\mb p(\cdot\mid\mb U)$ and $\mb W(\cdot\mid \mb U)$ for all $\mb U\in\Rcal$ satisfying \cref{eq:valid_promise,eq:valid_interim_promise}, where the interim future promise is defined as in \cref{eq:formula_interim_promise}.
\end{fact}

\subsection{A convenient choice of promised utility functions}
\label{subsec:convenient_choice_promised_utility}
Given an allocation function $\mb p(\cdot\mid\mb U)$, the remaining choice of the mechanism design is in the form of the promised utility function $\mb W(\cdot\mid\mb U)$ so that it satisfies \cref{eq:valid_interim_promise,eq:valid_promise}. We introduce a specific choice of form for these promised utility functions that will be useful for our mechanism design constructions.

The specific construction is somewhat more general than the present sequential allocation setting. Consider the following problem: given $n$ independent random variables $\tilde Z_1,\ldots,\tilde Z_n$ and a direction $\mb \alpha\in\Rbb^n$, we aim to construct a coupling $(Z_1,\ldots,Z_n)$ such that (1) for all $i\in[n]$, $\Ebb[Z_i\mid \tilde Z_i]=\tilde Z_i$, and (2) that is as far along the direction $\mb\alpha$ as possible in the worst-case sense. That is, we aim to solve the problem
\begin{equation}\label{eq:optimal_coupling}
    \max_{\mb Z} \min_{\omega\in\Omega} \mb\alpha^\top \mb Z(\omega),\;\; \text{s.t.} \;\; \forall i\in[n],\Ebb[Z_i\mid \tilde Z_i]=\tilde Z_i.
\end{equation}
Taking the expectation, any coupling $\mb Z$ satisfying the marginal constraints must satisfy
\begin{equation}\label{eq:easy_upper_bound}
    \min_{\omega\in\Omega} \mb\alpha^\top \mb Z(\omega) \leq \Ebb[\mb \alpha^\top \mb Z] = \mb\alpha^\top \Ebb[\tilde{\mb Z}].
\end{equation}
This value can also be reached for instance using the following coupling from \cite{bgs19}. For $i\in[n]$ such that $\alpha_i=0$, we can let $Z_i=\tilde Z_i$. Otherwise, we define 
\begin{equation}\label{eq:coupling_formula_old}
    Z_i = \tilde Z_i- \frac{1}{n-1}\sum_{j\neq i}\frac{\alpha_j}{\alpha_i} \paren{\tilde Z_j - \Ebb[\tilde Z_j]}.
\end{equation}
We can easily check that this coupling reaches the upper bound from \cref{eq:easy_upper_bound} since the coordinates of $\tilde{\mb Z}$ are independent. This coupling forces the vector $\mb Z$ to lie in the hyperplane $\{\mb x: \mb \alpha^\top (\mb x-\Ebb[\tilde{\mb Z}]) =0\}$.

\cite{bgs19} used this construction to specify the future promise functions: for any $\mb U$, letting $Z_i:=W_i(v_i\mid\mb U)$, they use the promise function $\mb W(\mb v\mid\mb U;\mb\alpha(\mb U))$ defined as the previous coupling for some pre-specified vector $\mb\alpha(\mb U) \in\Rbb^n\setminus\{\mb 0\}$. For our purposes, we will use a slightly different coupling. The term $\alpha_i$ in the denominator leads to instabilities that we can mitigate by using the coupling from \cref{eq:coupling_formula_old} only for the variables $i$ with largest value of $|\alpha_i|$. Let $i_1$ and $i_2$ be the indices of the largest and second-largest components of $\mb\alpha$ respectively. We pose
\begin{equation}
\label{eq:coupling_formula}
    Z_i = \begin{cases}
        \tilde Z_{i_1} - \sum_{i\neq i_1} \frac{\alpha_i}{\alpha_{i_1}} (\tilde Z_i-\Ebb[\tilde Z_i])& i=i_1,\\
        \tilde Z_{i_2} - \frac{\alpha_{i_1}}{\alpha_{i_2}}(\tilde Z_{i_1}-\Ebb[\tilde Z_{i_1}]) & i=i_2,\\
        \tilde Z_i &i\notin\{i_1,i_2\}.
    \end{cases}
\end{equation}
Both couplings defined in \cref{eq:coupling_formula_old,eq:coupling_formula} are optimal for the previous optimization problem.

\begin{lemma}
\label{lemma:optimal_coupling}
    For any $\mb \alpha\in\Rbb^n$, the couplings defined in \cref{eq:coupling_formula_old,eq:coupling_formula} are solutions of the problem in \cref{eq:optimal_coupling} with value $\mb\alpha^\top (\Ebb[\tilde{\mb Z}])_{i\in[n]}$.
\end{lemma}

With this construction at hand, to specify an allocation strategy for the central planner, we may only specify the allocation function $\mb p(\cdot\mid\mb U)$ and the directions $\mb\alpha(\mb U)$. By defining $\mb W(\cdot\mid\mb U)$ via \cref{eq:coupling_formula}, the constraint \cref{eq:valid_interim_promise} is automatically satisfied. The only constraint that remains to be checked to have a valid allocation is \cref{eq:valid_promise}, which asks that the future promises stay within the constructed region.

\begin{remark}
    We are now ready to state our main results. For ease of exposition, we state all our main results in the following \cref{sec:generic_mechanism,sec:main_tool_lower_bound,sec:infinite_horizon_paper,sec:links_to_finite_horizon,sec:examples} for equalitarian weights $\mb\alpha=\mb 1$. Note that these can easily be generalized to arbitrary weights $\mb\alpha\in\Rbb_+^n\setminus\{\mb 0\}$ by replacing the valuation $v_i$ of an agent $i\in[n]$ by $\alpha_i v_i$.
\end{remark}

\subsection{Characterization of scenarios without social welfare gap}
\label{subsec:trivial_cases}

For technical purposes, we need to treat separately trivial cases when the regions $\Ucal_0=\Vcal_1$ already contain an optimal utility vector. In words, these corresponds to cases when the optimal social welfare can be achieved with \emph{memoryless} mechanisms which allocate based solely on reports at the current iteration. In the following, we show that this may only happen in two scenarios (see \cref{fig:trivial_case} for an illustration). The first one corresponds to the case when one agent's utilities dominates all others, while the second one involves some slightly more complex hierarchy between agents. In both cases, we can give simple optimal mechanisms:

\begin{figure}[t]
\centering
\begin{tikzpicture}[scale=.75]
    \draw[-latex,line width=1pt] (-0.5,0) node[left]{(a)\;\;\;} -- (10,0) node[right]{\small Utility};
    \draw[line width=1pt] (0,0.1) -- (0,-0.1) node[below]{$0$};
    \draw[|-|,red,line width = 1.5pt] (6,0) -- (9,0);
    \draw[red,decorate,decoration={brace,amplitude = 5pt,raise=2ex}] (6,0) -- node[above,yshift=1.5em]{\small Support of $u_i$ }(9,0);
    \draw[decorate,decoration={brace,amplitude = 5pt,raise=2ex}] (0,0) -- node[above,yshift=1.5em]{\small Support for all $u_j$ with $j\neq i$}(5.8,0);

    \draw[-latex,line width=1pt] (-0.5,-2) node[left]{(b)\;\;\;} -- (10,-2) node[right]{\small Utility};
    \draw[line width=1pt] (0,-2+0.1) -- (0,-2.1) node[below]{$0$};
    \draw[red,line width=2pt] (0.03,-2+0.2) -- (0.03,-2.2);
    \draw[blue,line width=2pt] (-0.03,-2+0.2) -- (-0.03,-2.2) ;
    \draw[|-|,red,line width = 1.5pt] (7,-2) -- (9,-2);
    \draw[red,decorate,decoration={brace,mirror,amplitude = 5pt,raise=2ex}] (7,-2) -- node[below,yshift=-1.5em]{\small Support of $u_{i_1}$ }(9,-2);

    \draw[|-|,blue,line width = 1.5pt] (5,-2) -- (6.5,-2);
    \draw[blue,decorate,decoration={brace,amplitude = 5pt,raise=2ex}] (5,-2) -- node[above,yshift=1.5em]{\small Support of $u_{i_2}$ }(6.5,-2);

    \draw[|-|,green!80!black,line width = 1.5pt] (2,-2) -- (4,-2);
    \draw[green!80!black,decorate,decoration={brace,amplitude = 5pt,raise=2ex}] (2,-2) -- node[above,yshift=1.5em,xshift=-1em]{\small Support of $u_{i_3}$ }(4,-2);
    \draw[decorate,decoration={brace,mirror,amplitude = 5pt,raise=2ex}] (0.2,-2) -- node[below,yshift=-1.5em]{\small Support for all $u_j$ with $j\notin\{i_1,i_2,i_3\}$}(1.8,-2);
\end{tikzpicture}
\caption{Illustration of the two scenarios (a) and (b) from \cref{lemma:trivial_case} which correspond to cases in which there is already an optimal allocation strategy for $\gamma=0$ (resp. $T=1$) in the infinite-horizon (resp. finite-horizon) setting. Figure (a) corresponds to the scenario in which agent $i$ dominates all other agents. Figure (b) corresponds the scenario in which agent $i_1$ and $i_2$ have non-zero utility mass at $0$ and the support of $u_i$ for $i\in\{i_1,i_2,i_3\}$ follow a hierarchy.}
    \label{fig:trivial_case}
\end{figure}

\begin{lemma}
\label{lemma:trivial_case}
    First, we have $\Ucal_0=\Vcal_1$. Next, the following are equivalent.
    \begin{enumerate}
        \item There exists an optimal vector in $\Ucal_0$ ($\SWG(0)=0$).
        In particular, $\Ucal_\gamma$ and $\Vcal_T$ always have an optimal vector for all $\gamma\in[0,1)$ and $T\geq 1$.
        \item For all $i\in[n]$, either 
        \begin{enumerate}
            \item there exists $j\neq i$ such that $\Pbb(u_j \geq u_i)=1$,
            \item or there exists $[m_i,M_i]$ such that $\Pbb(u_i \in [m_i,M_i]\cup\{0\})=1$ but $\Pbb(u_j\in (m_i,M_i))=0$ for all $j\neq i$.
        \end{enumerate}
        \item Either one of these scenarios holds
        \begin{enumerate}
            \item There exists $i\in[n]$ such that for all $j\neq i$, $\Pbb(u_i \geq u_j)=1$. In which case, $\Ebb[u_i]\mb e_i\in\Ucal_0$ is optimal, and is reached by the mechanism that always allocates to agent $i$.
            \item There exists a subset of agents $S=\{i_1,\ldots,i_k\}\subseteq\{i\in[n]: \alpha_i>0\}$ and a sequence $m_0=\infty>m_1 \geq m_2\geq \ldots m_{k-1}>0$ for which: (1) If $l<k$, $\Dcal_{i_l}$ is supported on $\{0\}\cup [m_l,m_{l-1}]$ and $\Pbb_{u\sim\Dcal_{i_l}}(u=0)>0$. (2) $\Dcal_{i_k}$ is supported on $[0,m_{k-1}]$. (3) For $i\notin S$, we have $\Pbb(u_i \leq u_{i_k})=1$.
            
            In this case, $\mb U = \sum_{l\in[k]} \Pbb(u_{i_{l'}}=0,l'<l)\Ebb[u_{i_l}] \mb e_{i_l} \in\Ucal_0$ is optimal, and is reached by the following allocation mechanism for reported utilities $\mb v\in[0,\bar v]^n$: $\mb p(\mb v) = \mb e_{i_{l(\mb v)}}$ where $l(\mb v) = \min\{ l:v_{i_l}>0 \}\cup\{k\}$.
        \end{enumerate}
    \end{enumerate}
\end{lemma}
The proof is given in \cref{subsec:no_welfare_gap_characterization}.
In light of the second characterization of \cref{lemma:trivial_case}, and even when $\Ucal_0$ does not contain an optimal vector, we note for agents $i\in[n]$ satisfying either condition 2(a) or 2(b), ensuring incentive-compatibility while optimizing for social welfare is somewhat straightforward (see \cref{lemma:def_I_properties} for more details). Indeed, if condition 2(a) holds, then it is possible to ignore agent $i$, never allocating to them, and still yield an optimal utility vector by allocating to $j$ instead. On the other hand, if condition 2(b) holds, the reports of agent $i$ are also not really necessary to reach an optimal vector. Knowing the utilities $u_j$ for the other agents $j\neq i$ and whether $u_i>0$ is sufficient: if this is the case, we can allocate to agent $i$ when $\max_{j\neq i}u_j< M_i$ and to an agent in $\argmax_{j\neq i}u_j$ otherwise.

This motivates the definition of the complementary set of agents, for which more work is needed to ensure incentive-compatibility: $\tilde I:=\{i\in[n]:i\text{ does not satisfy 2(a) nor 2(b)} \}$. As it turns out, ties between different agents are especially convenient for the central planner since it can freely decide how to break ties without incurring social welfare cost. It will therefore be useful to define the set of agents $I$ augmented with these convenient agents, on which it suffices to focus.
\begin{equation}\label{eq:def_set_I}
    I:=\tilde I\cup\{i\in [n]:\Pbb(u_i=\max_{j\neq i} u_j>0)>0\}. 
\end{equation}

\section{A generic resource allocation mechanism}
\label{sec:generic_mechanism}

We are now ready to detail our generic resource allocation mechanism for the $\gamma$-discounted setting. The extension to the finite horizon-$T$ setting is discussed in \cref{sec:links_to_finite_horizon}. The mechanism aims to achieve any utility vector within a pre-specified ball $B(\mb x,r)$ which has sufficiently large margin within the full-information region $\Ucal^\star$, that is, $B(\mb x, r+\delta)\subseteq\Ucal^\star$ for some margin parameter $\delta>0$. We refer to \cref{fig:lemma_ball_in_region} for an illustration. The corresponding \emph{Promised Utility Mechanism} (PUM) requires an appropriate margin to successfully achieve utility vectors in the target ball, as detailed below.

\begin{figure}[t]
\centering
\begin{subfigure}{.5\textwidth}
  \centering
  \begin{tikzpicture}
\begin{axis}[width=7.5cm,height=7.5cm,xmin=0, xmax=0.5,
    ymin=0, ymax=0.5, xlabel={$U_1$},         ylabel={$U_2$},          tick label style={font=\tiny}]
 \addplot[domain=0:1/3, smooth, samples=501,line width=1pt]
    {1/2-3/2*x^2}; 
\addplot[domain=1/3:1/2, smooth, samples=501, line width=1pt]
    {(6*(1/2-x))^(1/2)/3}; 
\draw[ fill=gray!70] (0.16,0.3) circle (0.09);
\draw[dashed] (0.16,0.3) circle (0.1375);

\draw[<->] (0.16+0.09,0.3) -- node[below]{\small $\delta$} (0.16+0.1375,0.3);
\draw[<->] (0.165,0.3) -- node[below]{\small $r$} (0.16+0.09,0.3);

\draw[fill=black] (0.16,0.3) node[below]{\small\textbf{\textit x}} circle (0.005);

\draw (0.4,0.35) node{$\mathcal U^\star$};

\end{axis}
\end{tikzpicture}

\end{subfigure}%
\begin{subfigure}{.5\textwidth}
  \centering
  \begin{tikzpicture}
\begin{axis}[width=7.5cm,height=7.5cm,xmin=0, xmax=1/2,
    ymin=0, ymax=1/2, xlabel={$U_1$},         ylabel={$U_2$},          tick label style={font=\tiny}]

\draw[color = white,fill=gray] (1/4-1/24,1/2-1/24) -- (1/4-1/24 + 0.04,1/2-1/24-0.04)  --  (1/4-1/24 - 0.01,1/2-1/24-0.01) -- (1/4-1/24 - 0.06,1/2-1/24+0.06/5) -- (1/4-1/24,1/2-1/24);

\draw[line width=1 pt] (0,1/2) -- (1/4-1/24,1/2-1/24) -- (1/2-1/24,1/4-1/24) -- (1/2,0);

\draw (0.05,0.162) circle (0.32);
\draw[dashdotted] (0.05,0.162) circle (0.323);
\draw[dashed] (0.05,0.162) circle (0.34);

\draw[<->](0.05+0.32,0.162) --  (0.05+0.34,0.162)node[right]{\small $\delta$} ;
\draw[<->] (0.055,0.162) -- node[below]{\small $r$} (0.05+0.32,0.162);

\draw[fill=black] (0.05,0.162) node[below]{\small\textbf{\textit x}} circle (0.005);

\draw (0.43,0.3) node{$\mathcal U^\star$};

\end{axis}
\end{tikzpicture}
\end{subfigure}
\caption{Illustration of \cref{lemma:ball_in_region_no_assumptions} (left) and \cref{lemma:prove_upper_bounds} (right) with $2$ agents. (1) On the left, if the gray ball $B(\mb x, r)$ is within the full-information utility set $\Ucal^\star$ with a margin $\delta$ satisfying Eq~\eqref{eq:constraint_safe_boundary}, then $\text{PUM}(\mb x, r,\delta)$ is successful and $B(\mb x, r)$ is contained in $\Ucal_\gamma$. The region $\Ucal^\star$ here corresponds to agents having uniform utility distributions on $[0,1]$. 
(2) On the right, the ball $B(\mb x, r)$ separates the region $\Ucal^\star$ in two connected components, the one $\Ccal$ closest to the boundary being included within a margin $\delta$. If $\delta$ satisfies the constraints from \cref{lemma:prove_upper_bounds}, then the gray region is outside $\Ucal_\gamma$ (note that the region is slightly smaller than the connected component $\Ccal$ (corresponding to the dash-dotted ball). The region $\Ucal^\star$ here corresponds to agents having uniform utility distributions on $\{1/6,5/6\}$.}
    \label{fig:lemma_ball_in_region}
\end{figure}

\begin{lemma}\label{lemma:ball_in_region_no_assumptions}
    Fix $\gamma\in (0,1)$. Let $\mb x\in \Ucal^\star$ and $r,\delta>0$ such that $B(\mb x, r+\delta) \subseteq \Ucal^\star$. If
    \begin{equation}\label{eq:constraint_safe_boundary}
        \frac{r\gamma}{1-\gamma} \geq \delta \geq C\frac{1-\gamma}{\gamma r},
    \end{equation}
    then the mechanism $\text{PUM}(\mb x, r, \delta)$ (see \cref{alg:main_mechanism}) does not fail and achieves any utility vector within $B(\mb x,r)$. In particular, when this holds we have $B(\mb x,r) \subseteq \Ucal_\gamma$. Here, 
    \begin{equation*}
        C = 4n^2 \bar v^2 \paren{ 1 + \frac{\max_{i\in[n]} \Ebb[u_i]}{\min_{i\in[n]}(\Ebb[u_i] -x_i-r)}  }^2.
    \end{equation*}
\end{lemma}

As remark, we aim to use this mechanism for balls $B(\mb x, r)$ that are close to the social-welfare-maximizing boundary of $\Ucal^\star$. Hence, the margin $\delta$ to this boundary intuitively corresponds to the suboptimality incurred by \text{PUM} compared to monetary mechanisms. Effectively, this reduces the multi-horizon resource allocation problem into a purely geometric one: finding the ball satisfying the margin condition \cref{eq:constraint_safe_boundary} which is closest to the social-welfare-maximizing boundary of $\Ucal^\star$. Importantly, the required margin is smaller for larger balls---with larger radius $r$---which hints to the fact that smoothed boundaries for $\Ucal^\star$ will allow for smaller social-welfare gaps. This intuition will be formalized in the following sections.

We now describe the $\text{PUM}$ resource allocation strategy.
From the promised utility framework, one only needs to specify appropriate allocation and promised utility functions $\mb p(\cdot\mid\mb U)$ and $\mb W(\cdot\mid\mb U)$ for $\mb U\in B(\mb x, r)$. Ideally, for a point on the boundary $\mb U=\mb x+r\mb y$ with $\mb y\in S_{n-1}$, we aim to use the allocation function realizing the utility vector $\mb x+(r+\delta)\mb y$ in the full-information case and construct the promised utility function using the form described in \cref{subsec:convenient_choice_promised_utility} with $\mb\alpha(\mb U)=\mb x-\mb U$. The intuition is that this choice of allocation and promised utility function pushes the promised utility vector $\mb W(\mb u\mid\mb U)$ towards the center of the ball. In practice, as per \cref{eq:coupling_formula}, we need to lower bound the second-largest component of $\mb\alpha(\mb U)$, which will require considering a larger region $\Rcal\supseteq B(\mb x, r)$ on which the allocation and promised utility functions are defined.

\begin{algorithm}[t]

\caption{Promised Utility Mechanism $\text{PUM}(\mb x, r, \delta)$ for the $\gamma$-discounted setting}\label{alg:main_mechanism}

\LinesNumbered
\everypar={\nl}
\SetAlgoNoEnd
\setcounter{AlgoLine}{0}

\KwIn{Ball $B(\mb x,r)$, margin $\delta$ with $B(\mb x, r+\delta)\subseteq\Ucal^\star$, target utility $\mb U_0\in B(\mb x, r)$}

Initialize $\mb U\gets \mb U_0$

\For{$t\geq 1$}{
Decompose $\mb U= \mb s \odot \paren{\sum_{i\in[n]}q_i\Ebb[u_i]\mb e_i + q_0 (\mb x+r\mb y)}$ where $\mb q\geq 0,\sum_{i=0}^n q_i\leq 1$, $\mb y\in S_{n-1}$, $\mb s\in[0,1]^n$.

Fix an allocation function $\tilde {\mb p}(\cdot)$ realizing utility $\mb x+(r+\delta)\mb y$ in the full-information case

Compute interim functions $\tilde P_i(\cdot):=\Ebb_{\mb u_{-i}}[\tilde p_i(\cdot,\mb u_{-i})]$ and $\tilde W_i(\cdot)$ according to \cref{eq:formula_interim_promise} for $i\in[n]$

Observe reports $\mb v=(v_i)_{i\in[n]}$ at round $t$ and select allocation $\mb s\odot \paren{\sum_{i\in[n]}q_i\mb e_i + q_0  \tilde{\mb p}(\mb v)}$

Define $\tilde {\mb W}$ using \cref{eq:coupling_formula} for $\mb\alpha=-r\mb y$, and $\tilde Z_i=\tilde W_i(v_i)$ for $i\in[n]$

Update $\mb U\gets \mb s\odot \paren{\sum_{i\in[n]}q_i\Ebb[u_i]\mb e_i + q_0  \tilde{\mb W}}$

\lIf{$\mb U\notin \Rcal$ as defined in \cref{eq:def_enlarged_region}}{Mechanism fails. \textbf{break}}
}

\end{algorithm}
\paragraph{Formal construction of the mechanism.} Precisely, we construct these functions on the region
\begin{equation}\label{eq:def_enlarged_region}
    \Rcal:= \set{\mb y: \mb 0\leq \mb y\leq \mb z, \mb z\in \text{Conv}(\Ucal^{NI},B(\mb x,r))}.
\end{equation}
To give some intuition about this utility set, we have
$\text{Conv}(\Ucal^{NI},B(\mb x,r)) = \text{Conv}(\mb 0;\{ \Ebb[u_i]\mb e_i,i\in[n] \}; B(\mb x,r) )$. This is the convex hull of the target ball with the trivial allocations that always allocate the resource to a single agent $i\in[n]$. Additionally, we extend this region in all directions $-\mb e_i$ for $i\in[n]$ so long as points stay within the positive orthant $\Rbb_+^n$. For any $\mb U \in \Ucal^\star$, we fix $p(\cdot;\mb U)$ an allocation function which reaches the utility vector $\mb U$ when having access to the full information of agent's utilities.

We start by specifying the allocation strategy for extreme points $\mb U$ of $\Rcal$ that still belong to the ball $B(\mb x,r)$, for which we can use the strategy described above. Writing $\mb U = \mb x + r\mb y$ where $\mb y\in S_{n-1}$ is a unit vector, we define the allocation function via $p(\cdot\mid \mb U):= p(\cdot;\mb x + (r+\delta)\mb y)$ and $\mb \alpha(\mb U) = \mb x-\mb U = -r\mb y$. 
We then extend the definition of the allocation function to the whole domain $\Rcal$.
    To do so, note that any vector $\mb U\in \Rcal$ can be characterized by $\mb 0\leq \mb U\leq \mb y$ where $\mb y\in \text{Conv}(\Ucal^{NI},B(\mb x,r))$ and is also an extreme point of $\Rcal$. In particular, $\mb y$ can be written as the convex combination of $\mb 0$, vectors $\Ebb[u_i]\mb e_i$ and some extreme point $\tilde {\mb y}$ of $\Rcal$ within the ball $B(\mb x, r)$ (we only need one point from $B(\mb x, r)$ at most because the ball is strictly convex). Such a decomposition can be obtained via standard Carath\'eodory constructions. As a summary, we obtain
    \begin{equation*}
        \mb y = \sum_{i\in[n]} q_i \Ebb[u_i]\mb e_i + q_0\tilde{\mb y},\quad\text{where}\quad \mb q\geq \mb 0, \sum_{i\leq n} q_i\leq 1,
    \end{equation*}
    and $U_i = s_i y_i$ where $s_i\in[0,1]$ for all $i\in[n]$.
    We then define the allocation and promised utility function via
    \begin{align*}
        p_i(\cdot \mid \mb U) = s_i\paren{ \sum_{i\in[n]} q_i \mb e_i + q_0 p_i(\cdot\mid\tilde{\mb y})},\;\; i\in[n]. \\ 
        W_i(\cdot \mid \mb U) = s_i\paren{\sum_{i\in[n]} q_i \Ebb[u_i] \mb e_i + q_0  W_i(\cdot\mid\tilde{\mb y})},\;\; i\in[n].
    \end{align*}
    Note in particular that for all the extreme points $\mb U = \Ebb[u_i]\mb e_i$, the allocation function $\mb p(\cdot\mid\mb U) = \mb e_i$ always gives the resource to agent $i$, as should be expected. The complete $\text{PUM}$ strategy is summarized in \cref{alg:main_mechanism}.

\paragraph{Remarks on \cref{lemma:ball_in_region_no_assumptions}.}
As an important remark, the term $C$ from \cref{lemma:ball_in_region_no_assumptions} should be viewed as a constant. Indeed, since we aim to bound the social welfare gap $\SWG(\gamma)$ for non-monetary mechanisms, we can use \cref{lemma:ball_in_region_no_assumptions} with small balls locally close to an optimal utility vector $\mb U\in\Ucal^\star$. For sufficiently small balls and close to this optimum point, the term $C$ becomes
\begin{equation*}
    C \approx 4n^2 \bar v^2 \paren{ 1 + \frac{\max_{i\in[n]}  \Ebb[u_i]}{\min_{i\in[n]}(\Ebb[u_i] -U_i)}  }^2,
\end{equation*}
which is independent from $\gamma$. Further, the term $C$ from \cref{lemma:ball_in_region_no_assumptions} can be greatly simplified if one has additional structure in the problem, as exemplified by the following assumption.

\begin{assumption}\label{assumption:lower_bounded_valuation}
    The utility distributions of the agents $i\in[n]$ have support in $[\underline v , \bar v]$, where $\bar v \geq \underline v>0$.
\end{assumption}

\begin{lemma}\label{lemma:ball_in_region}
    Under \cref{assumption:lower_bounded_valuation}, we can choose $C = \bar v^2 ( 2n + \bar v/\underline v)^2$ in the statement of \cref{lemma:ball_in_region_no_assumptions}.
\end{lemma}

The sufficient condition from \cref{lemma:ball_in_region_no_assumptions} can be generalized by splitting agents into groups and applying the resource allocation mechanism to each group separately. For simplicity, we defer the complete statement of this lower bound tool on the achievable region $\Ucal_\gamma$ to the Appendix (\cref{lemma:ball_in_region_optimized}).

\section{A geometric tool for proving lower bounds on the social welfare gap}
\label{sec:main_tool_lower_bound}


In this section, we present our main tool for proving social welfare gaps between non-monetary and monetary mechanisms, for the $\gamma$-discounted infinite-horizon setting. Equivalently, we aim to show that some regions in the full-information region $\Ucal^\star$ cannot be achieved by non-monetary mechanisms.

As a reminder, \cref{lemma:ball_in_region_no_assumptions} showed that any ball $B(\mb x, r)$ with sufficient margin $\delta$ within the full-information region $\Ucal^\star$ is achievable by non-monetary mechanisms. 
Intuitively, we show a corresponding negative counterpart: if the ball $B(\mb x,r)$ is too close to the boundary of $\Ucal^\star$, then the region between the ball and this boundary cannot be achieved. More formally, we show that if the ball $B(\mb x,r)$ is so close to the social-welfare-maximizing utility vectors of $\Ucal^\star$ that it cuts the region $\Ucal^\star$ in (at least) 2 components, under appropriate margin conditions, the social-welfare-maximizing component cannot be achieved (see the right figure of \cref{fig:lemma_ball_in_region} for an illustration). The corresponding statement will therefore be \emph{local}, that is, we focus on the region of $\Ucal^\star$ that is close to being optimal for the social welfare: $\{\mb y: \mb 1^\top \mb y \geq \max_{\mb z\in \Ucal^\star} \mb 1^\top \mb z - c_1\}$ for some $c_1>0$:

\begin{lemma}\label{lemma:prove_upper_bounds}
    Suppose that $\Ucal_0$ does not already contain an optimal vector (see \cref{lemma:trivial_case} for a characterization). There exist two constants $c_1,c_2>0$ such that the following holds. 
    For $\gamma\in(0,1)$ and $r>0$, suppose that $\Ucal^\star \setminus B^\circ(\mb x, r)$ has a connected component $\Ccal\subseteq \{\mb y: \mb 1^\top \mb y \geq \max_{\mb z\in \Ucal^\star} \mb 1^\top \mb z - c_1\}$. If
    \begin{equation*}
        \Ccal \subseteq B(\mb x, r+\delta),\quad \delta = \min\paren{c_2\frac{1-\gamma}{\gamma r},r},
    \end{equation*}
    then the region $\tilde\Ccal:=\Ccal \setminus B^\circ\paren{\mb x,r+\frac{1-\gamma}{\gamma}\delta}$ is not achievable by non-monetary mechanisms: $\Ucal_\gamma\cap \tilde\Ccal=\emptyset$.
    In particular, for any $\mb U\in\Ccal$, writing $\mb U = \mb x+ \|\mb U-\mb x\| \mb d$,
    \begin{equation*}
        \max_{\mb y\in\Ucal_\gamma} \mb d^\top (\mb y-\mb x) < r+ \frac{1-\gamma}{\gamma}\delta .
    \end{equation*}
\end{lemma}
We make a few remarks about this result. First, the margin $\delta$ required ($\delta=\Ocal(\frac{1-\gamma}{\gamma r})$) has exactly the same form as the margin required for our positive results in \cref{lemma:ball_in_region_no_assumptions}. This will be essential to show that our mechanisms achieve near-optimal social welfare.
Second, the condition $\Ccal\subseteq \{\mb y: \mb 1^\top \mb y \geq \max_{\mb z\in \Ucal^\star} \mb 1^\top \mb z - c_1\}$, which localizes the region of interest, is mild since the constant $c_1>0$ is independent of $\gamma$ (especially since it is possible to reach points that are $c_1$-close to optimal for sufficiently small $\gamma$ from the folk theorem). Last, ideally, \cref{fig:lemma_ball_in_region} should prove that the full component $\Ccal$ is not achievable by non-monetary mechanisms, however, for technical reasons, we prove this result for a slightly smaller region $\tilde\Ccal$. Fortunately, this will not affect our near-optimality guarantees.


As for \cref{lemma:ball_in_region_no_assumptions}, the statement can be refined by considering a subset of agents. In particular, it is useful for future proofs to consider the setting in which a single agent $i\in[n]$ competes for the allocation against all other agents $[n]\setminus\{i\}$. The precise statement is deferred to the Appendix (\cref{lemma:upper_bounds_2D}).

\section{Social Welfare Gap Characterizations for the discounted infinite-horizon setting}
\label{sec:infinite_horizon_paper}

Using the tools developed in the previous section \cref{lemma:ball_in_region_no_assumptions,lemma:prove_upper_bounds} (and generalizations \cref{lemma:ball_in_region_optimized,lemma:upper_bounds_2D}), 
we now derive our main characterizations of the social welfare gap for the discounted infinite-horizon setting.

\subsection{A universal rate \texorpdfstring{$\Ocal(\sqrt{1-\gamma})$}{}}

As a first consequence of \cref{lemma:ball_in_region_no_assumptions}, we show that for appropriately chosen parameters, the mechanism $\text{PUM}$ always achieves social welfare gap at most $\Ocal(\sqrt{1-\gamma})$. As a remark, the folk theorem only ensures asymptotic decay of this social welfare gap. In comparison, to the best of our knowledge, this is the first result which achieves a quantitative universal convergence rate to first-best allocation as agents are more patient, irrespective of utility distributions or the number of agents. \footnote{\cite{gorokh2021monetary} also provide general bounds for the finite $T$-horizon setting but for the weaker notion of approximate PBE, and therefore are not comparable to our perfect PBE guarantees.}

\begin{theorem}\label{thm:universal_lower_bound_simplified}
    The optimal social welfare regret always satisfies $\SWG(\gamma) = \Ocal(\sqrt{1-\gamma})$.
\end{theorem}

The proof of this result proceeds by showing that any vector $\mb U\in\Ucal^\star$ is at distance at most $\Ocal(\sqrt{1-\gamma})$ from $\Ucal_\gamma$. In the standard case where $\mb U\in\Ucal^\star$ satisfies $0<U_i<\Ebb[u_i]$ for all $i\in[n]$, \cref{thm:universal_lower_bound_simplified} is a simple consequence of \cref{lemma:ball_in_region_no_assumptions} of which \cref{fig:proof_slow_rate} gives a visual proof. We first note that $\prod_{i\in[n]}[0,U_i]\subseteq \Ucal^\star$. To prove the desired result, it then suffices to construct a ball $B(\mb x, r)$ within this region and that satisfies the constraints Eq~\eqref{eq:constraint_safe_boundary} with $r=\delta=\sqrt{C\frac{1-\gamma}{\gamma}}$ as depicted in the left figure of \cref{fig:proof_slow_rate}.

\begin{figure}
\begin{subfigure}{.5\textwidth}
    
    \centering
\begin{tikzpicture}
\begin{axis}[width=7.5cm,height=7.5cm,xmin=0, xmax=1/2,
    ymin=0, ymax=1/2, xlabel={$U_1$},         ylabel={$U_2$},          tick label style={font=\tiny}]

\draw[line width=1 pt] (0,1/2) -- (1/4,5/12) -- (1/2,0);

\draw[fill=black] (1/4,5/12) node[above right]{\small\textbf{\textit U}} circle (0.005);

\draw[dotted] (0,5/12) -- (1/4,5/12) -- (1/4,0);

\draw[dashed] (1/4-.08,5/12-.08)  circle (0.08);

\draw[fill=gray!70] (1/4-.08,5/12-.08) circle (0.04);

\draw[<->](1/4 - 0.04,5/12 - 0.08) --  (1/4 ,5/12 - 0.08) node[right]{\small $\delta$} ;
\draw[<->] (1/4 - 0.08,5/12 - 0.08) -- node[below]{\small $r$} (1/4 - 0.04,5/12 - 0.08);

\draw (0.4,0.3) node{$\mathcal U^\star$};

\end{axis}
\end{tikzpicture}
\end{subfigure}
\begin{subfigure}{.5\textwidth}
  \centering
  \begin{tikzpicture}
\begin{axis}[width=7.5cm,height=7.5cm,xmin=0, xmax=0.5,
    ymin=0, ymax=0.5, xlabel={$U_1$},         ylabel={$U_2$},          tick label style={font=\tiny}]
 \addplot[domain=0:1/3, smooth, samples=501,line width=1pt]
    {1/2-3/2*x^2}; 
\addplot[domain=1/3:1/2, smooth, samples=501, line width=1pt]
    {(6*(1/2-x))^(1/2)/3}; 
\draw[ fill=gray!70] (.24,.24) circle (0.09);
\draw[dashed] (.24,.24) circle (0.1315);

\draw[<->] (.24+.063,.24+.063) -- node[above left]{\small $\delta$} (.24+.09,.24+.09);
\draw[<->] (.24+.002,.24+.002) -- node[above left]{\small $r$} (.24+.063,.24+.063);

\draw[fill=black] (.24,.24) node[left]{\small\textbf{\textit x}} circle (0.005);

\draw[<->] (.24,.24-0.015) -- node[below]{\small $r_0$} (.24+0.1315,.24-0.015);

\draw (0.28,0.45) node{$\mathcal U^\star$};

\draw[fill=black] (1/3,1/3) node[above right]{\textbf{\textit U}} circle (0.005);

\end{axis}
\end{tikzpicture}

\end{subfigure}%

    \caption{Visual proof of \cref{thm:universal_lower_bound_simplified} (left) and \cref{thm:simpler_geometric_fast_rate} (right) given an optimal vector $\mb U$ with $U_i>0$ for all $i\in[n]$. (1) On the left, the dotted region satisfies $\prod_{i\in[n]}[0,U_i]\subseteq \Ucal^\star$. We can then fit a ball $B(\mb x, r)$ as in the figure with $r=\delta=\sqrt{C\frac{1-\gamma}{\gamma}}$, satisfying Eq~\eqref{eq:constraint_safe_boundary}. Hence $B(\mb x,r)\subseteq \Ucal_\gamma$ and as a result $d(\mb U,\Ucal_\gamma)=\Ocal(\sqrt{1-\gamma})$. (2) On the right, if $B(\mb x,r_0)\subseteq \Ucal^\star$ is tangent at $\mb U$, we can choose $r,\delta$ satisfying $r+\delta=r_0$ and $\delta=C\frac{1-\gamma}{\gamma r} $. These satisfy Eq~\eqref{eq:constraint_safe_boundary}. Note that $\delta\sim C\frac{1-\gamma}{r_0}$ as $\gamma\to 1$, which gives $d(\mb U,\Ucal_\gamma)\leq \delta=\Ocal(1-\gamma)$.}
    \label{fig:proof_slow_rate}
\end{figure}

While this already shows that non-monetary mechanisms can have good non-asymptotic performance in general, the convergence rate of the social welfare gap may be significantly faster when the boundary of $\Ucal^\star$ is smooth. This is also captured by \cref{lemma:ball_in_region_no_assumptions}: the smoother the local boundary of $\Ucal^\star$, the larger the radius $r$ of the ball that can be fit close to that boundary, and the smaller the margin $\delta\approx (1-\gamma)/r$ that is needed. In particular, in ideal cases where the ball has radius $r$ independent of $\gamma$, \cref{lemma:ball_in_region_no_assumptions} shows that the mechanism \text{PUM} yields smaller social welfare gap $\Ocal(1-\gamma)$. We detail these scenarios in the following section.

\subsection{Sufficient conditions for faster rates \texorpdfstring{$\Ocal(1-\gamma)$}{}}
We now discuss situations when fast social welfare gap convergence rates $\Ocal(1-\gamma)$ can be achieved. Roughly speaking, these correspond to the ideal convergence rate: these are precisely the fastest rates achieved in \cite{bgs19} for the case of $n=2$ agents and well-behaved smooth utility distributions. Additionally, these will correspond to the fastest convergence rate that may carry over to the finite-horizon setting (see \cref{sec:infinite_horizon_paper} and \cref{lemma:no_faster_1/T}). From a practical perspective, understanding conditions for fast convergence rates is helpful to understand which distributional properties are most desirable for the implementation of non-monetary mechanisms.
As a warm-up, we start with a simple geometric condition. Intuitively, if the boundary of $\Ucal^\star$ is locally smooth at the neighborhood of optimal utility vectors, then one can achieve the faster rate:

\begin{theorem}\label{thm:simpler_geometric_fast_rate}
    Suppose that there exists an optimal vector $\mb U\in\Ucal^\star$ such that a ball tangent to $\mb U$ of radius $r_0>0$ is contained within $\Ucal^\star$, i.e., taking $\mb x = \mb U-r_0\mb 1$ we have $B(\mb x,r_0)\subseteq \Ucal^\star$. Then, $\SWG(\gamma) = \Ocal(1-\gamma)$.
\end{theorem}


For instance, \cref{thm:simpler_geometric_fast_rate} implies that whenever the boundary of $\Ucal^\star$ is twice-differentiable locally around an optimal point $\mb U$ then one can fit a ball tangent to $\mb U$ within the convex set $\Ucal^\star$ (see \cref{lemma:quick_proof} for a quick proof or \citet[Theorem 1.0.9]{barb2009topics} for a stronger statement).
This result is obtained by applying \cref{lemma:ball_in_region_no_assumptions} with the parameters $\mb x$, and $r,\delta>0$ such that $\delta = C\frac{1-\gamma}{\gamma r}$ (to satisfy \cref{eq:constraint_safe_boundary}) and $r+\delta=r_0$. This gives a social welfare gap of $\delta=\Theta(1-\gamma)$. This construction is depicted in the right figure of \cref{fig:proof_slow_rate}. 
While \cref{thm:simpler_geometric_fast_rate} gives a purely geometric condition, it can be translated into explicit conditions on the utility distributions: intuitively \cref{thm:simpler_geometric_fast_rate} requires that any two agents may simultaneously have near-optimal utilities with non-negligible probability. This condition can however be significantly generalized, as detailed below. 

Our main characterization for fast convergence rates $O(1-\gamma)$ converts this infinite-horizon continuous incentive-compatibility problem into a purely \emph{combinatorial} test: whether a certain directed graph $\Gcal$ admits a partition into \emph{strongly connected components}. We start by giving the main intuitions about this result then formally state it in \cref{thm:faster_rates_1-beta}. Intuitively, the graph $\Gcal$ contains an edge $i\to j$ when the central planner can use agent $i$ to enforce incentive-compatibility on agent $j$ through added rewards/penalties. This flexibility typically arises when $i$ and $j$ share similar utilities: the central planner can then freely choose which agent to allocate to, without incurring significant social welfare cost. On each strongly-connected component of $\Gcal$, we then show that the central planner can enforce incentive-compatibility at a low social welfare price. In turn, if $\Gcal$ admits a partition into strongly connected components, the central planner can enforce incentive-compatibility for all agents simultaneously and achieve fast convergence $O(1-\gamma)$.


We now give the formal construction for the graph $\Gcal$. As discussed in \cref{subsec:trivial_cases}, it suffices to focus on the agents in $I$ as defined in \cref{eq:def_set_I}. For any pairs of agents $i,j\in I$, we introduce the function
\begin{equation*}
    f(\eta;i,j):= \Ebb\sqb{u_i \1_{u_j =\max_{l\in[n]}u_l}\1_{ u_j\in[ u_i,(1+\eta) u_i]}}.
\end{equation*}
This function quantifies to what extent (1) agent $j$ has the highest utility among all agents and (2) agent $i$ has near-highest utility, simultaneously. To give some intuitions, notice that under this event, the central planner has the flexibility to allocate the resource either to agent $i$ or $j$ at a small social welfare cost proportional to $\eta$. On the other hand, because of the incentive-compatibility constraints \cref{eq:incentive_compatibility}, the central planner needs to be able to penalize agent $j$ on future rounds if they reported a high utility $v_j(t)$ at time $t$, and similarly, reward agent $j$ on future rounds if they reported a low utility $v_j(t)$ (notice that the interim promised utility $W_j(v_j(t)\mid \mb U)$ is non-increasing in $v_j(t)$ from \cref{eq:formula_interim_promise}).
Therefore, the higher the value of $f(\eta;i,j)$, the easier it is for the central planner to fulfill its future promises/penalties to agent $j$, leveraging agent $i$.

\begin{figure}
    \begin{subfigure}{.58\textwidth}
            \centering
    \begin{tikzpicture}[scale=2.5] Scale: x=12cm means [0,1] is 12cm wide
  \draw[-latex] (0,0) -- (1.1,0) node[right] {$x$};
  \draw[-latex] (0,0) --  (0,1.1);
  \node[rotate=90,anchor=south] at (-0.25,0.5) {Density / Mass};

  \filldraw[fill=blue!20,draw=blue,thick] (0,0) rectangle (0.25,1);
  \draw[blue,ultra thick] (0.75,0) -- (0.75,0.75);
  \fill[blue] (0.75,0.75) circle (0.03);

  \filldraw[fill=red!20,draw=red,thick] (0.5,0) rectangle (0.6667,1);
  \draw[red,ultra thick] (0.2,0) -- (0.2,5/6);
  \fill[red] (0.2,5/6) circle (0.03);

  \filldraw[fill=green!20,draw=green!50!black,thick] (0.75,0) rectangle (1,1);
  \draw[green!50!black,ultra thick] (0.5,0) -- (0.5,3/8);
  \fill[green!50!black] (0.5,3/8) circle (0.03);
  \draw[green!50!black,ultra thick] (0.125,0) -- (0.125,3/8);
  \fill[green!50!black] (0.125,3/8) circle (0.03);

  \draw[blue,ultra thick] (0.75,0) -- (0.75,0.75);
  \fill[blue] (0.75,0.75) circle (0.03);

  \node[blue] at (.15,1.1) { $\Dcal_1$};
  \node[red] at (.6,1.1) { $\Dcal_2$};
  \node[green!50!black] at (0.9,1.1) { $\Dcal_3$};

  \foreach \x/\label in {0,0.5,1}{
    \draw (\x,0) -- (\x,-0.02);
    \node[below] at (\x,-0.02) { $\label$};
  }
  \foreach \y in {0.5,1}{
    \draw (0,\y) -- (-0.015,\y);
    \node[left] at (-0.015,\y) { \y};
  }

\end{tikzpicture}
    \end{subfigure}
    \begin{subfigure}[t]{.4\textwidth}
    \begin{center}
        \begin{tikzpicture}[scale=1.2]
        \node (1) at (90:1cm) {\color{blue} $1$};   
        \node (2) at (210:1cm) {\color{red} $2$};  
        \node (3) at (330:1cm) {\color{green!50!black} $3$};

        \draw [-latex,line width=1pt] (1) -- (3);
        \draw [-latex,line width=1pt] (3) -- (2);
        \draw [-latex, bend right,line width=1pt] (1) to (2);
        \draw [-latex, bend right,line width=1pt] (2) to (1);

        \node at (0,-.8) {$\Gcal$};
    \end{tikzpicture}
    \end{center}
    
    \end{subfigure}

    \caption{Utility distributions with $n=3$ agents (left) and corresponding graph $\Gcal$ (right). (1) On the left, utility distributions are a mixture of atoms---represented with bullets---and uniform distributions on intervals---represented in rectangles. E.g., $\Dcal_1$ has an atom on $x=3/4$ of mass $3/4$ and a density of $1$ on $[0,1/4]$. Formally, $\Dcal_1=\frac{1}{4}\text{Unif}([0,\frac{1}{4}])+\frac{3}{4}\delta_{3/4}$, $\Dcal_2=\frac{1}{6}\text{Unif}([\frac12,\frac23])+\frac{5}{6}\delta_{1/5}$, and $\Dcal_3=\frac{1}{4}\text{Unif}([\frac34,1])+\frac{3}{8}\delta_{1/8}+\frac{3}{8}\delta_{1/2}$, where $\text{Unif}(I)$ and $\delta_x$ denotes the uniform over an interval $I$ and the Dirac at $x$ respectively. (2) On the right, the graph $\Gcal$ as defined in \cref{eq:def_graph_G}. Intuitively, $i\to j$ when the central planner may select agent $i$ while agent $j$ had highest utility, without incurring large social welfare cost. Since $\{1,2,3\}$ is strongly-connected, \cref{thm:faster_rates_1-beta} implies $\SWG(\gamma)=\Ocal(1-\gamma)$.}
    \label{fig:construction_G}
\end{figure}

We then consider the following graph $\Gcal$ on $I$ such that for $i\neq j\in I$,
    \begin{equation}\label{eq:def_graph_G}
        i\to j\quad\Longleftrightarrow\quad f(\eta;i,j)\underset{\eta\to 0}{=} \Omega(\eta).
    \end{equation}
    \cref{fig:construction_G} gives an illustration of $\Gcal$ for concrete utility distributions.
From the previous discussion, $i\to j$ may be interpreted as follows: the central planner can use agent $i$ as a lever to flexibly ensure incentive-compatibility for agent $j$. In turn, we can show that fast convergence rates can be achieved if $\Gcal$ is sufficiently connected.

\begin{theorem}\label{thm:faster_rates_1-beta}
    Suppose that there exists a partition of agents $I=I_1\sqcup I_2\sqcup\cdots \sqcup I_q$ such that for all $s\in[q]$, $|I_s|\geq 2$ and $I_s$ is strongly connected in $\Gcal$. Then, $\SWG(\gamma) = \Ocal(1-\gamma)$.
\end{theorem}



The proof of \cref{thm:faster_rates_1-beta} uses the more involved version \cref{lemma:ball_in_region_optimized} of \cref{lemma:ball_in_region_no_assumptions} which allows to group agents according to the suitable partition. At the high-level, within each strongly-connected component $I_i$ for $i\in[q]$, a central planner has enough flexibility to ensure incentive-compatibility for agents in $I_i$ leveraging other agents from $I_i$. In particular, this requires each component $I_i$ to have at least $2$ elements, as stated in \cref{thm:faster_rates_1-beta}.
Note that when $|I|\geq 5$, this strongly connected condition is stronger than simply assuming that for all $i\in I$, there exists $j,j'\in I$ such that $i\to j$ and $j'\to i$, as can be seen in \cref{fig:simple_example_bad_graph}. 

The conditions in \cref{thm:faster_rates_1-beta} hold under many standard distributional assumptions, as shown below.

\begin{corollary}\label{cor:simple_examples_fast_convergence}
    Denote by $m$ the infimum of the support of $\max_{i\in[n]} u_i$. Suppose that for any $i\in I$, there exists $j\in I$ with $j\neq i$ such that either
    \begin{itemize}
        \item there exists $m_i>0$ such that $\Pbb(u_i=m_i),\Pbb(u_j=m_i)>0$, and $\Pbb(u_k\leq m_i)>0$ for all $k\notin\{i,j\}$,
        \item or there exist $b_i>a_i> m$ such that both $u_i$ and $u_j$ have an absolutely-continuous part that has density bounded away from zero on $[a_i,b_i]$ by some constant.
    \end{itemize}
    Then, $\SWG(\gamma) = \Ocal(1-\gamma)$.
\end{corollary}

\begin{figure}[t]
    \centering
    \begin{tikzpicture}
        \graph {
          1/$1$ ->[bend left] 2/$2$ -> 3/$3$ -> 4/$4$ -> [bend left] 5/$5$ ;
          2 -> [bend left] 1;
          5 -> [bend left] 4
        };
    \end{tikzpicture}
    \caption{Example of graph for which each node has an incoming and outcoming edge but does not admit a partition into strongly connected components. For any partition of the nodes $\{1,\ldots,5\}$ into sets of size at least $2$, node $3$ is not strongly connected within its component.}
    \label{fig:simple_example_bad_graph}
\end{figure}


As a comparison, the main result from \citep{bgs19} implies that if the distributions $\Dcal_1,\ldots,\Dcal_n$ have a density on $[0,\bar v]$ that is bounded above and below by constants, one can achieve a $\Ocal(1-\gamma)$ convergence rate. In particular, \cref{cor:simple_examples_fast_convergence} shows that this holds even if one only has densities bounded below.

\subsection{Necessary conditions for faster rates $\Ocal(1-\gamma)$} 

We next also derive necessary conditions for having the faster rates $\Ocal(1-\gamma)$. Intuitively, we show that if $\Gcal$ has an isolated node i.e., there exists $i\in I$ with no inner edges and no outer edges, then $\Ocal(1-\gamma)$ rates are impossible. Formally, we introduce the following function
\begin{equation*}
    g(\eta):= \min_{i\in I} \sum_{j\in I\setminus\{i\} } \paren{f(\eta;i,j) + f(\eta;j,i)},
\end{equation*}
which quantifies the maximum ability of the central planner to fulfill its future promises/penalties for the worst agent $i\in I$ (leveraging any other agent). In particular, if this bound decays as $g(\eta)=_{\eta\to 0}o(\eta)$ then $\Gcal$ has an isolated node.\footnote{Note that converse also holds whenever $f$ is non-pathological, e.g. if for all $i\neq j\in I$, $f(\eta;i,j)=_{\eta\to0}\Theta(\eta^{\alpha_{i,j}})$ for some $\alpha_{i,j}\geq 0$.} We then show that the condition $g(\eta)=_{\eta\to 0}o(\eta)$ is necessary for achieving fast rates for social welfare. This formalizes the intuition that if no agent can be leveraged to flexibly fulfill the future promise of some agent $i$, then achieving the fast rate $\Ocal(1-\gamma)$ is impossible. 

\begin{theorem}\label{thm:no_faster_rate_1-beta}
    If $g(\eta) =_{\eta\to 0} o(\eta)$, then $\SWG(\gamma) =\omega(1-\gamma)$.
\end{theorem}


While this necessary condition does not exactly match the sufficient conditions from \cref{thm:faster_rates_1-beta}, these are written in similar format: enough flexibility to penalize/reward each agent, as quantified by the function $f$, is necessary to achieve faster rates.

\subsection{Convergence rate bounds for general distributions}

We are now ready to give our most general characterizations. While \cref{thm:faster_rates_1-beta} and \cref{thm:no_faster_rate_1-beta} focused on characterizing conditions for reaching the faster rate $\Ocal(1-\gamma)$, the methodology can be used to characterize any rate between $\sqrt{1-\gamma}$ and $1-\gamma$ more generally. 

Recall that the function $f(\eta;i,j)$  measures the central planner’s flexibility to meet future promises/penalties for agent $j$ by leveraging agent $i$. We will show that a larger value of this function leads to a faster convergence rate of the social welfare.
Specifically, when characterizing fast rates $\Ocal(1-\gamma)$, we focused on lower bounds on $f$ of the form $f(\eta;i,j) =\Omega(\eta)$. In comparison, when weaker lower bounds are available, weaker convergence rates are still achievable. 

\paragraph{Warm-up example with $2$ agents.} 
Before diving into the general multi-agent case, we start with a simpler example that illustrates important intuitions. We consider the $2$-agent scenario with $I=\{1,2\}$, and assume for simplicity that each agent gives similar amount of flexibility for the central planner to enforce incentive-compatibility on the other agent in the following manner:
\begin{equation*}
    f(\eta;1,2) , f(\eta;2,1) = \Theta(\eta^p),\quad \eta\in[0,1],
\end{equation*}
for some parameter $p\geq 1$. Here, larger flexibility $f(\eta;1,2),f(\eta;2,1)$ correspond to smaller values of $p$. Note that in this scenario, the quantity $g(\eta)=\Theta(\eta^p)$ also coincides with $f(\eta;1,2)$ and $f(\eta;2,1)$ up to constant factors. We refer to \cref{fig:construction_H} for a concrete example where $p=2$.

\begin{figure}
    \begin{subfigure}[t]{.58\textwidth}
    \centering
\begin{tikzpicture}[scale=2.5] 
  \draw[-latex, thick] (0,0) -- (3/2+.1,0) node[right] {$x$};
  \draw[-latex, thick] (0,0) --  (0,1.1);
  \node[rotate=90,anchor=south] at (-0.2,0.5) {Density / Mass};

\draw[red, thick, fill=red!20, domain=0:1, variable=\x] 
  plot ({\x}, {(1-2*\x)^2}) -- (1,0) -- (0,0) -- (0,1);

  \draw[blue, ultra thick] (1/2,0) -- (1/2,1/6);
  \fill[blue] (1/2,1/6) circle (0.03);

  \draw[blue, ultra thick] (3/2,0) -- (3/2,1/6);
  \fill[blue] (3/2,1/6) circle (0.03);

  \node[blue] at (0.5,.3) { $\Dcal_1$};
  \node[red] at (0.85,1) { $\Dcal_2$};

  \foreach \x/\label in {0,0.5,1,1.5}{
    \draw (\x,0) -- (\x,-0.02);
    \node[below] at (\x,-0.02) { $\label$};
  }
  \foreach \y in {1,2,3}{
    \draw (0,\y/3) -- (-0.015,\y/3);
    \node[left] at (-0.015,\y/3) { \y};
  }

\end{tikzpicture}
    \end{subfigure}
    \begin{subfigure}[t]{.4\textwidth}
    \begin{center}
        \begin{tikzpicture}[scale=1.8]
        \node (1) at (-1,0) {\color{blue} $1$}; 
        \node (2) at (1,0) {\color{red} $2$}; 

        \draw [-latex, bend left,line width=1pt] (1) to (2);
        \draw [-latex, bend left,line width=1pt] (2) to (1);
        \node at (0,.5) {\small $f(\eta;1,2) =\Theta(\eta^2)$};
        \node at (0,-.5) {\small $f(\eta;2,1) =\Theta(\eta^2)$};
        \node at (0,-.8) {$\Hcal$};
    \end{tikzpicture}
    \end{center}
    
    \end{subfigure}

    \caption{Utility distributions with $n=2$ agents (left) and corresponding graph $\Hcal$ (right). (1) On the left, $\Dcal_1$ is uniform on $\{1/2,3/2\}$, and $\Dcal_2$ has density $3(1-2x)^2$ on $[0,1]$. (2) By construction, the central planner incurs cost $\leq \eta$ when allocating to agent $1$ while agent $2$ had higher utility only with probability $\Theta(\eta^2)$, and vice versa: $f(\eta;1,2),f(\eta;2,1)=\Theta(\eta^2)$. In turn, this implies $f(\eta;\{1,2\}), g(\eta) = \Theta(\eta^2)$. Hence, \cref{thm:full_characterization} (or Prop.~\ref{prop:2_agent_case_example}) implies $\SWG(\gamma) = \Theta( (1-\gamma)^{3/4})$.}
    \label{fig:construction_H}
\end{figure}

The following result derives upper upper and lower bounds on convergence rates for this scenario. Specifically, it gives social welfare bounds between $\Theta(1-\gamma)$ and $\Theta(\sqrt{1-\gamma})$ depending on the flexibility of the central planner to enforce incentive-compatibility, as quantified by $f(\eta;1,2),f(\eta;2,1)$ through the parameter $p$. If the planner has larger flexibility to implement rewards/penalties---for smaller values of $p$---the convergence rate is faster.
\begin{proposition}\label{prop:2_agent_case_example}
    Fix $p\geq 1$. Suppose that $I=\{1,2\}$ and $f(\eta;1,2),f(\eta;2,1)=\Theta(\eta^p)$ for all $\eta\in[0,1]$. Then, $\SWG(\gamma)=\Theta((1-\gamma)^{(1+\frac{1}{p})/2})$.
\end{proposition}
This gives tight social welfare gap characterizations in the case when the flexibility functions $f(\eta;1,2),f(\eta;2,1)$ are polynomial, interpolating between the fast rate $O(1-\gamma)$ (when $p=1$) and the slow rate $O(\sqrt{1-\gamma})$ when the flexibility functions are smaller ($p\to\infty$).

\paragraph{General setting.} Going from the $2$-agent case to the general multi-agent setting requires a few notations. In particular, we first introduce a way to aggregate the flexibility functions $f(\eta;i,j)$ into a single quantity governing the social welfare convergence rate.
Formally, let $\Hcal$ be the graph on $I$ with capacities $f(\eta;i,j)$ for any pair $i\neq j\in I$. We denote by $\Hcal_{|S}$ this graph restricted to a subset of agents $S\subseteq I$. Now fix any partition $I=I_1\sqcup I_2\sqcup\cdots \sqcup I_q$ with $|I_s|\geq 2$ for all $s\in[q]$, which we denote $\Pcal$. For any $\eta\geq 0$, we define
\begin{equation*}
    f(\eta;\Pcal) := \min_{s\in[q]}\min_{\substack{i,j\in I_s\\i\neq j}} \set{\text{max-flow from $i$ to $j$ in }\Hcal_{|I_s} }.
\end{equation*}
We refer to \cref{fig:construction_H} for a visualization of the graph $\Hcal$ and how to derive $f(\eta;\Pcal)$ for a concrete example of utility distributions. 
Importantly, $f(\eta;\Pcal)$ measures the central planner’s ability to meet future promises/penalties using the partition $\Pcal$ to group agents, that is, only leveraging agents from the same group to meet their incentive-compatibility constraints. Hence, we will show that the higher this quantity, the faster the convergence rate for the social welfare. In comparison, in the $2$-player case, the only possible partition is to group all agents together $I_1=\{1,2\}$; accordingly we have $f(\eta;\Pcal)=\min(f(\eta;1,2),f(\eta;2,1))$.

\begin{remark}\label{remark:link_to_fast_rates}
    In the definition of $f(\eta;\Pcal)$, the constraint on the max-flow between any two elements of a group $I_s$ is analogous to the constraint that $I_s$ was strongly connected within $\Gcal$ in \cref{thm:faster_rates_1-beta}. Precisely, we can check that if a partition $\Pcal$ satisfies the constraints for \cref{thm:faster_rates_1-beta} then $f(\eta;\Pcal)=_{\eta\to 0}\Omega(\eta)$, since there is a path between any two nodes $i\neq j\in I_s$ within $I_s$ using only edges with capacities $\Omega(\eta)$.
\end{remark}

As in \cref{thm:faster_rates_1-beta}, the choice of the partition $\Pcal$ allows for further flexibility for the central planner. Nevertheless, we always have $f(\eta;\Pcal)\leq g(\eta)$
since the max-flow from or to some node is at most the total capacity of its incident edges. Hence, $g(\eta)$ intuitively upper bounds the ability of the central planner to ensure incentive-compatibility for agent $i$, leading to lower bounds on convergence rates.

We are now ready to derive general upper and lower bounds on convergence rates using $f(\eta;\Pcal)$ and $g(\eta)$. As in the $2$-agent scenario, the following result gives social welfare upper bounds between $\Theta(1-\gamma)$ and $\Theta(\sqrt{1-\gamma})$ depending on the flexibility of the central planner to enforce incentive-compatibility as quantified by $f(\eta;\Pcal)$: the larger the function $f(\eta;\Pcal)$---hence the more flexibility the planner has to implement rewards/penalties---the faster the resulting convergence. These upper bounds are complemented by corresponding lower bounds depending on $g(\eta)$. While upper and lower bounds do not match in general, they share a similar form and can be tight---e.g., see \cref{prop:2_agent_case_example} and \cref{sec:examples}. We refer to \cref{fig:construction_H} for a visualization of an explicit example when these yields tight rates strictly between $\Theta(1-\gamma)$ and $\Theta(\sqrt{1-\gamma})$.

\begin{theorem}\label{thm:full_characterization}
    There exist $C,C_\eta>0$ (see \cref{eq:def_constants} for their definition) such that the following holds. Fix any partition $I=I_1\sqcup I_2\sqcup\cdots \sqcup I_q$ with $|I_s|\geq 2$ for all $s\in[q]$, denoted $\Pcal$. Then, for any $\gamma\in[1/2,1)$,
    \begin{align}\label{eq:upper_bound_general}
        &\SWG(\gamma)\leq C(1-\gamma)  + C\sqrt{1-\gamma}\cdot\inf\, \{1\}\cup \notag\\
        &\set{\eta\in(0,1]:\forall \eta'\in[\eta,1], f(\eta';\Pcal) \geq C_\eta\sqrt{1-\gamma} \frac{\eta'}{\eta} }.
    \end{align}
    On the other hand, there exist $c_\eta,\eta_0,c>0$ depending only on $\Dcal_1,\ldots,\Dcal_n$ such that for any $\gamma\in[9/10,1)$,
    \begin{align}\label{eq:lower_bound_general}
        &\SWG(\gamma) \geq c\sqrt{1-\gamma} \cdot\notag\\ 
        &\sup\, \{0\}\cup \{\eta\in[0,\eta_0]: g(\eta) \leq c_\eta\sqrt{1-\gamma} \} .
    \end{align}
\end{theorem}


This general social welfare characterization recovers all previous results: the universal convergence rate $\Ocal(\sqrt{1-\gamma})$ from \cref{thm:universal_lower_bound_simplified}, the sufficient conditions from \cref{thm:faster_rates_1-beta} under which rates $\Ocal(1-\gamma)$ can be achieved,\footnote{From \cref{remark:link_to_fast_rates}, whenever $\Pcal$ satisfies the constraints from \cref{thm:faster_rates_1-beta} we have $f(\eta;\Pcal)=\Omega(\eta)$.} the corresponding necessary conditions from \cref{eq:lower_bound_general},\footnote{\cref{thm:full_characterization} shows that $g(\eta)=_{\eta\to0}\Omega(\eta)$ is necessary for achieving rates $\Ocal(1-\gamma)$.} and the $2$-agent scenario \cref{prop:2_agent_case_example}.


We now briefly discuss the main difference between the upper and lower bound from \cref{thm:full_characterization}. The lower bound \cref{eq:lower_bound_general} corresponds to a 2-agent subproblem where one agent $i\in I$ plays against the best utility among all other agents $\max_{j\neq i}u_j$; while the upper
bound \cref{eq:upper_bound_general} uses the full partition $\Pcal$, hence considers more complex interactions between groups of agents. Therefore, the bounds intuively coincide whenever there is no asymmetric advantage in clustering---see the $2$-agent scenario from \cref{prop:2_agent_case_example}---or under specific partition-symmetric structure of agents.

\subsection{Mechanism trajectory and efficiency}

\begin{figure}[t]
    \centering
    
    \begin{subfigure}{.8\textwidth}
        \centering
        \includegraphics[width=0.54\textwidth]{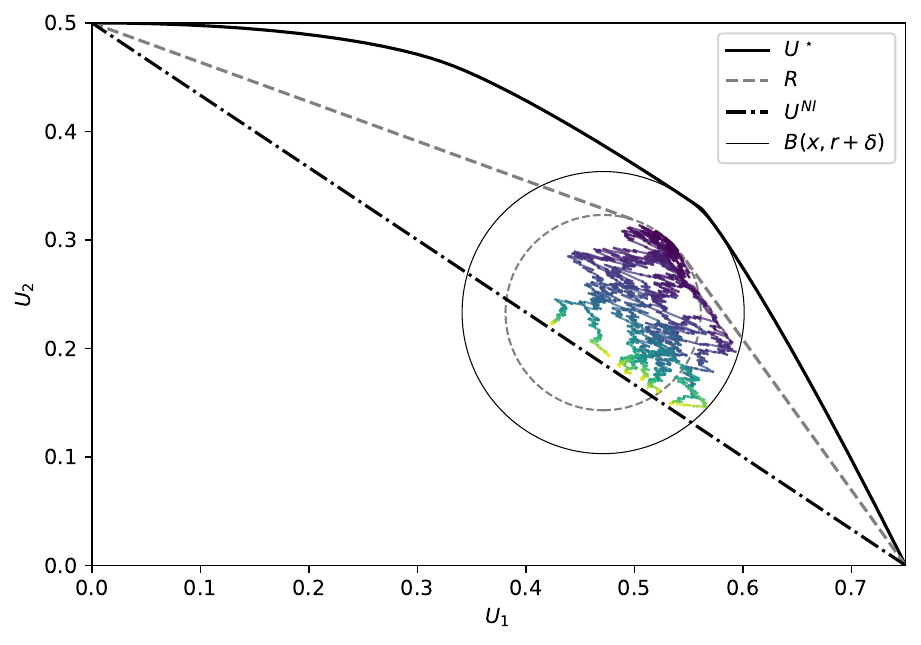}
        \includegraphics[width=0.45\textwidth]{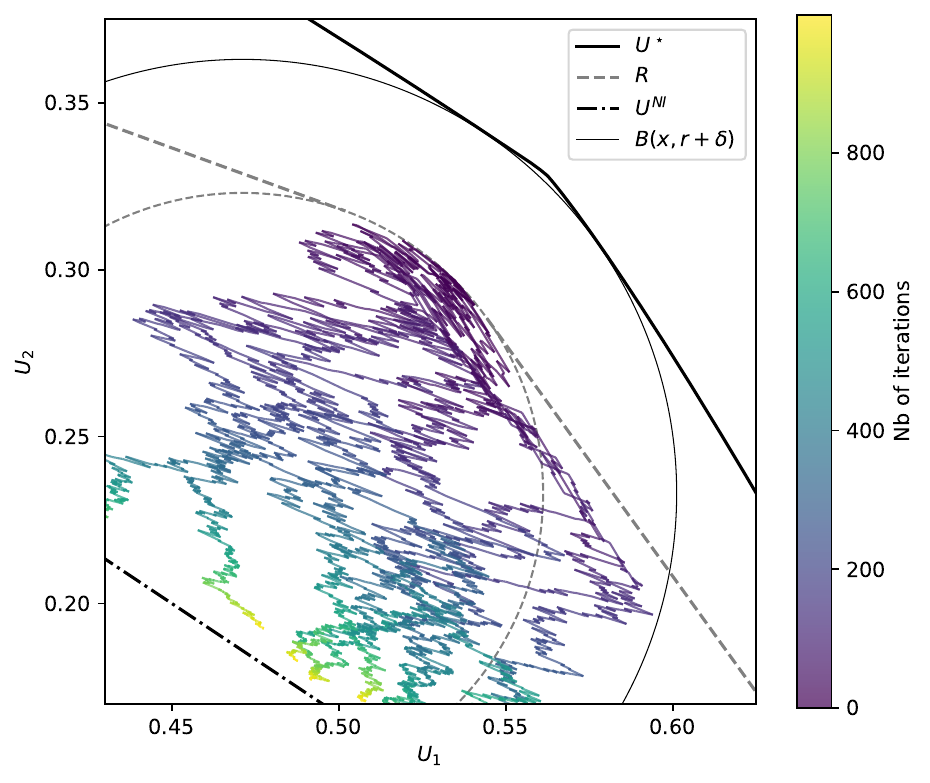}
        \caption{\small $\delta=0.04$ and $1000$ iterations}
        \label{fig:large_delta}
    \end{subfigure}

    \vspace{1em}

    \begin{subfigure}{.8\textwidth}
        \centering
        \includegraphics[width=0.54\textwidth]{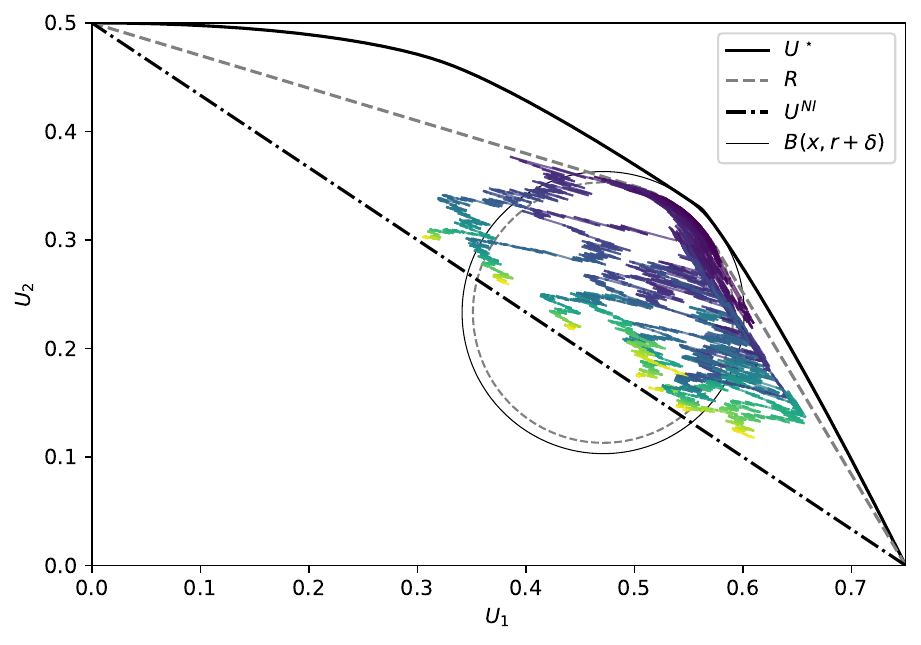}
        \includegraphics[width=0.45\textwidth]{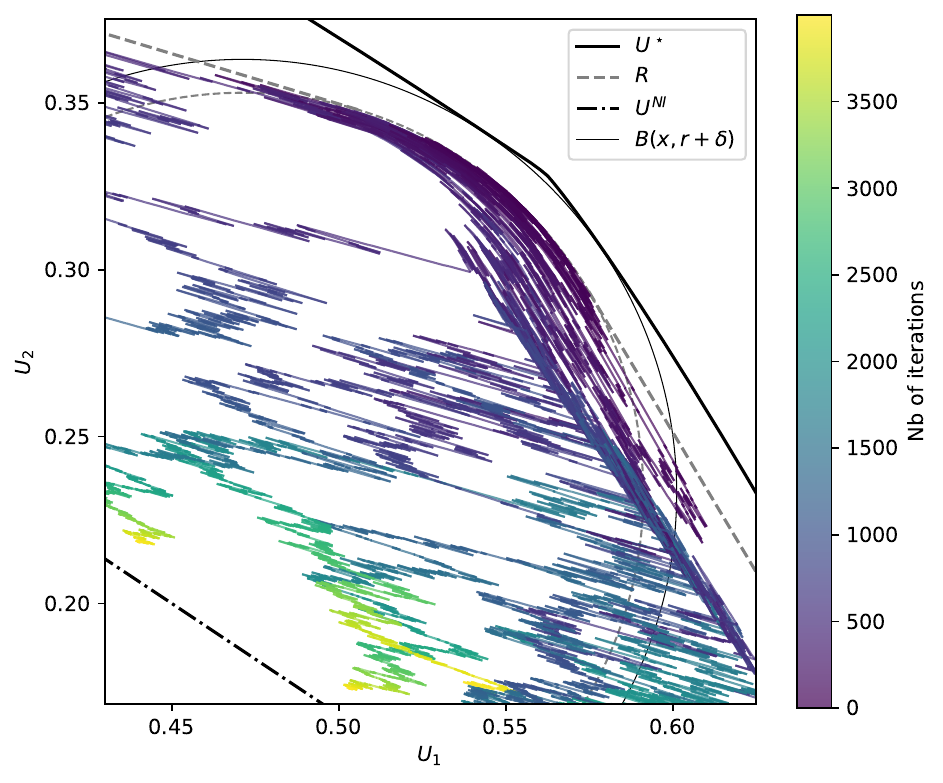}
        \caption{\small $\delta=0.01$ and $4000$ iterations}
        \label{fig:small_delta}
    \end{subfigure}

    \caption{Sample trajectories of promised utilities for $n=2$ agents and $\gamma=0.99$. The full-information region $\Ucal^\star$ (resp. no-information region $\Ucal^{NI}$ is drawn in bold solid (resp. dashed) black line. The ball $B(\mb x,r+\delta)\subseteq\Ucal^\star$ is drawn in regular black line.  The region $\Rcal$ (defined in \cref{eq:def_enlarged_region}) on which the mechanism is implemented is drawn in dashed gray line. \cref{fig:large_delta,fig:small_delta} each show 10 sample trajectories for a conservative margin $\delta=0.04$ and a tighter margin $\delta=0.01$ respectively. Plots on the right show a zoomed view of the same sample trajectories as their left plot. Crucially, with a tighter margin, trajectories in \cref{fig:large_delta,fig:small_delta} linger along the boundary, where social welfare is the largest.}
    \label{fig:sample_trajectories_all}
\end{figure}

To illustrate our proposed mechanism for reaching the convergence rates from \cref{thm:full_characterization}, we consider the same $n=2$ agent setting as in \cref{fig:construction_H}, in which agent $1$ has a discrete utility distribution and agent $2$ has a uniformly-continuous utility distribution.
Given a ball $B(\mb x,r)$ and an adequate margin $\delta$ such that $B(\mb x,r+\delta)$ remains within the full-information region $\Ucal^\star$, the mechanism implements the strategy described in \cref{sec:generic_mechanism} to fulfill promises within this ball $B(\mb x,r)$.

In \cref{fig:sample_trajectories_all} we show sample trajectories of promised utilities under this mechanism. These trajectories follow the same general behavior: starting from the initial promised utility on the boundary of $B(\mb x,r)$, they slowly drift away from this boundary and asymptotically converge towards the boundary the no-information region $\Ucal^{NI}$ where promised utilities are straightforward to fulfill. Specifically, \cref{fig:sample_trajectories_all} shows trajectories for the same choice of ball $B(\mb x,r+\delta)\subseteq \Ucal^\star$ but different margins $\delta$. Notably, if the mechanism converges for smaller values of this margin, its social welfare is increased since it may reach utility vectors closer to the boundary of $\Ucal^\star$.   
The margin $\delta$ for which our proofs ensure convergence are given by the constraints within \cref{lemma:ball_in_region_no_assumptions}. While the order of magnitude from these constraints cannot be improved in general given our proposed lower bounds, the specific constant $C$ therein may be overly conservative. In \cref{fig:large_delta} and \cref{fig:small_delta} we show the convergence of the mechanism for a conservative margin and a tighter margin respectively. For conservative margins $\delta$, the promised utilities easily drift towards $\Ucal^{NI}$. In comparison, for tighter margins as in \cref{fig:small_delta}, sample trajectories lingers close to the boundary of the region $B(\mb x,r)$ (first 500 iterations), where the algorithm achieves the highest social welfare efficiency.\footnote{This empirical behavior confirms the intuition from \cref{lemma:ball_in_region_no_assumptions}: the curvature of regions on which promised utility trajectories can be confined depends on the available margin $\delta$. For smaller margins, smoother region with larger curvature radius $r$ are required.}

As a remark, promised utilities are not guaranteed to remain within the ball $B(\mb x,r)$. Indeed, we recall that the mechanism \text{PUM} from \cref{sec:generic_mechanism} instead considers an enlarged region $\Rcal$ (see \cref{eq:def_enlarged_region}). 

\section{From the discounted infinite-horizon setting to the finite-horizon setting}
\label{sec:links_to_finite_horizon}

Up to minor changes, the mechanisms as well as the tools designed to characterize the social welfare gap in the discounted setting also carry to the finite-horizon $T$ setting. At the conceptual level, our main finding is that the difference between the infinite-horizon discounted and finite-horizon settings is essentially \emph{only} that in the finite-horizon case, the unavoidable last round $T$ allocation alone forces $\Omega(1/T)$ social welfare gap (see \cref{lemma:no_faster_1/T}). This is because on the last round, the central planner can no longer enforce future rewards/penalties: this corresponds to a one-shot resource allocation problem without money for which incentive-compatibility is impossible in general (see Myerson-Satterthwaite's impossibility theorem). In comparison, in the infinite-horizon setting, much faster rates including exponential ones are possible (see \cref{thm:discrete_case_full,thm:extra_fast_rates}): these are precluded by the last round allocation in the finite-horizon case.

A simple observation is that the utility regions $\Vcal_T$ are still linked by a dynamic programming formulation. In turn, the promised utility framework still applies up to modifying the discount factor to be horizon-dependent $\gamma(T)=1-1/T$. Naturally, for longer horizons, agents are more patient: $\gamma(T)$ is increasing in the horizon $T$. This is formalized by adapting Facts \ref{fact:promised_utility_1} and \ref{fact:promised_utility_2} to the finite-horizon case, which we defer to the Appendix (see \ref{fact:promised_utility_finite_horizon}). 


\subsection{Lower bounds on the social welfare gap}

We can now compare the regions $\Vcal_T$ for $T\geq 1$ to the region $\Ucal_{\gamma(T)}$. 

\begin{lemma}\label{lemma:upper_bound_finite_horizon}
    For any $T\geq 1$, we have
    $\Vcal_T \subseteq \Ucal_{\gamma(T)}$.
\end{lemma}

As a direct consequence of \cref{lemma:upper_bound_finite_horizon}, all techniques developed earlier to give upper bounds on $\Ucal_{\gamma(T)}$ also apply to $\Vcal_T$. In particular, \cref{lemma:prove_upper_bounds} (and \cref{lemma:upper_bounds_2D}), \cref{thm:no_faster_rate_1-beta}, and the second claim \cref{eq:lower_bound_general} from \cref{thm:full_characterization} all hold by changing $\Ucal_\gamma$ to $\Vcal_T$ and $\gamma$ to $\gamma(T)=1-1/T$.
In addition, we can easily prove that in the finite-horizon $T$ setting, one cannot reach faster rates than $\Ocal(1-\gamma(T))= \Ocal(1/T)$. Intuitively, this is because on the last allocation iteration, the reports of the agents carry no information, hence the central planner can only reach a utility vector within $\Ucal_0$ for this iteration, which incurs a non-zero (constant) regret.

\begin{lemma}\label{lemma:no_faster_1/T}
    Suppose that $\Ucal_0$ does not already contain an optimal vector (see \cref{lemma:trivial_case} for a characterization). Then, there exists $C>0$ such that for all $T\geq 1$, we have $\SWG(T) \geq \frac{C}{T}$.
\end{lemma}

\subsection{Mechanisms for the finite-horizon setting}

We now show that the positive results from the $\gamma$-discounted setting can be extended to the finite-horizon case, yielding the same social welfare gap guarantees up to logarithmic factors. The main argument is the following, which shows that utility vectors within a ball with almost the same margin as in \cref{lemma:ball_in_region_no_assumptions} within $\Ucal^\star$ are still achievable with an adjusted variant of the mechanism introduced in \cref{sec:generic_mechanism}.

\begin{lemma}\label{lemma:ball_in_region_finite_horizon}
    There exist $c_0\geq 1$ and $r_0>0$ (depending only on the utility distributions) such that the following holds. Fix $T\geq 2$, $\mb x\in\Ucal^\star$, and $r,\delta>0$ such that $B(\mb x, r+\delta)\subseteq \Ucal^\star$. If
    \begin{equation}\label{eq:finite_horizon_safe}
        r_0\geq r\geq \delta \geq c_0 C\frac{1-\gamma(T)}{\gamma(T)r}\paren{1+\ln\frac{r}{\delta}},
    \end{equation}
    where $C$ is as defined in \cref{lemma:ball_in_region_no_assumptions}, then a variant of $\text{PUM}(\mb x,r,\delta)$ can achieve any utility vector in $B(\mb x, r)$.

    In particular, we can simplify the statement as follows. Suppose that the assumptions of \cref{lemma:ball_in_region_no_assumptions} are satisfied with $\gamma=\tilde \gamma(T)=1-\ln T/T$ and the constant $C'=c_0C$, as well as $r_0\geq r\geq \delta$. Then, $B(\mb x, r)\subseteq \Vcal_T$.
\end{lemma}
As a result, all guarantees on the infinite-horizon mechanism---\cref{thm:universal_lower_bound_simplified,thm:faster_rates_1-beta}, \cref{cor:simple_examples_fast_convergence}, and the first claim \cref{eq:upper_bound_general} from \cref{thm:full_characterization}---hold for the finite-horizon case by changing $\Ucal_\gamma$ to $\Vcal_T$ and $\gamma$ to $1-\ln T/T$.

For the sake of conciseness, we give the main structure of the variant mechanism here and defer the complete description to the Appendix (see \cref{alg:mechanism_finite_horizon}). 
It constructs a ball $B_t:=B(\mb x^{(t)},r_t)\subseteq\Vcal_t$ with an appropriate margin $\delta_t$ within $\Ucal^\star$ and ensures that when $t$ iterations remain, it only needs to fulfill a utility vector within this ball $B_t$. When only $1$ iteration remains, the ball must start within the region $B_1\subseteq \Ucal^{NI}$; however, as the horizon $t\in[T]$ grows, the ball slowly converges to the target $B_T\approx B(\mb x, r)$. The specific trajectory depends on the parameters $r$ and $\delta$ and is illustrated in \cref{fig:finite_horizon}. At the high-level, it proceeds in two phases: (1) first decrease the radius of the ball and the margin together, $\delta_t\approx r_t=\Theta(1/\sqrt t)$, until the target radius $r$ is reached, (2) then decrease the margin via $\delta_t=\Theta(1/(rt))$ while keeping the radius constant. This trajectory of the achievable balls within $\Vcal_t$ for $t\in[T]$ illustrates the evolution of the strategy of the central planner throughout the sequential allocation problem.

\begin{figure}[t]
\centering
  \begin{tikzpicture}[scale=.9]
\begin{axis}[width=7.5cm,height=7.5cm,xmin=0, xmax=0.5,
    ymin=0, ymax=0.5, xlabel={$U_1$},         ylabel={$U_2$},          tick label style={font=\tiny}]

\tikzmath{\a=1/3;\p=1/20; }
\tikzmath{\b=1-2/3; \c=1+2/3;}
 \addplot [domain=1-2*\a:1+2*\a, smooth, samples=501,line width=1pt]
    ({\p*(x/4+1/4+\a/2)*(1/2+\a-x/2)/(2*\a) + (1/2+\a)*(1-\p)/2}, {\p*(x/2-1/2+\a)/(4*\a)+(1-\p)/4}); 

\draw[line width=1pt] (1/2,0) -- ({\p*(\b/4+1/4+\a/2)*(1/2+\a-\b/2)/(2*\a) + (1/2+\a)*(1-\p)/2}, {\p*(\b/2-1/2+\a)/(4*\a)+(1-\p)/4});

\draw[line width=1pt] (0,1/2) -- ({\p*(\c/4+1/4+\a/2)*(1/2+\a-\c/2)/(2*\a) + (1/2+\a)*(1-\p)/2}, {\p*(\c/2-1/2+\a)/(4*\a)+(1-\p)/4});
    

\draw[line width =1pt] (0,1/2)--(1/2,0);
\draw (0.09,0.33) node{$\mathcal U^{NI}$};

\draw[dotted] (0.391,0.228) circle (0.03);
\draw[fill=gray!30] (0.391,0.228) circle (0.025);

\draw[dashed] (0,0.228+0.03) -- (0.391,0.228+0.03);
\draw[dashed] (0.391+0.03,0) -- (0.391+0.03,0.228);


\draw[red, dotted] (0.391+0.03 - 0.05,0.228+0.03-0.05) circle (0.05);
\draw[red] (0.391+0.03 - 0.05,0.228+0.03-0.05) circle (0.025);


\draw[blue,dotted] (0.285,0.125) circle (0.125);
\draw[blue] (0.285,0.125) circle (0.125/2);

\draw[blue,<->] (0.285-0.125/2,0.125) -- node[below]{$r_0$} (0.285,0.125);
\draw[blue,<->] (0.285-0.125/2,0.125) -- node[below]{$r_0$} (0.285-0.125,0.125);

\draw[red,<-> ] (0.391+0.03 - 0.05,0.228+0.03-0.05) -- node[below]{\small $r$} (0.391+0.03 - 0.05-0.025,0.228+0.03-0.05) ;
\draw[red,<-> ] (0.391+0.03 - 2*0.05,0.228+0.03-0.05) -- node[below]{\small $r$} (0.391+0.03 - 0.05-0.025,0.228+0.03-0.05) ;

\draw[<->] (0.391,0.228) -- (0.391+0.025,0.228) node[right]{$r$};
\draw[<->] (0.391,0.228) -- (0.391+0.031/1.414,0.228+0.031/1.414) node[above]{\small \qquad $r+\delta$};



\draw (0.35,0.35) node{$\mathcal U^\star$};

\end{axis}
\end{tikzpicture}
    \caption{Illustration of the mechanism to approach the gray radius-$r$ ball in the finite-horizon $T$ setting (see \cref{lemma:ball_in_region_finite_horizon}). For all $t\in[T]$ we construct a ball within $\Vcal_t$. The mechanism has two phases. For $t=1$, we can take the blue radius-$r_0$ ball within $\Ucal^{NI}\subseteq\Vcal_1$. We first start by reducing the radius to $r$ until we reach the red ball. As $t$ grows, the radius reduces as $r_t=\sqrt{C\frac{1-\gamma(t)}{\gamma(t)}}=\sqrt{\frac{C}{t-1}}$, which requires conserving a margin $\delta_t=r_t$ depicted in the blue and red dotted circles. The second phase is to decrease the margin from $\delta_t=r$ (as shown in the red dotted circle) towards the desired margin $\delta$, all while staying within the dashed region which belongs to $\Ucal^\star$. In this phase, the margin decreases as $\delta_t=C\frac{1-\gamma(t)}{\gamma(t)r}$.}
    \label{fig:finite_horizon}
\end{figure}

In comparison to \cref{lemma:upper_bound_finite_horizon}, in general \cref{lemma:ball_in_region_finite_horizon} essentially requires a sub-optimal choice $\gamma=1-\ln T/T$; ideally, this should hold for $\gamma=\gamma(T)$ to match the social welfare gap lower bounds from \cref{lemma:upper_bound_finite_horizon}. We note that in some cases, \cref{lemma:ball_in_region_finite_horizon} only requires $\gamma=1-\Ocal(1/T)$, yielding tight results compared to the $\gamma$-discounted setting (the extra factor $\ln\frac{r}{\delta}$ vanishes when $\delta=\Theta(r)$, see \cref{cor:discrete_case_full_finite_horizon} for a concrete example). In other cases, however, this $\ln T$ term may not vanish, typically when utility distributions are somewhat smooth. While we believe that this logarithmic gap is essentially necessary, we leave this question open. The core of the problem is captured in the following simple scenario.
\begin{openpb}\label{open_question}
    Consider the $n=2$ agent finite-horizon $T$ resource allocation setting where the utility distributions are uniform $\Ucal([0,1])$ for both agents. Characterize the behavior (in $T$) of $\SWG(T)$.
\end{openpb}
Our results show that the social welfare gap is between $\Omega(1/T)$ and $\Ocal(\ln T/T)$. We conjecture that the answer is $\ln T/T$, which would imply that there is a separation between the infinite-horizon and finite-horizon settings beyond the fact that faster rates than $1/T$ are not achievable in the finite-horizon setting.


\section{Examples and special cases}
\label{sec:examples}

In this section, we apply the general methods described above to simple utility distributions of interest. 

\subsection{Utility distributions with density}

We start with smooth utility distributions. \cite{bgs19} shows that if the utility distributions have density upper and lower bounded by a constant on the domain $[0,\bar v]$ then the convergence rate is $\Theta(1-\gamma)$. Using the previous results, we can give further details about this convergence: in \cref{cor:simple_examples_fast_convergence} we showed that having densities bounded below is sufficient to have a convergence rate $\Ocal(1-\gamma)$, and \cref{thm:full_characterization} implies that having densities bounded above is sufficient to prove that the rate of convergence is $\Omega(1-\gamma)$.

\begin{corollary}\label{cor:super_smooth_case}
    Suppose that the utility distributions $\Dcal_1,\ldots,\Dcal_n$ admit a density upper bounded by some constant on $[0,\bar v]$ and $\Ucal_0$ does not already include an optimal vector. Then, $\SWG(\gamma)=\Omega(1-\gamma)$.
    
    If they admit a density lower bounded by some constant on $[0,\bar v]$, then $\SWG(\gamma) = \Ocal(1-\gamma)$.
\end{corollary}

For the finite-horizon setting, we have the following.

\begin{corollary}\label{cor:super_smooth_case_finite_horizon}
    Suppose that the utility distributions $\Dcal_1,\ldots,\Dcal_n$ admit a density upper bounded by some constant on $[0,\bar v]$ and $\Ucal_0$ does not already include an optimal vector. Then, for any $T\geq 1$, we have $\frac{c}{T} \leq \SWG(T) \leq \frac{C\ln(T+1)}{T}$ for some constants $c,C>0$.
\end{corollary}

\subsection{Discrete utility distributions}

We next turn to the other extreme case of discrete utility distributions. In this case, \cref{thm:full_characterization} implies that the social welfare gap is typically $\Omega(\sqrt{1-\gamma})$, which can be reached by \cref{thm:universal_lower_bound_simplified}. In some cases, however, when significant ties between agents arise, the convergence can be much faster (exponential) as detailed below. Geometrically, this corresponds to the case when the boundary of $\Ucal^\star$ is perfectly flat at social-welfare-maximizing utility vectors.

\begin{theorem}\label{thm:discrete_case_full}
    Suppose that all utility distributions $\Dcal_1,\ldots,\Dcal_n$ are discrete. Suppose that $\Ucal_0$ does not already contain an optimal vector.
    If there exists $i\in I$ such that for all $j\neq i$, $\Pbb(u_i=u_j=\max_{l\in[n]}u_l>0)=0$, then $\SWG(\gamma)= \Theta(\sqrt{1-\gamma})$.
    Otherwise, there is a constant $c>0$ such that $\SWG(\gamma) = \Ocal\paren{(1-\gamma) e^{- c/\sqrt{1-\gamma}}}$.
\end{theorem}

The second case of \cref{thm:discrete_case_full} corresponds to the case when with non-zero probability there are (significant) ties between different agents. Ties are especially advantageous for the central planner because it can therefore decide which agent to allocate to without incurring any social welfare cost. As a result, exponential rates may also arise beyond purely discrete distributions, as detailed below.

\begin{theorem}\label{thm:extra_fast_rates}
    If for all $i\in I$ there exists $j\neq i$ with $\Pbb(u_i=u_j=\max_{l\in[n]}u_l>0)>0$, then there exists a constant $c>0$ that only depends only on the utility distributions such that $\SWG(\gamma) = \Ocal\paren{(1-\gamma) e^{- c/\sqrt{1-\gamma}}}$.
\end{theorem}

Achieving these faster rates requires an additional tool compared to \cref{lemma:ball_in_region_no_assumptions}. Indeed, applying directly this result would only prove a convergence rate $\Ocal(1-\gamma)$: this is the fastest rate that can be proved with \cref{lemma:ball_in_region_no_assumptions} in its direct form. In practice, however, these can be combined with \emph{gluing} arguments to yield faster rates when the frontier of $\Ucal^\star$ is particularly flat. Roughly speaking, as per \cref{lemma:ball_in_region_no_assumptions}, achieving exponentially small social welfare gap requires using balls with larger radius, which is prohibited by the diameter of $\Ucal^\star$. However, this issue may be solved by truncating this large-radius ball and gluing it to smaller-radius balls whose achievability can directly be guaranteed by \cref{lemma:ball_in_region_no_assumptions}. We give an illustration of this gluing strategy in \cref{fig:gluing}. Repeatedly applying this process in turns yields exponentially small social welfare gaps.


\begin{figure}[t]
    \centering

    \begin{subfigure}[t]{0.32\textwidth}
        \centering
        \begin{tikzpicture}[scale=0.65]
\begin{axis}[width=8cm,height=8cm,xmin=0, xmax=0.55, ymin=0, ymax=0.55, xlabel={$U_1$},         ylabel={$U_2$},   tick label style={font=\tiny}]

\def\centerarc[#1](#2)(#3:#4:#5){ \draw[#1] ($(#2)+({#5*cos(#3)},{#5*sin(#3)})$) arc (#3:#4:#5); }

\draw[line width=1 pt] (0,0.3) -- (1/4,1/2) -- (1/2,1/4) -- (0.3,0);

\draw[dotted] (3/8 - 0.197/1.414 , 3/8 - 0.197/1.414) circle (0.197);
\draw[pattern = north west lines, pattern color=gray!30] (3/8 - 0.197/1.414 , 3/8 - 0.197/1.414) circle (0.197-0.013);

\draw[<->](3/8 - 0.197/1.414 - 0.184/1.414, 3/8 - 0.197/1.414 - 0.184/1.414) --  (3/8 - 0.197/1.414 - 0.197/1.414, 3/8 - 0.197/1.414 - 0.197/1.414) node[left] {\small $\delta_0$} ;

\draw (0.45,0.45) node{$\mathcal U^\star$};

\end{axis}
\end{tikzpicture}
        \caption{}
        \label{fig:sub1}
    \end{subfigure}
    \hfill
    \begin{subfigure}[t]{0.32\textwidth}
        \centering
        \begin{tikzpicture}[scale=0.65]
\begin{axis}[width=8cm,height=8cm,xmin=0, xmax=0.55, ymin=0, ymax=0.55, xlabel={$U_1$},         ylabel={$U_2$},   tick label style={font=\tiny}]
\def\centerarc[#1](#2)(#3:#4:#5){ \draw[#1] ($(#2)+({#5*cos(#3)},{#5*sin(#3)})$) arc (#3:#4:#5); }

\draw[line width=1 pt] (0,0.3) -- (1/4,1/2) -- (1/2,1/4) -- (0.3,0);

\draw (3/8 - 0.197/1.414 , 3/8 - 0.197/1.414) circle (0.197-0.013);

\draw[dotted] (3/8-0.062 - 0.197/2.828 , 3/8+0.062 - 0.197/2.828) circle (0.197/2);
\draw[pattern = north west lines, pattern color=blue!30] (3/8-0.062 - 0.197/2.828 , 3/8+0.062 - 0.197/2.828)  circle (0.197/2-0.026);
\draw[blue] (3/8-0.062 - 0.197/2.828 , 3/8+0.062 - 0.197/2.828)  circle (0.197/2-0.026);

\draw[dotted] (3/8+0.062 - 0.197/2.828 , 3/8-0.062 - 0.197/2.828) circle (0.197/2);
\draw[pattern = north west lines, pattern color=blue!30] (3/8+0.062 - 0.197/2.828 , 3/8-0.062 - 0.197/2.828)  circle (0.197/2-0.026);
\draw[blue] (3/8+0.062 - 0.197/2.828 , 3/8-0.062 - 0.197/2.828)  circle (0.197/2-0.026);


\draw[<->, blue](3/8-0.062 - 0.197/2.828 +0.0725/1.414 , 3/8+0.062 - 0.197/2.828 +0.0725/1.414 ) --  node[above right] {\small $\delta_1$} (3/8-0.062 - 0.197/2.828 +0.197/2.8284 , 3/8+0.062 - 0.197/2.828 +0.197/2.8284)  ;

\draw (0.45,0.45) node{$\mathcal U^\star$};

\end{axis}
\end{tikzpicture}
        \caption{}
        \label{fig:sub2}
    \end{subfigure}
    \hfill
    \begin{subfigure}[t]{0.32\textwidth}
        \centering
      \begin{tikzpicture}[scale=.65]
\begin{axis}[width=8cm,height=8cm,xmin=0, xmax=0.55, ymin=0, ymax=0.55, xlabel={$U_1$},         ylabel={$U_2$},   tick label style={font=\tiny}]
\def\centerarc[#1](#2)(#3:#4:#5){ \draw[#1] ($(#2)+({#5*cos(#3)},{#5*sin(#3)})$) arc (#3:#4:#5); }

\draw[line width=1 pt] (0,0.3) -- (1/4,1/2) -- (1/2,1/4) -- (0.3,0);

\draw (3/8 - 0.197/1.414 , 3/8 - 0.197/1.414) circle (0.197-0.013);

\draw[blue] (3/8-0.062 - 0.197/2.828 , 3/8+0.062 - 0.197/2.828)  circle (0.197/2-0.026);

\draw[blue] (3/8+0.062 - 0.197/2.828 , 3/8-0.062 - 0.197/2.828)  circle (0.197/2-0.026);

\centerarc[dotted](3/8 - 1.6*0.197/1.414 , 3/8 - 1.6*0.197/1.414)(10:80:1.6*0.197);
\centerarc[red](3/8 - 1.6*0.197/1.414 , 3/8 - 1.6*0.197/1.414)(8:82:0.307); 


\draw[draw=none,pattern = north east lines, pattern color=red!70](3/8 - 0.197/1.414 + 0.185/1.414, 3/8 - 0.197/1.414 + 0.185/1.414) arc (45:90+15.5:0.184)  arc (90+51:63:0.197/2-0.026) arc (66:23:0.307) arc (23:-50:0.197/2-0.026) arc (-15:45:0.184);

\draw[<->, red] (3/8 - 1.6*0.197/1.414 +0.305/1.414 , 3/8 - 1.6*0.197/1.414 +0.305/1.414) node[above right] {\small $\delta_2$}    -- (3/8 - 1.6*0.197/1.414 +1.6*0.198/1.414 , 3/8 - 1.6*0.197/1.414 +1.6*0.198/1.414)   ;


\draw (0.45,0.45) node{$\mathcal U^\star$};

\end{axis}
\end{tikzpicture}
        \caption{}
        \label{fig:sub3}
    \end{subfigure}

    \caption{Illustration of how the \emph{gluing} strategy achieves faster convergence rates when the frontier of the full-information region $\Ucal^\star$ is close to flat. The figure depicts a region $\Ucal^\star$ with an exaggerated piece-wise affine boundary for illustration purposes. In \cref{fig:sub1}, the dashed gray region corresponds to the region that can be shown to be achievable in $\Ucal_\gamma$ using \cref{lemma:ball_in_region_no_assumptions} directly, which is limited by the maximum-radius ball tangent to the region of optimal utility vectors that can be fit within $\Ucal^\star$ (represented with a dotted circle). 
    The necessary margin $\delta_0$ is inversely proportional to the radius of the corresponding black ball (\cref{eq:constraint_safe_boundary}). The gluing strategy artificially increases this radius as follows: In \cref{fig:sub2} we consider the dashed blue balls which can be proved to be within $\Ucal_\gamma$ directly using \cref{lemma:ball_in_region_no_assumptions}. Using the blue regions as anchor, we then glue the red arc in \cref{fig:sub3}, resulting in a smaller margin $\delta_2$. This results in the extra red-dashed region which we prove is within $\Ucal_\gamma$. The gluing strategy can be used with an arbitrary number of layers to obtain exponentially small social welfare gap.
    }
    \label{fig:gluing}
\end{figure}


By \cref{lemma:no_faster_1/T}, such exponential rates cannot be extended to the finite-horizon setting. In this case, we obtain the following social welfare gaps.

\begin{corollary}\label{cor:discrete_case_full_finite_horizon}
    Suppose that all utility distributions $\Dcal_1,\ldots,\Dcal_n$ are discrete and that $\Ucal_0$ does not already contain an optimal vector. 
    If there exists $i\in I$ such that for all $j\neq i$, $\Pbb(u_i=u_j=\max_{l\in[n]}u_l>0)=0$, then $\SWG(T)=\Theta(\frac{1}{\sqrt T})$. Otherwise,
    $\frac{c}{T} \leq \SWG(T) \leq \frac{C\ln(T+1)}{T}$ for $T\geq 1$ for constants $c,C>0$.
\end{corollary}

\section{Conclusion and future work}
\label{sec:conclusion}

We consider a sequential resource allocation problem without monetary transfers, for any number of agents and general utility distributions and aim to understand the social welfare gap between monetary and non-monetary mechanisms. Our results uncover a rather surprising richness of the social welfare landscape, depending on geometrical properties of the utility distributions of agents---essentially the smoothness of the Pareto-frontier of the full-information achievable utility set. At the high-level, our results show that scenarios for which approximate ties between utilities of agents are frequent are most favorable for non-monetary mechanisms; this intuition is further quantified by corresponding social welfare gap characterizations. On the algorithmic side, we propose unified resource allocation mechanisms (PUM) based on the promised utility framework. Conveniently, while the social welfare gap depends heavily on utility distributions, it suffices to use this generic mechanism for appropriately distribution-dependent parameters to achieve near-optimal social welfare gap guarantees among non-monetary mechanisms. Additionally, we unify the analysis of the infinite-horizon $\gamma$-discounted and finite $T$-horizon settings, which formalizes the links between both settings using analogous resource allocation mechanisms. 

These general characterizations uncover new quantitative insights on the social welfare gap for non-monetary mechanisms. First, our mechanisms can always achieve a social welfare gap $\Ocal(\sqrt{1-\gamma})$ and $\Ocal(1/\sqrt T)$ for the infinite-horizon and finite-horizon settings respectively. At the high-level this shows some quantitative efficiency of non-monetary mechanisms irrespective of the number of agents or their utility distributions; in comparison, the folk theorem ensures asymptotic convergence without specific rates. While this bound is typically tight for discrete distributions, considerably smaller social welfare gaps are possible depending on the geometry of the utility distributions: as small as $\tilde \Ocal(1/T)$ for finite $T$-horizon, and even exponentially small in the infinite-horizon setting. Our quantitative bounds show that the main driver for the convergence rate is whether the central planner has enough flexibility to penalize or reward agents in future rounds, where this flexibility corresponds to scenarios when agents have similar utility for the resource.

While we focused on the standard resource allocation setting, other models without monetary transfers could also be studied. In particular, one could consider the setting in which the central planner also has a utility for the resource but still aims to maximize social welfare. This is equivalent to the mixed strategic/non-strategic setting in which a subset of the agents is strategic and at least one of the agents is not: their utilities are directly revealed to the central planner. As it turns out, our results also directly apply to this allocation problem. All lower bounds on the achievable utility region can still be achieved in this setting since the central planner has more information and hence can implement the same mechanisms. On the other hand, as emphasized in the proofs, all of our upper bounds on the achievable utility regions are derived in the setting for which a single agent $i\in[n]$ is strategic and all other agents $[n]\setminus\{i\}$ are not: the only incentive-compatibility constraint used is that for agent $i$. Because this is a simpler problem for the central planner, our upper bounds on the achievable utility region (showing that some regions cannot be achieved) still hold. In summary, all our characterizations are still valid for the mixed strategic/non-strategic allocation model.

Among other potential extensions, one could consider multiple resources, constructing mechanisms that do not rely on lotteries (in our constructions, the allocation at each iteration is a lottery on the simplex $\Delta_n$), or considering correlations between the utilities of the agents for the resource. Our mechanisms also heavily rely on the assumption that the central planner has perfect knowledge of the geometry of the full-information utility set close to the optimal welfare boundary, determined by the agent utility distributions. Note that only the local geometry of the welfare-maximizing boundary is useful for our constructions. From our analysis, such perfect knowledge seems necessary to design near-optimal mechanisms given the intricate dependency of the optimal convergence rates on the boundary geometry. A natural question is to understand the trade-offs between maximizing social welfare efficiency and requiring minimal information from agent utilities for the central planner. As an initial remark, in light of the proof of \cref{thm:universal_lower_bound_simplified}, the universal rates $\Ocal(\sqrt{1-\gamma})$ and $1/\sqrt T$ only require the knowledge of a single near-optimal utility vector in the full-information setting. We recall that optimal allocations in the full-information setting simply correspond to first-best allocations. Hence, the only information that is required is an estimate of the utility gathered by each agent under a first-best allocation. This is similar in flavor to mechanisms from artificial currencies which require an estimate of the budget to allocate to each agent.

Another interesting research direction is to tighten the link between the infinite-horizon and finite-horizon settings. Our results show that there can be a clear separation between the two, since exponential convergence rates are possible for the infinite-horizon setting but not for the finite-horizon case: the gluing strategy we propose does not carry to the finite-horizon setting. Except for these exponential rates, however, our analysis shows that the main tools for the infinite-horizon setting can be used for the finite-horizon case up to logarithmic factors $\ln T$ by taking $\gamma(T)=1-1/T$. While we give some examples in which there is no logarithmic gap between these two settings, it is unclear whether such a gap is necessary in general. We conjecture that this is the case, and formalize in \cref{open_question} the simplest case for which such a gap may exist.

\section*{Acknowledgments.}
This work is being partly funded by AFOSR grant FA9550-23-1-0182.

\bibliographystyle{abbrvnat}
\bibliography{refs}

\appendix

\section*{Appendix}

In this Appendix, we give the proofs and further details on the results of the main paper. First, in \cref{sec:proof_facts}, we prove the preliminary results from \cref{sec:preliminaries} on the promised utility framework, including the proof of \cref{lemma:trivial_case} which characterizes situations when there is no social welfare gap between monetary and non-monetary mechanisms. Next, we prove the guarantees from \cref{sec:generic_mechanism} in \cref{sec:proof_main_mechanism} on our proposed allocation resource mechanism, together with a strengthened version of this guarantee \cref{lemma:ball_in_region_optimized}. In \cref{sec:tools_construct_upper_bounds}, we next prove our main geometric tool to prove lower bounds on the social welfare gap from \cref{sec:main_tool_lower_bound}, together with a strengthened version \cref{lemma:upper_bounds_2D}.
Next, we derive our main characterizations for the discounted infinite-horizon setting from \cref{sec:infinite_horizon_paper} in \cref{sec:infinite-horizon}, including the universal $\Ocal(\sqrt{1-\gamma})$ rate from \cref{thm:universal_lower_bound_simplified}, the characterizations for faster rates $\Ocal(1-\gamma)$ from \cref{thm:faster_rates_1-beta,thm:no_faster_rate_1-beta} and the full characterization from \cref{thm:full_characterization}. As a consequence of the previous results, we then prove the results on specific cases from \cref{sec:examples}. In particular, we detail when and how exponential convergence rates can be obtained via the gluing strategy for the infinite-horizon setting.
Last, we prove our results on the finite-horizon setting in \cref{sec:finite_horizon}. In particular, we prove results from \cref{sec:links_to_finite_horizon} showing how tools for the infinite-horizon setting can be used for the finite horizon. As a consequence of these derivations, we prove the results for the special cases in \cref{sec:examples} in the finite-horizon setting.

\noindent\textbf{Technical notations.} For convenience, we introduce a few useful notations. For any $S\subseteq[n]$, denote by $P_S:\Rbb^n\to\Rbb^S$ the projection onto the coordinates of $S$. For $\mb y\in\Rbb^S$, and $r>0$, we also denote by $B_S(\mb y,r)$ the ball $B(\mb y,r)$ in $\Rbb^S$.
Next, we denote $Z=\max_{i\in[n]}u_i$, $Z_{-i} = \max_{j\neq i}u_j$ for $i\in[n]$. Last, for $S\subseteq[n]$ we denote $Z_S=\max_{i\in S} u_i$ and $Z_{-S}=\max_{i\notin S} u_i$.

\section{Proofs from \cref{sec:preliminaries}}
\label{sec:proof_facts}

The results Facts \ref{fact:convexity}, \ref{fact:promised_utility_1}, and \ref{fact:promised_utility_2}, and \cref{lemma:optimal_coupling} can be considered as known in the literature, e.g.\ see \cite{myerson1981optimal} and \cite{bgs19} (Proposition 2.1, Proposition 2.2); these are given for completeness. First, we check that the utility regions are convex.

\begin{proof}[of \cref{fact:convexity}]
    Let $\theta\in[0,1]$. For any resource allocation setting, consider two realizable utilities $\mb U_1$ and $\mb U_2$. From the revelation principle there exist two incentive-compatible allocation strategies $\mb S_1$ and $\mb S_2$ realizing utilities $\mb U_1$ and $\mb U_2$ respectively. Consider the allocation strategy $\mb S$ which first samples a Bernoulli $B\sim\Bcal(\theta)$, then implements $\mb S_1$ if $B=1$ and $\mb S_2$ otherwise, for all times. By linearity, under truthful reporting, this allocation realizes the the utilities $\theta \mb U_1+ (1-\theta)\mb U_2$. It only remains to check that truthful reporting is still a PBE for $\mb S$, which is a consequence of the linearity of the value function $V_i$ as detailed below. Fix any agent $i$, time $t$, utility $u_t(i)$, and history $\mb h(t)$. For any reporting strategy $\sigma_{i,t}$ for time $t$, we denote by $\sigma_{i,t}\circ \textbf{truth}_i$ the concatenated strategy for agent $i$ which uses strategy $\sigma_{i,t}$ at time $t$ and truthful reporting for following rounds. We have
    \begin{align*}
        V_{i,t}&(\mb S,\sigma_{i,t}\circ \textbf{truth}_i,\textbf{truth}_{-i}\mid u_t(i),\mb h(t)) \\
        &= \Pbb[B=1\mid u_t(i),\mb h(t)]\cdot  V_{i,t}(\mb S_1,\sigma_{i,t}\circ \textbf{truth}_i,\textbf{truth}_{-i}\mid u_t(i),\mb h(t)) \\
        &\qquad + \Pbb[B=0\mid u_t(i),\mb h(t)] \cdot V_{i,t}(\mb S_0,\sigma_{i,t}\circ \textbf{truth}_i,\textbf{truth}_{-i}\mid u_t(i),\mb h(t))\\
        &\geq \Pbb[B=1\mid u_t(i),\mb h(t)]\cdot  V_{i,t}(\mb S_1,\textbf{truth}_i,\textbf{truth}_{-i}\mid u_t(i),\mb h(t)) \\
        &\qquad + \Pbb[B=0\mid u_t(i),\mb h(t)] \cdot V_{i,t}(\mb S_0,\textbf{truth}_i,\textbf{truth}_{-i}\mid u_t(i),\mb h(t)) \\
        &= V_{i,t}(\mb S, \textbf{truth}_i,\textbf{truth}_{-i}\mid u_t(i),\mb h(t)),
    \end{align*}
    since both $\mb S_1$ and $\mb S_2$ are incentive-compatible.
    In words, it is not advantageous to deviate from truthful reporting at a single round $t$ from any state. 
    Then, truthful reporting is a PBE by backwards induction (or e.g., using the performance difference lemma \citep{kakade2002approximately}).
\end{proof}

We next prove \cref{fact:promised_utility_1} which gives necessary and sufficient conditions for a utility region $\Rcal$ to be achievable in the $\gamma$-discounted setting.

\begin{proof}[of \cref{fact:promised_utility_1}]
    We start with showing that the condition is sufficient to show $\Rcal'\subseteq \Rcal\subseteq \Ucal_\gamma$. To do so, we use the allocation function $\mb p(\cdot\mid\mb U)$ for $\mb U\in\Rcal$ as the strategy at all times. Precisely, fix an arbitrary vector $\mb U_0\in \Rcal$. Suppose we have constructed $\mb U_{t-1}$ for $t\geq 1$, recalling the notation $\mb v(t)$ for the agent reports at time $t$, we pose
    \begin{equation*}
        \mb p(t) := \mb p(\mb v(t) \mid \mb U_{t-1}) \quad \text{and} \quad \mb U_t:= \mb W( \mb v(t) \mid \mb U_{t-1}).
    \end{equation*}
    It is well defined since by induction $\mb U_t\in\Rcal$ for all $t\geq 0$ from Eq~\eqref{eq:valid_promise}. We denote by $\mb S$ the constructed strategy. We now check that truthful reporting for all agents $\mb\sigma=\textbf{truth}$ forms a perfect Bayesian equilibrium.

    To do so, we check that for all $i\in[n]$, any time $t\geq 0$, the expected remaining utility of agent $i$ by reporting $v$ at time $t$ is exactly $W_i(v\mid \mb U_{t-1})$. Indeed, for any strategy $\sigma_i$ for agent $i$, we have
    \begin{align*}
        V_{i,t}&(\mb S,\textbf{truth}_{-i}, \sigma_i \mid u_i(t),\mb h(t)) \\
        &= \sum_{t'\geq t}(1-\gamma) \Ebb\sqb{\gamma^{t'-t}u_i(t')\Ebb_{\mb u_{-i}(t')}[p_i(\mb u_{-i}(t'),v_i(t')\mid\mb U_{t'-1}) ] \mid u_i(t),\mb h(t) } \\
        &= \sum_{t'\geq t}(1-\gamma) \Ebb\sqb{\gamma^{t'-t}u_i(t') P_i(v_i(t')\mid\mb U_{t'-1})  \mid u_i(t), \mb U_{t-1}}\\
        &\overset{(i)}{\leq} \sum_{t'\geq t}\gamma^{t'-t} \Ebb[(1-\gamma)u_i(t') P_i(u_i(t')\mid\mb U_{t'-1}) + \gamma W_i(u_i(t')\mid \mb U_{t'-1})\\
        &\qquad \qquad\qquad\qquad \qquad\qquad - \gamma W_i(v_i(t')\mid\mb U_{t'-1})  \mid u_i(t),  \mb U_{t-1} ]\\
        &\overset{(ii)}{=}(1-\gamma)u_i(t)P_i(u_i(t)\mid\mb U_{t-1})+\gamma ( W_i(u_i(t)\mid\mb U_{t-1}) - W_i(v_i(t)\mid\mb U_{t-1})) \\
        &\qquad \qquad\qquad\qquad \qquad\qquad + \sum_{t'>t}\gamma^{t'-t}\Ebb\sqb{ U_{t'-1,i}   - \gamma W_i(v_i(t')\mid\mb U_{t'-1}) \mid u_i(t),  \mb U_{t-1}  }
    \end{align*}
    In $(i)$ we applied the incentive-compatibility equation \cref{eq:incentive_compatibility}, and in $(ii)$ we took the expectation over $u_i(t')\sim\Dcal_i$ within each expectation and used \cref{eq:target_met}. Now note that by construction, for $t'\geq 1$ we have
    \begin{equation*}
        W_i(v_i(t')\mid\mb U_{t'-1}) = \Ebb_{\mb u_{-i}(t')}[W_i(\mb u_{-i}(t'),v_i(t')\mid\mb U_{t'-1}] = \Ebb_{\mb u_{-i}(t')}[U_{t,i}].
    \end{equation*}
    As a result, with a telescoping argument, we obtain
    \begin{equation*}
        V_{i,t}(\mb S,\textbf{truth}_{-i}, \sigma_i \mid u_i(t),\mb h(t)) \leq (1-\gamma)u_i(t)P_i(u_i(t)\mid\mb U_{t-1})+\gamma ( W_i(u_i(t)\mid\mb U_{t-1}),
    \end{equation*}
    where the only inequality used is that of $(i)$, which is an equality for $\sigma_i=\textbf{truth}$, that is when agent $i$ reports truthfully. This exactly shows that \cref{eq:perfect_bayesian_equilibrium} holds. We can now check that under truthful reporting,
    \begin{align*}
        V_i(\mb S,\textbf{truth}) &= \Ebb[V_i,1(\mb S,\textbf{truth}\mid u_i(1)) ] \\
        &\overset{(i)}{=} \Ebb[(1-\gamma)u_i(1)P_i(u_i(1)\mid\mb U_0)+\gamma ( W_i(u_i(1)\mid\mb U_0)] \overset{(ii)}{=} U_{0,i},
    \end{align*}
    where in $(i)$ we used the previous computations and in $(ii)$ we used \cref{eq:target_met}. Therefore we checked that $\mb U_0\in\Ucal_\gamma$ for any arbitrary $\mb U_0\in\Rcal$.

    We now prove that it is necessary. Let $\Rcal$ be an achievable region, that is $\Rcal\subseteq\Ucal_\gamma$. By definition of $\Ucal_\gamma$, for all $\mb U\in\Ucal_\gamma$ there is a strategy $\mb S(\mb U)$ that realizes $\mb U$. By the revelation principle, without loss of generality, truthful reporting is a perfect Bayesian equilibrium. We define $\mb p(\mb v\mid\mb U):=\mb p(1)$ where $\mb p(1)$ is the first allocation of the strategy $\mb S(\mb U)$ having received reports $\mb v$. Let $\tilde u_i(t)\overset{i.i.d.}{\sim}\Dcal_i$ be independent sequences for all $i\in[n]$. We then let
    \begin{equation*}
        \mb W(\mb v\mid\mb U):= (1-\gamma)\Ebb\sqb{\sum_{t\geq 2}\gamma^{t-2}\tilde u_i(t)p_i(t) \mid \mb v(1)=\mb v}, \quad \mb v\in[0,\bar v]^n.
    \end{equation*}
    That is, $\mb W(\mb v\mid\mb U)$ is the expected remaining utility knowing that the first reports were $\mb v$. By construction, since $\mb S(\mb U)$ realizes the utilities $\mb U$, \cref{eq:target_met} is satisfied. Next, note that for any first reports $\mb v(1)=\mb v$, the strategy $\mb S(\mb U)$ starting from time $2$ exactly realizes the utility vector $\mb W(\mb v\mid\mb U)$ (the allocation problem is time-invariant): \cref{eq:valid_promise} holds. Last, we assumed that truthful reporting satisfies \cref{eq:perfect_bayesian_equilibrium}. Consider any strategy $\sigma_i$ which reports an arbitrary value $v_i(1)=v(u_i(1))$ but truthfully reports at all times $t\geq 2$. Taking the expectation of \cref{eq:perfect_bayesian_equilibrium} for strategy $\sigma_i$ and $t=1$, over the realization of the first allocation $\mb p(1)$ shows that \cref{eq:incentive_compatibility} holds.
\end{proof}

We now turn to the second characterization from \cref{fact:promised_utility_2}, which characterizes the form of the interim utility promise, following an argument from \cite{myerson1981optimal}.

\begin{proof}[of \cref{fact:promised_utility_2}]
    Fix $\gamma\in(0,1)$. Given \cref{fact:promised_utility_1}, it suffices to show that \cref{eq:target_met,eq:valid_promise,eq:incentive_compatibility} are equivalent to \cref{eq:valid_promise,eq:valid_interim_promise}.

    We start by supposing that \cref{eq:target_met,eq:valid_promise,eq:incentive_compatibility} hold. We only need to show \cref{eq:valid_interim_promise}. Fix $\mb U\in\Rcal$. For readability we omit $\mid\mb U$ from all functions below. First, for any $u,v\in[0,\bar v]$ with $u\leq v$, applying \cref{eq:incentive_compatibility} to $u$ and $v$ implies that
    \begin{equation*}
        \frac{1-\gamma}{\gamma}(P_i(u)-P_i(v ))v\leq W_i(v )-W_i(u ) \leq \frac{1-\gamma}{\gamma}(P_i(u )-P_i(v ))u.
    \end{equation*}
    Note that the function $P_i(\cdot )$ takes values in $[0,1]$ since $\mb p(\cdot )$ takes values in $\Delta_n$. Also, the previous equation implies in particular that $P_i(\cdot )$ is non-decreasing and that $W_i(\cdot)$ is constant on any interval on which $P_i(\cdot)$ is constant. We can therefore sum the previous equations for $u=u_i$ and $v=u_{i+1}$ for regular partitions $0\leq u_1,\ldots,u_k=w$ for a fixed $w\in[0,\bar v]$ interpreting the sums as Riemann integrals (possible because $P_i$ is non-decreasing hence well-behaved) which implies
    \begin{align}\label{eq:formula_1}
        W_i(w)-W_i(0) = -\frac{1-\gamma}{\gamma} \int_{P_i(0)}^{P_i(w)} P_i^{<-1>}(y)dy
        =-\frac{1-\gamma}{\gamma} \paren{ wP_i(w) - \int_0^w P_i(x)dx }
    \end{align}
    This proves \cref{eq:formula_interim_initial}.
    We next use \cref{eq:target_met} to compute
    \begin{align*}
        U_i &= \Ebb_{u_i}[(1-\gamma) u_i P_i(u_i) + \gamma W_i(u_i)]\\
        &=\gamma W_i(0) + (1-\gamma)\Ebb_{u_i}\sqb{ \int_0^{u_i} P_i(x)dx }=\gamma W_i(0) + (1-\gamma)\Ebb_{u_i}\sqb{ \int_0^{\bar v} P_i(x) \1_{u_i\geq x}dx }\\
        &=\gamma W_i(0) + (1-\gamma)\int_0^{\bar v} \bar F_i(x)P_i(x)dx.
    \end{align*}
    Plugging this formula for $W_i(0)$ into \cref{eq:formula_1} gives the desired formula for the interim allocation \cref{eq:formula_interim_promise}. \cref{eq:valid_interim_promise} holds by definition of the interim promise $W_i(\cdot\mid\mb U)$.

    Now suppose that \cref{eq:valid_promise,eq:valid_interim_promise} hold. Using the previous computations, we note that \cref{eq:target_met} is satisfied. It only remains to check the incentive-compatibility constraint \cref{eq:incentive_compatibility}. Having fixed $\mb U\in\Rcal$, for any $i\in[n]$ and $u,v\in[0,\bar v]$, we have
    \begin{align*}
        (1-\gamma)&u P_i(v) +\gamma W_i(v) = U_i + (1-\gamma)\paren{ P_i(v) (u-v) + \int_0^v P_i(x)dx -\int_0^{\bar v}\bar F_i(x)P_i(x)dx }\\
        &= U_i + (1-\gamma)\paren{  \int_0^u P_i(x)dx -\int_0^{\bar v}\bar F_i(x)P_i(x)dx } + (1-\gamma)\int_u^v (P_i(x)-P_i(v))dx.
    \end{align*}
    Only the last term depends on $v$, hence the quantity is maximized for $v=u$ because $P_i(\cdot)$ is non-decreasing. This proves \cref{eq:incentive_compatibility} and ends the proof.
\end{proof}

Last, we prove \cref{lemma:optimal_coupling} which shows that the couplings from \cref{eq:coupling_formula_old,eq:coupling_formula} are optimal.

\begin{proof}[of \cref{lemma:optimal_coupling}]
    From \cref{eq:easy_upper_bound}, the optimal value of \cref{eq:optimal_coupling} is upper bounded by $\mb\alpha^\top \Ebb[\tilde{\mb Z}]$. For both couplings in \cref{eq:coupling_formula_old,eq:coupling_formula} we can easily check that for all realization $\omega$, $\mb Z\in\{\mb y: \mb\alpha^\top \mb y = \mb\alpha^\top \Ebb[\tilde{\mb Z}]\}$. Next, by independence of the variables $\tilde Z_i$ for $i\in[n]$, we can also prove that in both cases for all $i\in[n]$, $\Ebb[Z_i\mid\tilde Z_i]=\tilde Z_i]$. Hence these couplings satisfy the constraints and achieve the value $\mb\alpha^\top \Ebb[\tilde{\mb Z}]$ which ends the proof.
\end{proof}

\subsection{Characterization of cases with no social welfare gap}
\label{subsec:no_welfare_gap_characterization}
We now turn to the main result in this section, which is the \cref{lemma:trivial_case}. This gives a characterization of cases when $\Ucal_0=\Vcal_1$ already contains an optimal utility vector.

\begin{proof}[of \cref{lemma:trivial_case}]
    First note that the perfect Bayesian equations \cref{eq:perfect_bayesian_equilibrium} characterizing $\Vcal_1$ for $t=T=1$ are identical to those for $\Ucal_0$ for $t=1$, which is the only time step that matters for the total utility $\mb V(\mb S, \mb\sigma)$ since $\gamma=0$. Hence we directly have $\Ucal_0=\Vcal_1$.
    We now turn to the main claim of the result.
    \paragraph{Proving $1.\Rightarrow 2.$}
    We introduce the preorder $\succeq$ such that $i\succeq j$ if and only if $\Pbb(u_i \geq u_j)=1$ (it is reflexive and transitive, but not antisymmetric). Let $T = \{i\in[n],\nexists j\neq i, j\succeq i)$, which gives
    \begin{equation}\label{eq:hypothesis_non_dominated}
        \forall i\in T,\forall j\neq i,\quad \Pbb(u_i > u_j)>0.
    \end{equation}
    This implies in particular that $\Pbb(u_i=0)<1$. For every $i\in T$, recall the notation $\Supp(\Dcal_i)$ for the support of $\Dcal_i$. We define
    \begin{equation}\label{eq:def_m_M}
        m_i := \inf (\Supp(\Dcal_i)\setminus\{0\})\quad \text{and} \quad
        M_i := \sup (\Supp(\Dcal_i)\setminus\{0\}).
    \end{equation}
    Note that these are well-defined since $\Supp(\Dcal_i)\setminus \{0\}\neq \emptyset$, otherwise $\Pbb(u_i=0)=1$. In particular, $M_i>0$.
    By assumption, there exists an optimal vector $\mb U_0\in \Ucal_0$. Hence, by the revelation principle, there is an allocation function $\mb p(\cdot)$ that reaches $\mb U_0$ and is incentive-compatible as per \cref{eq:incentive_compatibility}.
    Now fix $i\in T$. Because $\mb U_0$ is an optimal vector, with probability one on $\mb u$, we have $p_i(\mb u)>0$ only if $i\in\argmax_{j\in[n]} u_j$. We also recall that $P_i(\cdot)$ is non-decreasing by hypothesis. As a result, for all $u\in \Supp(\Dcal_i)$ we have
    \begin{equation}\label{eq:constraint_1_bis}
        \Pbb(Z_{-i} < u) \leq P_i(u) \leq \Pbb(Z_{-i} \leq u).
    \end{equation}
    In particular, this holds for $u\in\{m_i ,M_i \}$.
    In our setting, because $\gamma=0$, the incentive-compatibility constraint implies that for all $u\in (0,\bar v]$, $u\in\argmax_{v\in[0,\bar v]} u P_i(v)$
    where $P_i(v_i) = \Ebb[p_i(v_i,\mb u_{-i})]$. This implies that $P_i$ is constant on $(0,\bar v]$ for all $i\in[n]$. Together with \cref{eq:constraint_1_bis}, this shows that
    \begin{equation}
        \Pbb(Z_{-i} < M_i)\leq P_i(M_i) = \lim_{x\to m_i^+} P_i(x) \leq \lim_{x\to m_i^+} P_i(Z_{-i}\leq x) = \Pbb(Z_{-i}\leq m_i) .
    \end{equation}
    Hence, either $m_i=M_i$, or $\Pbb(Z_{-i}\leq m_i) = \Pbb(Z_{-i}<M_i)$. In both cases, we obtained $\Pbb(Z_{-i} \in(m_i,M_i))=0$, while $\Supp(\Dcal_i)\subseteq[m_i,M_i]\cup\{0\}$ hence $\Pbb(u_i \in [m_i,M_i]\cup\{0\})=1$.
    For any $j\neq i$,
    \begin{align*}
        0=\Pbb(Z_{-i} \in(m_i,M_i)) &\geq \Pbb(u_j \in(m_i,M_i)\;\text{and}\;\,\forall k\notin\{i,j\}, u_k <M_i)\\
        &= \Pbb(u_j \in(m_i,M_i)) \prod_{k\notin\{i,j\}} \Pbb( u_k <M_i).
    \end{align*}
    However, \cref{eq:hypothesis_non_dominated} implies that for all $k\neq i$, we have $\Pbb(u_k<M_i)>0$. Together with the previous equation this gives $\Pbb(u_j \in(m_i,M_i))=0$, which holds for all $j\neq i$.

    \paragraph{Proving $2.\Rightarrow 3.$}
    We suppose that the second statement of \cref{lemma:trivial_case} holds. Further, suppose that scenario 3(a) does not hold, that is, $([n],\succeq)$ does not have a maximum element. We now construct the set $M$ as the set of maximal elements, that is
    \begin{equation*}
        M = \{i\in[n]: j\succeq i\Rightarrow i\succeq j\}.
    \end{equation*}
    On $M$, the preorder $\succeq$ is symmetric, hence is an equivalence relation and induces equivalence classes. We define $S$ as a set containing exactly one representative for each equivalence class of $\succeq$ on $M$. Note that for any $i\neq j\in S$ we have neither $i\succeq j$ nor $j\succeq i$. By assumption, we also have $|S|\geq 2$. For any $i\in M$, we have $\Pbb(u_i=0)<1$, otherwise for all $j\in M$, $j\succeq i$ and there would be only one equivalence class. 
    Let $T$ be the union of equivalence classes that are singletons. Note that $T$ coincides with the definition from the proof that $1.\Rightarrow 2.$, hence elements of $T$ do not satisfy assumption 2(a) as seen in \cref{eq:hypothesis_non_dominated}. Hence they must satisfy assumption 2(b): defining $m_i,M_i$ as in \cref{eq:def_m_M},
    \begin{equation}\label{eq:assumption_b_satisfied}
        \Pbb(u_i \in [m_i,M_i]\cup\{0\})=1\quad \text{and} \quad \Pbb(u_j \in(m_i,M_i))=0,\quad j\neq i.
    \end{equation}
    For $i\in M\setminus T$, there exists $j\in M$ with $j\neq i$ in its equivalence class. Then, $i\succeq j$ and $j\succeq i$, which implies that the distributions $\Dcal_i$ and $\Dcal_j$ are identical and are equal to a Dirac. We can define $m_i=M_i>0$ as the value such that $\Pbb( u_i=m_i)=1$. Note that with these values, \cref{eq:def_m_M,eq:assumption_b_satisfied} hold for all $i\in M$. 
    Note that $M\setminus T$ only contains one equivalence class, otherwise we would have $i,j\in M\setminus T$ with say $m_i<m_j$ which implies that $i$ is not maximal. In summary, $|S\setminus T|\leq 1$.

    We next prove that for all $i,j\in S$, $M_j\notin (m_i,M_i)$. Otherwise, by definition of $M_j$ in \cref{eq:def_m_M} for all $\epsilon>0$ we have $\Pbb(u_j \in (m_i,M_j))>0$ contradicting \cref{eq:assumption_b_satisfied}. Similarly, we have $m_j\notin (m_i,M_i)$.
    This implies that all intervals $(m_i,M_i)$ for $i\in S$ are disjoint. We order them by increasing order $S=\{i_1,\ldots,i_k\}$ where $M_{i_1}\geq m_{i_1} \geq M_{i_2} \geq m_{i_2}\geq\ldots \geq M_{i_k}\geq m_{i_k}$.

    For any $i\in T$ and $j\in S\setminus T$, we must have $m_j\leq m_i$, otherwise $m_j\geq M_i$ which implies $j\succeq i$ since $\Pbb(u_j=m_j)=1$. This shows that the element $j\in S\setminus T$ if it exists, appears last $j = i_k$. 
    For any $l\in[k-1]$, we have $i_l\in T$ so $\Supp(\Dcal_{i_l})\subseteq [m_{i_l},M_{i_l}]\cup\{0\})$. Then, we must have $\Pbb(u_{i_l}=0)>0$, otherwise we would have $i_l\succeq i_{l+1}$. We also note that $m_{i_{k-1}}\geq M_{i_k}>0$.
    By construction of $S$, for any $i\notin S$ there must exist $j\in S$ such that $j\succeq i$. If $j = i_l$ where $l\in[k-1]$ we must have $\Pbb(u_i=0)=1$ since we just proved that $\Pbb( u_{i_l}=0)>0$. Hence, in all cases $i_k\succeq i$.
    
    Up to renaming $m_{i_l}$ by $m_l$, this ends the proof of the characterization 3(b). It is straightforward to check that the proposed allocation function $\mb p(\cdot)$ is incentive-compatible and reaches the utility vector
    \begin{equation*}
        \mb U = \sum_{l\in[k]} \Pbb(u_{i_{l'}}=0,l'<l)\Ebb[u_{i_l}] \mb e_i.
    \end{equation*}
    In particular, $\mb U\in\Ucal_0$. We can easily check that it always allocates to the agent within $\{i_l,l\in[k]\}$ with maximum value, which is sufficient since $i_k\succeq i$ for all $i\notin S$. This ends the proof of $2.\Rightarrow 3.$

    \paragraph{Proving $3.\Rightarrow 1.$} This is immediate because the vector $\mb U\in\Ucal_0$ is optimal.    
\end{proof}

When $\Ucal_0$ does not contain an optimal vector, the characterization from \cref{lemma:trivial_case} does not hold: there exist agents $i\in[n]$ that do not satisfy assumptions 2(a) nor 2(b). However, we can still give a simple characterization of the distributions $\Dcal_i$ for indices $i\in[n]$ that satisfy one of these conditions, as detailed in the following lemma. We recall that $\tilde I$ denotes the set of agents that do not satisfy 2(a) nor 2(b).

\begin{lemma}\label{lemma:def_I_properties}
    Let $J=\{i\in[n]\setminus \tilde I:\forall j\neq i,\Pbb(u_i> u_j)>0\}$ and $K = \{i\in[n]\setminus (\tilde I\cup J):\forall j\neq i,\Pbb(u_i\geq u_j)>0\}$. There is an ordering of $J=\{j_1,\ldots,j_{|J|}\}$ and values $0\leq m\leq m_{j_1}\leq M_{j_1}\leq m_{j_2}\leq M_{j_2}\leq \ldots \leq m_{j_{|J|}}\leq M_{j_{|J|}}$ such that
    \begin{itemize}
        \item For $i\notin \tilde I\cup J\cup K$, $\Pbb(u_i=\max_{j\in[n]}u_j)=0$.
        \item For $k\in K$, $\Pbb(u_k\leq m)=1$ and $\Pbb(u_k=m)>0$. Further for each $k\in K$ there exists $j(k)\neq k$ such that $\Pbb(u_{j(k)}\geq m)=1$ and $\Pbb(u_{j(k)}> m)>0$.
        \item For $j\in J$, $\Pbb(u_j\in[m_j, M_j]\cup\{0\})=1$ and $\Pbb(u_i \in (m_i,M_i))=0$ for all $i\neq j$. Also, for $j\in J\setminus\{j_1\}$, $\Pbb(u_j=0)>0$.
        \item Either $\Pbb(u_{j_1}=0)>0$, $K=\emptyset$, or: $m_{j_1}=m$ and $\Pbb(u_{j_1}=m)>0$.
    \end{itemize}
\end{lemma}

\begin{proof}
    By construction of $K$, if $i\notin \tilde I\cup J\cup K$ there exists $j\in[n]$ such that $\Pbb(u_j>u_i)=1$. This proves the first claim.
    We now suppose that $K\neq\emptyset$. For $k\in K$ since $k\notin \tilde I\cup J$, there exists $j(k)\in[n]$ with $j(k)\neq k$ such that $\Pbb(u_k > u_{j(k)})=0$. Using the definition of $K$ we obtain $\Pbb(u_{j(k)} = u_k)>0$. Since $u_k$ and $u_{j(k)}$ are independent, the two previous equations imply that there exists a value $m_k$ such that $\Pbb(u_k\leq m_k)=1=\Pbb(u_{j(k)}\geq m_k)$ and $\Pbb(u_k=m_k),\Pbb(u_{j(k)}=m_k)>0$. Further, if $K$ contains two distinct elements $k\neq k'$, then by definition of $K$ we have $0<\Pbb(u_{k'} \geq u_{j(k)}) \leq \Pbb(u_{k'} \geq m_k)$.
    As a result, we have $m_{k'}\geq m_k$. By symmetry, we obtain $m_k=m_{k'}$. In other words, there is a single value $m$ such that for all $k\in K$:
    \begin{equation*}
        \Pbb(u_k\leq m) = 1\quad \text{and} \quad \Pbb(u_k = m) > 0,
    \end{equation*}
    and if $K\neq \emptyset$ there exists $l_K\in[n]$ such that $\Pbb(u_{l_K}\geq m)=1$ and $\Pbb(u_{l_K}=m)>0$.

    Next, we turn to elements of $J$. For $j\in J$, let $m_j:=\inf \Supp(\Dcal_j)\setminus\{0\}$ and $M_j:=\sup  \Supp(\Dcal_j)$. By definition of $J$, we know that for all $i\neq j$, $\Pbb(u_i\in(m_j,M_j))=0$. Using the same arguments as in \cref{lemma:trivial_case} we show that the intervals $(m_j,M_j)$ for $i\in J$ are all disjoint, which ensures that there is an ordering $J=\{j_1,\ldots,j_{|J|}\}$ satisfying the desired inequalities. The only remaining piece is to check that $m_{j_1} \geq m$, provided that $K\neq\emptyset$. Indeed, since $j_1\in J$ and $u_k=m$ with non-zero probability for $k\in K$, we have $m\notin (m_{j_1},M_{j_1})$. By definition of $J$, we also have $0<\Pbb(u_{j_1}> u_{l_K}) \leq  \Pbb(u_{j_1}>m)$. Hence, $M_{j_1}>m$ which implies $M_{j_1}\geq m_{j_1}\geq m$.

    By definition of $J$, for $l\in[|J|-1]$ one has $\Pbb(u_{j_l}>u_{j_{l+1}})>0$. From the previous characterization of the supports of $u_j$ for $j\in J$, this implies that $\Pbb(u_{j_{l+1}}=0)>0$. This proves the third claim.

    Suppose that $K\neq\emptyset$ and $J\neq\emptyset$. Then for $k\in K$ by definition of $K$ we also have $0<\Pbb(u_k \geq u_{j_1}) = \Pbb(u_{j_1}\leq m)$. This implies either $\Pbb(u_{j_1}=0)>0$ or $m_{j_1}=m$ and $\Pbb(u_{j_1}=m)>0$.
\end{proof}

\section{Proofs from \cref{sec:generic_mechanism} for the main resource allocation mechanisms}
\label{sec:proof_main_mechanism}


We now prove \cref{lemma:ball_in_region_no_assumptions} which gives conditions for a ball to be achievable within $\Ucal_\gamma$ by the mechanims \text{PUM}.

\begin{proof}[of \cref{lemma:ball_in_region_no_assumptions}]
    Fix parameters $\gamma$, $\mb x$, $r,\delta$ satisfying the assumptions. To show that $B(\mb x, r)\subseteq \Ucal_\gamma$, we construct an online allocation strategy that reaches these utilities. In fact, we give an allocation strategy to reach utilities 
    \begin{equation*}
        \Rcal:= \set{\mb y: \mb 0\leq \mb y\leq \mb z, \mb z\in Conv(\Ucal^{NI},B(\mb x,r))}.
    \end{equation*}

    \paragraph{Definition of the allocation function.}  We briefly recall the definition of the allocation function as defined in the main document. For convenience for any $\mb U \in \Ucal^\star$, we fix $p(\cdot;\mb U)$ an allocation function which reaches the utility vector $\mb U$ when having access to the full information of agent's utilities.
    
    We start by specifying the allocation strategy for extreme points $\mb U$ of $\Rcal$ that still belong to the ball $B(\mb x,r)$. Following the promised utility definition above, we only need to specify an allocation function $p(\cdot\mid\mb U)$ and a direction $\mb\alpha(\mb U)$ for these utility vectors $\mb U$. In particular, these can be written as $\mb U = \mb x + r\mb y$ where $\mb y\in S_{n-1}$ is a unit vector. We define the allocation function via
    \begin{equation*}
        p(\cdot\mid \mb U):= p(\cdot;\mb x + (r+\delta)\mb y) \quad \text{and} \quad \mb \alpha(\mb U) = \mb x-\mb U = -r\mb y.
    \end{equation*}

    We now extend the definition of the allocation function to the whole domain $\Rcal$.
    To do so, note that any vector $\mb U\in \Rcal$ can be characterized by $\mb 0\leq \mb U\leq \mb y$ where $\mb y\in Conv(\Ucal^{NI},B(\mb x,r))$ and is also an extreme point of $\Rcal$. In particular, $\mb y$ can be written as the convex combination of $\mb 0$, vectors $\Ebb[u_i]\mb e_i$ and some extreme point $\tilde {\mb y}$ of $\Rcal$ within the ball $B(\mb x, r)$ (we only need one point from $B(\mb x, r)$ at most because the ball is strictly convex). Such a decomposition can be obtained via standard Carath\'eodory constructions. As a summary, we obtain
    \begin{equation*}
        \mb y = \sum_{i\in[n]} q_i \Ebb[u_i]\mb e_i + q_0\tilde{\mb y},\quad\text{where}\quad \mb q\geq \mb 0, \sum_{i\leq n} q_i\leq 1,
    \end{equation*}
    and $U_i = s_i y_i$ where $s_i\in[0,1]$ for $i\in[n]$.
    We define the allocation and promised utility function by
    \begin{align}
        \mb p(\cdot \mid \mb y) = \sum_{i\in[n]} q_i \mb e_i + q_0 \mb p(\cdot\mid\tilde{\mb y})\quad &\text{and}\quad  p_i(\cdot\mid\mb U) = s_i p_i(\cdot\mid\mb y),\quad i\in[n].\label{eq:def_allocation_function_app} \\ 
        \mb W(\cdot \mid \mb y) = \sum_{i\in[n]} q_i \Ebb[u_i] \mb e_i + q_0 \mb W(\cdot\mid\tilde{\mb y})\quad &\text{and}\quad  W_i(\cdot\mid\mb U) = s_i W_i(\cdot\mid\mb y),\quad i\in[n].\label{eq:def_promised_utility_function_app}
    \end{align}

    \paragraph{Checking that this is a valid allocation strategy for region $\Rcal$.} Using the promised utility framework, to show that this is a valid strategy for utilities within $\Rcal$, we only need to show that \cref{eq:valid_promise,eq:valid_interim_promise} are satisfied. That is, we check that the future promised utilities remain within $\Rcal$ and their expectation with respect to other agents' utility matches the interim future promise defined as in \cref{eq:formula_interim_promise}. We start by proving that this is the case for the extreme points $\mb U$ of $\Rcal$ in the ball $B(\mb x, r)$. For these utility vectors, because we used the choice of allocation strategy from \cref{eq:coupling_formula} we only need to check \cref{eq:valid_promise}. We prove the stronger statement that these stay within the ball $B(\mb x, r)$.

    As above, we write $\mb U=\mb x + r\mb y$ with $\mb y\in S_{n-1}$. Note that $\mb\alpha(\mb U)$ is the normal vector to the boundary of $\Rcal$ at $\mb U$. Further, $\mb \alpha(\mb U)\in\Rbb_+^n$. Indeed, fix $i\in[n]$ and construct $\mb U^{(i)}$ such that $U^{(i)}_i=0$ and $U^{(i)}_j=U_j$ for all $j\neq i$. By construction of $\Rcal$ using the rectangles we still have $\mb U^{(i)}\in\Rcal$ so by convexity of $\Rcal$,
       $ 0\leq \mb\alpha(\mb U)^\top (\mb U-\mb U^{(i)}) = \alpha_i(\mb U) U_i.$
    Now because $B(\mb x, r+\delta)\subseteq \Ucal^\star\subseteq\Rbb_+^n$, we must have $U_i>0$. This proves that $\alpha_i(\mb U)\geq 0$.
    
    Intuitively, the choice of $\alpha(\mb U) = \mb x-\mb U$ pushes the promised utilities towards the center of the ball $B(\mb x, r)$. Further, we can also characterize the expectation of the promised utility vector which we denote $\bar{\mb W}:= \Ebb[\mb W(v_i\mid \mb U)]$. By the promise keeping equality \cref{eq:target_met},
    \begin{equation*}
        \gamma \bar{\mb W} = \gamma \Ebb[\mb W( \mb v\mid \mb U)] = \mb U - (1-\gamma) (\Ebb[u_i p_i(\mb v\mid \mb U)])_{i\in[n]}.
    \end{equation*}
    Now because the incentive-compatibility constraint is enforced and by definition of the allocation function, we have $\Ebb[u_i p_i(\mb v\mid \mb U)] = \Ebb[u_i p_i(\mb u;\mb x + (r+\delta)\mb y)] = x_i + (r+\delta)y_i$ for all $i\in[n]$. Combining this with the previous equation gives,
    \begin{equation}\label{eq:mean_towards_center_ball}
        \bar{\mb W} =  \mb U - \frac{1-\gamma}{\gamma}\delta \mb y = \mb x + \paren{r-\frac{1-\gamma}{\gamma}\delta}\mb y.
    \end{equation}
    Hence, the term in $\delta$ from \cref{eq:mean_towards_center_ball} also pushes future promise vectors towards the ball center.

    Using the definition of the promised utilities given the interim promise utility from \cref{eq:formula_interim_promise}, for any $i\in[n]$ since $W_i(v_i\mid\mb U)$ is non-increasing in $v_i$,
    \begin{align*}
        |W_i(v_i\mid\mb U) - \bar W_i|  &\leq W_i(0\mid \mb U) - W_i(\bar v \mid \mb U) = \frac{1-\gamma}{\gamma} \int_0^{\bar v} (P_i(\bar v\mid\mb U) - P_i(v\mid\mb U))dv\leq \frac{1-\gamma}{\gamma} \bar v.
    \end{align*}
    Now let $i_1,i_2$ be the indices of largest and second-largest value of $|\alpha_i(\mb U)|$ respectively. By the coupling formula from \cref{eq:coupling_formula} and because $\mb\alpha(\mb U)\geq \mb 0$, we have
    \begin{equation}\label{eq:W_deviation_bound}
        \| \mb W(\mb v\mid \mb U) - \bar {\mb W}\|_1 \leq \frac{1-\gamma}{\gamma} \bar v  \paren{n+\frac{\sum_{j\neq i_1}\alpha_j(\mb U)}{\alpha_{i_1}(\mb U)}  + \frac{\alpha_{i_1}(\mb U)}{\alpha_{i_2}(\mb U)}} 
        \leq \frac{1-\gamma}{\gamma} \bar v  \paren{2n + \frac{\alpha_{i_1}(\mb U)}{\alpha_{i_2}(\mb U)}} .
    \end{equation}
    We now give an upper bound on $\alpha_{i_1}(\mb U) / \alpha_{i_2}(\mb U)$. We recall that the region $\Rcal$ contains all extreme points $\Ebb[u_i]\mb e_i$ for $i\in[n]$. Hence, since $\mb\alpha(\mb U)$ is a supporting hyperplane of $\Rcal$ at $\mb U$, we have,
    \begin{equation*}
        0 \leq \mb\alpha(\mb U)^\top (\mb U - \Ebb[u_{i_1}]\mb e_{i_1}) \leq \alpha_{i_1}(\mb U) (U_{i_1}- \Ebb[u_{i_1}]) + \alpha_{i_2}(\mb U)\sum_{j\neq i_1} U_j 
    \end{equation*}
    In the second inequality, we used the fact that for all $j\neq i_1$, we have $0\leq \alpha_j(\mb U)\leq \alpha_{i_2}(\mb U)$. Because we can write $\mb U=\mb x+ r\mb y$ for $\mb y\in S_{n-1}$, we have $U_{i_1} \leq x_{i_1} + r$. Similarly, for any $i\in[n]$, we have $U_i \leq x_i+r \leq 2x_i$, where in the last inequality, we used the fact that $\mb x - r \mb e_i \in\Rcal\subseteq\Rbb_+^n$. Then,
    \begin{equation*}
        \alpha_{i_1}(\mb U) (\Ebb[u_{i_1}] - x_{i_1} - r) \leq 2\alpha_{i_2}(\mb U)\sum_{j\neq i_1} x_j \leq 2n\alpha_{i_2}(\mb U)\max_{i\in[n]} x_i,
    \end{equation*}
    which implies
    \begin{equation}\label{eq:bound_alpha_1/alpha_2}
        \frac{\alpha_{i_1}(\mb U)}{\alpha_{i_2}(\mb U)} \leq  \frac{2n\max_{i\in[n]} x_i}{\min_{i\in[n]} (\Ebb[u_i] - x_i - r)}.
    \end{equation}

    Putting this together with \cref{eq:W_deviation_bound} gives
    \begin{equation*}
        \|\mb W(\mb v\mid \mb U) - \mb W\|_2 \leq \|\mb W(\mb v\mid \mb U) - \mb W\|_1 \leq  \frac{1-\gamma}{\gamma}\cdot 2n\bar v \paren{1 + \frac{\max_{i\in[n]}  \Ebb[u_i]}{\min_{i\in[n]} (\Ebb[u_i] - x_i - r)}}.
    \end{equation*}
    For convenience, we will write $V := 2n\bar v \paren{1 + \frac{\max_{i\in[n]}  \Ebb[u_i]}{\min_{i\in[n]} (\Ebb[u_i] - x_i - r)}}$.
    We recall that by construction of the promised utilities, \cref{lemma:optimal_coupling} shows that $\mb W(\mb v\mid \mb U)$ lies in the hyperplane of equation $\mb\alpha(\mb U)^\top (\mb x-\bar{\mb W}) = 0$, that is $\mb y^\top (\mb x-\bar{\mb W}) = 0$. Given \cref{eq:mean_towards_center_ball}, to show that for all $\mb v$, we have $\mb W(\mb v\mid\mb U)\in B(\mb x, r)$, it suffices to show that
    \begin{equation*}
        \paren{r-\frac{1-\gamma}{\gamma}\delta,\sup_{\mb v} \|\mb W(\mb v\mid \mb U) - \mb W\|_2 } \in B(0,r).
    \end{equation*}
    To prove this, using $\delta \leq r\gamma/(1-\gamma)$ we compute
    \begin{align*}
       r^2 - \paren{r-\frac{1-\gamma}{\gamma}\delta}^2 - \paren{ \frac{1-\gamma}{\gamma}  V }^2 
        &= \frac{1-\gamma}{\gamma^2}\sqb{ 2\gamma \delta r  - (1-\gamma)\delta^2- (1-\gamma) V^2 }\\
        &\geq \frac{1-\gamma}{\gamma^2}\sqb{ \gamma \delta r- (1-\gamma) V^2 }
        \geq 0. 
    \end{align*}
    This ends the proof that whenever $\mb U$ is an extreme point of $\Rcal$ within the ball $B(\mb x, r)$, the future promised utilities $\mb W(\mb v\mid\mb U)$ stay within the ball $B(\mb x, r)\subseteq \Rcal$.

    It remains to check that \cref{eq:valid_promise,eq:valid_interim_promise} hold for all other utility vectors $\mb U\in\Rcal$. As in the definition of the allocation function, we write $U_i = s_i y_i$ where $s_i\in[0,1]$ for $i\in[n]$, and $\mb y = \sum_{i\in[n]} q_i \Ebb[u_i]\mb e_i + q_0\tilde{\mb y}$ where $\mb q\geq \mb 0$, $\sum_{i\leq n} q_i\leq 1$, and $\tilde{\mb y}$ is an extreme point of $\Rcal$ within the ball $B(\mb x, r)$. In the definition of the allocation function from \cref{eq:def_allocation_function_app}, the only component that is non-constant for $\mb v\in[0,\bar v]^n$ is the contribution of $\mb p(\cdot\mid\tilde{\mb y})$. By linearity,
    \cref{eq:valid_interim_promise} is then a consequence of the definitions \cref{eq:def_allocation_function_app,eq:def_promised_utility_function_app}. We turn to \cref{eq:valid_promise}.
    In the previous part, we showed that $\mb W(\cdot \mid\tilde{\mb y})\in B(\mb x, r)$ for all possible reports. As a result, by linearity this directly shows that the future promised utilities stay in the the corresponding ellipsoid $\Ecal(\mb U)$ defined below:
    \begin{equation*}
        \forall \mb v\in[0,\bar v]^n,\quad \mb W(\mb v\mid\mb U) \in \set{(s_i (q_i \Ebb[u_i] + q_0 x_i) )_{i\in[n]} + \mb z: q_0^2 \sum_{i\in[n]} (s_i z_i)^2 \leq r^2} := \Ecal(\mb U). 
    \end{equation*}
    We can easily show that $\Ecal(\mb U)\subseteq \Rcal$ by construction of the region $\Rcal$. This ends the proof.
\end{proof}

In \cref{lemma:ball_in_region}, we simplified the form of the constant $C$ from \cref{lemma:ball_in_region_no_assumptions} under \cref{assumption:lower_bounded_valuation} which asks that all valuations are lower bounded by some fixed value $\underline v>0$. To give some intuitions, the term $C$ in \cref{lemma:ball_in_region_no_assumptions} essentially measures how close the tentative ball $B(\mb x, r)$ is to the limiting hyperplanes $\{\mb x\in\Rbb^n: x_i =\Ebb[u_i]\}$ for $i\in[n]$. Under \cref{assumption:lower_bounded_valuation}, the following result gives geometrical properties of the optimal set $\Ucal^\star$, which intuitively imply that points in the optimal set $\Ucal^\star$ are sufficiently far from the hyperplanes $\{\mb x\in\Rbb^n: x_i =\Ebb[u_i]\}$ for $i\in[n]$.

\begin{lemma}\label{lemma:achievable_region}
    Under \cref{assumption:lower_bounded_valuation}, we have
    \begin{equation*}
        \Ucal^\star \subseteq \Rbb_+^n \cap \bigcap_{i\in[n]}\set{\mb x: \underline v \sum_{j\neq i} x_j \leq  \bar v (\Ebb[u_i] - x_i) }.
    \end{equation*}
\end{lemma}

\begin{proof}
    We already characterized the extreme points of $\Ucal^\star$ as $\mb 0$ together with the utilities reached by optimal allocation functions $\mb p(\cdot)$ for $\mb\alpha$-weighted objectives for $\mb\alpha\in\Rbb_+^n\setminus\{\mb 0\}$. 

    Now for $i\in[n]$, consider the weights $\mb \alpha$ such that $\alpha_i=\bar v$, and $\alpha_j = \underline v$ for all $j\neq i$. From \cref{assumption:lower_bounded_valuation} the valuations always lie in $[\underline v,\bar v]$, hence, the allocation function that always allocates the resource to agent $i$ is $\mb\alpha$-optimal. This allocation reaches the utility vector $\Ebb[u_i] \mb e_i$. Because this allocation is $\mb\alpha$-optimal, this proves the constraint $\Ucal^\star\subseteq \set{\mb x: \mb\alpha^\top \mb x \leq \bar v \Ebb[u_i]}$.
\end{proof}

We are now ready to give the proof of \cref{lemma:ball_in_region}.

\begin{proof}[of \cref{lemma:ball_in_region}]
    Going back to the proof of \cref{lemma:ball_in_region_no_assumptions}, we aim to improve the upper bound of $\mb\alpha_{i_1}(\mb U)/\mb\alpha_{i_2}(\mb U)$ from \cref{eq:bound_alpha_1/alpha_2} using \cref{lemma:achievable_region}. Up to rescaling, we suppose without loss of generality that $\alpha_{i_1}(\mb U)=\bar v$. Now suppose by contradiction that $\alpha_{i_2}(\mb U)<\underline v$. In particular, for all $j\neq i_1$, we have $\alpha_j(\mb U) \leq \alpha_{i_2}(\mb U)<\underline v$. Further, \cref{lemma:achievable_region}, since $\mb U\in\Rcal\subseteq \Ucal^\star$, we have
    \begin{equation*}
         \underline v \sum_{j\neq i_1} U_j \leq  \bar v (\Ebb[u_{i_1}] - U_{i_1})
    \end{equation*}
    Combining the previous remarks gives,
    \begin{equation*}
        \mb\alpha(\mb U)^\top (\mb U - \Ebb[u_{i_1}]\mb e_{i_1}) \leq (\alpha_{i_2}-\underline v)U_{i_2} + \underline v \sum_{j\neq i_1 } U_j - \bar v(\Ebb[u_{i_1}]-U_{i_1}) \leq (\alpha_{i_2}-\underline v)U_{i_2}.
    \end{equation*}
    We briefly argue that $U_{i_2}>0$. We recall that $\mb U = \mb x + r\mb y $ for some element $\mb y\in S_{n-1}$. Further, because $\mb\alpha(\mb U)\in\Rbb_+^n$, we must also have $\mb y\in \Rbb_+^n$ otherwise $\mb U$ wouldn't be $\mb\alpha$-optimal within $\Rcal$. Because $B(x,r+\delta)\subseteq\Rbb_+^n$ and $r+\delta>0$, we have $U_{i_2}\geq x_{i_2}>0$. This proves that
    $\mb\alpha(\mb U)^\top (\mb U - \Ebb[u_{i_1}]\mb e_{i_1}) < 0.$
    This contradicts the fact that $\mb\alpha(\mb U)$ is the normal of a supporting hyperplane at $\mb U$ of $\Rcal$, which contains in particular $ \Ebb[u_{i_1}]\mb e_{i_1}$. In conclusion, $\alpha_{i_1}(\mb U) / \alpha_{i_2}(\mb U) \leq \bar v / \underline v$. We can now use the same proof as for \cref{lemma:ball_in_region_no_assumptions} with $V := \bar v (2n + \bar v/\underline v)$.
\end{proof}

To understand the optimal policies for some social welfare, it may be useful to focus locally on this direction $\mb\alpha=\mb 1$. In particular, it may be possible to ignore some irrelevant agents in $[n]$. This leads to a stronger local version than the statement given in \cref{lemma:ball_in_region_no_assumptions}.

\begin{lemma}\label{lemma:ball_in_region_optimized}
    Fix $\gamma\in[0,1)$.
    Let $\mb U$ an optimal vector and $\mb x\in \Ucal^\star$ such that for all $i\notin I$, we have $x_i = U_i$. Let $r,\delta>0$ and suppose that there is a partition $ I =I_1\sqcup \ldots \sqcup I_q$ with $|I_s|\geq 2$ for all $s\in[q]$ such that
    \begin{equation}\label{eq:sets_proba_disjoints}
        \forall s\neq s'\in[q],\quad \Pbb\paren{Z_{I_s} = Z_{I_{s'}} = Z>0}=0,
    \end{equation}
    and for all $s\in[q]$,
    \begin{equation}\label{eq:constraint_ball_partition}
       B_{I_s}((x_i)_{i\in I_s}, r+\delta)  \subseteq P_{I_s}\paren{ \Ucal^\star \cap\{\mb y:\forall i\notin I_s,y_i=U_i\} }.
    \end{equation}
    Then, if \cref{eq:constraint_safe_boundary} holds with $C:=4n^2 \bar v^2\paren{1+\frac{\max_{i\in [n]} \Ebb[u_i]}{\min_{s\in[q],i\in I_s}(\Ebb[u_i]\Pbb(Z_{I_s}=Z>0) - x_i -r)}}^2$, we have
    \begin{equation*}
        B(\mb x, r)\cap \{\mb y:\forall i\notin  I ,y_i=x_i\} \subseteq \mb x + \{0\}^{[n]\setminus  I }\otimes \bigotimes_{s\in[q]} B_{I_s}(\mb 0, r)\subseteq \Ucal_\gamma.
    \end{equation*}
\end{lemma}

As a remark, note that while \cref{lemma:trivial_case} essentially shows that only agents in $\tilde I$ need to be carefully treated (all others can be treated perfectly even with $\gamma=0$), \cref{lemma:ball_in_region_optimized} takes all $ I $ agents into consideration. Precisely, we added agents $i\in[n]$ such that $\Pbb(u_i=Z_{-i}>0)>0$. Adding these agents can only help to satisfy the constraints from \cref{lemma:ball_in_region_optimized}. Indeed, by construction there is some other agent $j\neq i$ with $j\in I $ such that $\Pbb(u_i=u_j=Z)>0$ hence we can group $i$ and $j$ in the same set of the partition $ I =I_1\sqcup \ldots \sqcup I_q$. As a result, the set of optimal vectors of $\Ucal^\star$ is perfectly flat along the direction $(\1_{k=j}- \1_{k=i})_{k\in[n]}$, hence we can always fit a ball to satisfy Eq~\eqref{eq:constraint_ball_partition}. In summary, taking agents $ I \setminus \tilde I$ into account within \cref{lemma:ball_in_region_optimized} was without loss. On the other hand, these $ I \setminus \tilde I$ agents bring extra flexibility that can be beneficial for agents $\tilde I$ to satisfy Eq~\eqref{eq:constraint_ball_partition}. This flexibility will be useful in \cref{subsec:discrete_distributions} for discrete utility distributions.

\begin{proof}[of \cref{lemma:ball_in_region_optimized}]
    We define $J$ and $K$ as in \cref{lemma:def_I_properties}. Throughout, we use the characterizations in this lemma. First, note that if $k\in K$ and $m>0$, then $\Pbb(u_k=Z_{-k}>0)\geq \Pbb(u_k=u_{j(k)}=m)\Pbb(\forall j\notin\{k,j(k)\},u_j\leq m)>0$, where we bounded the second term using the definition of $k\in K$. Hence, this implies $k\in I$. Therefore, for any $i\notin  I $ we have either (1) $i\notin \tilde I\cup J\cup K$ or (2) $i\in K$ and $m=0$, or (3) $i\in J$. In all cases, either $x_i=U_i=0$ or $i\in J$.

    Then, for any $\mb V\in\{\mb y:\forall i\notin  I ,y_i=x_i=U_i\}$, and any allocation $\mb p(\cdot)$ that realizes $\mb V$ in the full-information setting, we have
    \begin{align}
        \forall i\notin I \cup J,\quad p_i(\mb u)&=0  \;(a.s.)\label{eq:allocation_1}\\
        \forall i\in J\setminus  I ,\quad p_i(\mb u)&=\1_{\forall j\neq i,u_j\leq m_i} \1_{u_i>0} \;(a.s.).\label{eq:allocation_2}
    \end{align}
    This is because by assumption, $\mb U$ is an optimal vector.
    For convenience, we pose for $i\notin  I $,
    \begin{equation*}
        p_i^{(0)}(\mb u) = \begin{cases}
            0 & i\notin I \cup J\\
            \1_{\forall j\neq i,u_j\leq m_i} \1_{u_i>0} &i\in J\setminus  I .
        \end{cases}
    \end{equation*}
    It will be useful later to view $[n]\setminus  I $ as part of the partition as well hence we pose $I_0:=[n]\setminus  I $. We can easily check that \cref{eq:sets_proba_disjoints} is also satisfied for all $r\neq r'\in\{0,\ldots,q\}$ since
    \begin{equation*}
        \Pbb\paren{Z_{I_0} = Z_{I}>0}\leq \sum_{i\in I_0} \Pbb\paren{u_i = Z>0}=0.
    \end{equation*}

    We first focus on coordinates within $I_s$ for some fixed $s\in[q]$. Let $\mb V\in \Ucal^\star \cap\{\mb y:\forall i\notin I_s,y_i=U_i\}$ and $\mb p(\cdot;\mb V)$ an allocation that realizes $\mb V$ in the full information setting. Because $\mb U$ is an optimal vector, we have
    \begin{equation*}
        \Ebb\sqb{Z_{-I_s} \1_{Z_{-I_s}>Z_{I_s}} } \leq \sum_{i\notin I_s} U_i \leq \Ebb\sqb{Z_{-I_s} \1_{Z_{-I_s}\geq Z_{I_s}}}
    \end{equation*}
    Because of \cref{eq:sets_proba_disjoints}, we have $\Pbb(Z_{-I_s}= Z_{I_s}>0)\leq \sum_{s'\neq s} \Pbb(Z_{I_{s'}}= Z_{I_s}=Z>0)=0$. Then,
    \begin{equation*}
        \sum_{i\notin I_s}  V_i =\sum_{i\notin I_s} U_i = \Ebb\sqb{Z_{-I_s} \1_{Z_{-I_s}\geq Z_{I_s}}}.
    \end{equation*}
    As a result, on the event $\Ecal_s:=\{Z_{-I_s}\geq Z_{I_s}\}$, without loss of generality, we can suppose that $\mb p$ allocates to agents in $[n]\setminus I_s$:
    \begin{equation}\label{eq:no_interactions}
        \forall i\in I_s,\quad \Pbb(p_i(\mb u)>0,\Ecal_s)=0.
    \end{equation}
    In the above, we used the fact that if $Z_{-I_s}=Z_{I_s}=0$, we have $u_i=0$ for all $i\in I_s$. Thus allocating to agents in $I_s$ in that case is useless. We suppose that \cref{eq:no_interactions} will always be satisfied by the allocations of the form $\mb p(\cdot;\mb V)$ from now.
    Note that $1-\Pbb(\Ecal_s)= \Pbb(Z_{I_s}>Z_{-I_s})=\Pbb(Z_{I_s}\geq Z_{-I_s},Z_{I_s}>0) = \Pbb(Z_{I_s}=Z>0)>0$. In the last inequality, we used $\emptyset\subsetneq I_s\subseteq  I $.
    
    \cref{eq:no_interactions} is also a characterization in the following sense. Consider the resource allocation problem with only agents $I_s$ but such that there is no allocation on the event $\Ecal_s$. Denote by $\Ucal_s^\star\in\Rbb_+^{I_s}$ the utility region that can be achieved in this setting. Then,
    \begin{equation}\label{eq:reachable_utilities_subgame}
        \Ucal_s^\star = P_{I_s}\paren{ \Ucal^\star \cap\{\mb y:\forall i\notin I_s,y_i=U_i\} }.
    \end{equation}
    Indeed, \cref{eq:no_interactions} shows the inclusion $\supseteq$ and for the inclusion $\subseteq$, any allocation from the game with agents $I_s$ can be completed on $\Ecal_s$ by allocating the resource to any agent in $\argmax_{i\notin I_s} u_i$. In particular, by assumption, we have $
        B_{I_s}((x_i)_{i\in I_s},r+\delta)\subseteq \Ucal_s^\star.$
    In this setting with $I_s$ agents, we can also define the region $\Ucal_{s}^{NI}$ that can be achieved without reports, that is $\Ucal_{s}^{NI}=Conv(\Ebb[u_i](1-\Pbb(\Ecal_s))\mb e_i,i\in I_s)$. 
    The only difference with a standard resource allocation problem between agents $I_s$ is the constraint on $\Ecal_s$. However, the proof of \cref{lemma:ball_in_region_no_assumptions} still holds in this setting as specified below.
    We construct the utility region
    \begin{equation*}
        \Rcal_s := \set{\mb y\in\Rbb^{I_s}:\mb 0\leq \mb y \leq \mb z, \mb z\in Conv(\Ucal_{s}^{NI} ,B_{I_s}((x_i)_{i\in I_s},r))}.
    \end{equation*}
    We then specify an allocation strategy for extreme points $\mb U$ of $\Rcal_s$ that still belong to the ball $B(\mb x, r)$. We write $\mb U=\mb x+r\mb y$ where $\|\mb y\|=1$ and pose
    \begin{equation*}
        \mb p^{(s)}(\cdot\mid \mb U):=\mb p^{(s)}(\cdot;\mb x+(r+\delta)\mb y)\quad \text{and}\quad \mb\alpha(\mb U) = \mb x-\mb U,
    \end{equation*}
    where $\mb p^{(s)}(\cdot;\mb z)$ is an allocation function that realizes $\mb z\in\Ucal_s^\star$ in the full information setting. We then extend the definition of $\mb p^{(s)}(\cdot\mid \mb U)$ and $\mb W^{(s)}(\cdot\mid\mb U)$ to the complete region $\Rcal_s$ as in \cref{lemma:ball_in_region_no_assumptions} in \cref{eq:def_allocation_function_app,eq:def_promised_utility_function_app}. The proof of \cref{lemma:ball_in_region_no_assumptions} shows that if
    $\frac{r\gamma}{1-\gamma} \geq \delta \geq C_s\frac{1-\gamma}{\gamma r},$
    where $C_s = 4|I_s|^2\bar v^2\paren{1+\frac{\max_{i\in I_s} \Ebb[u_i]}{\min_{i\in I_s}(\Ebb[u_i](1-\Pbb(\Ecal_s)) - x_i -r)}}^2$, then the allocation and promised utility functions constructed are valid on $\Rcal_s$ (they satisfy \cref{eq:valid_promise,eq:valid_interim_promise}). We can check that this is the case since with the choice of parameters, $C\geq\max_{s\in[q]}C_s$.

    We now pose $\Rcal_0:=(x_i)_{i\in I_0}$ and construct an allocation mechanism to reach the utilities
    \begin{equation*}
        \Rcal:= \bigotimes_{0\leq s\leq q} \Rcal_s \supseteq \mb x + \{0\}^{I_0 }\otimes \bigotimes_{s\in[q]} B_{I_s}(\mb 0, r),
    \end{equation*}
    by treating each cluster $I_s$ of agents separately. We pose for all $\mb U\in\Rcal$, $i\in [n]$ and $\mb u\in[0,\bar v]^n$,
    \begin{align*}
        \mb p(\mb u\mid\mb U) &:= \begin{cases}
            p^{(s)}_i(\mb u_{I_s}\mid\mb U_{I_s}) \frac{\1(Z_{I_s}>Z_{-I_s})}{1-\Pbb(\Ecal_s)} & i\in I_s,s\in [q]\\
            p^{(0)}_i(\mb u) & i\in I_0.
        \end{cases}\\
        \mb W(\mb u\mid\mb U) &:= \begin{cases}
            W^{(s)}_i(\mb u_{I_s}\mid\mb U_{I_s}) & i\in I_s,s\in [q]\\
            x_i & i\in I_0 .
        \end{cases}
    \end{align*}
    By design of the events $\Ecal_s$, we still have $\mb p(\mb u\mid\mb U)\in\Delta_n$ (there are no collisions between different partitions). By construction, \cref{eq:valid_promise} holds and \cref{eq:valid_interim_promise} is satisfied for all $i\in I $. It only remains to check that \cref{eq:valid_interim_promise} holds for $i\in I_0=[n]\setminus I $, which is equivalent to checking both \cref{eq:target_met} and the incentive-compatibility constraint \cref{eq:incentive_compatibility}. \cref{eq:target_met} is directly satisfied by \cref{eq:allocation_1,eq:allocation_2}. For the incentive-compatibility, note that for $i\notin  I $ the definition of $p_i^{(0)}(\mb u)$ only uses the information $\1_{u_i>0}$ about $u_i$. Hence the interim promise $P_i^{(0)}(u_i)$ is constant on $(0,\bar v]$. As a result, the allocation is also incentive-compatible for agent $i$ (this is always true if $u_i=0$).

    Altogether, this shows that $\Rcal\subseteq \Ucal_\gamma$, which ends the proof.
\end{proof}

\section{Proofs from \cref{sec:main_tool_lower_bound} to prove lower bounds on the social welfare gap}
\label{sec:tools_construct_upper_bounds}

We now prove \cref{lemma:prove_upper_bounds} that gives tools to prove upper bounds on the achievable region $\Ucal_\gamma$.
To prove this result, we first need to derive some properties of valid mechanisms for the central planner.

Using the promised utility framework, we suppose that we are given a utility region $\Rcal$, allocation functions $\mb p(\cdot\mid\mb U)$ and promised utility functions $\mb W(\cdot\mid\mb U)$ satisfying \cref{eq:valid_promise,eq:valid_interim_promise}. We start by showing that to achieve the utilities from a point $\mb U$ in the boundary of $\Ucal^\star$ in the full information setting, the allocation function needs to discriminate enough between low and high agent utilities. In practice, this will be helpful to derive a lower bound on the range of interim future promises if one aims to reach the boundary of $\Ucal^\star$. We recall the definition of $\tilde I$ as the set of indices that do not satisfy conditions 2(a) nor 2(b) from \cref{lemma:trivial_case}.

\begin{lemma}\label{lemma:Delta_W_boundary}
    Let $i\in \tilde I$. Define $q_i := \min\{v\in [0,\bar v]: \Pbb(u_i \leq v) \geq 1/2 \}$ the median of $\Dcal_i$. Then, there exists $\eta_i>0$ such that the following holds. Fix any allocation function $\mb p:[0,\bar v]^n\to \Delta_n$ that is non-decreasing (that is, for all $i\in[n]$, $P_i(u_i) = \Ebb_{\mb u_{-i}}[p_i(\mb u)]$ is non-decreasing) and such that
        $(\Ebb[u_i p_i(\mb u)])_{i\in[n]}$ is an optimal utility vector.
    Then,
    \begin{equation*}
        \Ebb_{u_i\sim\Dcal_i}[|P_i(u_i)-P_i(q_i)| \min(u_i,q_i)]\geq \eta_i.
    \end{equation*}
\end{lemma}

\begin{proof}
    Fix an index $i\in\tilde I$ satisfying the assumptions. We decompose $[0,\bar v]\setminus \Supp(\Dcal_i)$ as a union of disjoint intervals $(a_0,b_0=\bar v] \cup [a_1=0,b_1) \cup \bigcup_{k\geq 2} (a_k,b_k)$. We note that $u_i\sim\Dcal_i$ is not a constant random variable otherwise posing $m_i=M_i$ would satisfy 2(b), where $u_i=m_i$ almost surely, contradicting $i\in\tilde I$. This implies that $b_1<a_0$. 
    

    Now fix an optimal allocation $p$ for the social welfare. Because it is optimal, with probability one on $\mb u$, we have $p_i(\mb u)>0$ only if $i\in\argmax_{j\in[n]} u_j$. We also recall that $P_i(\cdot)$ is non-decreasing by hypothesis. As a result, for all $u\in \Supp(\Dcal_i)\setminus\{a_k,b_k,k\geq 0\}$ we have
    \begin{equation}\label{eq:constraint_1}
        \Pbb(Z_{-i} < u) \leq P_i(u) \leq \Pbb(Z_{-i} \leq u).
    \end{equation}
    Next, because $P_i$ is non-decreasing, we have $P_i(\bar v) \geq \Pbb(Z_{-i}<a_0)$ and
    \begin{equation}\label{eq:constraint_2}
        \begin{cases}
            \forall u\in[a_k,b_k],\quad \Pbb(Z_{-i} < a_k) \leq P_i(u) \leq \Pbb(Z_{-i} \leq b_k), &k\geq 1\\
            \forall u\in[a_0,\bar v],\quad \Pbb(Z_{-i} < a_0) \leq P_i(u) \leq P_i(\bar v)
        \end{cases}
    \end{equation}
    
    We now consider the problem of minimizing the desired quantity
    \begin{multline*}
        \Ebb_{u_i\sim\Dcal_i}[|P_i(u_i)-P_i(q_i)| \min(u_i,q_i)] \\
        = \Ebb_{u_i}[q_i P_i(u_i)\1_{u_i>q_i}] - \Ebb_{u_i}[u_i P_i(u_i)\1_{u_i<q_i}] + (\Ebb[u_i\1_{u_i<q_i}]-q_i\Pbb(u_i>q_i)) P_i(q_i), 
    \end{multline*}
    under the constraints \cref{eq:constraint_1,eq:constraint_2}. We can directly check that the optimum is achieved by setting $P_i^\star(u)$ equal to the upper bounds from \cref{eq:constraint_1,eq:constraint_2} if $u<q_i$, and set $P_i^\star(u)$ equal to their lower bounds if $u>q_i$. For $q_i$, we have $P_i^\star(q_i) =\Pbb(Z_{-i}< q_i)$ if $\Ebb[u_i\1_{u_i<q_i}]\leq q_i\Pbb(u_i>q_i)$ and $P_i^\star(q_i) = \Pbb(Z_{-i}\leq q_i)$ otherwise. We denote $\eta_i := \Ebb_{u_i\sim\Dcal_i}[|P_i(u_i)-P_i(q_i)| \min(u_i,q_i)]$ the corresponding value of $P_i^\star$. As a result, we obtained that for any optimal allocation function,
    \begin{equation*}
        \Ebb[|P_i(u_i)-P_i(q_i)| \min(u_i,q_i)] \geq \eta_i.
    \end{equation*}

    Note that to prove the desired result, it would suffice to show that $\eta_i>0$. The main point is that this lower bound $\eta_i$ is tight since the interim function $P_i^\star$ can be achieved by an optimal allocation function that either allocates to the agent with index $\argmax_{j\neq i} u_j$ or $i$ (only if $ u_i = Z_{-i}$). Hence, to end the proof it suffices to show that for any allocation function $p$ that is optimal for the social welfare, one has $\Ebb[|P_i(u_i)-P_i(q_i)| \min(u_i,q_i)] >0.$
    Suppose this is not the case by contradiction, hence $P_i(u) = P_i(q_i)$ for $\Dcal_i$-almost all $u\in (0,\bar v]$. We recall that $P_i$ is non-decreasing. Therefore, letting $m_i=\inf \Supp( u_i)\setminus \{0\}$ and $M_i=\sup \Supp(u_i)$, if $M_i > q_i$ we obtain that $P_i$ is constant on the interval $[q_i,M_i)$. Similarly, if $m_i<q_i$ then $P_i$ is constant on $(m_i,q_i]$. In all cases, we showed that $P_i$ is constant on $(m_i,M_i)$. Hence, \cref{eq:constraint_1} implies that
    \begin{equation*}
         \Pbb(Z_{-i}\leq m_i)\geq \lim_{x\to m_i^+}P_i(x) = \lim_{x\to M_i^-}P_i(x) \geq \Pbb(Z_{-i} <M_i).
    \end{equation*}
    Because $m_i\leq M_i$, the previous equation implies $\Pbb(Z_{-i}\in(m_i,M_i))=0$.
    On the other hand, by construction we have $\Pbb( u_i \in\{0\}\cup [m_i,M_i]) = 1$. Since condition 2(a) from \cref{lemma:trivial_case} does not hold because $i\notin\tilde I$, for all $j\neq i$, we have
    \begin{equation*}
        \Pbb(Z_{-i}<M_i) = \prod_{j\neq i}\Pbb(u_j < M_i) \geq \prod_{j\neq i}\Pbb(u_j < u_i) >0.
    \end{equation*}
    Combining this with the previous equation implies $\Pbb(Z_{-i}\leq m_i)>0$. We now fix $j\neq i$. We have
    \begin{align*}
        0=\Pbb(Z_{-i}\in(m_i,M_i)) &\geq \Pbb\paren{Z_{-\{i,j\}} \leq m_i } \Pbb( u_j\in(m_i,M_i)) \geq \Pbb(Z_{-i}\leq m_i) \Pbb( u_j\in(m_i,M_i)).
    \end{align*}
    Therefore, $\Pbb( u_j\in(m_i,M_i))=0$. This holds for all $j\neq i$, hence condition 2(b) of \cref{lemma:trivial_case} holds which contradicts the definition of $i\in \tilde I$. This ends the proof that $\eta_i>0$ and the desired result.
\end{proof}

We note that having $i\in \tilde I$ is necessary to obtain such a constant $\eta_i>0$. Indeed, suppose that condition 2(a) from \cref{lemma:trivial_case} holds. In that case, there are optimal allocations that never allocate the resource to agent $i$ (since there is an agent $j$ that always has at least the same utility), which results in having $P_i(u)=0$ for all $u\in[0,\bar v]$. On the other hand, if condition 2(b) holds, there is an interval $[m_i,M_i]$ such that $\Pbb( u_i \in[m_i,M_i]\cup\{0\})=1$ but $\Pbb(u_j \in(m_i,M_i))=0$ for all $j\neq i$. Then, taking into consideration the reports of agent $i$ is not really necessary: we can safely allocate the resource to agent $i$ if and only if $Z_{-i} \leq m_i$ and $u_i>0$ (if $u_i=0$, then truthful reporting is always an equilibrium for agent $i$). This optimal allocation then yields a constant interim allocation probability $P_i(u) = \Pbb(Z_{-i} \leq m_i)$ for all $u\in(0,\bar v]$.

We now use \cref{lemma:Delta_W_boundary} to derive a the lower bound on the range of the interim future promise $W_i$ even when the realized utility is not on the boundary of $\Ucal^\star$, but somewhat close to it. Having the realized utility vector $\mb U$ close to the boundary is necessary, otherwise, for instance if $\mb U\in\Ucal^{NI}$, we can use a constant allocation function $\mb p(\cdot)$ so that $W_i(v)$ is constant for $v\in [0,\bar v]$ for all $i\in[n]$.

\begin{lemma}\label{lemma:lower_bound_Delta_W}
    Fix $\gamma\in(0,1]$ and an index $i\in \tilde I$. Define $q_i\in[0,\bar v]$ as in \cref{lemma:Delta_W_boundary}. Then, there exists $\delta_i,\eta_i>0$ ($\eta_i$ is the same as in \cref{lemma:Delta_W_boundary}) such that the following holds. For any allocation function $\mb p:[0,\bar v]^n \to\Delta_n$ that is non-decreasing (that is, for all $i\in[n]$ $P_i:u_i\mapsto \Ebb_{\mb u_{-i}}[p_i(\mb u)]$ is non-decreasing) and such that
    \begin{equation*}
        \max_{\mb x\in\Ucal^\star} \mb 1^\top \mb x - \mb 1^\top  (\Ebb[u_i p_i(\mb u)])_{i\in[n]} \leq \delta_i,
    \end{equation*}
    the interim promised utility function $W_i(\cdot)$ for $i\in[n]$ constructed via \cref{eq:formula_interim_promise} using the interim allocation functions $P_i$ is non-increasing and satisfies
    \begin{equation*}
        \Ebb_{u_i\sim\Dcal_i}|W_i(u_i) - W_i(q_i)| \geq \frac{ 1-\gamma}{2\gamma}\eta_i.
    \end{equation*}
\end{lemma}

\begin{proof}
    The fact that the interim promise function $W_i(\cdot\mid\mb U)$ is non-increasing is direct from the definition in \cref{eq:formula_interim_promise}. We omit the terms $\mid\mb U$ in the rest of the proof for readability. From the same definition, for any $0\leq a\leq b\leq \bar v$,
    \begin{align*}
        W_i(a)-W_i(b) &= \frac{1-\gamma}{\gamma} \paren{\int_0^{b} ( P_i(b) - P_i(v))dv - \int_0^{a} ( P_i(a) - P_i(v))dv}\\
        &=\frac{1-\gamma}{\gamma} \paren{(P_i(b) - P_i(a))a + \int_{a}^{b} ( P_i(b) - P_i(v))dv} \geq \frac{1-\gamma}{\gamma} (P_i(b) - P_i(a))a.
    \end{align*}
    As a result,
    \begin{align}
        \Ebb|W_i(u_i) - W_i(q_i)| &= \Ebb[(W_i(q_i)-W_i(u_i))\1_{u_i\geq q_i}] +  \Ebb[(W_i(u_i)-W_i(q_i))\1_{u_i\leq q_i}]\\
        &\geq \frac{1-\gamma}{\gamma}\Ebb\sqb{|P_i(u_i) - P_i(q_i)|\min(u_i,q_i)}.\label{eq:from_W_to_int}
    \end{align}
    Hence, we aim to lower bound the expectation on the right-hand side. 
    First, recall that $         \max_{\mb x\in\Ucal^\star} \mb 1^\top \mb x = \Ebb [Z] = \Ebb \max(u_i,Z_{-i})$,
    and that the maximum is attained for any first-best allocation function. Next, for $u\in[0,\bar v]$, we have
    \begin{equation*}
         \mb 1^\top  (\Ebb[u_i p_i(\mb u)])_{i\in[n]} 
        \leq \Ebb\sqb{u_i p_i(\mb u) +  Z_{-i}(1-p_i(\mb u)) } =\Ebb\sqb{  u_i P_i(u_i) +  Z_{-i}(1-p_i(\mb u)) }.
    \end{equation*}
    Now for any value $u\in[0,\bar v]$, let $z_i(u) = \inf\{v: \Pbb(Z_{-i} \geq v) \leq 1- P_i(u)\}$. In particular, since $P_i(\cdot)$ is non-decreasing, so is $z_i(\cdot)$. We can check that
    \begin{equation*}
        \Ebb_{\mb u_{-i}}[Z_{-i}(1-p_i(\mb u))] \leq \Ebb_{\mb u_{-i}}[Z_{-i}\1_{Z_{-i}>  z_i(u_i)}] + z_i(u_i) \sqb{ \Pbb(Z_{-i} \geq z_i(u_i)) - (1-P_i(u_i))  }.
    \end{equation*}
    Because of the second term in the right-hand side, we define a variable $B_i\sim \Bcal(\Pbb(Z_{-i} \geq z_i(u_i)) - (1-P_i(u_i)))$ which is independent from $\mb u_{-i}$. Putting the previous equations together yields
    \begin{multline}\label{eq:base_ineq_eta}
        \max_{\mb x\in\Ucal^\star} \mb 1^\top \mb x - \mb 1^\top  (\Ebb[u_i p_i(\mb u)])_{i\in[n]}
        \geq\\
        \Ebb[\max(u_i ,Z_{-i}) - Z_{-i}\1_{Z_{-i}>z_i(u_i)} - u_i \1_{Z_{-i}<z_i(u_i)}- \1_{Z_{-i}=z_i(u_i)}(Z_{-i} B_i + u_i (1-B_i) )] =: A(P_i).
    \end{multline}
    This inequality is tight in the following sense. For any non-decreasing function $P_i$, there is a non-decreasing allocation function $\mb p(\cdot)$ with interim allocation function $P_i$ for agent $i$, and
    \begin{equation*}
        \max_{\mb x\in\Ucal^\star} \mb 1^\top \mb x - \mb 1^\top  (\Ebb[u_i p_i(\mb u)])_{i\in[n]} = A(P_i).
    \end{equation*}
    This allocation can directly be obtained from the definition of $A(P_i)$: when $Z_{-i}<z_i(u_i)$ we allocate to agent $i$, when $Z_{-i}>z_i(u_i)$ we allocate to some agent in $\argmax_{j\neq i} u_j$, and if $Z_{-i}=z_i(u_i)$ we allocate to one of these two cases depending on the value of $B_i$. 
    In summary, for any $\delta\geq 0$,
    \begin{multline}\label{eq:simplification_to_interim_allocations}
        \set{P_i(\cdot):\exists \text{ non-decreasing }\mb p(\cdot), \max_{\mb x\in\Ucal^\star} \mb 1^\top \mb x - \mb 1^\top  (\Ebb[u_i p_i(\mb u)])_{i\in[n]} \leq \delta, P_i(\cdot) = \Ebb_{\mb u_{-i}}[p_i(\cdot,\mb u_{-i})] }\\
        = \set{\text{non-decreasing }P_i:[0,\bar v]\to[0,1],\;A(P_i)\leq \delta}.
    \end{multline}
    As a result, we now only focus on non-decreasing interim allocation functions $P_i$ satisfying $A(P_i)\leq \delta$. The set of non-decreasing functions on $[0,\bar v]\to[0,1]$ is closed for the product measure (point-wise convergence) and so is the set of functions satisfying the constraint $A(P_i)\leq \eta$ by the dominated convergence theorem. Further, Tychonov's theorem implies that the set of functions $[0,\bar v]\to[0,1]$ is compact for the product measure. This therefore implies that the following optimization problem
    \begin{equation*}
        \inf_{\substack{\text{non-decreasing }P_i:[0,\bar v]\to[0,1] \\
        A(P_i)\leq \delta}} \Ebb_{u_i\sim\Dcal_i}[|P_i(u_i)-P_i(q_i)| \min(u_i,q_i)] =: B(\delta)
    \end{equation*}
    achieves its minimum, and further that $B(\delta)\to B(0)$ as $\delta\to 0$. Note that $\delta=0$ corresponds to the case when the allocation is exactly optimal for the social welfare. This is the case that was covered by \cref{lemma:Delta_W_boundary}, which shows that $B(0) \geq \eta_i$ for some constant $\eta_i>0$ (that only depends on $i$, and the distributions $\Dcal_1,\ldots,\Dcal_n$). In particular, there exists $\delta_i>0$ such that $B(\delta_i) \geq \eta_i/2$. Through \cref{eq:simplification_to_interim_allocations}, this exactly shows that for any non-decreasing allocation function $p$ satisfying the constraint $ \mb 1^\top \mb x - \mb 1^\top  (\Ebb[u_i p_i(\mb u)])_{i\in[n]} \leq \delta_i$, we have
    \begin{equation*}
        \Ebb\sqb{|P_i(u_i) - P_i(q_i)|\min(u_i,q_i)} \geq \frac{\eta_i}{2}.
    \end{equation*}
    Together with \cref{eq:from_W_to_int} this ends the proof.
\end{proof}

\cref{lemma:lower_bound_Delta_W} gives a lower bound on the range required for the interim future promised utilities. In fact it shows a stronger statement about their dispersion which is necessary for our proofs. We now make use of these range lower bounds to show that some regions of $\Ucal^\star$ cannot be achieved by $\Ucal_\gamma$. Intuitively, because future promised utilities need to remain within the achievable region $\Ucal_\gamma$ (see \cref{eq:valid_promise}) if interim promised utilities are somewhat scattered, this gives a local upper bound on the maximum curvature of $\Ucal_\gamma$.

\begin{proof}[of~\cref{lemma:prove_upper_bounds}]
    We start by specifying $c_1$ and $c_2$.
    By the second characterization of \cref{lemma:trivial_case}, we have $\tilde I\neq\emptyset$, otherwise, $\Ucal_0$ would already contain an optimal vector. Fix such an index $i\in \tilde I$. Let $\delta_i,\eta_i>0$ be the constants for which \cref{lemma:lower_bound_Delta_W} holds.
    We pose $c_1 := \delta_i$ and $c_2:=\eta_i^2/16$.

    Next, fix some parameters $\gamma,\mb x,r$ and $\delta$ satisfying the assumptions of the lemma. 
    Since $\Ucal^\star \setminus B^\circ(\mb x, r)$ is compact, so is the component $\Ccal$. By contradiction, we suppose that $\Ucal_\gamma \cap(\Ccal\setminus B^\circ(\mb x,r+\frac{1-\gamma}{\gamma}\delta))\neq \emptyset$.   
    We consider a solution $\mb U_0$ to the following optimization problem
    \begin{equation}\label{eq:problem_optim}
        \max_{\mb y\in \Ccal\cap \Ucal_\gamma} \|\mb y-\mb x\|.
    \end{equation}
    The maximum is attained because both $\Ccal$ and $\Ucal_\gamma$ are compact. Further, by assumption,
    \begin{equation*}
        \|\mb U_0-\mb x\| \geq r+\frac{1-\gamma}{\gamma}\delta.
    \end{equation*}
    Because $\mb U_0\in\Ucal_\gamma$, there exists a mechanism for the central planner that reaches this utility. Using the promised utility framework, we fix a non-decreasing allocation function $\mb p(\cdot\mid\mb U)$ and a promised utility function $\mb W(\cdot\mid \mb U)$ for all $\mb U\in\Ucal_\gamma$ that satisfy \cref{eq:valid_promise} (promises stay within $\Ucal_\gamma$) and \cref{eq:valid_interim_promise} (interim promises $W_i(v_i\mid\mb U)$ as defined in \cref{eq:formula_interim_promise}).

    \paragraph{Step 1.} Let $\mb U_1 := (\Ebb[u_i p_i(\mb v\mid \mb U_0)])_{i\in[n]} \in\Ucal^\star$. As a first step, we aim to show that
    \begin{equation}\label{eq:first_step_upper_bound}
        \mb 1^\top \mb U_1 \geq \max_{\mb z\in \Ucal^\star} \mb 1^\top \mb z - \delta_i.
    \end{equation}
    By \cref{eq:valid_promise} and the convexity of $\Ucal_\gamma$, we have
    \begin{equation}\label{eq:def_U_2}
        \mb U_2:=\Ebb[\mb W(\mb u\mid \mb U_0)] = \mb U_0 + \frac{1-\gamma}{\gamma}(\mb U_0 - \mb U_1) \in \Ucal_\gamma.
    \end{equation}
    In the equality, we used the form of the interim promises \cref{eq:formula_interim_promise}, or equivalently, \cref{eq:target_met}. 
    We next write $\mb U_0 = \mb x + r_0 \mb y_0$ where $r_0\geq r+\frac{1-\gamma}{\gamma}\delta$ and $\mb y_0\in S_{n-1}$. We argue that it suffices to show that $\mb y_0^\top (\mb U_1- \mb U_0) \geq 0$ to obtain \cref{eq:first_step_upper_bound}. Indeed, if this holds, then for any $\mb U\in[\mb U_0,\mb U_1]$, one has
    \begin{equation*}
        \|\mb U-\mb x\| \geq \mb y_0^\top (\mb U-\mb x) = \mb y_0^\top (\mb U-\mb U_0) + r_0 \geq r_0 \geq r.
    \end{equation*}
    Because $\Ucal^\star$ is convex, $[\mb U_0,\mb U_1]\subseteq\Ucal^\star$. Combining this with the previous equation shows that $[\mb U_0,\mb U_1] \in \Ucal^\star \setminus B^\circ(\mb x, r)$. Hence $\mb U_1$ belongs to the same component $\Ccal$ as $\mb U_0$. By assumption, $\Ccal\subseteq\{\mb y:\mb 1^\top \mb y\geq \max_{\mb z\in \Ucal^\star} \mb 1^\top \mb z - c_1\}$ and we posed $c_1=\delta_i$. This proves \cref{eq:first_step_upper_bound}.
    We now show
    \begin{equation}\label{eq:increase_radius_to_U_01}
        \mb y_0^\top (\mb U_1- \mb U_0) \geq 0.
    \end{equation}
    By contradiction suppose that this does not hold, then \cref{eq:def_U_2} implies that $\mb y_0^\top (\mb U_2- \mb U_0) > 0$. The same arguments as above then show that for any $\mb U\in[\mb U_0,\mb U_2]$,
    \begin{equation*}
        \|\mb U - \mb x\| \geq  \mb y_0^\top (\mb U-\mb U_0) + r_0 > r_0.
    \end{equation*}
    The previous arguments further show that $\mb U_2\in\Ccal$. Recall from \cref{eq:def_U_2} that $\mb U_2\in\Ucal_\gamma$. Hence, the previous inequality contradicts the definition of $\mb U_0$ as a maximizer of the problem in \cref{eq:problem_optim}.
    
    \paragraph{Step 2.} We are now ready to apply \cref{lemma:lower_bound_Delta_W} to the allocation $\mb p(\cdot\mid\mb U_0)$ since the resulting utility vector $\mb U_1$ satisfies the necessary conditions as checked in Step 1. 
    \comment{For more readability, we will drop the symbols $|\mb U_0$ in the rest of this proof for the allocations $\mb p(\cdot\mid\mb U_0)$, the future promised utilities $\mb W(\cdot\mid\mb U_0)$ as well as the interim promised utilities $W_i(u_i\mid\mb U_0)$.}
    We obtain
    \begin{equation}\label{eq:guarantee_dispersion_delta_W}
        \Ebb_{u_i\sim\Dcal_i}|W_i(u_i\mid \mb U_0) - W_i(q_i\mid\mb U_0)| \geq \frac{ 1-\gamma}{2\gamma}\eta_i,
    \end{equation}
    where $q_i := \min\{v\in [0,\bar v]: \Pbb(u_i \leq v) \geq 1/2 \}$. Define $p_B = \frac{1/2 - \Pbb(u_i<q_i)}{\Pbb(u_i=q_i)}$, with the convention $0/0=0$ and let $B\sim \Bcal(p_B)$ be a Bernoulli variable independent from all other variables. 
    Define
    \begin{align*}
        \mb U_+ &= 2\Ebb_{\mb u}[\mb W(u_i,\mb u_{-i} \mid \mb U_0)\1_{u_i<q_i} + B \mb W(u_i,\mb u_{-i} \mid \mb U_0)\1_{u_i=q_i}]\\
        \mb U_- &= 2\Ebb_{\mb u}[\mb W(u_i,\mb u_{-i} \mid \mb U_0)\1_{u_i>q_i} + (1-B) \mb W(u_i,\mb u_{-i} \mid \mb U_0)\1_{u_i=q_i}].
    \end{align*}
    From \cref{eq:valid_promise}, by convexity of $\Ucal_\gamma$, and because $\Pbb(u_i<q_i) + \Pbb(u_i=q_i,B=1) = 1/2$, both utility vectors satisfy $\mb U_+,\mb U_-\in\Ucal_\gamma$.
    The goal of this step is to show that either $\mb U_+$ or $\mb U_-$ have a better objective than $\mb U_0$ for the optimization problem \cref{eq:problem_optim}, to reach a contradiction. We first give a few properties about $\mb U_-$ and $\mb U_+$.
    We note that
    \begin{equation}\label{eq:U_2_midpoint}
        \mb U_2 = \Ebb[\mb W(\mb u\mid\mb U_0)] = \frac{\mb U_+ + \mb U_-}{2}.
    \end{equation}
    Next, recalling that the interim allocation function $W_i(\cdot\mid\mb U_0)$ is non-increasing, we have
    \begin{align*}
        &\|\mb U_+ - \mb U_-\| \\
        &\geq (\mb U_+)_i - (\mb U_-)_i \\
        &= (\mb U_+)_i - W_i(q_i\mid\mb U_0) +   W_i(q_i\mid\mb U_0) - (\mb U_-)_i\\
        &= 2\Ebb_{u_i}[(W_i(u_i \mid \mb U_0) - W_i(q_i\mid\mb U_0))\1_{u_i>q_i}]  + 2\Ebb_{u_i}[(W_i(q_i \mid \mb U_0) - W_i(u_i \mid \mb U_0))\1_{u_i<q_i}]\\
        & = 2\Ebb_{u_i}|W_i(u_i\mid \mb U_0) - W_i(q_i\mid\mb U_0)|.
    \end{align*}
    Combining with \cref{eq:guarantee_dispersion_delta_W} yields
    \begin{equation}\label{eq:U+_and_U-_far_appart}
        \|\mb U_+ - \mb U_-\| \geq  \frac{ 1-\gamma}{\gamma}\eta_i,
    \end{equation}
    We now lower bound $\|\mb U_2-\mb x\|$. By \cref{eq:def_U_2}, and recalling that $\mb U_0 = \mb x+r_0\mb y_0$, we obtain
    \begin{equation*}
        \mb y_0^\top (\mb U_2-\mb x) = r_0 -\frac{1-\gamma}{\gamma} \mb y_0^\top(\mb U_1-\mb U_0).
    \end{equation*}
    Within Step 1, we also showed that $\mb U_1\in\Ccal \subseteq B(\mb x,r+\delta)$, where in the last step we used one of the assumptions. Hence, $\mb y_0^\top (\mb U_1-\mb U_0) = \mb y_0^\top (\mb U_1-\mb x) - r_0 \leq (r+\delta)-r_0 \leq \delta.$
    Putting the two previous equations together yields
    \begin{equation}\label{eq:lower_bound_distance_U_2_to_x}
        \|\mb U_2-\mb x\| \geq \mb y_0^\top (\mb U_2-\mb x) \geq r_0-\frac{1-\gamma}{\gamma}\delta \geq r,
    \end{equation}
    where in the last inequality we used the assumption on $r_0$.
    We now argue that $\mb U_2\in\Ccal$. Indeed, since $\mb U_0,\mb U_2\in\Ucal^\star$, by convexity $[\mb U_0,\mb U_2]\subseteq\Ucal^\star$. Next, for any $\mb U = (1-\theta)\mb U_2 + \theta \mb U_0$ for $\theta\in[0,1]$,
    \begin{align*}
        \mb y_0^\top(\mb U-\mb x) = \mb y_0^\top(\mb U_2-\mb x) + \theta \mb y_0^\top (\mb U_0-\mb U_2)
        \geq r + \frac{1-\gamma}{\gamma} \theta \mb y_0^\top (\mb U_1-\mb U_0) \geq r.
    \end{align*}
    In the first inequality, we used \cref{eq:lower_bound_distance_U_2_to_x}, and in the second we used \cref{eq:increase_radius_to_U_01}. As a result, using similar arguments as before, we have $[\mb U_0,\mb U_2]\in \Ucal^\star\setminus B^\circ(\mb x, r)$, hence $\mb U_0\in\Ccal$ implies $\mb U_2\in \Ccal$ as well.
    Next, by \cref{eq:U_2_midpoint} we can write $\mb U_+ = \mb U_2 + \mb\Delta$ and $\mb U_-=\mb U_2-\mb\Delta$ where $\mb\Delta:=(\mb U_+-\mb U_-)/2$. 
    
    Suppose that $(\mb U_2-\mb x)^\top \mb \Delta \geq 0$. Then,
    \begin{align*}
        \|\mb U_+-\mb x\|^2  &= \|\mb U_2 - \mb x\|^2 + \|\mb\Delta\|^2 + 2(\mb U_2-\mb x)^\top \mb \Delta \\
        &\geq \|\mb U_2 - \mb x\|^2 + \paren{\frac{1-\gamma}{2\gamma}\eta_i}^2\\
        &\geq r_0^2 + \frac{1-\gamma}{\gamma}\paren{\frac{1-\gamma}{4\gamma}(\eta_i^2 + 4\delta^2) -2r_0 \delta} > r_0^2 + \frac{1-\gamma}{\gamma}\paren{\frac{1-\gamma}{4\gamma}\eta_i^2 -2r_0 \delta}
    \end{align*}
    where in the second inequality we used \cref{eq:U+_and_U-_far_appart} and in the third inequality we used \cref{eq:lower_bound_distance_U_2_to_x}. We now recall that by hypothesis $\delta = \min( c_2 \frac{1-\gamma}{\gamma r},r)$. Hence, from the choice of $c_2$,
    $2r_0 \delta \leq 4r \delta \leq \frac{1-\gamma}{4\gamma}\eta_i^2.$
    Putting the two previous inequalities together, we obtain $\|\mb U_+-\mb x\| > \|\mb U_0-\mb x\|=r_0$. We recall that $\mb U_2\in\Ccal$ and by assumption $(\mb U_2-\mb x)^\top (\mb U_+-\mb U_2)\geq 0$. Since we also have $\mb U_+\in \Ucal^\star$, the same arguments as in Step 1 show that $\mb U_+\in\Ccal$. This contradicts the definition of $\mb U_0$ as the maximizer of the problem in \cref{eq:problem_optim}.

    In the other case when $(\mb U_2-\mb x)^\top \mb \Delta \leq 0$, the same arguments show that $\|\mb U_- -\mb x\| > \|\mb U_0-\mb x\|$ and $\mb U_-\in\Ccal$, reaching the same contradiction. This ends the proof of the desired result:
    \begin{equation}\label{eq:desired_property}
        \Ucal_\gamma \cap \paren{\Ccal \setminus B^\circ\paren{\mb x,r+\frac{1-\gamma}{\gamma}\delta}} = \emptyset.
    \end{equation}

    \paragraph{Proof of the second claim.}
    We now prove the second claim of the lemma as a simple consequence of \cref{eq:desired_property}. First, note that because $\Ucal_\gamma$ is compact, the maximum in $\max_{\mb y\in\Ucal_\gamma} \mb d^\top (\mb y-\mb x)$ is well defined. Fix $\mb U = \mb x+\|\mb U-\mb x\| \mb d \in\Ccal$ such a maximizer. Consider any point $\mb y\in\Ucal_\gamma$ such that $\mb d^\top (\mb y-\mb x) \geq \|\mb U-\mb x\|$. Then, by convexity of $\Ucal^\star$, we have $[\mb U,\mb y]\subseteq\Ucal^\star$. Further,
    \begin{equation*}
        (\mb U-\mb x)^\top (\mb y-\mb U) = \|\mb U-\mb x\| \mb d^\top (\mb y-\mb x) - \|\mb U-\mb x\|^2 \geq 0.
    \end{equation*}
    Then, the same arguments as in Step 1 show that since $\mb U\in\Ccal$, we must also have $\mb y\in\Ccal$. Because we have $\mb y\in\Ucal_\gamma$, \cref{eq:desired_property} implies that $\mb y\in B^\circ(\mb x,r+\frac{1-\gamma}{\gamma}\delta)$. In particular,
    \begin{equation*}
        \mb d^\top(\mb y-\mb x) \leq \|\mb y-\mb x\| \,\|\mb d\| < \paren{ r+ \frac{1-\gamma}{\gamma}\delta } \|\mb d\|.
    \end{equation*}
    This ends the proof.
\end{proof}

In some cases, it may be useful to design lower bounds based on a subproblem instead of the full resource allocation problem as stated in \cref{lemma:prove_upper_bounds}. Intuitively, we may want to focus on the impact of a single agent $i$, which corresponds to a 2-agent setting in which one has utility $u_i$ and the other has utility $Z_{-i}$. In practice, this setup will be very useful, and we specify the corresponding lower bounds in the result below. This is a simple consequence of \cref{lemma:prove_upper_bounds}.  For any $\mb x=(x_i,x_{-i})\in\Rbb^2$ and $r>0$, we define the sets
\begin{align*}
    B(\mb x,r;i) &:= \{\mb y\in\Rbb^n: (y_i-x_i)^2 + (\mb 1_{-i}^\top \mb y_{-i} - x_{-i})^2 \leq r^2\}\\
    B(\mb x,r;i) &:= \{\mb y\in\Rbb^n: (y_i-x_i)^2 + (\mb 1_{-i}^\top \mb y_{-i} - x_{-i})^2 < r^2\}.\\
\end{align*}

\begin{lemma}\label{lemma:upper_bounds_2D}
    Fix an index $i\in \tilde I$.
    There exist two constants $c_1,c_2>0$ such that the following holds. 
    
    For $\gamma\in(0,1)$, $r>0$, and $(x_i,x_{-i})\in\Rbb^2$, suppose that $\Ucal^\star \setminus B^\circ(\mb x,r;i) $ has a connected component $\Ccal\subseteq \{\mb y: \mb 1^\top \mb y \geq \max_{\mb z\in \Ucal^\star} \mb 1^\top \mb z - c_1\}$. If
    \begin{equation*}
        \Ccal \subseteq B(\mb x,r+\delta;i),\quad \delta = \min\paren{c_2\frac{1-\gamma}{\gamma r},r},
    \end{equation*}
    then we have
    \begin{equation*}
        \Ucal_\gamma \cap \paren{\Ccal \setminus B^\circ\paren{\mb x,r+\frac{1-\gamma}{\gamma}\delta;i}} = \emptyset.
    \end{equation*}
    In particular, for any $\mb U\in \Ccal$, writing $(U_i,\mb 1_{-i}^\top \mb U_{-i}) = \mb x + \|(U_i,\mb 1_{-i}^\top \mb U_{-i}) -\mb x\|(d_i,d_{-i})$,
    \begin{equation*}
        \max_{\mb y\in\Ucal_\gamma} d_i (y_i-x_i) + d_{-i} (\mb 1_{-i}\mb y_{-i}^\top -x_{-i}) <  \paren{ r+ \frac{1-\gamma}{\gamma}\delta } \|\mb d\|.
    \end{equation*}
\end{lemma}

\begin{proof}
    In the proof of \cref{lemma:prove_upper_bounds}, we first used the characterization from \cref{lemma:trivial_case} to show that there exists some index $i\in \tilde I$ ($\tilde I\neq\emptyset$). From then, the only incentive-compatibility constraint that was used within the proof was that of agent $i$. Precisely, it only uses \cref{eq:valid_interim_promise} for agent $i$ to use the fact that the interim promised utility is given according to \cref{eq:formula_interim_promise} for which \cref{lemma:Delta_W_boundary} gives bounds.

    As a result, the lower bounds also hold in a setting in which agents $j\neq i$ are non-strategic, that is, the central planner observes all utilities $u_j$ for $j\notin i$. If we denote by $\tilde \Ucal_{\gamma,i}$ the utility region that can be achieved in this setting, we clearly have $\Ucal_\gamma\subseteq \tilde \Ucal_{\gamma,i}$ since there are fewer constraints. From the perspective of maximizing the total utility, this corresponds to the setting in which there are effectively two agents, agent $i$ with the same utility distribution $u_i\sim\Dcal_i$ and an agent $-i$ that has utilities $u_{-i}:=Z_{-i}$. We denote by $\Dcal_{-i}$ its distribution and introduce the linear mapping
    \begin{equation*}
        \Phi:\mb U\in \Rbb^n \mapsto (U_i,\mb 1_{-i}^\top \mb U_{-i}).
    \end{equation*}
    Let $\Ucal_i^*$ and $\Ucal_{\gamma,i}$ be the regions achievable with full information, and in the $\gamma$-discounted strategic setting for this 2-agent allocation problem. We first show that $\Ucal_i^\star = \Phi(\Ucal^\star).$
    Since these are both convex sets, it suffices to focus on extreme points: by definition of $\Dcal_{-i}$, for any $\mb d\in\Rbb_+^2\setminus \{\mb 0\}$,
    \begin{equation*}
        \max_{\mb x\in\Ucal_{\gamma,i}}\mb d^\top \mb x = \Ebb\max\paren{d_i u_i,d_{-i}Z_{-i}} = \max_{\mb x\in\Ucal^\star} (d_i,d_{-i}\mb 1_{-i})^\top \mb x = \max_{\mb x\in\Ucal^\star} \mb d^\top \Phi(\mb x).
    \end{equation*}
    Importantly, we also have
    \begin{equation}\label{eq:inclusion_new_game}
        \Phi(\Ucal_\gamma)\subseteq \Phi(\tilde\Ucal_{\gamma, i})=\Ucal_{\gamma,i}.
    \end{equation}
    The first inequality is immediate from $\Ucal_\gamma\subseteq\tilde \Ucal_{\gamma,i}$ and the equality comes from the definition of $\Dcal_{-i}$ and $\Phi$. Before applying the proof of \cref{lemma:prove_upper_bounds}, we only need to check that $\Dcal_i$ still does not satisfy the conditions 2(a) nor 2(b) from \cref{lemma:trivial_case} in the new 2-agent setting for the direction $\mb d:=(1,1)$. Using the assumption $i\in \tilde I$, for all $j\in[n]$ we have $\Pbb(u_i >  u_j)>0$ hence since $u_i$ and $u_j$ are independent, there exists $m_j$ such that $\Pbb(u_i>m_j),\Pbb(u_j\leq m_j)>0$. Then,
    \begin{align*}
        \Pbb(u_i>u_{-i}) &\geq \Pbb(u_i>\max_{j\neq i}m_j,\forall j\neq i, u_j \leq m_j) = \prod_{j\neq i}\Pbb( u_j \leq m_j)\cdot \min_{j\neq i}\Pbb(u_i >m_j)>0.
    \end{align*}
    Hence 2(a) does not hold. Now suppose by contradiction that 2(b) holds and let $[m_i,M_i]$ the guaranteed interval such that $\Pbb(u_i \in[m_i,M_i]\cup\{0\})=1$ and
    \begin{equation*}
        0=\Pbb(u_{-i} \in(m_i,M_i)) = \Pbb(Z_{-i} \in (m_i ,M_i)).
    \end{equation*}
    In the proof of \cref{lemma:trivial_case} (within the proof of 1.$\Rightarrow$ 2.), we showed that this implies $\Pbb(u_j\in(m_i,M_i))=0$ for all $j\neq i$, which contradicts the hypothesis $i\in \tilde I$.

    Therefore, we are ready to apply the proof of \cref{lemma:prove_upper_bounds} for the 2-agent game for direction $\mb 1=(1,1)$ and agent $i$. Indeed, we can check that in the 2-agent setting, $B(\mb x, r) = \Phi(B(\mb x, r;i))$ and similarly for their open counterparts. Hence, using $\Ucal_i^\star = \Phi(\Ucal^\star)$,
    \begin{equation*}
        \Ucal_i^\star \setminus B^\circ(\mb x, r) = \Phi(\Ucal^\star \setminus B^\circ(\mb x, r;i)),
    \end{equation*}
    and as a result $\Phi(\Ccal)$ is a connected component of $\Ucal_i^\star \setminus B^\circ(\mb x, r)$, which satisfies
    \begin{align*}
        \Phi(\Ccal)&\subseteq \Phi(\{\mb y: \mb 1^\top \mb y \geq \max_{\mb z\in \Ucal^\star} \mb 1^\top \mb z - c_1\}) = \{\mb y\in\Rbb^2:\mb 1^\top \mb y \geq \max_{\mb z\in \Ucal_i^\star} \mb 1^\top \mb z - c_1\},\\
        \Phi(\Ccal)&\subseteq \Phi(B(\mb x,r+\delta;i)) = B(\mb x, r+\delta).
    \end{align*}
    Thus, the proof of \cref{lemma:prove_upper_bounds} gives
    \begin{equation*}
        \Phi(\Ucal_\gamma)\cap \paren{\Phi(\Ccal) \setminus B^\circ\paren{\mb x,r+\frac{1-\gamma}{\gamma}\delta}} \subseteq \Ucal_{\gamma,i} \cap \paren{\Phi(\Ccal) \setminus B^\circ\paren{\mb x,r+\frac{1-\gamma}{\gamma}\delta}} = \emptyset,
    \end{equation*}
    where in the inclusion we used \cref{eq:inclusion_new_game}. This gives the desired result. The second claim is a direct translation of the second claim from \cref{lemma:prove_upper_bounds} that we obtained on this 2-agent game.  
\end{proof}

To simplify the use of \cref{lemma:ball_in_region_no_assumptions,lemma:prove_upper_bounds}, we give simple equivalent computational constraints to prove that a ball belongs to $\Ucal^\star$, directly in terms of the distributions $\Dcal_1,\ldots,\Dcal_n$.

\begin{lemma}\label{lemma:on_balls}
    Let $\mb x$ and $r>0$. $B(\mb x, r)\subseteq \Ucal^\star$ if and only if $\min_{i\in[n]} x_i\geq r$ and
    \begin{equation*}
        \forall \mb\alpha\in\Rbb_+^n\setminus\{\mb 0\},\quad \Ebb\sqb{\max_{i\in[n]}\alpha_i u_i} \geq \mb\alpha^\top\mb x + \|\mb\alpha\|r. 
    \end{equation*}
\end{lemma}

\begin{proof}
    Any convex set $C$ is the intersection of its supporting hyperplanes. Hence,
    \begin{align*}
        B(\mb x,r)\subseteq \Ucal^\star &\Longleftrightarrow \forall \mb\alpha\neq \mb 0,\quad \max_{\mb y\in\Ucal^\star} \mb\alpha^\top \mb y \geq \max_{\mb y\in B(\mb x, r)} \mb\alpha^\top \mb y\\
        &\Longleftrightarrow B(\mb x, r)\subseteq \Rbb_+^n \quad\text{and}\quad \forall \mb\alpha\in\Rbb_+^n\setminus\{\mb 0\},\quad \Ebb\sqb{\max_{i\in[n]}\alpha_i u_i} \geq  \mb\alpha^\top\mb x + \|\mb\alpha\|r. 
    \end{align*}
    This ends the proof.
\end{proof}

\section{Characterizations in the discounted infinite-horizon setting}
\label{sec:infinite-horizon}

In this section, we prove our main results concerning lower bounds for the reachable region $\Ucal_\gamma$ compared to the full-information reachable set $\Ucal^\star$. Before doing so, we need to introduce a pruning procedure that will help ``pre-condition'' the allocation problem.

\subsection{A pruning procedure}

For a given vector $\mb U\in\Ucal^\star$, we consider the following recursive pruning process. First let $I(\mb U) = \{i\in[n]: U_i>0\}$ and initialize $\mb U^{(0)} = \mb U$. For any $k\geq 0$, if there exists $i\in I(\mb U)$ with $U^{(k)}_i\geq \Ebb[u_i]$, we let $i_{k+1}:=i$. If $\Pbb(u_i=0)=0$, pose $\mb U^{(k+1)}:=\mb 0$, otherwise
\begin{equation*}
    \mb U^{(k+1)} := \frac{\mb U^{(k)} - \Ebb[u_{i_k}]\mb e_{i_k}}{\Pbb(u_{i_k}=0)}.
\end{equation*}
If for all $i\in I(\mb U)$, $U_i<\Ebb[u_i]$, the process stops. We denote by $K$ the index of the last constructed index $i_K$ and denote $\tilde{\mb U}:= \mb U^{(K+1)}$ the final constructed vector. We pose $J(\mb U) = \{i\in I(\mb U): \tilde U_i>0\}$.

Properties of this pruning process are given below in \cref{lemma:pruning_process}. The last claim shows that the pruning procedure preserves the regions $\Ucal_\gamma$ while deleting agents. Essentially, this shows that without loss of generality, we can assume that $0<U_i<\Ebb[u_i]$ for all $i\in[n]$ (those are properties satisfied by $(\tilde U_j)_{j\in J(\mb U)}$), which will be useful to derive general lower bounds.

\begin{lemma}\label{lemma:pruning_process}
    Let $\mb U\in\Ucal^\star$ and construct construct $i_1,\ldots,i_K$, $\tilde{\mb U}$ and $J(\mb U)$ according to the pruning process above. 
    \begin{enumerate}
        \item For all $k\in [K]$, $\mb U^{(k)}\in\Ucal^\star$. In particular, $U^{(k)}_i\leq \Ebb[u_i]$ for all $i\in I(\mb U)$. Also, for $k\in[K-1]$, $\Pbb(u_{i_k}=0)>0$.
        \item
        \begin{equation*}
            \forall k\in[K],\; U_{i_k} = \Ebb[u_{i_k}] \prod_{l<k}\Pbb(u_{i_l}=0)\quad \text{and}\quad  \mb U = (U_i\1_{i\notin J(\mb U)})_{i\in [n]} + \prod_{k\in[K]}\Pbb(u_{i_k}=0) \cdot \tilde{\mb U}.
        \end{equation*}
        \item $\tilde{\mb U}\in \Ucal^\star$. Equivalently, $(\tilde U_j)_{j\in J(\mb U)}$ belongs to the achievable set $\tilde\Ucal^\star$ for the full-information game in which only agents $J(\mb U)$ are present.
        \item Fix $\gamma\in[0,1)$ and denote by $\tilde\Ucal_\gamma$ the achievable set for the $\gamma$-discounted game in which only agents $J(\mb U)$ are present. Then, $\mb U\in\Ucal_\gamma$ if and only if $(\tilde U_j)_{j\in J(\mb U)}\in \tilde\Ucal_\gamma$.
    \end{enumerate} 
\end{lemma}

\begin{proof}
    Suppose that for some $k\in\{0,1,\ldots,K\}$ we have $\mb U^{(k)}\in\Ucal^\star$. By construction, $U^{(k)}_{i_k} \geq \Ebb[u_{i_k}]$. However, since $\mb U^{(k)}\in\Ucal^\star$ we also have $U^{(k)}_{i_k} \leq \Ebb[u_{i_k}]$. This proves $U^{(k)}_{i_k} = \Ebb[u_{i_k}]$. Also, $i_k\in I(\mb U)$ hence $\Ebb[u_{i_k}]\geq U_{i_k}>0$. As a result, $\Pbb(u_{i_k}>0)>0$ and any allocation function $\mb p(\cdot)$ that realizes $\mb U^{(k)}$ allocates to agent $i_k$ whenever $u_{i_k}>0$, $\Dcal_{i_k}$-almost surely. This implies that for any $i\neq i_k$,
    \begin{equation*}
        U^{(k)}_i = \Ebb[u_i p_i(\mb u)] = \Ebb[u_i p_i(\mb u)\1_{u_{i_k}=0}] =  \Pbb(u_{i_k}=0) \Ebb[u_i p_i(0,\mb u_{-i_k})].
    \end{equation*}
    If $\Pbb(u_{i_k}=0)=0$, then we had $\mb U^{(k)} = \Ebb[u_{i_k}]\mb e_{i_k}$ and the process ends with $k=K$. Otherwise, the previous equation already shows that the allocation in which we force $u_{i_k}=0$, that is 
    \begin{equation}\label{eq:def_new_allocation}
        \mb p':\mb u\in[0,\bar v]^n \mapsto (p_i(0,\mb u_{-i_k}) \1_{i\neq i_k})_{i\in[n]},
    \end{equation}
    realizes $\mb U^{(k+1)}$. Indeed, it only suffices to check $\Ebb[u_{i_k} p'_i(\mb u)]=0=U^{(k+1)}_{i_k}$. As a result, $\mb U^{(k+1)}\in\Ucal^\star$.
    This proves the first claim of the lemma as well as $\tilde {\mb U} = \mb U^{(K+1)}\in\Ucal^\star$, which is the third claim.
    The above arguments also show $U^{(k)}_{i_k} = \Ebb[u_{i_k}]$ for all $k\in[K]$, and that $U^{(k)}_i=0$ for all $i\notin I(\mb U)$ and $i\in\{i_1,\ldots,i_{k-1}\}$. We can then prove the second claim by simple induction.

    \paragraph{Proving that $\mb U\in\Ucal_\gamma\Rightarrow\tilde{\mb U}\in \Ucal_\gamma$.}
    To prove the last claim, further suppose that $\mb U^{(k)}\in\Ucal_\gamma$ for $\gamma\in[0,1)$ and some $k\in\{0,1,\ldots,K\}$. If $\Pbb(u_{i_k}=0)=0$, we know that $\mb U^{(k+1)}=\mb 0\in\Ucal_0\subseteq\Ucal_\gamma$. From now, we suppose that
    \begin{equation*}
        P_k:=\Pbb(u_{i_k}=0)>0.
    \end{equation*}
    Let $\mb p(\cdot\mid \mb V)$ and $\mb W(\cdot\mid\mb V)$ be valid incentive-compatible allocation and promised utility functions for all $\mb V\in\Ucal_\gamma$ (that is, satisfying \cref{eq:target_met,eq:valid_promise,eq:incentive_compatibility}). Let $I_k = \{i\in[n],U^{(k)}_i>0\}$. We denote $\Rcal_k = \Ucal_\gamma\cap \{\mb V:V_{i_k}=\Ebb[u_{i_k}], \;\forall i\notin I_k, V_i=0\}$. In particular, $\mb U^{(k)}\in\Rcal_k$. For any $\mb V\in\Rcal_k$,
    \begin{align*}
        \Ebb[u_{i_k}] = V_{i_k} &= (1-\gamma)\Ebb[u_{i_k}p_{i_k}(\mb u\mid \mb V)]+ \gamma \Ebb[W_{i_k}(\mb u\mid \mb V)]\\
        &\leq (1-\gamma)\Ebb[u_{i_k}] + \gamma \Ebb[u_{i_k}] = \Ebb[u_{i_k}].
    \end{align*}
    In the inequality, we used the fact that $\mb W(\mb v\mid\mb U^{(k)})\in\Ucal_\gamma\subseteq \Ucal^\star$, hence $W_{i_k}(\mb v\mid \mb U^{(k)}) \leq \Ebb[u_{i_k}]$. The above computations show that before, $\Dcal_{i_k}$-almost surely, $\mb p(\cdot\mid \mb U^{(k)})$ allocates to $i_k$ whenever $u_{i_k}>0$. Also, almost-surely $\mb W(\mb u\mid \mb U^{(k)}) \in\Rcal_k$ as well. Without loss of generality, we can assume that this the case deterministically for all $\mb u\in[0,\bar v]^n$ (up to modification of the promised utility function on zero-measure regions). We now define
    \begin{equation*}
        \Rcal'_k := \set{ \frac{\mb V-\Ebb[u_{i_k}]\mb e_{i_k}}{P_k},\;\mb V\in\Rcal_k}.
    \end{equation*}
    We construct the following allocation and promised utility functions similarly as in \cref{eq:def_new_allocation},
    \begin{align*}
        \mb p'(\mb v \mid \mb V') &:= (p_i(0,\mb v_{-i_k}\mid \Ebb[u_{i_k}]\mb e_{i_k} + P_k \mb V')\1_{i\neq i_k})_{i\in[n]},\\
        \mb W'(\mb v \mid \mb V') &:= \frac{1}{P_k} (\Ebb_{u_{i_k}\sim\Dcal_{i_k}} [W_i(u_{i_k},\mb v_{-i_k}\mid \Ebb[u_{i_k}]\mb e_{i_k} + P_k \mb V')]\1_{i\neq i_k})_{i\in[n]},
    \end{align*}
    for all $\mb V'\in\Rcal_k'$ and $\mb v\in[0,\bar v]^n$. We can check that
    \begin{equation*}
        \mb W'(\mb v\mid \mb V') = \frac{\mb V-\Ebb[u_{i_k}]\mb e_{i_k}}{P_k},\quad \text{where }\mb V=\Ebb_{u_{i_k}}[\mb W(u_{i_k},\mb v_{-i_k}\mid \Ebb[u_{i_k}]\mb e_{i_k} + P_k \mb V')] \in\Rcal_k.
    \end{equation*}
    Here we used the fact that $\Rcal_k$ is convex since $\Ucal_\gamma$ is convex.
    Thus, \cref{eq:valid_promise} is satisfied, that is all promised utilities stay within $\Rcal'_k$. Then, the same arguments as for \cref{eq:def_new_allocation} show that for any $\mb V'\in \Rcal'_k$, letting $\tilde{\mb V}:=\Ebb[u_{i_k}]\mb e_{i_k} + P_k \mb V'$ we have
    \begin{align*}
        (\Ebb[ u_i p_i'(\mb u \mid \mb V')])_{i\in[n]}=\frac{(\Ebb[ u_i p_i(\mb u \mid \tilde{\mb V})])_{i\in[n]} - \Ebb[u_{i_k}]\mb e_{i_k}}{P_k}.
    \end{align*}
    On the other hand, 
    \begin{align*}
        \Ebb[\mb W'(\mb u\mid \mb V')] = \frac{ \Ebb_{\mb u} \Ebb_{u_{i_k}'}[\mb W(u_{i_k}',\mb u_{-i_k}\mid \tilde{\mb V})] - \Ebb[u_{i_k}]\mb e_{i_k}}{P_k}
        = \frac{ \Ebb_{\mb u}[\mb W(\mb u\mid \tilde{\mb V})] - \Ebb[u_{i_k}]\mb e_{i_k}}{P_k}
    \end{align*}
    Combining the two previous equations we obtain
    \begin{align*}
         (1-\gamma)(\Ebb[ u_i p_i'(\mb u \mid \mb V')])_{i\in[n]} &+ \gamma \Ebb[\mb W'(\mb u\mid \mb V')]\\
         &=\frac{(1-\gamma)(\Ebb[ u_i p_i(\mb u \mid \tilde{\mb V})])_{i\in[n]} + \gamma \Ebb [\mb W(\mb u\mid \tilde{\mb V})]-\Ebb[u_{i_k}]\mb e_{i_k}}{P_k}\\
         &= \frac{\tilde{\mb V}  -\Ebb[u_{i_k}]\mb e_{i_k}}{P_k} = \mb V',
    \end{align*}
    where in the second equality we used the fact that $\mb p$ and $\mb W$ satisfy \cref{eq:target_met}. The previous equation shows that $\mb p'$ and $\mb W'$ also satisfy \cref{eq:target_met}. Last, for any $i\neq i_k$ and $u,v\in[0,\bar v]$ we have
    \begin{align*}
        (1-\gamma)u P'_i(v\mid \mb V') + \gamma W'_i(v\mid \mb V')  &= \frac{(1-\gamma)u P_i(v\mid \tilde{\mb V}) + \gamma W_i(v\mid \tilde{\mb V})}{P_k}.
    \end{align*}
    Hence, because $\mb p,\mb W$ is incentive-compatible (\cref{eq:incentive_compatibility}), so is $\mb p',\mb W'$ (we recall that for $i_k$, the utility of agent $i_k$ is always 0). This ends the proof that $\mb U^{(k+1)}\in\Rcal'_k\subseteq \Ucal_\gamma$. The induction then shows that $\tilde{\mb U}\in \Ucal_\gamma$, and hence $(\tilde U_j)_{j\in J(\mb U)}\in\tilde\Ucal_\gamma$.

    \paragraph{Proving that $\tilde{\mb U}\in \Ucal_\gamma \Rightarrow \mb U\in\Ucal_\gamma$.}
    Now suppose that $(\tilde U_j)_{j\in J(\mb U)}\in\tilde\Ucal_\gamma$, which implies in particular $\tilde{\mb U}\in \Ucal_\gamma$. Let $\Rcal'=\Ucal_\gamma\cap\{\mb V:\forall i\notin J(\mb U),V_i=0\}$. Let $\mb p'$ and $\mb W'$ be valid incentive-compatible allocation and promised utility functions for $\mb V\in\Ucal_\gamma$. We can easily check that for any $\mb V'\in \Rcal'$, then almost surely $W'_i(\mb u\mid\mb V')=0$ for all $i\notin J(\mb U)$, which implies $\mb W'(\mb u\mid\mb V')\in\Rcal'$. Without loss of generality, we suppose this is the case (up to modifying $\mb W$ on zero-measure regions).

    We next define $P_k:= \prod_{k\in[K]}\Pbb(u_{i_k}=0)$ and $\mb V_0:=\sum_{k\in[K]}\Ebb_{u_{i_k}} \prod_{l<k}\Pbb(u_{i_l}=0)\mb e_{i_l} $, we pose
    \begin{equation*}
        \Rcal := \{ \mb V_0 + P_k \mb V',\mb V'\in\Rcal' \}.
    \end{equation*}
    We consider the allocation $\mb p'$ which always tries to allocate to agents $i_1,i_2,\ldots,i_K$ with this priority order whenever they have non-zero utility, then resorts to using the allocation $\mb p$. Formally, for all $\mb v\in[0,\bar v]^n$ and $\mb V\in\Rcal$, letting $\mb V'=(\mb V-\mb V_0)/P_k\in\Rcal'$ we pose
    \begin{align*}
        \mb p(\mb v\mid\mb V)&:= \begin{cases}
            \mb e_{i_l} & \text{if } \max_{k\in[K]} v^{(t)}_{i_k}>0,\quad l=\min\{k\in[K]: v^{(t)}_{i_l}>0\}\\
            \mb p'(\mb v \mid \mb V') &\text{otherwise},
        \end{cases}\\
        \mb W(\mb v\mid \mb V)&:= \mb V_0 + P_k \mb W'(\mb v\mid\mb V').
    \end{align*}
    We now check that $\mb p,\mb W$ are valid for utility vectors in $\Rcal$. By construction of $\mb W$, it directly satisfies \cref{eq:valid_promise}. Next, for any $\mb V\in\Rcal$, with the same notation $\mb V'=(\mb V-\mb V_0)/P_k\in\Rcal'$ we have
    \begin{multline*}
        (1-\gamma)(\Ebb[u_ip_i(\mb u\mid \mb V)])_{i\in[n]} + \gamma \Ebb[\mb W(\mb v\mid\mb V)]\\
        =(1-\gamma)(\mb V_0 + P_k(\Ebb[u_ip'_i(\mb u\mid \mb V')])_{i\in[n]}) + \gamma(\mb V_0 + P_k\Ebb[\mb W'(\mb v\mid\mb V')])= \mb V_0 + P_k \mb V' = \mb V.
    \end{multline*}
    In the second equality, we used the fact that $\mb p',\mb W'$ satisfy \cref{eq:target_met}. This shows that $\mb p,\mb W$ also satisfy it. All that remains to check is the incentive-compatibility constraint \cref{eq:incentive_compatibility}. For any $k\in[K]$, assuming all other agents are truthful, agent $i_k$ is allocated the good whenever $u_{i_l}=0$ for all $l<k$ and its report satisfies $v_{i_k}>0$. Hence we only need to check that incentive-compatibility when $u_{i_K}=0$, which is immediate. For any $i\notin \{i_k,k\in[K]\}$ and $u,v\in[0,\bar v]$, we directly have
    \begin{equation*}
        (1-\gamma)u P_i(v\mid \mb V) + \gamma W_i(v\mid \mb V) = P_k\paren{(1-\gamma)u P'_i(v\mid \mb V') + \gamma W'_i(v\mid \mb V')},
    \end{equation*}
    hence the incentive-compatibility is guaranteed by that of $\mb p',\mb W'$. This ends the proof.
\end{proof}

\subsection{A universal $\Ocal(\sqrt{1-\gamma})$ convergence rate}
\label{subsec:universal_slow_rate}

With the pruning procedure and \cref{lemma:ball_in_region_no_assumptions}, we now  show that without any assumptions, the distance between the boundary of $\Ucal_\gamma$ and the optimal region $\Ucal_\star$ is at most $\Ocal(\sqrt{1-\gamma})$. This implies \cref{thm:universal_lower_bound_simplified}. 

\begin{theorem}\label{thm:universal_lower_bound}
    For any $\mb U\in\Ucal^\star$, construct $i_1,\ldots,i_K$, $\tilde{\mb U}$ and $J(\mb U)$ according to the pruning process. If $|J(\mb U)|\leq 1$, we have $\mb U\in\Ucal_0$. Otherwise, there exist $\gamma_0\in[1/2,1)$ and $C>0$ such that
    \begin{equation*}
        \forall \gamma\in[\gamma_0,1), \quad d(\mb U;\Ucal_\gamma) \leq C \sqrt{1-\gamma}.
    \end{equation*}
    Precisely, introducing the constant $\tilde C = 4|J(\mb U)|^2\bar v^2\paren{1+\frac{\max_{j\in J(\mb U)} \Ebb[u_i]}{\min_{j\in J(\mb U)}(\Ebb[u_j] - \tilde U_j)}}^2<\infty$, we have $C=2\tilde C \sqrt{|J(\mb U)|} \prod_{k\in[K]}\Pbb(u_{i_k}=0)$ and $\gamma_0=\max\paren{1/2,1- \min_{j\in J(\mb U)}\tilde U_i^2/(32\tilde C)}<1$.
\end{theorem}

\begin{proof}
    Fix $\mb U\in\Ucal^\star$ and define $i_1,\ldots,i_K$, $\tilde{\mb U}$, and $J(\mb U)$ according to the pruning process. If $|J(\mb U)|\leq 1$, it is clear that $(\tilde U_j)_{j\in J(\mb U)}\in\tilde\Ucal_0$, and \cref{lemma:pruning_process} ensures that $\mb U\in\Ucal_0$. We suppose that this is not the case from now. 
    By construction of the pruning process, for all $j\in J(\mb U)$, we have $\tilde U_j \in(0, \Ebb[u_j])$, and $\tilde U_i=0$ for all $i\notin J(\mb U)$. We therefore consider the reduced problem in which only agents in $J(\mb U)$ are present. By \cref{lemma:pruning_process}, $\tilde{\mb U}\in\Ucal^\star$, hence the utility vector $(\tilde U_j)_{j\in J(\mb U)}$ can still be reached in this simplified setting with full information. We denote by $\tilde\Ucal^\star = \{(x_j)_{j\in J(\mb U)}:\mb x\in\Ucal^\star\}$ the reachable utilities in this setting. By slight abuse of notation, we will still denote by $\tilde{\mb U}$ the vector $(\tilde U_j)_{j\in J(\mb U)}$.

    Note that $\Rcal:=\prod_{j\in J(\mb U)}[0,\tilde U_j]\subseteq \tilde\Ucal^\star$. Let $\tilde C =  4|J(\mb U)|^2\bar v^2\paren{1+\frac{\max_{j\in J(\mb U)} \Ebb[u_i]}{\min_{j\in J(\mb U)}(\Ebb[u_j] - \tilde U_j)}}^2$ and pose
    $r:=\delta := \sqrt{\tilde C\frac{1-\gamma}{\gamma}}.$
    Next, using $\gamma\geq \gamma_0$, we have
    \begin{equation*}
        2(r+\delta) \leq  4\sqrt{2\tilde C(1-\gamma)} \leq \min_{j\in J(\mb U)}\tilde U_i.
    \end{equation*}
    As a result, with $\mb x = \tilde{\mb U} - (r+\delta)\mb 1$, we have $B(\mb x, r+\delta)\subseteq \Rcal\subseteq\tilde\Ucal^\star$. Since $\gamma\geq 1/2$, the choice of parameters $\mb x,r,\delta$ satisfies the assumptions of \cref{lemma:ball_in_region_no_assumptions}. Thus, $B(\mb x, r)\subseteq \tilde \Ucal_\gamma$. In turn, \cref{lemma:pruning_process} implies that with $P:=\prod_{k\in [K]}\Pbb(u_{i_l}=0)$, we have
    \begin{equation*}
        (U_i\1_{i\notin J(\mb U)})_{i\in[n]} + P \cdot B_{J(\mb U)}(\mb x, r) \subseteq \Ucal^\star,
    \end{equation*}
    where $B_{J(\mb U)}(\mb x,r) = \{\mb y: (y_j)_{j\in J(\mb U)}\in B(\mb x,r),\,\forall i\notin J(\mb U),y_i=0\}$. Hence,
    \begin{equation*}
        d(\mb U,\Ucal^\star) \leq P\cdot d(\tilde{\mb U},\tilde\Ucal^\star) \leq P\cdot d(\tilde{\mb U},B(\mb x, r)) = P(\sqrt{|J(\mb U)|}(r+\delta)-r) \leq 2P\sqrt{|J(\mb U)|}r.
    \end{equation*}
    This ends the proof of the theorem.
\end{proof}

\subsection{Improved rates for smooth full-information regions $\Ucal^\star$}
\label{subsec:improved_rates_smooth_regions}

We first prove the following lemma.
\begin{lemma}\label{lemma:quick_proof}
    If $\Ucal^\star$ is twice-differentiable locally around an optimal point $\mb U$, then it contains a ball tangent to $\mb U$ of radius $\mathcal O(1/\|H(\mb U)\|)$ where $\|H(\mb U)\|$ is the operator norm of the directional Hessian $H(\mb U)$ of the boundary at $\mb U$.
\end{lemma}
\begin{proof}
    Up to restricting to a subset of agents, suppose without loss of generality that $U_i>0$ for all $i\in[n]$. In particular, $\Ucal^\star$ has non-empty interior. Consider the Taylor expansion of the boundary of $\Ucal^\star$ at $\mb U$ in coordinates given by the tangent hyperplane at $\mb U$. Then, the boundary of $\Ucal^\star$ is locally upper bounded by $\frac{1}{2} x^\top H(\mb U) x+o(\|x\|^2)\leq \|H(\mb U)\| \cdot \|x\|^2$ where $H(\mb U)$ is the directional Hessian at $\mb U$ and $\|H(\mb U)\|$ is its operator norm. Recalling that $\Ucal^\star$ is convex and has non-empty interior, this shows that $\Ucal^\star$ contains a ball tangent at $\mb U$ of radius $\mathcal O(1/\|H(\mb U)\|)$.
\end{proof}

Next, we prove the geometric result \cref{thm:simpler_geometric_fast_rate}, which is a direct application of \cref{lemma:ball_in_region_no_assumptions}.

\begin{proof}[of \cref{thm:simpler_geometric_fast_rate}]
    Fix $B(\mb x,r)$ a ball tangent at $\mb U=\mb x+r\mb 1$ to $\Ucal^\star$. 
    Let $C>0$ be the corresponding constant in \cref{eq:constraint_safe_boundary} from \cref{lemma:ball_in_region_no_assumptions}. The constraint \cref{eq:constraint_safe_boundary} is satisfied for $r(\gamma)=1-\delta(\gamma)$ and $\delta(\gamma) = 2C\frac{1-\gamma}{\gamma r}$ for $1-\gamma$ sufficiently small. Therefore, \cref{lemma:ball_in_region_no_assumptions} implies that $B(\mb x, r(\gamma))\subseteq\Ucal_\gamma$. We conclude by noting $\SWG(\gamma)\leq \sqrt n\cdot d(\mb U,B(\mb x, r(\gamma))=\sqrt n\delta(\gamma)=\Ocal(1-\gamma)$.
\end{proof}

We next prove \cref{thm:faster_rates_1-beta}, which gives more flexible and combinatorial conditions under which faster rates $\Ocal(1-\gamma)$ can be achieved. The most complete statement of this result is given below (and in particular implies \cref{thm:faster_rates_1-beta}, see condition (SC3) below). We recall the notation $P_S:\Rbb^n\to\Rbb^S$ for the projection on coordinates $S$.

\begin{theorem}\label{thm:faster_rates_1-beta_long_version} 
    Suppose that there exists a partition $ I =I_1\sqcup I_2\sqcup\cdots \sqcup I_q$ with $|I_s|\geq 2$ for all $s\in[q]$, an optimal vector $\mb U\in\Ucal^\star$ and $r>0$, such that (SC1) \cref{eq:sets_proba_disjoints} holds and taking $\mb x := (U_i - r\sum_{s\in[q]}\frac{1 }{\|\mb 1_{I_s}\|}\1_{i\in I_s})_{i\in[n]}$, we have for all $s\in[q]$,
    \begin{equation*}
       B_{I_s}(\mb x_{I_r}, r)  \subset P_{I_s}\paren{ \Ucal^\star \cap\{\mb y:\forall i\notin I_s,y_i=U_i\} }.
    \end{equation*}
    Then $\SWG(\gamma) = \Ocal(1-\gamma)$.
    
    Condition (SC1) is equivalent to the following condition (SC2): \cref{eq:sets_proba_disjoints} holds, and for all $s\in[q]$ and $\emptyset\subsetneq B\subsetneq I_s$,
    \begin{equation*}
        \Ebb\sqb{Z_B \1_{ Z_B\leq Z_{I_s\setminus B}=Z\leq  (1+\eta) Z_B}}\underset{\eta\to 0}{=} \Omega(\eta).
    \end{equation*}

   Last, (SC1) or (SC2) are implied by the following condition (SC3): there exists a partition $ I =I_1\sqcup I_2\sqcup\cdots \sqcup I_q$ such that for all $s\in[q]$, $|I_s|\geq 2$ and $I_s$ is strongly connected in $\Gcal$.
\end{theorem}

\begin{proof}[of \cref{thm:faster_rates_1-beta_long_version}]
    For $s\in[q]$, we define the function $\phi:\Rbb^{I_s}\to\Rbb$ via
    \begin{equation}\label{eq:def_phi}
        \phi^{(s)}(\mb\beta):=  \max_{\mb y\in P_{I_s}\paren{ \Ucal^\star \cap\{\mb y:\forall i\notin I_s,y_i=U_i\} }} \mb\beta^\top \mb y =    \Ebb\sqb{\1_{Z_{I_s}=Z>0}\cdot  \max_{i\in I_s} \beta_i u_i },\quad \mb\beta\in\Rbb^{I_s}.
    \end{equation}
    In the equality, we use the characterizations from \cref{lemma:ball_in_region_optimized}. We also recall that by \cref{eq:sets_proba_disjoints}, we have $\Pbb(Z_{I_s}=Z_{[n]\setminus I_s}>0)=0$.
    Note that $\phi$ is convex as the maximum of linear functions.

    \paragraph{Proving that the (SC1) is sufficient.} Suppose that (SC1) holds and fix parameters satisfying the assumptions. \cref{lemma:ball_in_region_optimized} implies that for $1-\gamma$ sufficiently small and $\delta = 2C(1-\gamma)/(\gamma r)$ for some constant $C>0$,
    $B(\mb x, r-\delta)\cap \{\mb y:\forall i\notin  I ,y_i=U_i\} \subseteq \Ucal_\gamma.$
    In turn, we obtain the desired result $\SWG \leq \sqrt n\cdot d(\mb U,\Ucal_\gamma) \leq \sqrt n\delta = \Ocal(1-\gamma)$.

    \paragraph{Proving that (SC3)$\Rightarrow$(SC2)$\Rightarrow$(SC1).}
     Suppose that (SC3) holds and denote $ I = J_1\sqcup  J_2\sqcup\cdots \sqcup  J_{\tilde q}$ the guaranteed partition. To prove (SC2) and (SC1) we will use a slightly different partition by merging some sets. Precisely, consider the undirected graph $\Fcal$ on $[\tilde q]$ such that $s\neq s'\in[q]$ are connected if and only if $\Pbb(Z_{ J_s}=Z_{ J_{s'}}=Z>0)>0$. We let $q$ be the number of connected components $C_1,\ldots,C_q$ of $\Fcal$. For $s\in[q]$ we let $I_s:=\bigcup_{t\in C_s}  J_t$, that is we merge the partition according to the graph $\Fcal$. Note that if $s\neq s'$ are connected within $\Fcal$ then
     \begin{equation*}
         0<\Pbb(Z_{ J_s}=Z_{ J_{s'}}=Z>0) \leq \sum_{i\in  J_s,j\in  J_{s'}} \Pbb(u_i=u_j=Z>0).
     \end{equation*}
     Hence, there exists $i\in  J_s$ and $j\in  J_{s'}$ such that $\Pbb(u_i=u_j=Z>0)$. This implies that $i\to j$ and $j\to i$ in the graph $\Gcal$, and hence $ J_s$ is strongly connected to $  J_{s'}$ via this double edge. This proves that the sets of the constructed partition $ I =I_1\sqcup  I_2\sqcup\cdots \sqcup  I_q$ are still strongly connected within $\Gcal$.

     We can directly check that \cref{eq:sets_proba_disjoints} is satisfied by construction of $I_1,\ldots,I_q$: for any $s\neq s'\in [q]$
     \begin{equation*}
         \Pbb(Z_{I_s}=Z_{I_{s'}}=Z>0)\leq \sum_{t\in C_s,t'\in C_{s'}} \Pbb(Z_{ J_t}=Z_{ J_{t'}}=Z>0)=0.
     \end{equation*}
     In the last equality we use the fact that $C_s$ and $C_{s'}$ are disconnected in $\Fcal$.
     
     We next define an allocation function $\mb p$ such that $\mb p(\mb u)$ is the uniform distribution on $\argmax_{j\in [n]}u_j$. We then define $\mb U=(\Ebb[u_i p_i(\mb u)])_{i\in[n]}$ as the utility vector realized by $\mb p$. Note that by construction, $\mb U$ is optimal and we have $U_i>0$ for all $i\in  I $. In particular, for all $s\in[q]$ we have $\phi^{(s)}(\mb 1_{I_s})=\mb 1_{I_s}^\top \mb U_{I_s}$.

     Suppose that there exists $c_1,c_2>0$ such that for any $s\in[q]$ and $\mb\beta\in\Hcal_s:=\{\mb\delta\in\Rbb^{I_s}:\mb 1_{I_s}^\top \mb\delta=0\}$, with $\|\mb\beta\|\leq c_1$, we have
    \begin{equation}\label{eq:intermediary_assumption}
        \phi^{(s)}(\mb 1_{I_s} + \mb\beta) -\phi^{(s)}(\mb 1_{I_s} ) - \mb\beta^\top \mb U_{I_s} \geq   c_2\|\mb\beta\|^2.
    \end{equation}
    Because the function $\mb\beta\mapsto \phi^{(s)}(\mb 1_{I_s} + \mb\beta) -\phi^{(s)}(\mb 1_{I_s} ) - \mb\beta^\top \mb U_{I_s} $ is convex in $\gamma$ and has value $0$ for $\mb\beta=\mb 0$, this implies that for all $\mb\beta\in\Hcal_s$,
    \begin{equation}\label{eq:intermediary_assumption_bis}
        \phi^{(s)}(\mb 1_{I_s} + \mb\beta) -\phi^{(s)}(\mb 1_{I_s} ) - \mb\beta^\top \mb U_{I_s} \geq   c_2 \min(\|\mb\beta\|^2,c_1\|\mb\beta\|) \geq  c_3 \min ( \|\mb\beta\|^2, \|\mb 1_{I_s}\| \|\mb\beta\|) ,
    \end{equation}
    for $c_3 = \min(c_2,c_1c_2/\|\mb 1_{I_s}\|)$.
    Let $r=\min( 2c_3 \min_{s\in[q]}\|\mb 1_{I_s} \|,\min_{i\in  I } U_i)$. We note that $\mb x_{I_s} = \mb U_{I_s}-r\frac{\mb 1_{I_s}}{\|\mb 1_{I_s}\|}$. Since $\mb\beta^\top\mb 1_{I_s}=0$, we have
    \begin{align*}
        (\mb 1_{I_s}+\mb\beta)^\top \mb x_{I_s}+ \|\mb 1_{I_s}+\mb\beta\| r &=(\mb 1_{I_s}+\mb\beta)^\top\mb U_{I_s} + r(\|\mb 1_{I_s}+\mb\beta\| - \|\mb 1_{I_s}\|)\\
        &= \phi^{(s)}(\mb 1_{I_s}) + \mb\beta^\top \mb U_{I_s} + r  \|\mb 1_{I_s}\| \paren{\sqrt{1+(\|\mb\beta\|/\|\mb 1_{I_s}\|)^2} - 1}\\
        &\leq \phi^{(s)}(\mb 1_{I_s}) + \mb\beta^\top \mb U_{I_s} + \frac{2r}{\|\mb 1_{I_s}\|}\min ( \|\mb\beta\|^2, \|\mb 1_{I_s}\| \|\mb\beta\|)  \leq \phi^{(s)}(\mb 1_{I_s} + \mb\beta).
    \end{align*}
    In the last inequality, we used \cref{eq:intermediary_assumption_bis}. As a result, \cref{lemma:on_balls} shows that we have $B_{I_s}(\mb x_{I_s}, r)\subseteq P_{I_s}\paren{ \Ucal^\star \cap\{\mb y:\forall i\notin I_s,y_i=U_i\} }$. We can also easily check that it contains $\mb U$. Hence, this proves (SC1).
    
    We now show that \cref{eq:intermediary_assumption} holds. We proceed independently for each $s\in[q]$. 
    Fix $\mb\beta \in\Hcal_s\setminus\{\mb 0\}$. We first show that there is a subset $\emptyset\subsetneq B\subsetneq I_s$ such that
    \begin{equation}\label{eq:reduce_to_subsets}
        \min_{i\in B} \beta_i - \max_{i\in I_s\setminus B} \beta_i \geq \frac{\|\mb\beta\|} {2n^2} \comment{ \quad\text{and} \quad  \min_{i\in B} \beta_i\geq \frac{1} {2n^2} \|\mb\beta\|}.
    \end{equation}
    Let $\epsilon =\frac{1} {2n^2}\|\mb\beta\|$. We consider the partition of $\Rbb$ into $(-\infty, 0]$, $(n\epsilon,\infty)$ and $n$ length-$\epsilon$ intervals partitioning $[0,n\epsilon)$. To prove \cref{eq:reduce_to_subsets} it suffices to show that there is one of these intervals of length $\epsilon$ that does not contain an element of the form $\beta_i$ for $i\in I_s$, but has elements below and above it. Suppose that this is not the case. Then, $\{\beta_i\geq 0,i\in I_s\}$ occupy consecutive intervals of the partition. Since $\mb 1_{I_s}^\top \mb\beta=0$, $\mb\beta$ has both positive and negative entries, this shows that the consecutive intervals occupied by $\{\beta_i,i\in I_s\}$ must start at $[0,\epsilon)$. There are at most $|I_s|-1\leq n-1$ such intervals (one point must belong to $(-\infty,0)$), hence we proved that all elements from $\{\beta_i\geq 0,i\in I_s\}$ lie consecutively within the $n$ intervals within $[0,(n-1)\epsilon)$. Then, since $\mb 1_{I_s}^\top \mb\beta=0$, we have
    \begin{equation*}
        \sum_{i\in I_s:\beta_i<0} |\beta_i| =  \sum_{i\in I_s:\beta_i\geq 0}\beta_i\leq  |I_s| (n-1)\epsilon <n^2\epsilon.
    \end{equation*}
    Therefore, we obtain the following contradiction
    \begin{equation*}
        \|\mb\beta\|\leq \|\mb\beta\|_1 \leq \sum_{i\in I_s} |\beta_i| < 2 n^2\epsilon = \|\mb\beta\|.
    \end{equation*}

    In the rest of the proof, we fix a subset $\emptyset\subsetneq B\subsetneq I_s$ satisfying \cref{eq:reduce_to_subsets}. We let $\gamma_1:= \min_{i\in B} \beta_i$ and $\gamma_2:= \max_{i\in I_s\setminus B} \beta_i.$
    By assumption, we have $\gamma_1\geq\gamma_2+\epsilon$.
    Provided that $\|\mb\beta\|\leq 1/2$, we have $|\gamma_2|\leq 1/2$. We suppose that this is the case from now.
    Recall that $\mb p$ is the allocation that realizes $\mb U$ in the full information setting. We also introduce the notation $\hat i$ for the agent that effectively receives the allocation (that is, $\hat i\sim\mb p(\mb u)$). Using \cref{eq:def_phi}, we have
    \comment{We recall that the allocation $\mb p$ and all the allocations in the characterization of $\phi$ in \cref{eq:def_phi} coincide with $\mb p^{(0)}$ on indices $i\notin I$. Therefore,}
    \begin{align}
        \phi^{(s)}(\mb 1_{I_s} + \mb\beta) &-\phi^{(s)}(\mb 1_{I_s}) - \mb\beta^\top \mb U_{I_s}\\
        &= \Ebb\sqb{\paren{\max_{i\in I_s}(1+\beta_i) u_i -(1+\beta_{\hat i})u_{\hat i}\1_{\hat i\in I_s}}\1_{Z=Z_{I_s}} } \notag \\
        &\geq \Ebb\sqb{\paren{\max_{i\in I_s}(1+\beta_i) u_i -(1+\beta_{\hat i})u_{\hat i}\1_{\hat i\in I_s} } \1_{Z_B\leq  Z_{I_s\setminus B}=Z\leq \frac{1+\gamma_1}{1+\gamma_2} Z_B}} \notag \\
        &\overset{(i)}{\geq}\frac{1}{n} \Ebb\sqb{\paren{(1+\gamma_1)Z_B -(1+\gamma_2)Z_{I_s\setminus B} } \1_{Z_B\leq Z_{I_s\setminus B}=Z\leq  \frac{1+\gamma_1}{1+\gamma_2} Z_B}} \notag \\
        &\overset{(ii)}{\geq} \frac{1+\gamma_2}{n}\Ebb\sqb{\paren{(1+\epsilon/2)Z_B -Z_{I_s\setminus B} } \1_{ Z_B\leq Z_{I_s\setminus B}=Z\leq  (1+\epsilon/4) Z_B}} \notag \\
        &\overset{(iii)}{\geq}   \frac{\epsilon}{8n}\Ebb\sqb{Z_B \1_{Z_B\leq Z_{I_s\setminus B}=Z\leq  (1+\epsilon/4) Z_B}}. \label{eq:upper_bound_deviation_ZB}
    \end{align}
    In (i), the additional factor $1/n$ comes from the fact that if we have a tie $Z=Z_{I_s\setminus B}$, there is at least a probability $1/n$ that $\hat i\in I_s\setminus B$ by definition of $\mb p$. In (ii) and (iii) we used $|\gamma_2|\leq 1/2$ so that $(1+\gamma_1)\geq (1+\gamma_2)(1+\epsilon/2)$ and $1+\gamma_2\geq 1/2$. We compute for any $\eta\geq 0$,
    \begin{align*}
        \Ebb\sqb{Z_B \1_{ Z_B\leq Z_{I_s\setminus B} = Z\leq  (1+\eta) Z_B}} &\geq \max_{j\in I_s\setminus B} \Ebb\sqb{Z_B \1_{ Z_B\leq Z_{I_s\setminus B}\leq  (1+\eta) Z_B} \1_{u_j = Z_{I_s\setminus B}=Z}}\\
        &\geq  \max_{i\in B,j\in I_s\setminus B} \Ebb\sqb{u_i \1_{u_j= Z}  \1_{ u_j \in[u_i,(1+\eta) u_i]}}.
    \end{align*}
    
    By assumption, because $\Gcal$ is strongly connected, there exists an outgoing edge from $B$. This implies that the right-hand side of the previous equation is $\Omega(\eta)$.
    This proves the condition (SC2).
    Because there are only a finite number of subsets $\emptyset\subsetneq B\subsetneq I_s$ and $s\in[q]$, this shows that there exists $c_3,c_4>0$ such that for all $s\in[q]$ and $\emptyset\subsetneq B\subsetneq I_s$,
    \begin{equation*}
         \Ebb\sqb{Z_B \1_{ Z_B\leq Z_{I_s\setminus B} = Z\leq  (1+\eta) Z_B}}  \geq c_3\eta,\quad \eta\in [0,c_4].
    \end{equation*}
    Together with the previous estimates, this shows that for $s\in[q]$, and any $\mb\beta\in\Hcal_s$ satisfying $\|\mb\beta\|\leq \min(1/2,c_4 n^{3/2})$, we have $\epsilon\leq c_4$ hence we can prove \cref{eq:intermediary_assumption} as follows
    \begin{align*}
        \phi^{(s)}(\mb 1_{I_s} + \mb\beta) -\phi^{(s)}(\mb 1_{I_s}) - \mb\beta^\top \mb U_{I_s} 
        &\geq \frac{c_3}{32n} \epsilon^2 = \frac{c_3}{32 n^4} \|\mb\beta\|^2.
    \end{align*}

    \paragraph{Proving that (SC1)$\Rightarrow$(SC2).}
    Suppose that we are given $\mb U$ and a radius $r>0$ satisfying (SC1). We use similar notations as before: let $\mb p(\cdot)$ be an allocation function that realizes $\mb U$ and denote by $\hat i\sim \mb p(\mb u)$ the agent that receives the resource with this allocation. By \cref{lemma:on_balls}, and using the same arguments as before, for $s\in[q]$ and every $\mb\beta\in\Hcal_s$, we have
    \begin{align*}
        \phi^{(s)}(\mb 1_{I_s}+\mb\beta) &\geq  \phi^{(I_s)}(\mb 1_{I_s}) + \mb\beta^\top \mb U_{I_s} + r  \|\mb 1_{I_s}\| \paren{\sqrt{1+(\|\mb\beta\|/\|\mb 1_{I_s}\|)^2} - 1}.
    \end{align*}
    Using the inequality $\sqrt{1+x}-x\geq (\sqrt 2-1)x$ for $x\in[0,1]$, this shows that for $\mb\beta\in\Hcal_s$ with $\|\mb\beta\|\leq \|\mb 1_{I_s}\|$,
    \begin{equation*}
        \phi^{(s)}(\mb 1_{I_s}+\mb\beta) - \phi^{(s)}(\mb 1_{I_s}) - \mb\beta^\top \mb U_{I_s} \geq \frac{(\sqrt 2-1)r}{\|\mb 1_{I_s}\|} \, \|\mb\beta\|^2.
    \end{equation*}
    Next fix any $\emptyset\subsetneq B\subsetneq I_s$. We consider the value $\mb\beta_B = (\eta\1_{i\in B} - \eta'\1_{i\in I_s\setminus B})_{i\in I_s}$ where $\eta\geq 0$ and we posed $\eta' := \eta |B|/| I_s\setminus B|$, so that $\mb 1_{I_s}^\top \mb\beta=0$. For sufficiently small $\eta\geq 0$, we have $\|\mb\beta\|=\eta \sqrt{2|B|}\leq \|\mb 1_{I_s}\|$.
    Then,
    \begin{align*}
         &\phi^{(s)}(\mb 1+\mb\beta)  - \phi^{(s)}(\mb 1) - \mb\beta^\top \mb U = \Ebb\sqb{\paren{\max_{i\in I_s}(1+\beta_i) u_i -(1+\beta_{\hat i})u_{\hat i}\1_{\hat i\in I_s}}\1_{Z=Z_{I_s}} }\\
         &\leq \Ebb\sqb{\paren{\max((1+\eta)Z_B,(1-\eta')Z_{I_s\setminus B}) - (1+\eta)Z_B\1_{\hat i\in B} - (1-\eta')Z_{I_s\setminus B}\1_{\hat i\in I_s\setminus B}} \1_{Z=Z_{I_s}}}\\
         &\leq \Ebb\sqb{\paren{(1+\eta)Z_B-(1-\eta')Z_{I_s\setminus B} } \1_{Z_B \leq Z_{I_s\setminus B}=Z\leq \frac{1+\eta}{1-\eta'}Z_B}}\\
         &\leq (\eta+\eta') \Ebb\sqb{Z_B\1_{Z_B \leq Z_{I_s\setminus B}=Z\leq \frac{1+\eta}{1-\eta'}Z_B}}.
    \end{align*}
    For $\eta\geq 0$ sufficiently small, we have $\frac{1+\eta}{1-\eta'}\leq 1+2(\eta+\eta')$. Therefore, for $\zeta\geq 0$ sufficiently small, taking $\eta\geq $ such that $\zeta=\eta+\eta'$, we obtained
    \begin{equation*}
        \Ebb\sqb{Z_B\1_{Z_B \leq Z_{I_s\setminus B}=Z\leq (1+2\zeta)Z_B}} \geq  \frac{(\sqrt 2-1)r |B|}{2\|\mb 1_{I_s}\| \paren{1+|B|/|I_s\setminus B|}^2} \,\zeta.
    \end{equation*}
    This proves (SC2).
\end{proof}

We now prove \cref{cor:simple_examples_fast_convergence} which gives simpler conditions for reaching the fast rates $\Ocal(1-\gamma)$.

\begin{proof}[of \cref{cor:simple_examples_fast_convergence}]
    We show that condition (SC3) from \cref{thm:faster_rates_1-beta} is satisfied. For any $i\in  I $, let $j\in  I $ be the corresponding index satisfying the assumption for $i$. In the first case, we obtain directly 
    \begin{equation*}
        \Pbb(u_i=u_j=Z=m_i) \geq \Pbb(u_i=m_i)\Pbb(u_j=m_i)\prod_{k\notin\{i,j\}} \Pbb(u_k\leq m_i)>0.
    \end{equation*}
    Hence $m_i>0$ implies $i\to j$ and $j\to i$.
    In the second case, let $f_i$ be the lower bound on the density of the absolutely-continuous part of $u_i$ and $u_j$ on $[a_i,b_i]$. Since $a_i>m$, we have $\Pbb(Z\leq a_i)>0$. This shows that for all $k\in[n]$, $\Pbb(u_k \leq a_i)>0$. Then,
    \begin{align*}
        \Ebb\sqb{u_i \1_{u_j =Z}\1_{ u_j\in[ u_i,(1+\eta) u_i]}}         &\geq a_i \Pbb(\forall k\notin\{i,j\},u_k\leq a_i) \Pbb(a_i\leq u_i\leq  u_j \leq  u_i+\eta a_i)\\
        &\geq a_i \prod_{k\notin\{i,j\}} \Pbb(u_k\leq a_i) \cdot f_i^2 (b_i-a_i-\eta a_i)\eta a_i =\Omega(\eta).
    \end{align*}
    This shows that $i\to j$. The same arguments also show that $j\to i$.

    In summary, $\Gcal$ contains a subgraph $\Gcal'$ in which $i\to j$ implies $j\to i$ and such that every node has an outgoing edge. We then use the partition of $ I $ given by the connected components of $\Gcal'$, that all have at least 2 elements and are strongly connected. Hence (SC3) is satisfied.
\end{proof}

\subsection{General rates for arbitrary utility distributions}
\label{subsec:general_rates_arbitrary_distributions}

\cref{lemma:ball_in_region_no_assumptions,lemma:ball_in_region_optimized}, and \cref{lemma:prove_upper_bounds,lemma:upper_bounds_2D} can be used more generally to prove any convergence rates between $\sqrt{1-\gamma}$ and $1-\gamma$, as characterized in \cref{thm:full_characterization}. Before doing so, we introduce an additional notation. Fix a partition $ I =I_1\sqcup I_2\sqcup\cdots \sqcup I_q$ with $|I_s|\geq 2$ for all $s\in[q]$, which we denote $\Pcal$. In the main text, we introduced the quantity $f(\eta;\Pcal)$ to quantify the central planner's ability to meet future promises using the partition $\Pcal$ to group agents. For technical purposes, we introduce another related quantity as follows:
\begin{equation*}
    \tilde f(\eta;\Pcal) := \min_{s\in[q]} \min_{0\subsetneq B\subsetneq I_s} \Ebb\sqb{Z_B \1_{ Z_B\leq Z_{I_s\setminus B}=Z\leq  (1+\eta) Z_B}}.
\end{equation*}

These functions are identical up to constant factors as checked below.

\begin{lemma}\label{lemma:f_vs_f_p}
    Fix a partition $ I =I_1\sqcup\ldots\sqcup I_q$ with $|I_s|\geq 2$ for all $s\in[q]$, which we denote $\Pcal$. For any $\eta\geq 0$, we have 
    \begin{equation*}
        \frac{1}{n^2} f(\eta;\Pcal)\leq \tilde f(\eta;\Pcal) \leq f(\eta;\Pcal).
    \end{equation*}
\end{lemma}

\begin{proof}
    Fix $\eta\geq 0$ and $s\in[q]$. Using the min-cut/max-flow duality theorem, we have
    \begin{align*}
         \min_{i\neq j\in I_s} \paren{ \text{max flow from $i$ to $j$ on $\Hcal_{|I_s}$} }  &= \min_{i\neq j\in I_s}\min_{\{i\}\subseteq B\subseteq I_s\setminus\{j\}} \sum_{k\in B, l\in I_s\setminus B} f(\eta;k,l)\\
         &= \min_{\emptyset \subsetneq B\subsetneq I_s } \sum_{i\in B,\, j\in I_s\setminus B} f(\eta;i,j).
    \end{align*}

    We next note that for any $\emptyset\subsetneq B\subsetneq I_s$, we have
    \begin{align*}
         \Ebb\sqb{Z_B \1_{ Z_B\leq Z_{I_s\setminus B}=Z\leq  (1+\eta) Z_B}} &= \Ebb\sqb{\max_{i\in B,\,j\in I_s\setminus B} u_i  \1_{u_i=Z_B} \1_{u_j=Z }  \1_{ u_j\in[u_i,(1+\eta) u_i] }}.
    \end{align*}
    Therefore,
    \begin{align*}
        \frac{1}{n^2}\sum_{i\in B,\,j\in I_s\setminus B} f(\eta;i,j) 
        \leq
        \max_{i\in B,\,j\in I_s\setminus B}  f(\eta;i,j)
        &\leq \Ebb\sqb{Z_B \1_{ Z_B\leq Z_{I_s\setminus B}=Z\leq  (1+\eta) Z_B}}\\
         &\leq \sum_{i\in B,\, j\in I_s\setminus B} f(\eta;i,j)
         \leq \sum_{i\in B,\, j\in I_s\setminus B} f(\eta;i,j).
    \end{align*}
    Taking the minimum over all $s\in[q]$ and $\emptyset\subsetneq B\subsetneq I_s$ then gives the desired result.
\end{proof}

We are now ready to prove the complete characterization from \cref{thm:full_characterization}

\begin{proof}[of \cref{thm:full_characterization}] We prove the upper and lower bound separately.
\paragraph{Proof of the first claim.}
    Fix a partition $\tilde \Pcal$ of $I = J_1\sqcup J_2\sqcup\cdots \sqcup J_{\tilde q}$ with $|J_2|\geq 2$ for $s\in[\tilde q]$. As a first step, we construct a new partition $ I =I_1\sqcup I_2\sqcup\cdots \sqcup I_q$ which we denote $\Pcal$ by merging together parts $J_s$ and $J_{s'}$ satisfying  $\Pbb(Z_{J_s}=Z_{J_{s'}}=Z>0)>0$. This is the same procedure as within the proof of (SC3)$\implies$ (SC2) for \cref{thm:faster_rates_1-beta_long_version}. This proof then shows that the resulting partition $\Pcal$ satisfies \cref{eq:sets_proba_disjoints}. Further, if two parts $J_s$ and $J_{s'}$ are merged, there exists $j_s\in J_s$ and $j_{s'}\in J_{s'}$ such that $\Pbb(u_{j_s} = u_{j_{s'}}=Z>0)>0$. In particular, $f(\eta;j_s,j_{s'}) = \Omega(1)$ as $\eta\to 0$: in words, there is a link between the two merged parts with capacity is independent of $\eta$ in $\Hcal$. Since there are finitely-many merged parts, this implies that there is a constant $\zeta >0$ (dependent only on the graph $\Hcal$) such that for all $\eta\geq 0$,
    \begin{equation}\label{eq:reduction_to_merged_parts}
        f(\eta;\Pcal) \geq \zeta\cdot f(\eta;\tilde\Pcal).
    \end{equation}
    
    We may now focus on the partition $\Pcal$ only. We will later use \cref{eq:reduction_to_merged_parts} to translate the obtained guarantee to $f(\eta;\tilde\Pcal)$. Precisely, we aim to apply \cref{lemma:ball_in_region_optimized} with the partition $\Pcal$, which now satisfies \cref{eq:sets_proba_disjoints}. We follow similar arguments as in the proof of \cref{thm:faster_rates_1-beta}. In particular, we define the function $\phi^{(s)}:\Rbb^{I_s}\to\Rbb$ as in \cref{eq:def_phi}. We consider the allocation function $\mb p(\mb u)$ that allocates uniformly on $\argmax_{j\in[n]} u_j$ and let $\mb U$ be the utility vector realized by $\mb p$. 
    
    We next introduce the parameters used. Let 
    \begin{equation}\label{eq:def_constants}
        C_2:=16n^2 \bar v^2\paren{1+\frac{\max_{i\in [n]} \Ebb[u_i]}{\min_{s\in[q],i\in I_s}(\Ebb[u_i]\Pbb(Z_{I_s}=Z>0) - U_i )}}^2
    \end{equation}
    and $r_0 = \min_{s\in[q],i\in I_s}(\Ebb[u_i]\Pbb(Z_{I_s}=Z>0) - U_i )/2$. Next, for convenience, we pose $c_0=\frac{1} {16n^5}$ and $c'=\frac{1}{8n^2}$. Let $C_r:=c'\sqrt{\frac{C_2}{2}}$, $C_\eta= 4c_0 c'C_r $, and $C_\delta = \frac{4C_r}{c'^2 }=\frac{2C_2}{C_r}.$ Last, we pose $f_0=f(c'/2;\Pcal)$. We define
    \begin{equation*}
        \eta(\gamma) := \inf\, \{1\}\cup \set{\eta\in\sqb{\frac{C_r}{\min(r_0,c_0f_0/4)}\sqrt{1-\gamma},1}:\forall \eta'\in[\eta,1], f(\eta';\Pcal) \geq C_\eta \sqrt{1-\gamma} \frac{\eta'}{\eta} }.
    \end{equation*}
    First, if $\eta(\gamma)\geq \frac{c'}{2\sqrt 2}>0$, by \cref{thm:universal_lower_bound} we have
    $\SWG(\gamma) \leq C_1\sqrt{1-\gamma}$,
    for the corresponding constant $C_1$.
    Hence ensuring that $C\geq \frac{2\sqrt 2\,C_1}{c'}$ gives the desired result. We now suppose that we have $\eta(\gamma)<  \frac{c'}{2\sqrt 2}$. Note that $f$ is right-continuous, hence we have $f(\eta';\Pcal)\geq C_\eta\sqrt{1-\gamma}\frac{\eta'}{\eta(\gamma)}$ for all $\eta'\in[\eta(\gamma),1]$. 
    In the following, we check that the assumptions of \cref{lemma:ball_in_region_optimized} are satisfied for $\gamma\in[1/2,1)$ with the parameters:
    \begin{equation*}
        r = C_r \frac{\sqrt{1-\gamma}}{\eta(\gamma)}, \quad \delta = C_\delta \sqrt{1-\gamma}\,\eta(\gamma), \quad \text{and} \quad \mb x = \paren{U_i - (r+2\delta) \frac{\1_{i\in I_s}}{\|\mb 1_{I_s}\|}}_{i\in[n]}.
    \end{equation*}
    First note that by assumption we have $\eta(\delta) \leq \frac{c'}{2\sqrt 2} = \sqrt{\frac{C_r}{2C_\delta}}$ so that $r\geq 2\delta$. In particular, since $\gamma\geq 1/2$ the constraint $\frac{r\gamma}{1-\gamma}\geq \delta$ from \cref{eq:constraint_safe_boundary} is satisfied.
    
    We now fix $s\in[q]$ and let $\Hcal_s:=\{\mb\delta\in\Rbb^{I_s}:\mb 1_{I_s}^\top \mb\delta=0\}$. From the proof of \cref{thm:faster_rates_1-beta_long_version} (see \cref{eq:upper_bound_deviation_ZB}) for any $\mb\beta\in\Hcal_s$ with $\|\mb\beta\| \leq 1/2$, we have
    \begin{align*}
        \phi^{(s)}(\mb 1_{I_s} + \mb\beta) -\phi^{(s)}(\mb 1_{I_s}) - \mb\beta^\top \mb U_{I_s} &\geq n^2 c_0\, \Ebb\sqb{Z_B \1_{Z_B\leq Z_{I_s\setminus B}=Z\leq  (1+c'\|\mb\beta\|) Z_B}} \|\mb\beta\| \\
        &\geq n^2 c_0 \tilde f(c'\|\mb\beta\|;\Pcal)\|\mb\beta\| \overset{(i)}{\geq} c_0 f(c'\|\mb\beta\|;\Pcal)\|\mb\beta\|,
    \end{align*}
    where in $(i)$ we used \cref{lemma:f_vs_f_p}.
    Because the function $\mb\beta\mapsto \phi^{(s)}(\mb 1_{I_s} + \mb\beta) -\phi^{(s)}(\mb 1_{I_s} ) - \mb\beta^\top \mb U_{I_s} $ is convex in $\gamma$ and has value $0$ for $\mb\beta=\mb 0$, this implies that for all $\mb\beta\in\Hcal_s$,
    \begin{equation*}
         \phi^{(s)}(\mb 1_{I_s} + \mb\beta) -\phi^{(s)}(\mb 1_{I_s}) - \mb\beta^\top \mb U_{I_s} \geq c_0 \min(f(c'\|\mb\beta\|;\Pcal),f_0)\cdot \|\mb\beta\| .
    \end{equation*}
    As a result, for any $\mb\beta\in\Hcal_s$,
    \begin{align*}
        A_s(\mb\beta)&:=\phi^{(s)}(\mb 1_{I_s} + \mb\beta) -  (\mb 1_{I_s} +  \mb\beta)^\top \mb x_{I_s}-  (r+\delta)\|\mb 1_{I_s}+\mb\beta\|  \\
        &\overset{(i)}{\geq} c_0 \min(f(c'\|\mb\beta\|;\Pcal),f_s) \|\mb\beta\| - (r+\delta)(\|\mb 1_{I_s}+\mb\beta\|-\|\mb 1_{I_s}\|) + \delta\|\mb 1_{I_s}\| \\
        &\overset{(ii)}{\geq} c_0 \min(f(c'\|\mb\beta\|;\Pcal),f_s) \|\mb\beta\|   -\frac{4r}{\|\mb 1_{I_s}\|}\min ( \|\mb\beta\|^2, \|\mb 1_{I_s}\| \|\mb\beta\|) + \delta\|\mb 1_{I_s}\|.
    \end{align*}
    In $(i)$ we used $\phi^{(s)}(\mb 1_{I_s})=\mb 1_{I_s}^\top \mb U_{I_s}$ (see proof of \cref{thm:faster_rates_1-beta_long_version}), $\mb 1_{I_s}^\top \mb\beta=0$, and the previous inequality. In $(ii)$ we used $r\geq 2\delta$ and $\mb 1_{I_s}^\top \mb\beta=0$.
    We now distinguish between three cases. First suppose that $\|\mb\beta\| \leq \eta(\gamma)/c'$. Then,
    \begin{align*}
        A_s(\mb\beta)
        \geq \delta\|\mb 1_{I_s}\| -\frac{4r}{c'^2\|\mb 1_{I_s}\|} \eta(\gamma)^2
        = \sqrt{1-\gamma}\eta(\gamma)\paren{C_\delta\|\mb 1_{I_s}\| - \frac{4C_r}{c'^2\|\mb 1_{I_s}\|}} \geq 0.
    \end{align*}

    Next suppose that $\eta(\gamma)/c'\leq \|\mb\beta\| \leq 1/2$.
    Note that by assumption, we have $\eta(\gamma)/c'\leq 1/2$. Therefore, since $f(\cdot;\Pcal)$ is non-decreasing we have $f(c'\|\mb\beta\|;\Pcal)\leq f_0$. Further, since $c'\|\mb\beta\|\geq \eta(\gamma)$, we also have $f(c'\|\mb\beta\|)\geq C_\eta\sqrt{1-\gamma} c'\|\mb\beta\|/\eta(\gamma)$. Hence,
    \begin{align*}
        A_s(\mb\beta)
        \geq c f(c'\|\mb\beta\|) \|\mb\beta\| -  \frac{4r}{\|\mb 1_{I_s}\|} \|\mb\beta\|^2
        \geq \paren{c_0c'C_\eta - \frac{4C_r}{\|\mb 1_{I_s}\|}} \frac{\sqrt{1-\gamma}}{\eta(\gamma)}\|\mb\beta\|^2 \geq 0.
    \end{align*}
    In the last inequality, we used the definition of $C_\eta$.
     Last, for $\|\mb\beta\| \geq 1/2$, we obtain
     $A_s(\mb\beta) \geq (c_0 f_0 - 4r)\|\mb\beta\|$. Now by construction of $\eta(\gamma)$, we have
     \begin{equation*}
         r = C_r\frac{\sqrt{1-\gamma}}{\eta(\gamma)} \leq \min(r_0,cf_0/4).
     \end{equation*}
     This therefore shows that for all $\mb\beta\in\Hcal_s$ we have $A_s(\mb\beta)\geq 0$. Finally, we check that the second constraint from \cref{eq:constraint_safe_boundary} is satisfied. By construction and because $\gamma\geq 1/2$,
     \begin{equation*}
         \delta= 2C_2\frac{1-\gamma}{r}\geq C_2\frac{1-\gamma}{\gamma r}.
     \end{equation*}
     Further, since $r\leq r_0$, we can check that
     \begin{equation*}
         4n^2 \bar v^2\paren{1+\frac{\max_{i\in [n]} \Ebb[u_i]}{\min_{s\in[q],i\in I_s}(\Ebb[u_i]\Pbb(Z_{I_s}=Z>0) - x_i -r)}}^2 \leq C_2.
     \end{equation*}

     Finally, we have checked that all assumptions to apply \cref{lemma:ball_in_region_optimized} are satisfied. Therefore,
    \begin{equation*}
        \mb x+ \{0\}^{[n]\setminus I }\otimes \bigotimes_{s\in[q]} r\frac{\mb 1_{I_s}}{\|\mb 1_{I_s}\|} \in\Ucal_\gamma.
    \end{equation*}
    As a result,
    \begin{equation*}
        \SWG(\gamma) = \max_{\mb x\in\Ucal^\star} \mb 1^\top \mb x- \max_{\mb x\in\Ucal_\gamma} \mb 1^\top \mb x \leq 2\delta \sum_{s\in[q]}\|\mb 1_{I_s}\| \leq n\sqrt n C_\delta \sqrt{1-\gamma}\eta(\gamma).
    \end{equation*}
    Ensuring that $C\geq n\sqrt n C_\delta\max\paren{1,\frac{C_r}{\min(r_0,c_0 f_0/4)}}$ and up to multiplying $C_\eta$ by a factor $1/\zeta$, the previous equation together with \cref{eq:reduction_to_merged_parts} this implies the desired result. Note that the term $C(1-\gamma)$ takes care of the case when $\eta(\gamma) = \frac{C_r}{\min(r_0,c_0 f_0/4)}\sqrt{1-\gamma}$.

\paragraph{Proof of the second claim.}
    For $i\in I $, we define the functions $g_i^+$ and $g_i^-$ on $\Rbb_+$ via
    \begin{align*}
        g_i^+(\eta):=\Ebb\sqb{u_i \1_{u_i\leq Z_{-i} \leq  (1+\eta) u_i}}  \quad \text{and} \quad
        g_i^-(\eta):=\Ebb\sqb{u_i \1_{Z_{-i} < u_i \leq (1+\eta)Z_{-i}} }.
    \end{align*}
    Last, we define $g_i(\eta) = g_i^+(\eta) + g_i^-(\eta)$.
    Let $\tilde c_\eta = \min\{1\}\cup\{\Ebb[u_i\1_{u_i=Z_{-i}}]/2:i\in I , \Pbb(u_i=Z_{-i}>0)>0\}>0$. Ensuring $c_\eta\leq \tilde c_\eta$ implies that for any $i\in I $ such that $\Pbb(u_i=Z_{-i}>0)>0$, we have (for any $\eta_0>0$)
    \begin{equation*}
         \sup\, \{0\}\cup \{\eta\in[0,\eta_0]: g_i(\eta) \leq c_\eta\sqrt{1-\gamma} \} \leq \sup\, \{0\}\cup \{\eta\in[0,\eta_0]: g_i(\eta) \leq \tilde c_\eta \} =0.
    \end{equation*}
    Hence, the desired inequality is trivial in that case. In the rest of the proof, we therefore focus on an index $i\in I $ with $\Pbb( u_i=Z_{-i}>0)=0$. In particular, we necessarily have $i\in \tilde I$. 
    We now introduce the function
    \begin{equation*}
        \phi(\mb y):= \max_{\mb x\in\Ucal^\star} \mb y^\top (x_i, \mb 1_{-i}^\top \mb x_{-i}) = \Ebb\sqb{\max\paren{y_iu_i,y_{-i} Z_{-i}}},\quad \mb y\in\Rbb^2\setminus\{\mb 0\}.
    \end{equation*}
    We fix $\mb d:=(1,1)$ and $\mb f:=(1,-1)$ so that $\mb d^\top \mb f=0$. Let $\mb U$ be an optimal vector and let $\mb p(\cdot)$ be an optimal allocation that realizes $\mb U$. In the 2-agent game it induces the utility vector $\mb V=(U_i,\mb 1_{-i}\mb U_{-i})$. We denote $\hat i\sim\mb p(\mb u)$ the agent that receives the allocation. Then, for any $\eta\geq 0$, using the fact that $\phi(\mb d) = \mb d^\top\mb V$ and the same arguments as in the proof of \cref{thm:faster_rates_1-beta_long_version}, we have
    \begin{align*}
        \phi(\mb d+\eta \mb f)-\phi(\mb d)-\eta \mb f^\perp \mb V
        &= \Ebb\sqb{\max\paren{(1+\eta)u_i, (1-\eta)Z_{-i}}  - (1+\eta)u_i\1_{\hat i=i} -  (1-\eta) Z_{-i} \1_{\hat i\neq i} }\\
        &= \Ebb\sqb{\paren{(1+\eta)u_i - (1-\eta)  Z_{-i}}\1_{\hat i\neq i}\1_{(1+\eta)u_i \geq (1-\eta) Z_{-i}}}\\
        &\overset{(i)}{=}  \Ebb\sqb{\paren{(1+\eta)u_i - (1-\eta ) Z_{-i}} \1_{u_i < Z_{-i} \leq \frac{1+\eta}{1-\eta}u_i}}.
    \end{align*}
    In $(i)$ we used the fact that $\Pbb(u_i=Z>0)=0$ hence almost surely, $\mb p$ only allocates to $i$ if $u_i >Z_{-i}$ or if $u_i=Z_{-i}=0$ in which case $u_i=0$ and the contribution is null. For $\eta\geq 0$ sufficiently small (independently of $\gamma$), we have $\frac{1+\eta}{1-\eta}\leq 1+3\eta$. Then,
    \begin{align*}
        \phi(\mb d+\eta \mb f)-\phi(\mb d)-\eta \mb f^\perp \mb V &\leq 2\eta \, \Ebb\sqb{u_i  \1_{u_i < Z_{-i} \leq (1+3\eta)u_i}} = 2\eta g_i^+(3\eta).
    \end{align*}
    In the last inequality we used the fact that $\Pbb(u_i=Z_{-i}>0)=0$.
    Similarly, for $\eta \leq 0$, with $|\eta|$ sufficiently small, we have
    \begin{align*}
         \phi(\mb d+\eta \mb f) -\phi(\mb d)-\eta \mb f^\perp \mb V &= \Ebb\sqb{\paren{(1-\eta) Z_{-i} - (1+\eta)u_i }\1_{\hat i=i}\1_{(1+\eta)u_i \leq (1-\eta) Z_{-i}}}\\
         &=  \Ebb\sqb{\paren{(1-\eta) Z_{-i} - (1+\eta)u_i } \1_{Z_{-i} < u_i \leq \frac{1-\eta }{1+\eta } Z_{-i} }}\\
         &\leq 2 |\eta| \, \Ebb\sqb{u_i  \1_{Z_{-i}<u_i \leq (1+3|\eta|)Z_{-i}}} =  2|\eta| g_i^-(3|\eta|).
    \end{align*}
    
    We next introduce $c_1,c_2>0$ the constants guaranteed from \cref{lemma:upper_bounds_2D}. We now pose $c_\eta = \min\paren{\tilde c_\eta,\frac{\sqrt{3c_2}}{48}}$, $c_r = 3\sqrt{3c_2}$, and fix $0<\eta_0\leq \min(1,\sqrt{2c_r^2/(3c_2)},c_1/(\frac{c_2 }{\sqrt{2}c_r}+\frac{c_r}{9\sqrt 2}))$ sufficiently small such that all the previous estimates hold for $|\eta|\leq \eta_0$. Now let $\gamma\in[9/10,1)$ and define 
    \begin{equation*}
        \eta_i'(\gamma) := \sup\, \{0\}\cup \set{\eta\in[0,\eta_0]: g_i(\eta)=g_i^+(\eta) + g_i^-(\eta) \leq c_\eta\sqrt{1-\gamma} }.
    \end{equation*}
    If $\eta_i'(\gamma)=0$, we always have $\SWG(\gamma) \geq c\sqrt{1-\gamma} \cdot \eta_i'(\gamma)$ for any choice of constants. Suppose that is not the case and fix $\lambda_i(\gamma)\in(0,\eta_i'(\gamma))$. By construction, since $g_i^+$ and $g_i^-$ are non-decreasing, we have
    $g_i(\lambda_i(\gamma)) \leq c_\eta\sqrt{1-\gamma}$. We then pose
    \begin{equation*}
        r:= c_r\frac{ \sqrt{1-\gamma}}{\lambda_i(\gamma)},\quad \xi :=\frac{c_2}{2c_r}\sqrt{1-\gamma }\,\lambda_i(\gamma),  \quad \text{and} \quad \mb x := \mb V-(r+\xi)\frac{\mb d}{\|\mb d\|} .
    \end{equation*}
    For convenience, we introduce the function
    \begin{equation*}
        \psi(\mb y):=\max_{\mb z\in B(\mb x,r)} \mb y^\top \mb z = \mb y^\top \mb x +  r \|\mb y\|
    \end{equation*}
    We then compute for any $\eta$,
    \begin{align*}
        \psi(\mb d+\eta\mb f) -  \phi(\mb d) - \eta\mb f^\top\mb V
          =r( \|\mb d+\eta\mb f\|-\|\mb d\|)-\xi \|\mb d\| =r\|\mb d\| \paren{\sqrt{1+\eta^2}-1} -\xi \|\mb d\| 
    \end{align*}
    In the first equality, we used the fact that $\mb U$ is optimal, hence $\mb d^\top\mb V=\phi(\mb d)$. Next, using $1+x/2\geq \sqrt{1+x}\geq 1+x/3$ for all $x\in[0,1]$, we obtain that for $\eta$ with $|\eta|\leq 1$,
    \begin{equation*}
        \frac{1}{3} r\|\mb d\| \eta^2 -\xi \|\mb d\| \leq \psi(\mb d+\eta\mb f) -  \phi(\mb d) - \eta\mb f^\top\mb V \leq  \frac{1}{2} r\|\mb d\| \eta^2 -\xi \|\mb d\|.
    \end{equation*}
    Putting this with the previous equations, for any $\eta$ with $|\eta|\leq \eta_0$, we obtained
    \begin{align*}
        \phi(\mb d+\eta\mb f)-\psi(\mb d+\eta\mb f) 
        \leq |\eta|\paren{2 g_i(3|\eta|) - \frac{\|\mb d\|}{3} |\eta| r} + \xi\|\mb d\|.
    \end{align*}
    In particular, for $\eta\in\{\pm \lambda_i(\gamma)/3)\}$, we have
    \begin{align*}
        \phi(\mb d+\eta\mb f)-\psi(\mb d+\eta\mb f) &\leq \frac{\lambda_i(\gamma)}{3}\paren{2 c_\eta \sqrt{1-\gamma} - \frac{c_r\|\mb d\|}{9}\sqrt{1-\gamma}} + \frac{c_2}{2c_r} \|\mb d\|\lambda_i(\gamma) \sqrt{1-\gamma}\\
        &\overset{(i)}{=}\lambda_i(\gamma)\sqrt{1-\gamma} \paren{\frac{2c_\eta}{3} - \frac{\sqrt{c_2}\|\mb d\|}{6\sqrt 3}} \overset{(ii)}{=} -\lambda_i(\gamma)\sqrt{1-\gamma} \frac{c_\eta}{2}<0.
    \end{align*}
    In $(i)$ we used the definition of $c_r$ and in $(ii)$ we used the definition of $c_\eta$ and $\|\mb d\|=\sqrt 2$.
    As a result, $\mb x + r\frac{\mb d+\eta\mb f}{\|\mb d+\eta \mb f\|}\notin \{(u_i,\mb 1_{-i}^\top \mb u_{-i}),\mb u\in \Ucal^\star\}$ for $\eta\in\{\pm \lambda_i(\gamma)/3\}$.
    We can then define $\Ccal'$ as the union of all connected components of $\Ucal^\star\setminus B^\circ(\mb x, r;i)$ that are included within
    \begin{equation*}
        \set{\mb u: \mb d^\top (u_i,\mb 1_{-i}^\top \mb u_{-i}) > \mb d^\top \paren{\mb x + r\frac{\mb d+\eta\mb f}{\|\mb d+\eta \mb f\|}} ,\; \eta = \frac{\lambda_i(\gamma)}{3} }.
    \end{equation*}
    Note that for $|\eta|$ sufficiently small, we have
    \begin{equation*}
        \mb d^\top \paren{\mb x + r\frac{\mb d+\eta\mb f}{\|\mb d+\eta \mb f\|}}= \max_{\mb z\in\Ucal^\star}\mb 1^\top \mb z -\xi\|\mb d\|-r\|\mb d\|\paren{1-\frac{1}{\sqrt{1+\eta^2}}}.
    \end{equation*}
    Now by construction of $\eta_0$, for $|\eta|=\lambda_i(\gamma)3$, since $\lambda_i(\gamma)<\eta_i(\gamma)\leq \eta_0$, we have
    \begin{equation*}
        \xi\|\mb d\|+r\|\mb d\|\paren{1-\frac{1}{\sqrt{1+\eta^2}}} \leq \xi\|\mb d\|+\frac{1}{2}r\|\mb d\|\eta^2\leq  \frac{c_2}{2c_r}\|\mb d\|\lambda_i(\gamma) + \frac{c_r}{18}\|\mb d\| \lambda_i(\gamma) \leq c_1.
    \end{equation*}
    This proves that $\Ccal'\subseteq \set{\mb u:\mb 1^\top \mb u\geq  \max_{\mb z\in\Ucal^\star}\mb 1^\top \mb z - c_1}$.
    Pose $\delta=c_2\frac{1-\gamma}{\gamma r}=\frac{2}{\gamma}\xi\geq 2\xi$. Since $\gamma\geq 9/10$,
    \begin{equation*}
        \frac{\delta}{r} \leq \frac{3\xi}{r} = \frac{3c_2}{2c_r^2}\lambda_i(\gamma)^2\leq \frac{3c_2}{2c_r^2}\eta_0^2\leq 1.
    \end{equation*}
    That is, $\delta\leq r$.
    We next prove $\Ccal'\subseteq B(\mb x,r+\delta;i)$. To do so, it suffices to check that
    \begin{equation*}
        \phi(\mb d+\eta\mb f) \leq (\mb d+\eta\mb f)^\top \mb x + (r+\delta)\|\mb d+\eta\mb f\|,\quad \eta\in \sqb{-\frac{\lambda_i(\gamma)}{3},\frac{\lambda_i(\gamma)}{3} }.
    \end{equation*}
    Using the previous computations, for any $\eta$ such that $|\eta|\leq \lambda_i(\gamma)/3$, we have
    \begin{align*}
        \phi(\mb d+\eta\mb f) - &(\mb d+\eta\mb f)^\top \mb x - (r+\delta)\|\mb d+\eta\mb f\| \\
        &\leq 2 |\eta| g_i(3|\eta|) - \frac{1}{3}(r+\delta)\|\mb d\| \eta^2-(\delta-\xi)\|\mb d\|\\
        &\overset{(i)}{\leq}  \frac{2 c_\eta \lambda_i(\gamma)}{3}\sqrt{1-\gamma} -\xi\|\mb d\| \overset{(ii)}{\leq} 0,
    \end{align*}
    where in $(i)$ we used $|\eta|\leq \lambda_i(\gamma)/3$ and $\delta\geq 2\xi$, and in $(ii)$ we used the definition of $c_\eta$ and $\xi$. This proves $\Ccal'\subseteq B(\mb x,r+\delta;i)$.
    We have now checked all the assumptions for \cref{lemma:upper_bounds_2D} which implies
    \begin{equation*}
        \Ucal_\gamma \cap \paren{\Ccal' \cap B^\circ\paren{\mb x,r+\frac{1-\gamma}{\gamma}\delta;i}} = \emptyset.
    \end{equation*}
    Next, note that by construction, $\mb U\in\Ccal$ and $(U_i,\\mb 1_{-i}^\top \mb U_{-i}) = \mb x+(r+\xi)\frac{\mb d}{\|\mb d\|}$. Hence, by \cref{lemma:upper_bounds_2D},
    \begin{align*}
        \max_{\mb y\in\Ucal_\gamma} \mb 1^\top \mb y  < \mb d^\top \mb x - \paren{ r+ \frac{1-\gamma}{\gamma}\delta } \|\mb d\| =  \max_{\mb y\in\Ucal^\star} \mb 1^\top \mb y  - \paren{\xi - \frac{1-\gamma}{\gamma}\delta}\|\mb d\|
    \end{align*}
    Hence, for $1-\gamma\leq 1/10$, we obtain
    \begin{equation*}
        \SWG(\gamma) = \max_{\mb y\in\Ucal^\star} \mb 1^\top \mb y - \max_{\mb y\in\Ucal_\gamma} \mb 1^\top \mb y \geq \frac{\xi}{2}\|\mb d\| = \frac{c_2\|\mb d\|}{2c_r} \sqrt{1-\gamma}\,\lambda_i(\gamma).
    \end{equation*}
    Because this holds for all $\lambda_i(\gamma) \in(0,\eta_i'(\gamma))$, this also holds for $\eta_i'(\gamma)$.

    In summary, we proved that for each $i\in I$, there exists constants $c_\eta(i),\eta_0(i),c(i)>0$ depending only on the utility distributions such that for $1-\gamma\leq 1/10$,
    \begin{equation*}
        \SWG(\gamma) \geq c(i)\sqrt{1-\gamma} \cdot \sup\, \{0\}\cup \{\eta\in[0,\eta_0(i)]: g_i(\eta) \leq c_\eta(i)\sqrt{1-\gamma} \} 
    \end{equation*}
    We take $c_\eta, \eta_0,c$ to be the minimum of these quantities for all $i\in I$, which shows that the above equation holds by replacing $g_i$ with $\min_{i\in I} g_i$, and $c_\eta(i),\eta_0(i),c(i)$ with $c_\eta,\eta_0,c$.
    The last step of the proof is to relate the functions $g$ and $\min_{i\in I}g_i$. For a given $\eta\in[0,\eta_0]$, there is $i\in I$ such that
    \begin{align*}
        g(\eta) &= \sum_{j\in I \setminus\{i\}} f(\eta;i,j)+f(\eta;j,i)\\
        &\geq  \Ebb\sqb{u_i \1_{Z_{-i}=Z}\1_{Z_{-i}\in[u_i,(1+\eta)u_i]} } +\Ebb\sqb{ Z_{-i} \1_{u_i=Z}\1_{u_i\in[Z_{-i},(1+\eta)Z_{-i}]}}\\
        &\geq g_i^+(\eta) + \frac{g_i^-(\eta)}{1+\eta} \geq \frac{1}{2} \min_{i\in I} g_i(\eta),
    \end{align*}
    where in the last inequality we used $\eta_0\leq 1/2$. Therefore, up to halving the parameter $c_\eta$, this proves the desired result.
\end{proof}

We now derive some immediate consequences from \cref{thm:full_characterization}. 
First, we prove \cref{thm:no_faster_rate_1-beta} that gives necessary conditions for having rates $\Ocal(1-\gamma)$.

\begin{proof}[or \cref{thm:no_faster_rate_1-beta}]
    If $g(\eta)\underset{\eta\to 0}{=}o(\eta)$, then
    $\sup\, \{0\}\cup \{\eta\in[0,\eta_0]: g(\eta) \leq c_\eta\sqrt{1-\gamma} \}  = \omega(\sqrt{1-\gamma}),$
    which gives the desired lower bound $\omega(1-\gamma)$ on the convergence rate.
    \comment{
    We now show that (NC2) implies $g_i(\eta)\underset{\eta\to 0}{=}o(\eta)$ (NC1). Using the same notations as in \cref{thm:full_characterization}, for any $\eta\geq 0$,
    \begin{align*}
        &g_i(\eta) = g_i^+(\eta) + g_i^-(\eta)\\
        &\leq\sum_{j\neq i} \Ebb\sqb{u_i \1_{ \alpha_i u_i\leq Z_{-i}=\alpha_ju_j \leq  (1+\eta) \alpha_i u_i}} + \Ebb\sqb{u_i \1_{Z_{-i}=\alpha_ju_j \leq \alpha_i u_i \leq (1+\eta)\alpha_ju_j} }\\
        &\leq \sum_{j\neq i,\alpha_j>0} \Ebb\sqb{u_i \1_{\alpha_ju_j=Z} \1_{\alpha_ju_j\in [ \alpha_i u_i, (1+\eta) \alpha_i u_i]}} 
        + \frac{\alpha_j}{\alpha_i}(1+\eta)\Ebb\sqb{u_j \1_{Z_{-i}=\alpha_ju_j \leq \alpha_i u_i \leq (1+\eta)\alpha_ju_j} }\\
        &\leq \sum_{j\neq i,\alpha_j>0} \Ebb\sqb{u_i \1_{\alpha_ju_j=Z} \1_{\alpha_ju_j\in [ \alpha_i u_i, (1+\eta) \alpha_i u_i]}} 
        + \frac{\alpha_j}{\alpha_i}(1+\eta)\Ebb\sqb{u_j \1_{\alpha_iu_i=Z}\1_{\alpha_iu_i\in[\alpha_ju_j , (1+\eta)\alpha_ju_j]} }.
    \end{align*}
    Hence, (NC2) implies (NC1), which ends the proof of the result.}
\end{proof}

Next, we can prove the $2$-agent scenario characterization from \cref{prop:2_agent_case_example}.

\begin{proof}[of \cref{prop:2_agent_case_example}]
    Since $I=\{1,2\}$ and $f(\eta;1,2),f(\eta;2,1)=\Theta(\eta^p)$, we have $g(\eta)=\Theta(\eta^p)$. Also, using the partition $\Pcal=\{\{1,2\}\}$ which groups all agents together we also obtain $f(\eta;\Pcal)=\Theta(\eta^p)$, where $p\geq 1$. A direct computation then shows that \cref{thm:full_characterization} implies $\SWG(\eta)=\Theta((1-\gamma)^{(1+\frac{1}{p})/2})$ as desired.
\end{proof}

Next, we prove \cref{cor:super_smooth_case}.

\begin{proof}[of \cref{cor:super_smooth_case}]
    We use \cref{thm:full_characterization}. Suppose that we have density upper bounds. From the distributional assumptions, for any $i,j\in I$ we have $f(\eta;i,j)\underset{\eta\to 0}{=}\Ocal(\eta)$. Hence, we also have $g(\eta)\underset{\eta\to 0}{=}\Ocal(\eta)$. This implies that $g(\eta)\underset{\eta\to 0}{=}\Ocal(\eta)$. Hence $\sup \{\eta\in[0,\eta_0]: g(\eta) \leq c_\eta\sqrt{1-\gamma} \}=\Omega(\sqrt{1-\gamma})$ as $\gamma\to 1$ and \cref{thm:full_characterization} provides the desired first claim. The second claim is taken directly from \cref{cor:simple_examples_fast_convergence}.
\end{proof}

\subsection{Remaining proofs for discrete distributions}
\label{subsec:discrete_distributions}

In this section, we give the remaining proof of \cref{thm:discrete_case_full} from \cref{sec:examples}. The tools developed until now already imply the following.

\begin{corollary}\label{cor:discrete_case}
    Suppose that all $\Dcal_1,\ldots,\Dcal_n$ are discrete and that $\Ucal_0$ does not already contain an optimal vector (see \cref{lemma:trivial_case} for a characterization). 
    If there exists $i\in I $ such that for all $j\neq i$, $\Pbb( u_i= u_j=Z>0)=0$, then $\SWG(\gamma) = \Theta(\sqrt{1-\gamma})$. Otherwise, $\SWG(\gamma) = \Ocal(1-\gamma)$.
\end{corollary}

\begin{proof}
    We let $m:=\max_{i\in[n]}\min ( \Supp(\Dcal_i))$ be the minimum of the support of $Z$.
    We start with the first case. The assumption implies that for some $i\in I$, $\Pbb( u_i=Z_{-i}>0)=0$. As a result, there exists $\eta_1>0$ such that
    \begin{equation*}
        \Pbb(u_i>0, Z_{-i}\in [ u_i/(1+\eta_1),(1+\eta_1) u_i])=0.
    \end{equation*}
    This implies that for all $\eta\in[0,\eta_1]$, we have $g_i(\eta)=0$. Then, \cref{thm:full_characterization} implies 
    \begin{equation*}
        \SWG(\gamma) \geq c\sqrt{1-\gamma} \min(\eta_0,\eta_1).
    \end{equation*}
    The other inequality is already guaranteed by \cref{thm:universal_lower_bound}, which ends the first claim.

    We now consider the second alternative. In this case, the first bullet point from \cref{cor:simple_examples_fast_convergence} is satisfied for all $i\in I $. This implies the second claim of the corollary.
\end{proof}

We now use a gluing method to improve the convergence rate in the second case of \cref{cor:discrete_case}. In this case, the frontier of $\Ucal^\star$ at optimal points is flat, which will allow to fit local caps from larger radius balls close to this flat surface. Here, the form of the resulting frontier can be explicitly constructed, which allows for a simpler analysis. The intuition is however the same as that from the gluing approach described in \cref{sec:examples} and gives the same convergence rates up to constants (potentially in exponents). The proof of \cref{thm:extra_fast_rates} is detailed below. In particular, it implies the second claim of \cref{thm:discrete_case_full}.

\begin{proof}[of \cref{thm:extra_fast_rates}] Suppose that the distributions satisfy the assumptions that for all $i\in I$, there exists $j\neq i$ with $\Pbb(u_i=u_j=Z>0)>0$. We denote the set of optimal utility vectors by
\begin{equation*}
    \Ucal^\star(\mb 1):=\set{\mb x\in\Ucal^\star, \mb 1^\top \mb x = \max_{\mb y\in\Ucal^\star} \mb 1^\top \mb y}.
\end{equation*}
We next construct the (undirected) graph $\Gcal$ on $ I $ such that for any $i\neq j\in I $, $i$ is connected to $j$ if and only if $\Pbb(u_i= u_j=Z>0)>0$. By assumption, for all $i\in I $ there exists $j\neq i$ such that $\Pbb( u_i= u_j=Z>0)>0$. In particular, this implies $j\in I $. This shows that $\Gcal$ does not have any isolated node. We then consider the partition $ I =I_1\sqcup \ldots \sqcup I_q$ by connected components of $\Gcal$. Since there is no isolated node, we have $|I_s|\geq 2$ for all $s\in[q]$. By construction, since two components are not connected, \cref{eq:sets_proba_disjoints} is satisfied. 
Following the same arguments as in \cref{lemma:ball_in_region_optimized}, it suffices to focus on each set $I_s$ independently for all $s\in[q]$. Precisely, we focus on the game where only agents $I_s$ are present but the central planner cannot allocate on the event $\Ecal_s :=\{Z_{-I_s}\geq Z_{I_s}\}$. Suppose we showed that the region $\Rcal_s\subseteq\Rbb_+^{I_s}$ is achievable in this setting for all $s\in[q]$, then the proof of \cref{lemma:prove_upper_bounds} implies that for any optimal vector $\mb U$, the region
\begin{equation}\label{eq:combining_together}
    \Rcal:=\mb U_{[n]\setminus  I } \otimes \bigotimes_{s\in[q]} \Rcal_s
\end{equation}
is achievable in the original setting with $[n]$ agents: $\Rcal\subseteq \Ucal_\gamma$.

We therefore focus on a single connected component $I_s$ for $s\in[n]$ from now. We fix any optimal vector $\mb U$ and let $\Hcal_s=\{\mb y:\forall i\notin I_s,y_i=U_i\}$. For any $\mb x\in\Rbb^{I_s}$ and $r>0$, we also denote $B_{I_s}(\mb x, r)$ the corresponding ball on the space $\Rbb^{I_s}$. We first show that $\Ucal^\star(\mb 1)\cap \Hcal_s$ has (full) dimension $|I_s|-1$. 
Note that since $I_s$ is connected, there exist values $m_1>\ldots>m_T>0$ such that letting
\begin{equation*}
    J_t:=\set{i\in I_s: \Pbb(u_i= m_t=Z)>0},\quad t\in[T],
\end{equation*}
we have (1) $|J_t|\geq 2$ for $t\in[T]$ and (2) no set $J_t$ is disjoint from the others, that is $J_t\cap \bigcup_{t'\neq t}J_{t'}\neq\emptyset$ for all $t\in[T]$.
We now define the event $\Fcal_t=\{\forall i\in J_t,u_i=m_t=Z\}$. By construction, $\Pbb(\Fcal_t)>0$ and all events $\Fcal_1,\ldots,\Fcal_T$ are disjoint. On each event $\Fcal_t$, an optimal allocation can choose to allocate to any agent in $J_t$. Let $\mb p^{(0)}(\cdot)$ an optimal allocation such that $\Ebb[u_ip^{(0)}_i(\mb u)]=U_i$ for all $i\notin I_s$ and that for any $t\in[T]$, allocates uniformly among agents $J_t$ on the event $\Fcal_t$. Denote $\mb U^{(0)}$ the utility vector realized by $\mb p^{(0)}$. The previous arguments imply
\begin{equation*}
    \Scal:=\mb U^{(0)} + \sum_{t\in[T]}\Pbb(\Fcal_t) \set{ \paren{ m_t\paren{q_i-\frac{1}{|J_t|}} \1_{i\in J_t}}_{i\in[n]}, \mb q\in\Delta_{J_t}} \subseteq \Ucal^\star(\mb 1),
\end{equation*}
where the sum between sets are Minkowski sums. Here each sum corresponds to the freedom that an optimal allocation has on the event $\Fcal_t$ to use any allocation within $\Delta_{J_t}$. From the assumptions (1) and (2) on the sets $J_1,\ldots,J_T$, with $\tilde r_0=\min_{t\in[T]}\Pbb(\Fcal_t) m_t/n$, we have
\begin{equation}\label{eq:starting_assumption}
    \mb U^{(0)} + \set{\paren{y_i\1_{i\in I_s}}_{i\in[n]}, \mb y\in B_{I_s}(0,\tilde r_0),\mb 1_{I_s}^\top \mb y=0 }  \subseteq \Ucal^\star(\mb 1) \cap\Hcal_s.
\end{equation}

We recall that for any $\mb x\in\Ucal^\star$, any $\mb 0\leq \mb y\leq \mb x$ satisfies $\mb y\in\Ucal^\star$. This also holds in the setting where only agents $I_s$ are present and there is no allocation on the event $\Ecal_s$. We denote by $\Ucal_s^\star$ the corresponding achievable region. In the proof of \cref{lemma:ball_in_region_optimized}, we showed (\cref{eq:reachable_utilities_subgame}) that $
        \Ucal_s^\star = P_{I_s}\paren{ \Ucal^\star \cap\Hcal_s },$
where $P_{I_s}:\Rbb^n\to\Rbb^{I_s}$ is the projection on the $I_s$ coordinates. For convenience, we also write $\mb U_s:=\mb U^{(0)}_{I_s}$.
As a result, from \cref{eq:starting_assumption} we can check that with $r_0 = \tilde r_0/(8\|\mb 1\|)$, for any $\mb z\in B_{I_s}(\mb U_s,r_0)\cap \{\mb y: \mb 1_{I_s}^\top (\mb y-\mb U_s)=0\}:=S_s$,
\begin{equation*}
    B_{I_s}\paren{\mb z - r_0\frac{\mb 1_{I_s}}{\|\mb 1_{I_s}\|}, r_0} \subseteq P_{I_s}(\Ucal^\star\cap\Hcal_s)= \Ucal_s^\star.
\end{equation*}

Note that for all $i\in I_s$, we have $\Ebb[u_i]\Pbb(Z_{I_s}=Z>0)-(\mb z - r_0\frac{\mb 1_{I_s}}{\|\mb 1_{I_s}\|})_i-r_0\geq \Pbb(\Fcal_t) m_t (1-1/|J_t|)-2r_0 - r_0 \geq 8r_0-3r_0\geq  2r_0 $, where $t$ is such that $i\in J_t$.
    Then, with 
    \begin{equation*}
    C=4n^2\bar v^2\paren{1+\frac{\max_{i\in[n]}  \Ebb[u_i] + r_0}{r_0 }}^2
    \end{equation*}
    and $\delta=4C\frac{1-\gamma}{r_0}$, provided that $\delta\leq r_0/2$ then \cref{eq:constraint_safe_boundary} is satisfied with $r=r_0-\delta$ and $\delta$:
    \begin{equation*}
        \frac{r\gamma}{1-\gamma}\geq r\geq \frac{r_0}{2}\geq \delta =4C\frac{1-\gamma}{r_0}\geq C \frac{1-\gamma}{\gamma r}.
    \end{equation*}
    Here, we used $\gamma\geq 1/2$ and $r_0-\delta\geq r_0/2$. Let $C_1:=5n\bar v$. 
    The inequality $\delta\leq r_0/4$ holds whenever $\gamma\geq \gamma_1:=\max(1-\min(\frac{r_0^2}{16C}, (\frac{r_0}{8C\|\mb 1\|})^2, \frac{r_0}{4C_1}),1/2)$. Then, for $\gamma\in[\gamma_1,1)$, the proof of \cref{lemma:ball_in_region_optimized} shows that
    \begin{equation}\label{eq:safe_ball}
        B_{I_s}\paren{\mb z - r_0\frac{\mb 1_{I_s}}{\|\mb 1_{I_s}\|}, r_0-\delta}\subseteq\Ucal_{s,\gamma},
    \end{equation}
    where $\Ucal_{s,\gamma}$ is the achievable region for the setting with only agents $I_s$ and in which the central planner cannot allocate on the event $\Ecal_s$, with discount factor $\gamma$.
    We define the following function $f:\Rbb\to\Rbb$,
    \begin{equation*}
        f(r) := \delta \, \frac{\cosh\paren{\frac{c_f r}{\sqrt{1-\gamma} } } }{\cosh\paren{\frac{c_f r_0}{\sqrt{1-\gamma}} }} = 4C\frac{1-\gamma}{r_0} \, \frac{\cosh\paren{\frac{c_f r}{\sqrt{1-\gamma} } } }{\cosh\paren{\frac{c_f r_0}{\sqrt{1-\gamma}} }},\quad r\geq 0.
    \end{equation*}
    where $c_f= \min\paren{ \frac{1}{4C_1},1}$. We then pose $\gamma_0=\max(\gamma_1,1-\min(r_0^2 c_f^2,(\frac{c_f C}{C_1})^2))$ and consider $\gamma\in[\gamma_0,1)$.
    We focus on the following region
    \begin{equation*}
        \Rcal_s = \set{\mb y\in\Rbb^{I_s}:\mb 0\leq \mb y\leq \mb z-f(r)\frac{\mb 1_{I_s}}{\|\mb 1_{I_s}\|}, \|\mb z-\mb U_{I_s}\|=r,\mb z\in S_s}
    \end{equation*}
    and now construct an allocation and promised utility function on $\Rcal_s$. For any $\mb z\in S_s$, we denote by $\mb p(\cdot;\mb z)$ an allocation that realizes $\mb z$ in the full information setting (within the game when only agents $I_s$ are present). We start by specifying the strategy on the boundary points of $\Rcal_s$. For $\mb z\in S_s$, let $r=\|\mb z-\mb U_s\|$ and $\mb U=\mb z-f(r)\mb 1_{I_s}/\|\mb 1_{I_s}\|$. We pose
    \begin{equation*}
        \mb p(\cdot\mid\mb U):=\mb p(\cdot;\mb z) \quad \text{and} \quad \mb\alpha(\mb U):=-\frac{\mb 1_{I_s}}{\|\mb 1_{I_s}\|} -f'(r)\frac{\mb z-\mb U_s}{\|\mb z-\mb U_s\|}
    \end{equation*}
    Note that $\mb\alpha(\mb U)$ was selected (up to the sign) as the normal to the frontier manifold $\Mcal_s:=\{\mb z'-f(\|\mb z'-\mb U_s\|)\mb 1_{I_s}/\|\mb 1_{I_s}\|,\mb z'\in S_s\}$. We then extend the definition of the allocation and promised utility functions to all $\mb U\in \Rcal_s$ as follows. Let $\mb V\in\Mcal_s$ such that $\mb 0\leq \mb U\leq \mb V$. We let
    \begin{equation*}
        p_i(\cdot\mid\mb U) := \frac{U_i}{V_i}p_i(\cdot\mid \mb V)\quad \text{and} \quad W_i(\cdot\mid\mb U) := \frac{U_i}{V_i}W_i(\cdot\mid \mb V),\quad i\in I_s,
    \end{equation*}
    with the convention $0/0=0$ when $U_i=V_i=0$. To prove that $\Rcal_s\subseteq\Ucal_{s,\gamma}$ it suffices to show that for all $\mb U\in\Rcal_s$ and $\mb v\in[0,\bar v]^{I_s}$,
    \begin{equation}\label{eq:goal_valid_promise_v1}
        \mb W(\mb v\mid\mb U)\in \Rcal_s \cup \bigcup_{\mb z\in S_s} \set{\mb y:\mb 0\leq \mb y\leq \mb x, \mb x\in B_{I_s}\paren{\mb z - r_0\frac{\mb 1_{I_s}}{\|\mb 1_{I_s}\|}, r_0-\delta}  },
    \end{equation}
    and \cref{eq:valid_interim_promise} is satisfied for all $\mb U\in\Rcal_s$. Here, we used \cref{eq:safe_ball} that already guarantees that the second term in the right-hand side of \cref{eq:goal_valid_promise_v1} is contained within $\Ucal_\gamma$. By the linearity of the definitions of the allocation and promised utility function, it suffices to check that for all boundary utility vectors $\mb U\in\Mcal_s$, \cref{eq:valid_interim_promise} is satisfied (this is guaranteed by \cref{eq:coupling_formula}) and to check that the following holds:
    \begin{equation}\label{eq:goal_valid_promise}
        \mb W(\mb v\mid\mb U)\in \Rcal_s \cup \bigcup_{\mb z\in S_s}  B_{I_s}\paren{\mb z - r_0\frac{\mb 1_{I_s}}{\|\mb 1_{I_s}\|}, r_0-\delta} ,\quad \mb U\in\Mcal_s,\mb v\in[0,\bar v]^{I_s},
    \end{equation}
    
   Fix such a vector $\mb U\in\Mcal_s$ and write $\mb U=\mb z - f(r)\frac{\mb 1_{I_s}}{\|\mb 1_{I_s}\|}$ where $r=\|\mb z-\mb U_s\|$. By the promise-keeping equality \cref{eq:target_met} (which is a consequence of the construction),
    \begin{equation}\label{eq:forumla_W_bar_gluing}
        \bar{\mb W}:=\Ebb[\mb W(\mb u\mid \mb U)] = \mb U - \frac{1-\gamma}{\gamma} f(r)\frac{\mb 1_{I_s}}{\|\mb 1_{I_s}\|}.
    \end{equation}
    By construction, we also have $\mb\alpha(\mb U)^\top (\mb W(\mb v\mid\mb U) - \bar{\mb W})=0$ for $v\in[0,\bar v]^{I_s}$. Last, the proof of \cref{lemma:ball_in_region_no_assumptions} (\cref{eq:W_deviation_bound}) shows that
    \begin{equation*}
        \|\mb W(\mb v\mid\mb U) - \bar{\mb W} \| 
        \leq \frac{1-\gamma}{\gamma} \bar v  \paren{2n + \frac{\alpha_{i_1}(\mb U)}{\alpha_{i_2}(\mb U)}}   ,\quad \mb v\in[0,\bar v]^{I_s},
    \end{equation*}
    where $\alpha_{i_1}(\mb U)$ and $\alpha_{i_2}(\mb U)$ are the first and second largest components in absolute value of $\mb\alpha(\mb U)$. Note that from $\gamma\geq \gamma_1$ and the definition of $\gamma_1$, we have
    \begin{equation*}
        \norm{\mb\alpha(\mb U)-\frac{\mb 1_{I_s}}{\|\mb 1_{I_s}\|}} = |f'(r)| = c_f\frac{\delta}{\sqrt{1-\gamma}} \frac{|\sinh(c_f r/\sqrt{1-\gamma})|}{\cosh(c_fr_0/\sqrt{1-\gamma})}\leq \frac{1}{2\|\mb 1\|}.
    \end{equation*}
    As a result, this shows that for all $i\in I_s$ we have $\alpha_i(\mb U) \|\mb 1_{I_s}\| \in [1/2,3/2]$.  Therefore, we obtained
    \begin{equation}\label{eq:bound_delta_W_gluing}
        \|\mb W(\mb v\mid\mb U) - \bar{\mb W} \| 
        \leq \frac{1-\gamma}{\gamma} \bar v  (2n + 3\|\mb 1\|) \leq 2C_1(1-\gamma)   ,\quad \mb v\in[0,\bar v]^{I_s}.
    \end{equation}
    Now let $d(\mb v) := \|P_{\mb 1_{I_s}^\perp}(\mb W(\mb v\mid\mb U)-\mb U_s)\| - \|P_{\mb 1_{I_s}^\perp}(\bar{\mb W}-\mb U_s)\| $, where $P_{\mb 1_{I_s}^\perp}$ denotes the projection onto the orthogonal space to $\mb 1_{I_s}\Rbb$. Because $\mb\alpha(\mb U)^\top(\mb W(\mb v\mid\mb U)-\bar{\mb W})=0$, we can write
    \begin{equation*}
        \mb W(\mb v\mid\mb U) -\bar{\mb W} = d(\mb v) \frac{\mb z-\mb U_s}{\|\mb z-\mb U_s \|} + d(\mb v) f'(r)\frac{(-\mb 1_{I_s})}{\|\mb 1_{I_s}\|} + \mb W_0(\mb v),
    \end{equation*}
    where $\mb W_0(\mb v)\in Span(\mb 1_{I_s},\mb z-\mb U_s)^\perp$. Using \cref{eq:forumla_W_bar_gluing} we have
    \begin{align*}
        \frac{-\mb 1_{I_s}^\top}{\|\mb 1_{I_s}\|} (\mb W(\mb v\mid\mb U) - \mb U_s)  &= f(r) + f(r)\frac{1-\gamma}{\gamma} +  \frac{-\mb 1_{I_s}^\top}{\|\mb 1_{I_s}\|} (\mb W(\mb v\mid\mb U) - \bar{\mb W})  \\
        &= f(r) + f(r)\frac{1-\gamma}{\gamma} + d(\mb v) f'(r).
    \end{align*}
    Using Taylor expansion's theorem, there exists $\tilde r\in[\min(r,r+d(\mb v)),\max(r,r+d(\mb v))]$ such that
    \begin{align*}
         f(r)\frac{1-\gamma}{\gamma}\geq \frac{-\mb 1_{I_s}^\top}{\|\mb 1_{I_s}\|} (\mb W(\mb v\mid\mb U) - \mb U_s) - f(r+d(\mb v)) &= f(r)\frac{1-\gamma}{\gamma} - \frac{f''(\tilde r)}{2} d(\mb v)^2\\
        &\overset{(i)}{=} f(r)(1-\gamma) - \frac{c_f^2 f(\tilde r)}{2(1-\gamma)} d(\mb v)^2\\
        &\overset{(ii)}{\geq} (1-\gamma) \paren{f(r) - 2c_f^2 C_1^2 f(\tilde r)}\\
        &\overset{(iii)}{\geq} (1-\gamma) \paren{f(r) - \frac{f(r+|d(\mb v)|)}{2}}
    \end{align*}
    In $(i)$ we used the definition the fact that $\cosh''(x)=\cosh(x)$ and in $(ii)$ we used $|d(\mb v)|\leq \|\mb W(\mb v\mid\mb U)-\bar{\mb W}\| $ together with \cref{eq:bound_delta_W_gluing}. Last, in $(iii)$ we used the definition of $c_f$ together with the fact that $f(\tilde r)\leq f(r+|d(\mb v)|)$. Now note that
    \begin{equation}\label{eq:bound_f_increase}
        f(r+|d(\mb v)|) \leq f(r) e^{\frac{c_f |d(\mb v)|}{\sqrt{1-\gamma}}} \leq f(r) e^{2c_fC_1\sqrt{1-\gamma}} \leq 2f(r). 
    \end{equation}
    In summary, since $\gamma\geq 1/2$, we showed that
    \begin{equation}\label{eq:useful_both_sides_bounds}
        f(r+d(\mb v)) \leq \frac{-\mb 1_{I_s}^\top}{\|\mb 1_{I_s}\|} (\mb W(\mb v\mid\mb U) - \mb U_s)  \leq f(r+d(\mb v)) + 2f(r)(1-\gamma).
    \end{equation}
    First suppose that $|r+d(\mb v)|\leq r_0$. We let $\mb z(\mb v) := \mb U_s+P_{\mb 1_{I_s}^\top} (\mb W(\mb v\mid\mb U)-\mb U_s)\in S_s$. The previous equation implies that $\mb W(\mb v\mid\mb U)\leq \mb z(\mb v)-f(\|\mb z(\mb v)-\mb U_s\|)\frac{\mb 1_{I_s}}{\|\mb 1_{I_s}\|}$. Also, $2f(r)(1-\gamma)\leq f(r_0)=\delta$, hence, we also obtain $\mb 0\leq  \mb W(\mb v\mid\mb U)$. Together with the previous inequalities, this gives $\mb W(\mb v\mid\mb U)\in\Rcal_s$.

    We now treat the remaining case when $|r+d(\mb v)|\geq r_0$. We first show $d(\mb v)\geq 0$. Otherwise, we have $2C_1(1-\gamma)\geq d(\mb v) \geq r_0$, which is absurd since $\gamma\geq \gamma_1$. For this boundary case, we let
    \begin{equation*}
        \mb z(\mb v) := \mb U_s+r_0\frac{P_{\mb 1_{I_s}^\top} (\mb W(\mb v\mid\mb U)-\mb U_s)}{\|P_{\mb 1_{I_s}^\top} (\mb W(\mb v\mid\mb U)-\mb U_s)\|},
    \end{equation*}
    and aim to show that
    \begin{equation}\label{eq:boundary_within_ball}
        \mb W(\mb v, \mb U)\in B_{I_s}\paren{\mb z(\mb v) - r_0\frac{\mb 1_{I_s}}{\|\mb 1_{I_s}\|}, r_0-\delta}.
    \end{equation}
    Note that $f(r_0)=\delta$ and by \cref{eq:useful_both_sides_bounds} there exists $\tilde f\in[0,2f(r_0)(1-\gamma)]\subseteq[0,f(r_0)]$ such that
    \begin{align*}
        R(\mb v):=&\norm{\mb W(\mb v, \mb U) -\paren{ \mb z(\mb v) - r_0\frac{\mb 1_{I_s}}{\|\mb 1_{I_s}\|} } }^2 = (r+d(\mb v)-r_0)^2 + (f(r+d(\mb v))+\tilde f - r_0)^2\\
        &\overset{(i)}{\leq} (r+d(\mb v)-r_0)^2 + (r_0-f(r+d(\mb v)))^2 \\
        &=(r_0-\delta)^2 +  (r+d(\mb v)-r_0)^2 -(2r_0 +f(r_0)-f(r+d(\mb v)))(f(r+d(\mb v))-f(r_0)) \\
        &\overset{(ii)}{\leq} (r_0-\delta)^2 +  (r+d(\mb v)-r_0)^2 -r_0(f(r+d(\mb v))-f(r_0)).
    \end{align*}
    In $(i)$ we used \cref{eq:bound_f_increase} to have $\delta=f(r_0)\leq f(r+d(\mb v))\leq 2f(r)\leq 2\delta$ which implies in particular $r_0-f(r+d(\mb v))-\tilde f\geq r_0-3\delta\geq r_0/4$ since $\delta\leq r_0/4$, and in $(ii)$ we again used $f(r+d(\mb v))\leq 2\delta$ and $r_0\geq \delta$. Now because $f$ is convex, we have $f(r+d(\mb v)) - f(r_0) \geq f'(r_0) (r+d(\mb v)-r_0)$. Further, by definition of $\gamma\geq \gamma_0$, we have $\frac{c_f r_0}{\sqrt{1-\gamma}}\geq 1$. As a result, we obtain $f'(r_0)\geq \frac{c_f\delta}{\sqrt{1-\gamma}} \cdot\frac{e-1/e}{e+1/e} \geq \frac{c_f\delta}{2\sqrt{1-\gamma}} $. Altogether, this implies
    \begin{align*}
        R(\mb v) -(r_0-\delta)^2 &\leq  (r+d(\mb v)-r_0) (r+d(\mb v)-r_0-r_0f'(r_0))\\
        &\overset{(i)}{\leq}  (r+d(\mb v)-r_0) (2C_1(1-\gamma)-2c_f C \sqrt{1-\gamma} )\overset{(ii)}{\leq} 0
    \end{align*}
    where in $(i)$ we used $r+d(\mb v)-r_0\leq d(\mb v)\leq 2C_1(1-\gamma)$ and in $(ii)$ we used the definition of $\gamma_0$. This ends the proof of \cref{eq:boundary_within_ball}. 
    
    In both cases, we proved that \cref{eq:goal_valid_promise} holds. Therefore, we have $\Rcal_s\subseteq\Ucal_{s,\gamma}$. This holds for all $s\in[q]$, hence the tensored region $\Rcal$ defined in \cref{eq:combining_together} is achievable in the original setting with $[n]$ agents: $\Rcal\subseteq \Ucal_\gamma$. In particular, with $\mb U_1:=\mb U_{[n]\setminus I }\otimes\bigotimes_{s\in[q]} \paren{\mb U_s-f(0)\frac{\mb 1_{I_s}}{\|\mb 1_{I_s}\|} }$, we obtain
    \begin{equation*}
        \SWG(\gamma) = \max_{\mb x\in\Ucal^\star}\mb 1^\top \mb x - \max_{\mb x\in \Ucal_\gamma}\mb 1^\top \mb x \leq f(0)\sum_{s\in[q]}\|\mb 1_{I_s}\| \leq \frac{4n\|\mb 1\|(1-\gamma) }{r_0 \cosh\paren{\frac{c_f r_0}{\sqrt{1-\gamma}}}} = \Ocal\paren{(1-\gamma) \,e^{-c_f r_0/\sqrt{1-\gamma}}},   
    \end{equation*}
    which ends the proof of the theorem.
\end{proof}

\section{Extension to the finite-horizon setting}
\label{sec:finite_horizon}

In this section, we show how the techniques developed in previous sections for the $\gamma$-discounted infinite-horizon setting can be used for the finite-horizon setting.

We start by proving the following finite-horizon counterpart to Fact~\ref{fact:promised_utility_1} and \ref{fact:promised_utility_2} which allows to use the promised utility framework for the finite-$T$ horizon setting also.

\begin{fact}\label{fact:promised_utility_finite_horizon}
    Fix $\Rcal\subseteq\Ucal^\star$ and $T\geq 2$. Then $\Rcal\subseteq\Vcal_T$ if and only if there exists functions $\mb p(\cdot\mid\mb U)$ and $\mb W(\cdot\mid \mb U)$ for all $\mb U\in\Rcal$ satisfying \cref{eq:target_met,eq:incentive_compatibility} for $\gamma = \gamma(T)$, and instead of the original constraint \cref{eq:valid_promise} they satisfy
    \begin{equation}\label{eq:valid_interim_promise_finite_horizon}
        \forall \mb U\in\Rcal,\forall\mb v\in[0,\bar v]^n,\quad \mb W(\mb v\mid\mb U)\in\Vcal_{T-1}.
    \end{equation}

    Equivalently, $\Rcal\subseteq\Vcal_T$ if and only if the promised utility function satisfies \cref{eq:valid_interim_promise} where the interim promised utility was defined via \cref{eq:formula_interim_promise} with $\gamma=\gamma(T)$, and satisfies \cref{eq:valid_interim_promise_finite_horizon}.
\end{fact}

\begin{proof}
    Suppose that \cref{eq:target_met,eq:incentive_compatibility} for $\gamma = \gamma(T)$, and \cref{eq:valid_interim_promise_finite_horizon} are satisfied. Then, for any $\mb U\in\Rcal$, we can use the allocation function $p(\mb v(1)\mid\mb U)$ for the first time $t=1$ and reports $\mb v(1)$, then realize the utility vector $\mb W(\mb v(1)\mid\mb U)$ since it belongs to the achievable region $\Vcal_{T-1}$ by \cref{eq:valid_interim_promise_finite_horizon}. This strategy is incentive-compatible at least for time $t=1$ hence from \cref{eq:target_met}, it realizes the vector $\mb U$, that is, $\mb U\in\Vcal_T$. On the other hand, if $\mb U$ can be realized, fix an incentive-compatible strategy to realize $\mb U$. Let $p(\cdot\mid\mb U)$ be the allocation function for the first time $t=1$ and define $\mb W(\mb v\mid\mb U) = \frac{1}{T}\Ebb\sqb{\sum_{t\geq 2}\tilde u_i(t)p_i(t)\mid\mb v(1)=\mb v}$ the remaining utility for times in $[2,T]$. We can easily check that these satisfy all the desired equations.

    The second claim can be proved with the same arguments as in the proof of \cref{fact:promised_utility_2}.
\end{proof}

We next prove \cref{lemma:upper_bound_finite_horizon} which shows that for $T\geq 2$, we have $\Vcal_T\subseteq \Ucal_{\gamma(T)}$, where $\gamma(T)=1-1/T$.

\begin{proof}[of \cref{lemma:upper_bound_finite_horizon}]
    For $t\geq 2$, we let $\mb p^{(t)}(\cdot\mid\mb U)$ and $\mb W^{(t)}(\cdot\mid\mb U)$ be allocation and promised utility functions for $\mb U\in\Vcal_t$ satisfying \cref{eq:target_met,eq:incentive_compatibility} for $\gamma(t)$ and \cref{eq:valid_interim_promise_finite_horizon} for $t$.
    
    For $T=1$, we have directly $\Vcal_1=\Ucal_0=\Ucal_{\gamma(1)}$. For $T\geq 2$, note that $\mb p^{(T)}$ and $\mb W^{(T)}$ already satisfy almost all the equations to show that the region $\Vcal_T$ belongs to $\Ucal_{\gamma(T)}$. The only difference is in \cref{eq:valid_interim_promise_finite_horizon} which implies that the promised utilities stay in $\Vcal_{T-1}$. Precisely, to show $\Vcal_T\subseteq\Ucal_{\gamma(T)}$ it suffices to show that $\Vcal_{T-1}\subseteq\Vcal_T$. This is also immediate if $T=2$ since $\Vcal_1=\Ucal_0\subseteq\Vcal_t$ for all $t\geq 1$. We now suppose that $T\geq 3$ and that we showed $\Vcal_{T-2}\subseteq\Vcal_{T-1}$.

    We introduce an allocation function and promised utility function on $\Vcal_{T-1}$ via
    \begin{equation*}
        \mb p(\cdot\mid\mb U):=\mb p^{(T-1)}(\cdot\mid\mb U) \quad\text{and}\quad \mb W(\cdot\mid\mb U) :=\frac{T-2}{T-1}\mb W^{(T-1)}(\cdot\mid\mb U) + \frac{1}{T-1}\mb U ,\quad \mb U\in\Vcal_{T-1}.
    \end{equation*}
    We now check that this is a valid allocation strategy. Indeed, with $\mb U\in\Vcal_{T-1}$, for all $i\in[n]$,
    \begin{align*}
        (1-\gamma(T))&\Ebb[u_i p_i(\mb u)] + \gamma(T)\Ebb[\mb W(\mb u\mid\mb U)] \\
        &= \frac{\mb U}{T} + \frac{T-1}{T}\paren{(1-\gamma(T-1))\Ebb[u_i p^{(T-1)}_i(\mb u)] + \gamma(T-1)\Ebb[\mb W^{(T-1)}(\mb u\mid\mb U)]} = \mb U.
    \end{align*}
    Hence \cref{eq:target_met} holds. Here, in the last inequality we used the fact that $\mb p^{(T-1)}$ and $\mb W^{(T-1)}$ satusfy \cref{eq:target_met} for $\gamma=\gamma(T-1)$. Similarly, for any $i\in[n]$ and $u,v \in[0,\bar v]$,
    \begin{align*}
        (1-\gamma(T)) &u P_i(v\mid\mb U) + \gamma(T) W_i(v\mid\mb U) \\
        &= \frac{U_i}{T} + \frac{T-1}{T}\paren{(1-\gamma(T-1))uP^{(T-1)}_i(v\mid\mb U) + \gamma(T-1)W^{(T-1)}_i(v\mid\mb U)},
    \end{align*}
    hence the incentive-compatibility \cref{eq:incentive_compatibility} also holds from that of $\mb p^{(T-1)}$ and $\mb W^{(T-1)}$. Last, for any $\mb u\in[0,\bar v]^n$, we have $\mb W^{(T-1)}(\mb v\mid \mb U)\in\Vcal_{T-2}\subseteq\Vcal_{T-1}$ and $\mb U\in\Vcal_{T-1}$. Because $\Vcal_{T-1}$ is convex, this implies $\mb W(\mb v\mid\mb U)\in\Vcal_{T-1}$. That is, \cref{eq:valid_interim_promise_finite_horizon} also holds, which ends the inductive proof that $\Vcal_{T-1}\subseteq\Vcal_T$. This proves the desired result.
\end{proof}

As a consequence of \cref{lemma:upper_bound_finite_horizon}, all techniques developed earlier to give upper bounds on $\Ucal_{\gamma(T)}$ also apply to $\Vcal_T$. 
We next show the simple universal $1/T$ lower bound for the finite-horizon setting.

\begin{proof}[of \cref{lemma:no_faster_1/T}]
    Let $\mb U\in\Vcal_T$ and consider an allocation strategy that realizes $\mb U$. For any $t\in[T]$, we recall the notation $\mb p(t)\in\Delta_n$ and $\mb u(t)$ for the allocation distribution and the agent utilities respectively at time $t$. The main remark is that the suboptimality of $\mb U$ is at least that of the first allocation $\mb p(1)$ and this corresponds to an allocation from $\Vcal_1$:
    \begin{align*}
        \max_{\mb x\in\Ucal^\star}\mb 1^\top \mb x - \mb 1^\top \mb U &= \Ebb\sqb{\frac{1}{T}\sum_{t\in[T]} \max_{\mb x\in\Ucal^\star}\mb 1^\top \mb x - \mb 1^\top (p_i(t)u_i(t))_{i\in[n]} } \\
        &\geq \frac{1}{T}\Ebb\sqb{ \max_{\mb x\in\Ucal^\star}\mb 1^\top \mb x - \mb 1^\top (p_i(1)u_i(1))_{i\in[n]} } \overset{(i)}{\geq} \frac{1}{T}\paren{\max_{\mb x\in\Ucal^\star}\mb 1^\top \mb x - \max_{\mb x\in\Ucal_0}\mb 1^\top \mb x}.
    \end{align*}
    In $(i)$ we used the fact that $(\Ebb[p_i(1)u_i(1)])_{i\in[n]}\in\Vcal_1=\Ucal_0$. By assumption, the term in the bracket in the last inequality is nonzero, otherwise $\Ucal_0$ would contain an optimal vector.
\end{proof}

We next turn to the lower bound tools on $\Ucal_\gamma$ and show that they still hold in the finite-horizon $T$ setting up to logarithmic factors. We start by showing that \cref{lemma:ball_in_region_no_assumptions} still holds with a variant of the mechanism \text{PUM}. The proof proceeds by giving a trajectory of achievable regions for all times in $t\in[T]$, illustrated in \cref{fig:finite_horizon}: one can approximate balls within $\Ucal^\star$ by constructing balls within $\Vcal_t$ converging to the original ball as $t\in [T]$ grows. The corresponding mechanism \text{FH-PUM} (Finite-Horizon Promised Utility Mechanism) is described formally within the following proof and summarized in \cref{alg:mechanism_finite_horizon}.

    \begin{algorithm}[t]

\caption{Promised Utility Mechanism $\text{FH-PUM}(\mb x, r, \delta)$ for the $T$-horizon setting}\label{alg:mechanism_finite_horizon}

\LinesNumbered
\everypar={\nl}
\SetAlgoNoEnd
\setcounter{AlgoLine}{0}

\KwIn{Ball $B(\mb x,r)$, margin $\delta$ with $B(\mb x, r+\delta)\subseteq\Ucal^\star$, target utility $\mb U_0\in B(\mb x, r)$, horizon $T$}

\textbf{Initialization:}
$r_0:=\frac{\min_{i\in[n]}\Ebb[u_i]}{6n}$, $\mb x^{(0)}:=3r_0\mb 1$, $\tilde\delta:=\frac{\delta}{2}$, $\mb z=\mb x+\tilde\delta\frac{\mb U_0-\mb x}{\|\mb U_0-\mb x\|}$, constant $\tilde C:=4(1+\frac{\max_{i\in[n]}\Ebb[u_i]}{r_0})^2 C$ where $C$ is defined as in \cref{lemma:ball_in_region_no_assumptions}.

Let $\mb U\gets \mb U_0$, $T_0=\max\{t\in[T]: 2\sqrt{\frac{\tilde C}{t-1}}\geq r_0\}$, and $\mb x^{(T_0)}=z^{(0)}$

\For{$t=T_0+1,\ldots,T$}{
    \lIf{$r<\sqrt{\frac{\tilde C}{t-1}} $}{
        Define $r_t=\delta_t:=\sqrt{\frac{\tilde C}{t-1}}$
    }
    \lElse{
        Define $r_t:=r$ and $\delta_t:=\max(\frac{\tilde C}{r(t-1)},\tilde\delta)$
    }
    Define $\mb y^{(t)}:= \mb z+\frac{r_t+\delta_t-(r+\tilde\delta)}{2r_0-(r+\tilde\delta)}(\mb x^{(0)}-\mb z)$ and $\mb x^{(t)}:=(1-\frac{1}{t})\mb x^{(t-1)}+\frac{1}{t}\mb y^{(t)}$
}

\vspace{3pt}

\textbf{Main mechanism:}
\For{$t=T,T-1,\ldots,1$}{
\lIf{$\mb U\in \Ucal^{NI}$}{
    Observe new reports $\mb v=(v_i)_{i\in[n]}$ and select allocation $(\frac{v_i}{\Ebb[u_i]})_{i\in[n]}$
}
\Else{
    Decompose $\mb U= \mb s \odot \paren{\sum_{i\in[n]}q_i\Ebb[u_i]\mb e_i + q_0 (\mb x^{(t)}+r_t\mb y)}$ where $\mb q\geq 0,\sum_{i=0}^n q_i\leq 1$, $\mb y\in S_{n-1}$, $\mb s\in[0,1]^n$.

    Fix an allocation function $\tilde {\mb p}(\cdot)$ realizing utility $\mb y^{(t)}+(r_t+\delta_t)\mb y$ for full-information case

    Compute interim functions $\tilde P_i(\cdot):=\Ebb_{\mb u_{-i}}[\tilde p_i(\cdot,\mb u_{-i})]$ and $\tilde W_i(\cdot)$ according to \cref{eq:formula_interim_promise} for $i\in[n]$
    
    Observe new reports $\mb v=(v_i)_{i\in[n]}$ and select allocation $\mb s\odot \paren{\sum_{i\in[n]}q_i\mb e_i + q_0  \tilde{\mb p}(\mb v)}$
    
    Define $\tilde {\mb W}$ using \cref{eq:coupling_formula} for $\gamma=1-\frac{1}{t}$, $\mb\alpha=-r_t\mb y$, and $\tilde Z_i=\tilde W_i(v_i)$ for $i\in[n]$
    
    Update $\mb U\gets \mb s\odot \paren{\sum_{i\in[n]}q_i\Ebb[u_i]\mb e_i + q_0  \tilde{\mb W}}$
    
    \lIf{$\mb U\notin \Ucal^{NI}$ and $\mb U\notin \Rcal_{t-1}$ as defined in \cref{eq:def_enlarged_region_finite_case}}{Mechanism fails. \textbf{break}}
}

}

\end{algorithm}

\begin{proof}[of \cref{lemma:ball_in_region_finite_horizon}]
    Let $\mb x$ and $r,\delta>0$ satisfying the requirements. 
    The assumptions imply that $0<x_i\leq \Ebb[u_i]$ for all $i\in[n]$. Now let $r_0:=\min_{i\in[n]}\Ebb[u_i]/(6n)$ and let $\mb x^{(0)}:= 3r_0\mb 1$. We can note that
    \begin{equation*}
        B(\mb x^{(0)},3r_0)\subseteq [0,6r_0]^n\subseteq\{\mb y:\mb 0\leq \mb y\leq (\Ebb[u_i]/n)_{i\in[n]}\} \subseteq \Ucal_0.
    \end{equation*}
    Also, up to renaming the constant $r_0$ to $r_0/2$, the assumption gives $r+\delta\leq 2r\leq r_0$. We next introduce the constant $\tilde c_0 :=(1+\frac{\max_{i\in[n]}\Ebb[u_i]}{r_0})^2$.

    \paragraph{Step 1.} We start with the case in which the assumption that \cref{eq:constraint_safe_boundary} holds with the same constant $\tilde C:=\tilde c_0 C$ where $C$ is the same constant as in the original statement of \cref{lemma:ball_in_region_no_assumptions}, and we have $\delta\leq r\leq r_0/2$. We will later see that from there we can easily extend to the general case.

    We define the set $\Tcal = \set{t\in\{2,\ldots,T\},r \geq \sqrt{\frac{\tilde C(1-\gamma(t))}{\gamma(t)}}=\sqrt{\frac{\tilde C}{t-1}} }$. Note that by hypothesis, we have $2r\geq r+\delta \geq r+\tilde C\frac{1-\gamma(T)}{\gamma(T)r} \geq  2\sqrt{\frac{\tilde C(1-\gamma(T))}{\gamma(T)}}$. Thus, $T\in\Tcal$. Because $\frac{1-\gamma(t)}{\gamma(t)}$ is decreasing in $t$, there exists $T_1\in[T]$ such that $\Tcal = \{T_1,T_1+1,\ldots,T\}$. We next let
    \begin{equation*}
        T_0 = \max\set{t\in[T],2\sqrt{\frac{\tilde C(1-\gamma(t))}{\gamma(t)}}\geq  r_0  },
    \end{equation*}
    which is well defined since $\gamma(1)=0$ and hence the set always contains time $t=1$.
    We note that for $T_1$, we have
    \begin{equation*}
        2\sqrt{\frac{\tilde C(1-\gamma(T_1))}{\gamma(T_1)}}< 2r\leq r_0,
    \end{equation*}
    Therefore, $T_0<T_1$ (by construction $T_1\geq 2$).
    We now specify our allocation mechanism.

    \paragraph{Definition of the allocation strategy.}
    We use a trivial allocation for times in $[T_0]$: we will only use the fact that $B(\mb x^{(0)},3r_0)\subseteq \Ucal_0\subseteq \Vcal_{T_0}$. 
    For times in $(T_0,T]$, we let
    \begin{equation*}
        r_t:= \begin{cases}
             \sqrt{\frac{\tilde C(1-\gamma(t))}{\gamma(t)}} = \sqrt{\frac{\tilde C}{t-1}} & t\in \{T_0,\ldots,T_1-1\}\\
            r & t\in\{T_1,\ldots,T\}.
        \end{cases}
    \end{equation*}
    To prove the desired result, we inductively prove that $B(\mb x^{(t)},r_t)\subseteq \Vcal_t$ for appropriately chosen parameters $\mb x^{(t)}$. Precisely, we first introduce the parameter
    \begin{equation*}
        \delta_t:= \begin{cases}
             r_t & t\in \{T_0+1,\ldots,T_1-1\}\\
            \max\paren{\tilde C\frac{1-\gamma(t)}{\gamma(t)r_t},\delta} & t\in\{T_1,\ldots,T\}.
        \end{cases}
    \end{equation*}
    We pose $\mb x^{(t)}:=\mb x^{(0)}$ for all $t\in[T_0]$ and for $t> T_0$ we introduce
    \begin{equation*}
        \mb y^{(t)} := \mb x + \frac{r_t+\delta_t-(r+\delta)}{2r_0-(r+\delta)} (\mb x^{(0)}-\mb x).
    \end{equation*}
    Note that by construction, $r_t+\delta_t\geq r+\delta$: this is immediate for $t\in[T_1,T]$ and for $t\in(T_0,T_1)$ since $t<T_1$ one has $r_t+\delta_t = 2r_t >2r\geq r+\delta$.
    We then construct the sequence of vectors $\mb x^{(t)}$ via
    \begin{equation}\label{eq:def_recursion_x}
        \mb x^{(t)}:= \gamma(t)\mb x^{(t-1)} + (1-\gamma(t)) \mb y^{(t)},\quad t\in\{T_0+1,\ldots,T\}.
    \end{equation}
    To prove that $B(\mb x^{(t)},r_t)\subseteq \Vcal_t$, we construct an allocation mechanism on the region
    \begin{equation}\label{eq:def_enlarged_region_finite_case}
        \Rcal_t:=\{\mb y:\mb 0\leq \mb y\leq \mb z,\mb z\in Conv(\Ucal^{NI},B(\mb x^{(t)},r_t)  \}.
    \end{equation}
    Next, by construction, note that
    \begin{equation}\label{eq:ponderated_sum_balls}
        B(\mb y^{(t)},r_t+\delta_t) = \frac{r_t+\delta_t-(r+\delta)}{2r_0-(r+\delta)} B(\mb x^{(0)},2r_0) +  \frac{2r_0-(r_t+\delta_t)}{2r_0-(r+\delta)} B(\mb x,r+\delta).
    \end{equation}
    By convexity of $\Ucal^\star$, since we have $B(\mb x^{(0)},2r_0),B(\mb x,r+\delta)\subseteq\Ucal^\star$ this also implies that $B(\mb y^{(t)},r_t+\delta_t) \subseteq\Ucal^\star$. This is precisely the reason behind the definition of $\mb y^{(t)}$.
    
    \comment{
    This is immediate from $B(\mb x,r+\delta)\subseteq\Ucal^\star$ if $t\geq T_1$ and $\delta_t=\delta$. Suppose that this is not the case. We already checked that $r_t+\delta_t-r-\delta\geq 0$. Therefore, by construction of $\mb y^{(t)}$ it suffices to check that $B(\mb y^{(t)},r_t+\delta_t)\subseteq\Rbb_+^n$. Since $t>T_0$, we have $4r_t\leq \min_{i\in[n]}x_i +r+\delta$. This already proves the claim for $t\in(T_0,T_1)$. For $t\in[T_1,T]$, because $r\geq \sqrt{\frac{C(1-\gamma(t))}{\gamma(t)}}$ we obtain $\delta_t=C\frac{1-\gamma(t)}{\gamma(t)r}\leq r=r_t$, hence $2(r_t+\delta_t)\leq 4r_t \leq \min_{i\in[n]}x_i +r+\delta$. This ends the proof that $B(\mb y^{(t)},r_t+\delta_t)\subseteq \Ucal^\star$.
    }

    We now construct an allocation function $\mb p^{(t)}(\cdot\mid\mb U)$ and a promised utility function $\mb W^{(t)}(\cdot\mid\mb U)$ for $\mb U\in\Rcal_t$ in a similar fashion to those constructed in the proof of \cref{lemma:ball_in_region_no_assumptions}. We first specify the allocation strategy when $\mb U$ is an extreme point of $\Rcal_t$ that still belongs to $B(\mb x^{(t)},r_t)$. We write $\mb U=\mb x^{(t)}+r_t\mb y$ where $\mb y\in S_{n-1}$ and let
    \begin{equation}\label{eq:def_allocation_promised_finite_horizon}
        \mb p^{(t)}(\cdot\mid\mb U):= \mb p(\cdot;\mb y^{(t)}+(r_t+\delta_t) \mb y) \quad \text{and} \quad \mb\alpha(\mb U) = \mb x^{(t)}-\mb U = -r_t\mb y,\quad t\in(T_0,T],
    \end{equation}
    where $\mb p(\cdot;\mb z)$ refers to an allocation that realizes $\mb z$ in the full information setting (we checked earlier that we always used an allocation $\mb p(\cdot;\mb z)$ with $\mb z\in\Ucal^\star$). We then extend the definition of the allocation mechanism to the complete region $\Rcal_t$ exactly as in the original proof using \cref{eq:def_allocation_function_app,eq:def_promised_utility_function_app}.

    \paragraph{Checking that the allocation strategy is valid.}
    We now check that this is a valid allocation. Using the same linearity arguments as in the proof of \cref{lemma:ball_in_region_no_assumptions}, it suffices to show that for extreme points $\mb U$ of $\Rcal_t$ within $B(\mb x^{(t)},r_t)$, the promised utilities lie in $B(\mb x^{(t-1)},r_{t-1})$ to show that for all $\mb U\in \Rcal_t$, \cref{eq:valid_interim_promise} is satisfied, and
    \begin{equation}\label{eq:recursive_valid_promised_utility}
        \forall\mb v\in[0,\bar v]^n,\quad \mb W^{(t)}(\mb v\mid\mb U)\in \Rcal_{t-1}.
    \end{equation}
    
    Fix such an extreme point $\mb U=\mb x^{(t)}+r_t\mb y$ where $\mb y\in S_{n-1}$.
    First, recall that $B(\mb y^{(t)},r_t+\delta_t)\subseteq\Ucal^\star$. We now check that \cref{eq:constraint_safe_boundary} is also satisfied. Note that by construction the sequences $r_t$ and $\delta_t$ are both non-increasing on $(T_0,T]$. This is clear by definition of $T_1$ for the sequence $(r_t)_t$. For $\delta_t$, we only need to check that $\delta_{T_1-1}=r_{T_1-1} \geq \delta_{T_1}$. To do so, we use $\delta_{T_1}\leq r_{T_1} \leq r_{T_1-1}$. By \cref{eq:ponderated_sum_balls}, we have $
        y_i^{(t)} + r_t+\delta_t \leq \max(x^{(0)}_i + 2r_0,x_i+r+\delta).$
    Now by hypothesis \cref{eq:constraint_safe_boundary} is satisfied for the parameters $\mb x, r,\delta$ and $\gamma=\gamma(T)$, and the constant $\tilde C$ instead of $C$. Hence, $\delta\geq \tilde C\frac{1-\gamma(T)}{\gamma(T)r}$ which implies $\delta_T=\delta$. Hence, because $\delta_t$ is non-increasing in $t$, we have $\delta_t\leq \delta$ for all $t\in[T_0,T_1]$, which implies
    \begin{equation*}
        y_i^{(t)} + r_t \leq \max(x^{(0)}_i + 2r_0,x_i+r) \leq \max(\Ebb[u_i]-r_0 ,x_i+r),
    \end{equation*}
    where in the last inequality we used the fact that $B(\mb x^{(0)},3r_0)\subseteq \Ucal_0$. The previous remarks imply that
    \begin{equation*}
        \tilde C = \tilde c_0 4n^2 \bar v^2 \paren{ 1 + \frac{\max_{i\in[n]}  \Ebb[u_i]}{\min_{i\in[n]}(\Ebb[u_i] -x_i-r)}  }^2 \geq 4n^2 \bar v^2 \paren{ 1 + \frac{\max_{i\in[n]}  \Ebb[u_i]}{\min_{i\in[n]}(\Ebb[u_i] -y^{(t)}_i-r_t)}  }^2.
    \end{equation*}
    Last, we check that
    \begin{equation*}
        \frac{\gamma(t) r_t}{1-\gamma(t)} \overset{(i)}{\geq} r_t \geq \delta_t \geq \tilde C \frac{1-\gamma(t)}{\gamma(t)r_t}.
    \end{equation*}
    where in $(i)$ we used the fact that $t\geq T_0+1\geq 2$, hence $\gamma(t) \geq 1/2$. Altogether, the previous arguments show that \cref{eq:constraint_safe_boundary} is satisfied for the parameters $\mb y^{(t)},r_t,\delta_t$ and $\gamma=\gamma(t)$.
    
    Next, compared to the definition of \cref{lemma:ball_in_region_no_assumptions}, these would be identical if the allocation was for $\mb U':=\mb y^{(t)}+r_t\mb y$ instead of $\mb U=\mb x^{(t)}+r_t\mb y$. To clarify this comparison, we introduce the allocation function $\mb p'(\cdot\mid\mb U'):=\mb p(\cdot;\mb U)$ and denote by $\mb W'(\cdot\mid\mb U')$ the promised utility function resulting from this choice of allocation function and the same direction choice $\mb\alpha(\mb U)$ as in \cref{eq:def_allocation_promised_finite_horizon}. Having checked that all conditions apply, the proof of \cref{lemma:ball_in_region_no_assumptions} implies that for all $\mb v\in[0,\bar v]^n$, $\mb W'(\mb v\mid\mb U')\in B(\mb y^{(t)},r_t)$.
    Now by definition of the promised utility functions (see \cref{eq:formula_interim_promise,eq:coupling_formula}), for any $\mb v\in[0,\bar v]^n$,
    \begin{equation*}
        \mb W^{(t)}(\mb v\mid \mb U) = \mb W'(\mb v\mid \mb U') + \frac{\mb U-\mb U'}{\gamma(t)} = \mb W'(\mb v\mid \mb U') - \frac{\mb y^{(t)}-\mb x^{(t)}}{\gamma(t)}.
    \end{equation*}
    As a result, using $r_t\leq r_{t-1}$, we obtain
    \begin{equation*}
        \mb W^{(t)}(\mb v\mid \mb U) \in B\paren{\mb y^{(t)} + \frac{\mb x^{(t)}-\mb y^{(t)}}{\gamma(t)}  ,r_t} = B(\mb x^{(t-1)},r_t) \subseteq B(\mb x^{(t-1)},r_{t-1}).
    \end{equation*}
    This ends the proof of \cref{eq:recursive_valid_promised_utility} hence $\Rcal_t\subseteq\Vcal_t$. As a summary, we obtained the induction: if $B(\mb x^{(t-1)},r_{t-1})\subseteq \Vcal_{t-1}$, then $B(\mb x^{(t)}, r_t)\subseteq \Vcal_t$ where $\mb x^{(t)}$ is defined in \cref{eq:def_recursion_x}. We recall that the initialization at $t=T_0$ is immediate from $B(\mb x^{(0)},r_0)\subseteq\Ucal_0\subseteq \Vcal_{T_0}$.

    \paragraph{Putting the recursive construction together.} Using the induction formula \cref{eq:def_recursion_x}, we can check that $\mb x^{(t)}=a_t\mb x^{(0)} + b_t(\mb x^{(0)}-\mb x) + c_t\mb x$ for $t\in[T_0,T]$ where $a_{T_0}=1$, $b_{T_0}=c_{T_0}=0$ and
    \begin{equation*}
        \begin{cases}
            a_t = \gamma(t) a_{t-1}\\
            b_t = \gamma(t) b_{t-1} + (1-\gamma(t))\frac{r_t+\delta_t-(r+\delta)}{2r_0-(r+\delta)}\\
            c_t = \gamma(t)c_{t-1} + (1-\gamma(t)),
        \end{cases} \quad t\in[T_0+1,\ldots,T_1-1].
    \end{equation*}
    Using the convexity inequality $\ln(1-x)\leq -x$ for $x\leq 1$, we obtain
    \begin{equation}\label{eq:ineq_a_c_T}
        0\leq a_{T}=1-c_T = \prod_{t=T_0+1}^{T} \gamma(t) \leq e^{-\sum_{t=T_0+1}^{T} \frac{1}{t}} \leq \frac{T_0+1}{T+1}.
    \end{equation}
    The recursion for $b_t$ requires more work. First, recall that $r+\delta\leq r_0$ and $r_t+\delta_t\geq r+\delta$. Therefore, if we define the inductive sequence $\tilde b_{T_0}:=0$ and
    \begin{equation*}
        \tilde b_t = \gamma(t) \tilde b_{t-1} + (1-\gamma(t))\frac{r_t+\delta_t-(r+\delta)}{r_0},
    \end{equation*}
    we have $0\leq b_t\leq \tilde b_t$ for all $t\in[T_0,T]$. Note that the induction can be rewritten as follows,
    \begin{equation*}
        r_0 t\tilde b_t = r_0 (t-1)\tilde b_{t-1} + r_t+\delta_t-(r+\delta).
    \end{equation*}
    As a result, letting $T_2=\min\{t\geq T_1,\delta_t=\delta\}$, we obtain
    \begin{align*}
        r_0 T\tilde b_T = \sum_{t=T_0+1}^T (r_t+\delta_t-r-\delta)
        &= 2\sum_{t=T_0+1}^{T_1-1} r_t - (T_1-T_0-1)(r+\delta) + \sum_{t=T_1}^{T_2-1}\delta_t- (T_2-T_1)\delta\\
        &\overset{(i)}{\leq} 4\sqrt{\tilde C(T_1-2)} + \frac{\tilde C}{r}\paren{1+\ln\frac{T_2-2}{T_1-1}\1_{T_2>T_1}},
    \end{align*}
    where in $(i)$ we used $r_t=\sqrt{\tilde C/(t-1)}$ for $t\in(T_0,T_1)$ and $\delta_t=\tilde C/((t-1)r)$ for $t\in[T_1,T_2)$, and used sum-integral inequalities. Now by definition of $T_1$ and $T_2$, we have $\sqrt{\frac{\tilde C}{T_1-2}}>r\geq \sqrt{\frac{\tilde C}{T_1-1}}$ and if $T_2>T_1$, we have $\delta< \tilde C\frac{1-\gamma(T_2-1)}{\gamma(T_2-1)r}=\frac{\tilde C}{r(T_2-2)}$. As a result, we obtained
    \begin{equation}\label{eq:ineq_b_T}
        0\leq b_T \leq \frac{4\sqrt{\tilde C(T_1-2)}}{r_0T} + \frac{\tilde C}{r_0 rT}\paren{1+\ln\frac{r}{\delta}} \leq \frac{\tilde C}{r_0rT}\paren{5 +\ln\frac{r}{\delta} }.
    \end{equation}

    \paragraph{Step 2.} We now consider the original case in which $B(\mb x,r+\delta)\subseteq \Ucal^\star$, $\delta\leq r\leq r_0/2$, and \cref{eq:finite_horizon_safe} holds with the constant $c_0=8\tilde c_0+20\tilde c_0/r_0 + 4(T_0+1)r_0/C$. We denote by $C$ the term in the statement of \cref{lemma:ball_in_region_no_assumptions} for parameters $\mb x$ and $r$.
    Fix any $\mb z\in B(\mb x,\delta/2)$. Then, with $\tilde \delta=\delta/2$, we still have $B(\mb z,r+\tilde \delta)\subseteq\Ucal^\star$, and
    \begin{equation*}
         r \geq \tilde\delta\geq \frac{c_0C}{2}\frac{1-\gamma(T)}{\gamma(T)r} \geq 4\tilde c_0 C \frac{1-\gamma(T)}{\gamma(T)r}.
    \end{equation*}
    Note that the constant corresponding to the parameters $\mb z,r$ in the statement of \cref{lemma:ball_in_region_no_assumptions} is at most $4C$ (note that $\|\mb z-\mb x\|\leq \delta/2$ and since $B(\mb x,r+\delta)\subseteq \Ucal^\star$, this still leaves a margin at least $\delta/2$ to the boundary of $\Ucal^\star$ for $B(\mb z,r)$). Therefore, the previous step shows that $B(\mb z^{(T)},r)\subseteq \Vcal_T$ where
    \begin{equation*}
        \mb z^{(T)} = a_T\mb x^{(0)}+b_T(\mb x^{(0)}-\mb z) + c_T\mb z.
    \end{equation*}
    Because this holds for all $\mb z\in B(\mb x,\delta/2)$,we obtained
    \begin{equation}\label{eq:result_step_1}
        B\paren{ a_T\mb x^{(0)}-b_T\mb 1 + c_T\mb x , r+\frac{\delta}{2}} \subseteq \Vcal_T.
    \end{equation}
    Now using \cref{eq:ineq_a_c_T,eq:ineq_b_T}, we have
    \begin{align*}
        \|a_T\mb x^{(0)}-b_T\mb 1 + c_T\mb x - \mb x\| &\leq \frac{T_0+1}{T+1}\|\mb x-\mb x^{(0)}\| + \frac{\tilde C(5+\ln \frac{r}{\delta})}{r_0rT} \|\mb x-\mb x^{(0)}\|\\
        &\leq \frac{(T_0+1)r_0^2 + \tilde C(5+\ln \frac{r}{\delta})}{r_0 rT}\sqrt n\max_{i\in[n]}\Ebb[u_i]
    \end{align*}
    As a result, recalling the definition of $c_0$, we obtain
    \begin{equation*}
        \|a_T\mb x^{(0)}-b_T\mb 1 + c_T\mb x - \mb x\| \leq \frac{c_0 C(1+\ln \frac{r}{\delta})}{2r(T+1)} \leq \frac{\delta}{2}.
    \end{equation*}
    Together with \cref{eq:result_step_1} this implies the desired result $B(\mb x,r)\subseteq \Vcal_T$. This proves the first claim. The corresponding mechanism $\text{FH-PUM}(\mb x, r, \delta)$ is summarized in \cref{alg:mechanism_finite_horizon}.

    For the second claim, we simply note that by assumption, we have $r\geq \delta\geq \frac{c_0C}{r(T-1)}$. In particular, $\frac{r}{\delta}\leq \frac{r^2}{c_0C}(T-1)\leq \frac{r_0^2}{c_0C}T$. Hence, up to modifying the constant $c_0$, the result holds if we replace the lower bound from \cref{eq:finite_horizon_safe} with $c_0 C\frac{1-\tilde \gamma(T)}{\tilde\gamma(T)}$ where $\tilde \gamma(T):=1-(1+\ln T)/T$.
\end{proof}

As a remark, note that when the margin $\delta$ is of the same order as $r$ within \cref{lemma:ball_in_region_finite_horizon}, then the logarithmic term $\ln\frac{r}{\delta}$ is a constant, which essentially means that we can use the same techniques to prove lower bounds on $\Vcal_T$ as for $\Ucal_{\gamma(T)}=\Ucal_{1-1/T}$, up to worsening constants. This is significant since from \cref{lemma:upper_bound_finite_horizon}, we also have $\Vcal_T\subseteq\Ucal_{\gamma(T)}$: in these cases, characterizing $\Vcal_T$ or $\Ucal_{\gamma(T)}$ with our tools is essentially equivalent.
On the other hand, as per the second claim of \cref{lemma:ball_in_region_finite_horizon}, the extra factor $\ln\frac{r}{\delta}$ can be of order $\ln T$, which potentially induces additional logarithmic factors to translate results from the inifinite-horizon setting to the finite-horizon setting.

Similarly as for \cref{lemma:ball_in_region_no_assumptions}, the optimized version \cref{lemma:ball_in_region_optimized} also holds in the finite-horizon setting up to worsening the constants.

\begin{lemma}\label{lemma:ball_in_region_optimized_finite_horizon}
    Fix $T\geq 2$. There exist constants $c_0\geq 8$ and $r_0>0$ such that the following holds. Suppose that all assumptions from \cref{lemma:ball_in_region_optimized} are satisfied with the exception that \cref{eq:constraint_safe_boundary} is replaced with \cref{eq:finite_horizon_safe}. Then, using the same notations,
    \begin{equation*}
        B(\mb x, r)\cap \{\mb y:\forall i\notin  I ,y_i=x_i\} \subseteq \mb x + \{0\}^{[n]\setminus  I }\otimes \bigotimes_{s\in[q]} B_{I_s}(\mb 0, r)\subseteq \Vcal_T.
    \end{equation*}

    Alternatively, suppose that all assumptions from \cref{lemma:ball_in_region_optimized} are satisfied with $\gamma=\tilde\gamma(T)=1-\ln T/T$ and the constant $C'=c_0C$, as well as $r_0\geq r\geq \delta$. Then the previous conclusion also holds.
\end{lemma}

\begin{proof}
    The same proof as in \cref{lemma:ball_in_region_optimized} carries to the finite-horizon setting: we can construct an allocation mechanism that treats separately agents by clusters $I_1,\ldots,I_q$ as described in the proof. The only difference is that instead of applying \cref{lemma:ball_in_region_no_assumptions} we apply \cref{lemma:ball_in_region_finite_horizon}, which required exactly the assumptions made within \cref{lemma:ball_in_region_optimized_finite_horizon}.
\end{proof}

As a conclusion, up to logarithmic factors, all lower bounds developed in the previous sections apply to the finite-horizon setting choosing $\gamma=\gamma(T)=1-1/T$.
We now prove \cref{cor:super_smooth_case_finite_horizon,cor:discrete_case_full_finite_horizon} as examples of how the results from the $\gamma$-discounted setting can help for the finite-horizon setting.

\begin{proof}[of \cref{cor:super_smooth_case_finite_horizon}]
    It suffices to prove the result for $T\geq 2$. The lower bound is directly taken from \cref{lemma:no_faster_1/T} and the upper bound is a consequence of the second claim of \cref{cor:super_smooth_case} together with \cref{lemma:ball_in_region_optimized_finite_horizon}.
\end{proof}

\begin{proof}[of \cref{cor:discrete_case_full_finite_horizon}]
    It suffices to prove the result for $T\geq 2$. The first claim is a consequence of the first result from \cref{cor:discrete_case} together with \cref{lemma:upper_bound_finite_horizon} for the lower bound and \cref{lemma:ball_in_region_finite_horizon} for the upper bound. Note that there is no extra $\ln T$ factor (we can take directly $\gamma=\gamma(T)=1-1/T$) in the upper bound because the proof of the convergence rate $\Ocal(\sqrt{1-\gamma})$ comes from \cref{thm:universal_lower_bound}, which only uses \cref{lemma:ball_in_region_no_assumptions} with parameters $r=\delta$. As a result, the extra term $\ln\frac{r}{\delta}$ from \cref{lemma:ball_in_region_finite_horizon} vanishes.
    The second claim is a consequence of the second claim of \cref{cor:discrete_case} together with \cref{lemma:ball_in_region_optimized_finite_horizon} for the upper bound. The lower bound is directly \cref{lemma:no_faster_1/T}.
\end{proof}

\end{document}

%% file: shortcuts.tex
\newcommand{\Bcal}{\mathcal{B}}
\newcommand{\Ccal}{\mathcal{C}}
\newcommand{\Dcal}{\mathcal{D}}
\newcommand{\Ecal}{\mathcal{E}}
\newcommand{\Fcal}{\mathcal{F}}
\newcommand{\Gcal}{\mathcal{G}}
\newcommand{\Hcal}{\mathcal{H}}

\newcommand{\Mcal}{\mathcal{M}}

\newcommand{\Pcal}{\mathcal{P}}
\newcommand{\Scal}{\mathcal{S}}
\newcommand{\Tcal}{\mathcal{T}}
\newcommand{\Ucal}{\mathcal{U}}
\newcommand{\Vcal}{\mathcal{V}}

\newcommand{\Ocal}{\mathcal{O}}

\newcommand{\Rcal}{\mathcal{R}}

\newcommand{\Ebb}{\mathbb{E}}

\newcommand{\Pbb}{\mathbb{P}}

\newcommand{\Rbb}{\mathbb{R}}

\newcommand{\1}{\mathbbm{1}}

\newcommand{\x}{\mathsf{x}}
\newcommand{\y}{\mathsf{y}}

\definecolor{dark_red}{rgb}{0.2,0,0}

\newcommand{\mb}[1]{\ensuremath{\boldsymbol{#1}}}

\DeclareMathOperator*{\argmax}{arg\,max}

\newcommand{\paren}[1]{\left( #1 \right)}
\newcommand{\sqb}[1]{\left[ #1 \right]}
\newcommand{\set}[1]{\left\{ #1 \right\}}

\newcommand{\norm}[1]{\left\|#1\right\|}

\newcommand{\comment}[1]{}
\newcommand{\test}[1]{}

\newcommand{\Supp}{\textnormal{Supp}}

\newcommand{\SWG}{\textnormal{SWG}}